\documentclass[10pt]{article}

\usepackage{amsfonts}
\usepackage{amsmath}
\usepackage{amssymb}
\usepackage{compozition}
\usepackage{graphicx}
\usepackage[all]{xy}
\usepackage{MnSymbol}
\usepackage{comment}

\usepackage{dsfont} 
\usepackage{undertilde}
\makeatletter
\newcommand\extleftrightarrow[2][]{\ext@arrow 0099{\longleftrightarrowfill@}{#1}{#2}}
\def\longleftrightarrowfill@{\arrowfill@\leftarrow\relbar\rightarrow}
\makeatother

\usepackage{tensor}

\usepackage{url}
\usepackage{verbatim}
\usepackage{tabu}

\usepackage{xstring}  

\usepackage{tscustomcommands}

\title{\textbf{Time Symmetry in Operational Theories}}

\author{Lucien Hardy\footnote{lhardy@perimeterinstitute.ca}\\
\textit{Perimeter Institute,}\\
\textit{31 Caroline Street North,}\\
\textit{Waterloo, Ontario N2L 2Y5, Canada}}
\date{}

\begin{document}

\pagestyle{empty}

\begin{titlepage}

\maketitle

\vskip 3.2cm

\[
\scalebox{1.3}{
\begin{Compose}{0}{0} \setdefaultfont{\hat} \setsecondfont{\mathtt} \setthirdfont{\mathsfb}
\crectangledmark[-1]{B}{3}{2.5}{0,0} \csymbol{V}
\thispoint{Bd}{0,-7} \thispoint{Bu}{0,7}
\jointbnoarrow[right]{Bd}{0}{B}{0} \csymbolthird{a}  \jointbnoarrow[left]{B}{0}{Bu}{0} \csymbolthird{b}
\jointbnoarrow[right]{Bd}{-2}{B}{-2} \csymbolthird{x}  \jointbnoarrow[left]{B}{2}{Bu}{2} \csymbolthird{y}
\end{Compose}
}
~~=~~
\scalebox{1.3}{
\begin{Compose}{0}{0}\setdefaultfont{} \setsecondfont{\mathsfb}
\crectangle[thin]{coef}{3.2}{2}{-12,0} \csymbol{\frac{\alpha_\mathsfb{xa}\sqrt{N_\mathsfb{xa}N_\mathsfb{by}}}{\alpha_\mathsfb{by}} }
\crectangledouble{Iup}{1.5}{1.5}{0,8}\csymbolthird{\hat{I}} \crectangledouble{Idown}{1.5}{1.5}{0,-8} \csymbolthird{\hat{I}}
\crectangledleft[-1]{A}{1.5}{2.5}{-6,0} \csymbol{V}
\relpoint{A}{0,-7}{Ad} \jointbleft[left]{Ad}{0}{A}{0} \csymbolalt[-8,0]{xa}
\relpoint{A}{0,7}{Cd} \jointbleft[left]{A}{0}{Cd}{0} \csymbolalt[-8,0]{by}
%
\crectangledleft{XA}{0.8}{0.8}{-2,3} \crectangledleft{BY}{0.8}{0.8}{-2,-3}
\relpoint{XA}{0.8,-1.8}{XAA} \joinrrnoarrow{XA}{0}{XAA}{0}
\joinrlnoarrow[below]{A}{1.2}{XAA}{0} \csymbol{xa}
\relpoint{BY}{0.8,1.8}{BYY} \joinrrnoarrow{BY}{0}{BYY}{0}
\joinrlnoarrow[above]{A}{-1.2}{BYY}{0} \csymbol{by}
\jointbleft[left]{XA}{0}{Iup}{-1} \csymbolalt[-8,6]{xa} \jointbleft[left]{Idown}{-1}{BY}{0} \csymbolalt[-8,-6]{by}
\crectangledright[-1]{AA}{1.5}{2.5}{6,0} \csymbol{V}
\relpoint{AA}{0,-7}{Ad} \jointbright[right]{Ad}{0}{AA}{0} \csymbolalt[8,0]{xa}
\relpoint{AA}{0,7}{Cd} \jointbright[right]{AA}{0}{Cd}{0} \csymbolalt[8,0]{by}
%
\crectangledright{XA}{0.8}{0.8}{2,3} \crectangledleft{BY}{0.8}{0.8}{2,-3}
\relpoint{XA}{-0.8,-1.8}{XAA} \joinllnoarrow{XA}{0}{XAA}{0}
\joinlrnoarrow[below]{AA}{1.2}{XAA}{0} \csymbol{xa}
\relpoint{BY}{-0.8,1.8}{BYY} \joinllnoarrow{BY}{0}{BYY}{0}
\joinlrnoarrow[above]{AA}{-1.2}{BYY}{0} \csymbol{by}
\jointbright[right]{XA}{0}{Iup}{1} \csymbolalt[8,6]{xa} \jointbright[right]{Idown}{1}{BY}{0} \csymbolalt[8,-6]{by}
\end{Compose}
}
\]

\thispagestyle{empty}
\end{titlepage}

\newpage

\begin{abstract}
The standard operational probabilistic framework (within which we can formulate Operational Quantum Theory) is time asymmetric.  This is clear because the conditions on allowed operations are time asymmetric.  It is odd, though, because Schoedinger's equation is time symmetric and probability theory does not care about time direction.  In this work we provide a time symmetric framework for operational theories in general and for Quantum Theory in particular.

The clearest expression of the time asymmetry of standard Operational Quantum Theory is that the deterministic effect is unique - meaning there is only one way to ignore the future - while deterministic (i.e.\ normalised) states are not unique.  In this paper, this time asymmetry is traced back to a time asymmetric understanding of the most basic elements of an operational theory - namely the operations (or boxes) out of which circuits are built.  We modify this allowing operations to have classical \emph{incomes} as well as classical outcomes on these operations.  We establish a time symmetric operational framework for circuits built out of operations. In particular, we demand that the probability associated with a circuit is the same whether we calculate it forwards in time or backwards in time. We do this by imposing various \emph{double properties}.  These are properties wherein a forward in time and a backward in time version of the same property are required. In this paper we provide new causality condition which we call \emph{double causality}.

To apply these operational ideas to Quantum Theory we adapt the duotensor and operator tensor approaches to the time symmetric setting.   We formulate physicality conditions consisting of (i) a condition that ensures the probability for any circuit is non-negative and (ii) the double causality condition.   Time symmetric operational Quantum Theory is then stated by the single axiom: Every physical operator has an operation corresponding to it.   We prove an extension theorem (this is a time symmetric version of the Stinespring extension theorem) enabling physical operators to be built from unitaries.

Curiously, we find that gauge parameters need to be introduced when we provide a concrete representation of the operator tensors associated with operations. They drop out when we calculate probabilities for circuits.  The particular gauge choice that is adopted in standard treatments of Quantum Theory goes naturally with a forward in time point of view.  We can, though, chose any values for these parameters whilst maintaining a time symmetric point of view.

As a future research direction, we propose the idea of a conditional frame of reference and elaborate on three such frames of reference - a time symmetric one, a forward one, and a backwards one.
\end{abstract}

\newpage

\pagestyle{plain}

\pagenumbering{roman} 
\setcounter{page}{1}

{~}

\tableofcontents

\newpage

\pagenumbering{arabic} 
\setcounter{page}{1}

\section{Introduction}

Quantum Theory is a probability calculus.   The usual tools of probability theory, appropriately applied, must enable us to calculate probabilities in the quantum case as in any other case.  Operational probabilistic theories concern the calculation of probabilities associated with circuits. These can be applied to Quantum Theory giving Operational Quantum Theory. In the standard formulation, Operational Quantum Theory is time asymmetric.  This is elucidated in the work of Chiribella, D'Ariano, and Perinotti \cite{chiribella2010informational}.  Future choices should not influence probabilities in the present and  corresponding to this is the property the mathematical property that deterministic effect is unique.  However, past choices do effect probabilities in the present and so the deterministic preparation is not unique.

This is rather strange since
\begin{enumerate}
\item Abstract probability theory concerns calculation of such objects as $\text{prob}(x|y)$ and, as such, knows nothing of time.
\item Quantum Theory, at the level of the Schr\"odinger equation, is time symmetric.
\item The Quantum Theory of measurement can be treated very simply without reference to the second law, for example with the von Neumann model of measurement.
\end{enumerate}
It is worth amplifying on the last point.  Although real measurement situations involve macroscopic instruments and, therefore, invariably entropic considerations are relevant, these simple models capture all the essential properties of measurement without invoking any entropic properties.

We will see it is possible to provide a time-symmetric operational formulation of quantum theory.  We do this by revisiting the notion of an operation and make it time symmetric. Then we set up a framework whereby circuits have the same probability whether the probability is calculated forwards or backwards in time.  We obtain this by imposing, and deriving various double properties (wherein there is a forward and a backward property).  The most important of these properties is double causality. This replaces the causality postulate of Chribella, D'Ariano, and Perinotti.   We further how to formulate Quantum Theory in the operator tensor framework.   

An extended introduction is given below in Sec.\ \ref{sec:extendedintroduction}.

\section{Previous work}\label{sec:previouswork}

The projection postulate of quantum theory, if taken literally as a physical process, is apparently time asymmetric since it implies that the wavefunction collapses after measurement.  However, the work of Aharonov, Bergmann and Lebowitz \cite{aharonov1964time} showed that it is possible to formulate pure state quantum theory, including the theory of measurement, in a time symmetric fashion.  For years this was thought to have dealt with the problem of time asymmetry in Quantum Theory.  The gauntlet was thrown down (again) with the work of Chribella, D'Ariano, and Perinotti \cite{chiribella2010informational} (see also \cite{d2018causality}) whose causality postulate - as discussed in the introduction above - revealed a deep time asymmetry in the standard operational formalism of Quantum Theory (where we now treat mixed states).   This set the stage for much thought this issue over the last decade or so.

The present work, in its inception, was strongly influenced by a presentation by Andrea Di Biagio  on preliminary ideas on time symmetry including \lq\lq a time-reversal invariant subtheory" of Quantum Theory.  Subsequently,  Di Biagio, in collaboration with Don\`a and Rovelli \cite{di2020quantum} showed how, it is possible to formulate operational Quantum Theory so that the mathematical machinery for prediction tasks - wherein we calculate $p(\text{future}|\text{past})$ - is of the same form as the mathematical machinery for postdiction tasks - wherein we calculate $p(\text{past}|\text{future})$.  This shows that there is a time reversal symmetry between prediction and postdiction.  The approach in \cite{di2020quantum} is similar to and clearly consistent with the present work.  There are some key differences however.   First, much of the discussion here is at the level of general operational theories.  Second, here we consider joint probabilities over past and future quantities of the form $p(\text{future},\text{past})$.   Then we set up mathematical machinery wherein we get the same value for these joint probabilities whether we calculate them in the forward time direction or the backwards time direction.  This perspective enables us to establish the double causality condition (discussed in Sec.\ \ref{sec:doublebisummationfromdoublecausality}) and the extension theorem (discussed in Sec.\ \ref{sec:extensiontheorem}.  Conditional probabilities can be obtained from these joint probabilities by application of the conditional probability formula.  In fact, we can \lq\lq transform" to the prediction or postdiction point of view as discussed in Sec.\ \ref{sec:proofthatformulationistimesymmetric} and Sec.\ \ref{sec:generalisedtheoryofconditionalframes}. Di Biagio et al.\ contains some interesting discussion on interpretation much of which is also pertinent to the approach in this paper. This covers the interpretation of having a unique deterministic  preparation (as well as a unique deterministic effect). They also discuss how different interpretations of Quantum Theory look in the light of their considerations.  Rovelli has other work on the issue of time symmetry.  For example, in  \cite{rovelli2020memory} he provides a model for memory (which plays a role in measurement) and discusses the connection with the second law.

A different take on the issue is provided in the paper \lq\lq Operational formulation of time reversal in quantum theory" by Oreshkov and Cerf \cite{oreshkov2016operational}.  They provide what looks, on the face of it, like an alternative time-symmetric operational formulation of Quantum Theory to that presented here or in DiBiagio et al.  It will be interesting to see if this approach is actually equivalent to the approach in the present paper.

There has been a tremendous amount of work on time-symmetry in Quantum Theory.  Indeed, it this is too vast a field to summarize here.  A few papers that are of particular interest are as follows. First, in 1964 Aharonov, Bergmann and Lebowitz \cite{aharonov1964time} provided a time-symmetric formulation of pure state quantum theory in which there are pre-selections (preparations) post-selections, and intermediate measurements.  A multi-time version appears in \cite{aharonov2009multiple}.    A mixed state version of this was provided by Silva {\it et al.}\ \cite{silva2014pre} (though there is no discussion of time-symmetry in this paper).

The connection to taking the adjoint and time reversal is discussed in Coecke and Kissinger \cite{coecke2018picturing} (see also \cite{quant-ph/0402014,coecke2014logic}).
In \cite{coecke2012time} Coecke and Lal  provide a discussion of reversing the time direction of a measurement process.

A recent paper by Chribella and Liu \cite{2012.03859} discusses the \emph{time flip operator}.

In this paper we will be primarily interested in getting to a time symmetric operational formulation of Quantum Theory.   There has been much interesting work which focus not only on the operational realm, but also on ontological questions arising from time symmetry.  See for example Price \cite{price2012does}, Wharton \cite{wharton2018new, wharton2020colloquium}, Leifer and Pusey \cite{leifer2017time},  Argaman \cite{argaman2021quantum}.   Also see the books by Price \cite{price1997time} by Schulman \cite{schulman1997time}.  It is hoped that the approach in this paper will provide a framework in which to further those discussions.

The approach taken here extends frameworks I have developed previously and makes them time symmetric.  These frameworks are duotensor framework in \cite{hardy2013formalism} and the operator tensor formulation of Quantum Theory   \cite{hardy2011reformulating, hardy2012operator} (these were the development of a perspectives from \cite{hardy2001quantum} and \cite{hardy2005probability, hardy2007towards}).  The duotensor and operator tensor approaches are built on the idea that we can regard states and processes as arising via a tomographic procedure using fiducial elements.

The duotensor and operator tensor approaches are diagrammatic and strongly inspired diagrammatic approaches due to Coecke and collaborators.  See the extraordinary book of Coecke and Kissinger \cite{coecke2017picturing} for a extensive discussion of this approach.  In Sec.\ \ref{sec:inputandoutputhilbertspaces} I provide a summary of an approach to Hilbert spaces with input and output structure.  This is essentially a summary of the approach of Coecke \cite{coecke2007linearizing} and Selinger \cite{selinger2007dagger} and summarized in \cite{coecke2017picturing}.  The doubling up process described in these works provides an alternative route (to the tomographic procedure mentioned above) to what I call operator tensors.  The notation, both diagrammatic and symbolic, adopted here is different from that in Coecke and Kissinger and chosen so as to match with my previous notation for operator tensors (which are represented by double border boxes).  The symbolic notation, in particular, is tensor-like and intended to represent faithfully the graphical aspect of the diagrammatic notation.

The general probability framework approach of \cite{hardy2001quantum} has predecessors (for example, see \cite{mackey1963mathematical, ludwig1985axiomatic, davies1970operational, gunson1967algebraic, mielnik1969theory, araki1980characterization, gudder1999convex, foulis1979empirical, fivel1994interference}) and has been a topic of considerable research in recent years (for example, see \cite{barrett2007information, barnum2011information, chiribella2010probabilistic, chiribella2010informational, hardy2009foliable, hardy2009operational, hardy2009operational2, oeckl2016local}).

The approach due to Chiribella, D'Ariano, and Perinotti \cite{chiribella2010probabilistic, chiribella2010informational} was also inspired by the diagrammatic approach of Coecke and others. It is in this context that the causality postulate is introduced.  The causality postulate is also at work in the iterative conditions on quantum combs discussed in \cite{chiribella2008quantum, chiribella2009theoretical}. One of the research directions discussed at the end of this paper concerns updating these iterative conditions in the light of the double causality principle from this paper.

A large part of my motivation for the present work (though not extensively discussed in this paper) is to deal with the challenge of indefinite causal structure which we would expect to arise in constructing a theory of Quantum Gravity.  Time asymmetry weds us more strongly to a background causal structure.  There has been much work on indefinite causal structure \cite{hardy2005probability, hardy2007towards, hardy2009quantum, chiribella2013quantum, oreshkov2012quantum, procopio2015experimental, hardy2020implementation} (see the nice article by Natalie Wolchover \cite{wolchover2021quantum} for an introduction).

A different strand of work has been pursued by Leifer and Spekkens on a formulating Quantum Theory as a causally neutral theory of inference \cite{leifer2013towards}.  Part of the idea there is to generalise the chain rule for conditional probabilities to Quantum Theory.  The duotensor approach discussed in Part \ref{part:duotensorformulation} also invokes this chain rule in the classical context and what might be regarded as a generalisation for general systems (that might be quantum).

A recent iteration of the inferential approach is due to Schmid, Selby, and Spekkens \cite{schmid2020unscrambling}.  This adds classical inferential structure to the quantum structure.  This is also done in the present work.   The approach of Schmid et al.\ is more general though not specifically adapted to a time symmetric formulation.

The program of Oeckl,  which starts from more of a Quantum Gravity perspective with the general boundary formalism \cite{oeckl2003general} and has moved in the direction of general probability theories with his positive formalism \cite{oeckl2013positive, oeckl2016local}, is of particular interest.  This approach does not assume up front any time or metric.

\section{Extended Introduction}\label{sec:extendedintroduction}

Since this is a long paper I will provide an extended introduction summarizing the main ideas.  Throughout the paper, I will state \emph{double properties}. These consist of a \emph{forward property} and a \emph{backward property} (and the two are the time reverse of one another).  I use several types of diagrammatic notation.   Generally wires that run up the page pertain to systems (quantum systems for example).   Wires that run across the page can pertain to results of measurements (this is classical information) or to indices that are summed over in some mathematical expression.

\subsection{Time symmetric operations}

In Sec.\ \ref{sec:timesymmetricoperations} we argue that the usual notion of an operation is not fit for purpose when it comes to formulating operational theories in a time symmetric manner.  We propose modifying this so that, in addition to outcomes, there also incomes.  A time symmetric operation looks like this
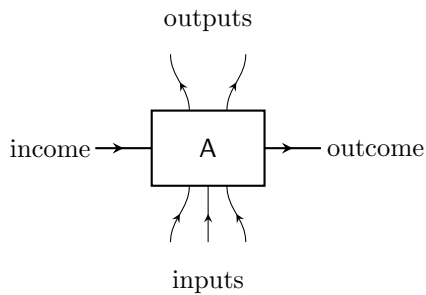
\[
\begin{Compose}{0}{0}
\Crectangle{A}{3}{2}{0,0}
\thispoint{AL}{-6,0} \thispoint{AR}{6,0}
\thispoint{A1in}{-2,-5} 
\thispoint{A2in}{0,-5} 
\thispoint{A3in}{2,-5} 
\thispoint{A1out}{-2,5}
 \thispoint{A2out}{0,5}
 \thispoint{A3out}{2,5} 
\Thistexthere{inputs}{0,-7.1} \Thistexthere{outputs}{0,6.8}  
\joinrlthick{AL}{0}{A}{0}   \Thistexthere{income}{-8.4,0}
\joinrlthick{A}{0}{AR}{0}   \Thistexthere{outcome}{8.9,0}
\jointb{A1in}{0}{A}{-1} \jointb{A2in}{0}{A}{0} \jointb{A3in}{0}{A}{1}
\jointb{A}{-1}{A1out}{0} \jointb{A}{1}{A3out}{0}
\end{Compose}
\]
Outcomes are available after the operation has been applied but not before. Incomes are available before the operation has been applied but not afterwards (an example is an ancilla prepared in some initial state, or the initial position of a pointer).  The wires associated with incomes and outcomes are referred to as \emph{pointers types}.  The inputs and outputs are for physical systems such as electrons, photons, etc.  We will refer to these as \emph{system types}.  We can also accommodate \emph{settings} (corresponding to some knob positions for example). In the above diagram thoughthese are suppressed for ease of presentation.

\subsection{Circuits}

Operations can be wired together. If there are no open wires left over then we have a circuit. For example
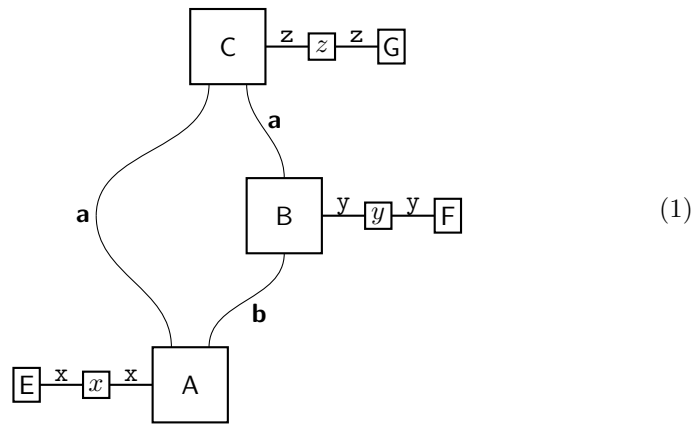
\begin{equation} 
\begin{Compose}{0}{-2.4} \setsecondfont{\mathsfb}
\crectangle{A}{2}{2}{0,0} \csymbol{A}  \crectangle{B}{2}{2}{5,9} \csymbol{B} \crectangle{C}{2}{2}{2,18} \csymbol{C}
\thispoint{amid}{-5,9}  \csymbolalt[-20,0]{a}
\jointbnoarrow[left]{A}{-1}{amid}{0}\jointbnoarrow{amid}{0}{C}{-1}
\jointbnoarrow[below right]{A}{1}{B}{0}  \csymbolalt[5,-5]{b}
\jointbnoarrow[above right]{B}{0}{C}{1}  \csymbolalt{a}
\GenxBoxincome{A}{2}{0}{x}{E}
\GenxBoxoutcome{B}{2}{0}{y}{F}
\GenxBoxoutcome{C}{2}{0}{z}{G}
\end{Compose}
\end{equation}
The small boxes with $x$, $y$, \dots are called \emph{readout boxes}.  They readout the income or outcome.  We demand that the probability, $p(x,y,z)$, is given by the circuit description and does not depend on extraneous factors.  We also demand that we get the same answer for this probability whether we do the calculation forwards or backwards in time.   Boxes like $\mathsf E$, viewed forward in time, correspond to a distribution over income values.  Boxes like $\mathsf F$, viewed backwards in time correspond to a distribution over outcome values.

\subsection{A simple classical situation}

In Part \ref{sec:asimpleclassicalsituation} we will see that we can view circuits as pertaining to a simple classical situation in which we have classical incomes and classical outcomes.  We can impose various double properties to this classical situation.  Given a few basic assumptions, we will see that time symmetry forces on us that distributions over incomes and outcomes must be flat.  Throughout the paper, we represent flat distributions by boxes with an $\mathsf R$ in them.   Thus, circuits will look like
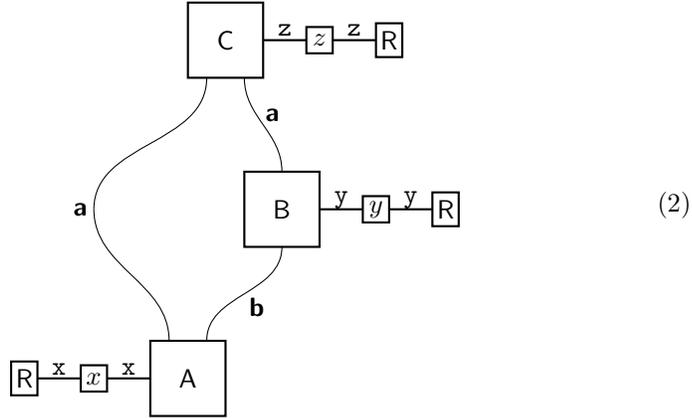
\begin{equation} 
\begin{Compose}{0}{-2.4} \setsecondfont{\mathsfb}
\crectangle{A}{2}{2}{0,0} \csymbol{A}  \crectangle{B}{2}{2}{5,9} \csymbol{B} \crectangle{C}{2}{2}{2,18} \csymbol{C}
\thispoint{amid}{-5,9}  \csymbolalt[-20,0]{a}
\jointbnoarrow[left]{A}{-1}{amid}{0}\jointbnoarrow{amid}{0}{C}{-1}
\jointbnoarrow[below right]{A}{1}{B}{0}  \csymbolalt[5,-5]{b}
\jointbnoarrow[above right]{B}{0}{C}{1}  \csymbolalt{a}
\RxBoxincome{A}{2}{0}{x}{E}
\RxBoxoutcome{B}{2}{0}{y}{F}
\RxBoxoutcome{C}{2}{0}{z}{G}
\end{Compose}
\end{equation}
Correspondingly, we argue that circuits should be doubly summing. Thus, if we sum over the incomes then the we get a flat distribution over the outcomes. And, if we sum over the outcomes, we get a flat distribution over the incomes.  This property is a generalization of bistochasticity.

\subsection{Equivalence, linearity, and physicality}

In Part \ref{part:equivalencelinearityandphysicality} we introduce a linearly extended notion of equivalence between operations. This makes it possible to regard expressions consisting of linearly weighted sums of operations as being equivalent to one another.  Armed with this mathematical machinery, we can define \emph{ignore} operations and introduce the notion of \emph{physicality}.

Ignore operations are written as
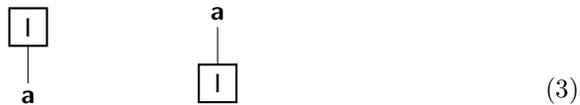
\begin{equation}
\begin{Compose}{0}{0}\setsecondfont{\mathsfb}
\Crectangle{I}{1}{1}{0,3} \thispoint{p}{0,0} \csymbolalt[0,-20]{a} \joinbtnoarrow{I}{0}{p}{0}
\end{Compose}
~~~~~~~~~~~~~~~~~
\begin{Compose}{0}{0}\setsecondfont{\mathsfb}
\Crectangle{I}{1}{1}{0,0} \thispoint{p}{0,3} \csymbolalt[0,20]{a} \jointbnoarrow{I}{0}{p}{0}
\end{Compose}
\end{equation}
They correspond to the situation where we have no incomes or outcomes.  In the case on the left, we have no output - we call this the \emph{ignore result}.  In the case on the right we have no input - we call this the \emph{ignore preparation}.  We show, employing some assumptions, that these ignore operations are unique (so, for example, all ignore preparations for a given system type are equivalent).

Physicality consists of constraints on operations such that the probabilities associated with circuits (i) are positive and (ii) respect the double summing property.  Positivity of probabilities for circuits is guaranteed by imposing $T$-\emph{positivity} on each operation. This is written
\begin{equation}
0~\underset{T}{\leq}
\begin{Compose}{0}{0}\setdefaultfont{}\setsecondfont{\mathsfb}\setthirdfont{\mathtt}
\crectangle{B}{1.5}{1.5}{0,0} \csymbol{\mathsf{B}}
\thispoint{d}{0,-3} \csymbolalt[0,-20]{a}
\thispoint{u}{0,3}  \csymbolalt[0,20]{b}
\thispoint{l}{-3,0} \csymbolthird[-20,0]{x}
\thispoint{r}{3,0}  \csymbolthird[20,0]{y}
\jointbnoarrow{d}{0}{B}{0}
\jointbnoarrow{B}{0}{u}{0}
\joinrlnoarrowthick{l}{0}{B}{0}
\joinrlnoarrowthick{B}{0}{r}{0}
\end{Compose}
\end{equation}
The double summing property is guaranteed by imposing \emph{double causality}.  Double causality is written
\begin{equation}
\Funitalxyab{1}{1}{1}{1} ~~~~~~~~~~~~~~~~  \Bunitalyxba{1}{1}{1}{1}
\end{equation}
The condition on the left is forward causality. The condition on the right is backwards causality. We show how double causality is implied by more basic assumptions.  The usual causality condition \cite{chiribella2010informational} in time asymmetric operational theories can be understood as an application of backwards causality.

Note that often the earth symbol is used to represent objects like $\mathsf R$ and $\mathsf I$. So we might, instead, write
\begin{equation}
\begin{Compose}{0}{0}
\crectangle{A}{0.8}{0.8}{0,-2} \csymbol{I}\thispoint{p}{0,2} \jointbnoarrow{A}{0}{p}{0}
\end{Compose}
~~=~~
\begin{Compose}{0}{0}
\supterminal{A}{0,-2} \thispoint{p}{0,2} \jointbnoarrow{A}{0}{p}{0}
\end{Compose}
~~~~~~~~~~~~~~~~~~~
\begin{Compose}{0}{0}
\crectangle{A}{0.8}{0.8}{0,2} \csymbol{I} \thispoint{p}{0,-2} \jointbnoarrow{p}{0}{A}{0}
\end{Compose}
~~=~~
\begin{Compose}{0}{0}
\sdownterminal{A}{0,2} \thispoint{p}{0,-2} \jointbnoarrow{p}{0}{A}{0}
\end{Compose}
\end{equation}
and
\begin{equation}
\begin{Compose}{0}{0}
\crectangle{A}{0.8}{0.8}{-2,0} \csymbol{R}\thispoint{p}{2,0} \joinrlnoarrowthick{A}{0}{p}{0}
\end{Compose}
~~=~~
\begin{Compose}{0}{0}
\sleftterminal{A}{-2, 0} \thispoint{p}{2,0} \joinrlnoarrowthick{A}{0}{p}{0}
\end{Compose}
~~~~~~~~~~~~~~~~~~~
\begin{Compose}{0}{0}
\crectangle{A}{0.8}{0.8}{2,0} \csymbol{R} \thispoint{p}{-2,0} \joinrlnoarrowthick{p}{0}{A}{0}
\end{Compose}
~~=~~
\begin{Compose}{0}{0}
\srightterminal{A}{2,0} \thispoint{p}{-2, 0} \joinrlnoarrowthick{p}{0}{A}{0}
\end{Compose}
\end{equation}
This notation is not used here however. 

\subsection{Duotensor formalism}

In Part \ref{part:duotensorformulation} we discuss the duotensor formalism. This acts as an intermediary between the operation picture and the operator picture.   I have discussed this in previous papers \cite{hardy2010formalism, hardy2011reformulating}. However, I introduce new structure here so that both the pointer types and the system types have duotensor structure associated with them.  The idea is that we introduce fiducial elements
\begin{equation}
\begin{Compose}{0}{0} \setsecondfont{\mathtt}
\sleftdatafid{F}{0,0} \blackdotsq{FL}{-2.5,0} \csymbolthird[-24,0]{x}
\relpoint{F}{2.5,0}{FR} \csymbolalt[15,0]{x}
\joinrlnoarrowthick[above]{F}{0}{FR}{0}  \joinrlnoarrow[above]{FL}{0}{F}{0}
\end{Compose}
~~=~~
\begin{Compose}{0}{0}
\Crectangle{R}{0.7}{0.7}{-3,0}
\crectangle{x}{0.6}{0.6}{0,0} \csymbolthird{x}
\relpoint{x}{3,0}{xR}
\joinrlnoarrowthick[above]{R}{0}{x}{0} \csymbol[0,5]{x}
\joinrlnoarrowthick[above]{x}{0}{xR}{0} \csymbol[0,5]{x}
\end{Compose}
~~~~~~~~~~~~~
\begin{Compose}{0}{0} \setsecondfont{\mathtt}
\srightdatafid{F}{0,0} \relpoint{F}{-2.5,0}{FL} \csymbolalt[-18,0]{x}
\blackdotsq{FR}{2.5,0} \csymbolthird[21,0]{x}
\joinrlnoarrow[above]{F}{0}{FR}{0}  \joinrlnoarrowthick{FL}{0}{F}{0}
\end{Compose}
~~=~~
\begin{Compose}{0}{0}
\Crectangle{R}{0.7}{0.7}{3,0}
\crectangle{x}{0.6}{0.6}{0,0} \csymbolthird{x}
\relpoint{x}{-3,0}{xR}
\joinlrnoarrowthick[above]{R}{0}{x}{0} \csymbol[0,5]{x}
\joinlrnoarrowthick[above]{x}{0}{xR}{0} \csymbol[0,5]{x}
\end{Compose}
\end{equation}
for pointers and
\begin{equation}
\begin{Compose}{0}{0}\setsecondfont{\mathsfb} \setthirdfont{\mathpzc}
\supsysfid{A}{0,0}  \thispoint{p}{0,2.5} \csymbolalt[0,20]{x} \jointbnoarrow{A}{0}{p}{0} \blackdot{d}{0,-2.5} \csymbolthird[0,-23]{x}\joinbtnoarrow{A}{0}{d}{0}
\end{Compose}
~~~~~~~~~~~~~~
\begin{Compose}{0}{0}\setsecondfont{\mathsfb} \setthirdfont{\mathpzc}
\sdownsysfid{A}{0,0}  \thispoint{p}{0,-2.5} \csymbolalt[0,-20]{x} \joinbtnoarrow{A}{0}{p}{0} \blackdot{d}{0,2.5} \csymbolthird[0,23]{x}\jointbnoarrow{A}{0}{d}{0}
\end{Compose}
\end{equation}
for systems.  Using linearity and equivalence, as discussed in Part \ref{part:equivalencelinearityandphysicality}, we can represent an operation as a sum over fiducial elements as follows
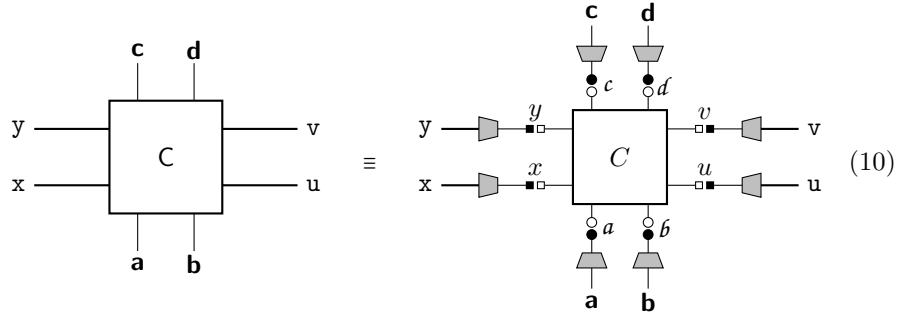
\begin{equation}\label{introGeneralOperationExpansion}
\begin{Compose}{0}{0}\setsecondfont{\mathsfb} \setthirdfont{\mathpzc}  \setdefaultfont{\mathtt}
\crectangle{T}{3}{3}{0,0} \csymbol{\mathsf{C}}
\thispoint{y7}{-7,1.5}   \csymbol[-24,0]{y} \joinrlnoarrowthick[above]{y7}{0}{T}{1.5}
\thispoint{x1}{-7,-1.5}   \csymbol[-24,0]{x} \joinrlnoarrowthick[above]{x1}{0}{T}{-1.5}
\thispoint{w8}{7,1.5}   \csymbol[24,0]{v} \joinrlnoarrowthick[above]{T}{1.5}{w8}{0}
\thispoint{y2}{7,-1.5}   \csymbol[24,0]{u} \joinrlnoarrowthick[above]{T}{-1.5}{y2}{0}
\upwire{T}{-1.5}{c}  \upwire{T}{1.5}{d}
\downwire{T}{-1.5}{a} \downwire{T}{1.5}{b}
\end{Compose}
~~~\equiv~~~
\begin{Compose}{0}{0}\setsecondfont{\mathsfb} \setthirdfont{\mathpzc} \setdefaultfont{\mathtt}
\crectangle{T}{2.5}{2.5}{0,0} \csymbolfourth{C}
\srightdatafid{y7}{-7,1.5}   \joinrlnoarrowbw[above]{y7}{0}{T}{1.5}\csymbolfourth[0,10]{y}
\thispoint{yL7}{-9.5,1.5}   \csymbol[-24,0]{y}  \joinrlnoarrowthick{yL7}{0}{y7}{0}
\srightdatafid{x1}{-7,-1.5}  \joinrlnoarrowbw[above]{x1}{0}{T}{-1.5} \csymbolfourth[0,10]{x}
\thispoint{xL1}{-9.5,-1.5} \csymbol[-24,0]{x} \joinrlnoarrowthick[above]{xL1}{0}{x1}{0}
\sleftdatafid{w8}{7,1.5}  \joinrlnoarrowwb[above]{T}{1.5}{w8}{0} \csymbolfourth[0,10]{v}
\thispoint{wR8}{9.5,1.5} \csymbol[24,0]{v}  \joinrlnoarrowthick[above]{w8}{0}{wR8}{0}
%
%
\sleftdatafid{y2}{7,-1.5} \joinrlnoarrowwb[above]{T}{-1.5}{y2}{0}  \csymbolfourth[0,10]{u}
\thispoint{yR2}{9.5,-1.5} \csymbol[24,0]{u}  \joinrlnoarrowthick[above]{y2}{0}{yR2}{0}
\fiddyupwire{T}{-1.5}{c}  \fiddyupwire{T}{1.5}{d}
\fiddydownwire{T}{-1.5}{a} \fiddydownwire{T}{1.5}{b}
\end{Compose}
\end{equation}
so long as we have tomographic locality.  Whenever there is an adjacent black and white dot we sum over the corresponding variable.  In this equation we have expanded the operation in terms of fiducials using a duotensor with all white dots.   Using the ideas of linearity and equivalence discussed in Part \ref{part:equivalencelinearityandphysicality} we see that we can represent an operation as a sum over fiducial elements.

\subsection{The operator tensor formalism}

In Part \ref{part:operatortensorformalism} we discuss the operator tensor formalism \cite{hardy2011reformulating, hardy2012operator}.  As with duotensors, I introduce new structure to allow both pointer and system types to be represented.    An operator tensor can be represented diagrammatically as follows
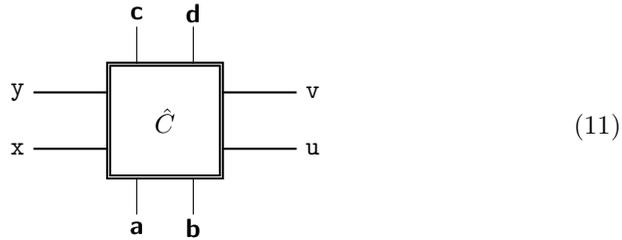
\begin{equation}
\begin{Compose}{0}{0}\setsecondfont{\mathsfb} \setthirdfont{\mathpzc}  \setdefaultfont{\mathtt}
\crectangledouble{T}{3}{3}{0,0} \csymbolfourth{\hat{C}}
\thispoint{y7}{-7,1.5}   \csymbol[-24,0]{y} \joinrlnoarrowthick[above]{y7}{0}{T}{1.5}
\thispoint{x1}{-7,-1.5}   \csymbol[-24,0]{x} \joinrlnoarrowthick[above]{x1}{0}{T}{-1.5}
\thispoint{w8}{7,1.5}   \csymbol[24,0]{v} \joinrlnoarrowthick[above]{T}{1.5}{w8}{0}
\thispoint{y2}{7,-1.5}   \csymbol[24,0]{u} \joinrlnoarrowthick[above]{T}{-1.5}{y2}{0}
\upwire{T}{-1.5}{c}  \upwire{T}{1.5}{d}
\downwire{T}{-1.5}{a} \downwire{T}{1.5}{b}
\end{Compose}
\end{equation}
Operator tensors play the role of density operators and completely positive maps in more standard approaches (though they do so in such a way that a calculation for the probability associated with a circuit is represented by a picture that looks the same as that circuit).  Before coming to operator tensors we provide a treatment of Hilbert spaces equipped with input and output structure which is, pretty much, equivalent to that presented in Coecke and Kissinger \cite{coecke2017picturing} though presented in new notation so it melds with previously adopted notation for operator tensors.  For example, an example of an operator tensor is
\begin{equation}
\begin{Compose}{0}{0}\setdefaultfont{} \setsecondfont{\mathsfb}
\crectangledouble{A}{2.5}{2.3}{0,0} \csymbol{\hat{B}}
\relpoint{A}{0,-4.5}{Ad} \jointbnoarrow[right]{Ad}{0}{A}{0} \csymbolalt{a}
\relpoint{A}{1,4.5}{Cd} \jointbnoarrow[right]{A}{1}{Cd}{0} \csymbolalt{a}
\relpoint{A}{-1,4.5}{Ed} \jointbnoarrow[right]{A}{-1}{Ed}{0} \csymbolalt{b}
\end{Compose}
~~=~~
\begin{Compose}{0}{0}\setdefaultfont{} \setsecondfont{\mathsfb}
\crectangledleft{A}{1.5}{2.3}{-3,0} \csymbol{B}
\relpoint{A}{0,-4.5}{Ad} \jointbleft[right]{Ad}{0}{A}{0} \csymbolalt{a}
\relpoint{A}{1,4.5}{Cd} \jointbleft[right]{A}{1}{Cd}{0} \csymbolalt{a}
\relpoint{A}{-1,4.5}{Ed} \jointbleft[right]{A}{-1}{Ed}{0} \csymbolalt{b}
\crectangledright{AA}{1.5}{2.3}{3,0} \csymbol{B}
\relpoint{AA}{0,-4.5}{Ad} \jointbright[right]{Ad}{0}{AA}{0} \csymbolalt{a}
\relpoint{AA}{1,4.5}{Cd} \jointbright[right]{AA}{1}{Cd}{0} \csymbolalt{b}
\relpoint{AA}{-1,4.5}{Ed} \jointbright[right]{AA}{-1}{Ed}{0} \csymbolalt{a}
\joinrlnoarrow[above]{A}{1.7}{AA}{1.7}\csymbol{a}
\joinrlnoarrow[above]{A}{0}{AA}{0}\csymbol[0,3]{b}
\joinrlnoarrow[above]{A}{-1.7}{AA}{-1.7}\csymbol{a}
\end{Compose}
\end{equation}
Each of the two halves are Hilbert space objects.  The horizontal wires connecting the two halves on the right represent summation over the associated indices.
This input output Hilbert space machinery is very useful as it allows us to see \lq\lq under the hood" and prove theorems.

We can restrict to the case of Hermitian operator tensors for which there exists an expansion in terms of fiducial operators having the form
\begin{equation}\label{introGeneralOperatorExpansion}
\begin{Compose}{0}{0}\setsecondfont{\mathsfb} \setthirdfont{\mathpzc}  \setdefaultfont{\mathtt}
\crectangledouble{T}{3}{3}{0,0} \csymbolfourth{\hat{C}}
\thispoint{y7}{-7,1.5}   \csymbol[-24,0]{y} \joinrlnoarrowthick[above]{y7}{0}{T}{1.5}
\thispoint{x1}{-7,-1.5}   \csymbol[-24,0]{x} \joinrlnoarrowthick[above]{x1}{0}{T}{-1.5}
\thispoint{w8}{7,1.5}   \csymbol[24,0]{v} \joinrlnoarrowthick[above]{T}{1.5}{w8}{0}
\thispoint{y2}{7,-1.5}   \csymbol[24,0]{u} \joinrlnoarrowthick[above]{T}{-1.5}{y2}{0}
\upwire{T}{-1.5}{c}  \upwire{T}{1.5}{d}
\downwire{T}{-1.5}{a} \downwire{T}{1.5}{b}
\end{Compose}
~~~=~~~
\begin{Compose}{0}{0}\setsecondfont{\mathsfb} \setthirdfont{\mathpzc} \setdefaultfont{\mathtt}
\crectangle{T}{2.5}{2.5}{0,0} \csymbolfourth{C}
\srightvecfid{y7}{-7,1.5}   \joinrlnoarrowbw[above]{y7}{0}{T}{1.5}\csymbolfourth[0,10]{y}
\thispoint{yL7}{-9.5,1.5}   \csymbol[-24,0]{y}  \joinrlnoarrowthick{yL7}{0}{y7}{0}
\srightvecfid{x1}{-7,-1.5}  \joinrlnoarrowbw[above]{x1}{0}{T}{-1.5} \csymbolfourth[0,10]{x}
\thispoint{xL1}{-9.5,-1.5} \csymbol[-24,0]{x} \joinrlnoarrowthick[above]{xL1}{0}{x1}{0}
\sleftvecfid{w8}{7,1.5}  \joinrlnoarrowwb[above]{T}{1.5}{w8}{0} \csymbolfourth[0,10]{v}
\thispoint{wR8}{9.5,1.5} \csymbol[24,0]{v}  \joinrlnoarrowthick[above]{w8}{0}{wR8}{0}
\sleftvecfid{y2}{7,-1.5} \joinrlnoarrowwb[above]{T}{-1.5}{y2}{0}  \csymbolfourth[0,10]{u}
\thispoint{yR2}{9.5,-1.5} \csymbol[24,0]{u}  \joinrlnoarrowthick[above]{y2}{0}{yR2}{0}
\fidopupwire{T}{-1.5}{c}  \fidopupwire{T}{1.5}{d}
\fidopdownwire{T}{-1.5}{a} \fidopdownwire{T}{1.5}{b}
\end{Compose}
\end{equation}
Thus, duotensors act as a bridge between operations and operator tensors in the following sense. From the expansion in \eqref{introGeneralOperationExpansion} we obtain a duotensor.  Then, we can use this duotensor to form an operator tensor using \eqref{introGeneralOperatorExpansion}.   As long as certain conditions are met on the fiducials from the operation theory and the fiducials from the operator theory, we are guaranteed that the circuits built from operator tensors will give the correct probabilities.

In Part \ref{part:operatortensorformalism} we see that there are gauge parameters, $\alpha_\mathsf{x}$ associated with providing a concrete representation of operator tensors.  For example, the ignore operators can be written
\begin{equation}
\begin{Compose}{0}{-0.07}\setdefaultfont{\hat}\setsecondfont{\mathtt}\setthirdfont{\mathsfb}
\Crectangledouble{I}{1}{1}{0,0}
\thispoint{p}{0,3.5}  \csymbolthird[0,20]{x} \jointbnoarrow{I}{0}{p}{0}
\end{Compose}
~=~ \frac{\alpha_\mathsfb{x}}{\sqrt{N_\mathsfb{x}}} \hat{\mathds{1}}^{\mathsfb{x}_1}
~~~~~~~~~~~~
\begin{Compose}{0}{-0.07}\setdefaultfont{\hat}\setsecondfont{\mathtt}\setthirdfont{\mathsfb}
\Crectangledouble{I}{1}{1}{0,0}
\thispoint{p}{0,-3.5}  \csymbolthird[0,-20]{x} \joinbtnoarrow{I}{0}{p}{0}
\end{Compose}
~=~ \frac{1}{\alpha_\mathsfb{x}\sqrt{N_\mathsfb{x}}} \hat{\mathds{1}}_{\mathsfb{x}_1}
\end{equation}
We are free to choose any value for $\alpha_\mathsf{x}$.  The standard choice in the usual (time asymmetric) formulation of Quantum Theory is $\alpha_\mathsfb{a}=\frac{1}{\sqrt{N_\mathsfb{a}}}$.  This corresponds to a forwards point of view.  Other choices are discussed in Sec.\ \ref{sec:chasingdowntheoneoverN}.  This story corresponds to physical systems.   There is a similar story corresponding to pointer types and the associated gauge parameter is $\beta_\mathtt{x}$.

\subsection{Time symmetric operational Quantum Theory}

In Part \ref{TSOQT} we finally come to the time symmetric operational formulation of Quantum Theory given by a single mathematical \emph{quantum axiom}.  To state this axiom we require two notions (a) correspondence and (b) physicality.  We say we have \emph{correspondence} from operations to operators if there is an operator associated with every operation such that if we replace operations by operators in a circuit then this operator circuit gives the probability for the circuit.  We prove the correspondence theorem that if
\begin{equation}
\begin{Compose}{0}{0} \setsecondfont{\mathtt}
\blackdotsq{FL}{-2.5,0} \csymbolthird[-24,0]{x}
\sleftdatafid{F}{0,0}
\srightdatafid{rF}{3,0}
\blackdotsq{rFR}{5.5,0} \csymbolthird[24,0]{x}
 \joinrlnoarrow[above]{FL}{0}{F}{0}
\joinrlnoarrowthick[above]{F}{0}{rF}{0} \csymbolalt{x}
\joinrlnoarrow[above]{rF}{0}{rFR}{0}
\end{Compose}
~~\equiv~~
\begin{Compose}{0}{0} \setsecondfont{\mathtt}
\blackdotsq{FL}{-2.5,0} \csymbolthird[-24,0]{x}
\sleftvecfid{F}{0,0}
\srightvecfid{rF}{3,0}
\blackdotsq{rFR}{5.5,0} \csymbolthird[24,0]{x}
 \joinrlnoarrow[above]{FL}{0}{F}{0}
\joinrlnoarrowthick[above]{F}{0}{rF}{0} \csymbolalt{x}
\joinrlnoarrow[above]{rF}{0}{rFR}{0}
\end{Compose}
~:=~ \begin{Compose}{0}{0}
\bbmatrixnoarrowsq{h}{0,0} \csymbolthird{x}
\end{Compose}
\end{equation}
and
\begin{equation}
\begin{Compose}{0}{0} \setsecondfont{\mathsfb} \setfourthfont{\mathpzc}
\blackdot{u}{0,3.5} \csymbolfourth[0,20]{x}\sdownsysfid{up}{0,1.5} \supsysfid{do}{0,-1.5} \blackdot{d}{0,-3.5}\csymbolfourth[0,-20]{x}
\jointbnoarrow[left]{d}{0}{do}{0} \jointbnoarrow[left]{do}{0}{up}{0} \csymbolalt{x} \jointbnoarrow[left]{up}{0}{u}{0}
\end{Compose}
~~\equiv~~
\begin{Compose}{0}{0} \setsecondfont{\mathsfb} \setfourthfont{\mathpzc}
\blackdot{u}{0,3.5} \csymbolfourth[0,20]{x}\sdownopfid{up}{0,1.5} \supopfid{do}{0,-1.5} \blackdot{d}{0,-3.5}\csymbolfourth[0,-20]{x}
\jointbnoarrow[left]{d}{0}{do}{0} \jointbnoarrow[left]{do}{0}{up}{0} \csymbolalt{x} \jointbnoarrow[left]{up}{0}{u}{0}
\end{Compose}
~~:=~~
\begin{Compose}{0}{0} \setfourthfont{\mathpzc}
\blackdot{u}{0,1.5}\blackdot{d}{0,-1.5} \jointbnoarrow[left]{d}{0}{u}{0}\csymbolfourth{x}
\end{Compose}
\end{equation}
then operations correspond to operators whose expansion is in terms of the same duotensor (compare \eqref{introGeneralOperationExpansion} with \eqref{introGeneralOperatorExpansion}).

Next we translate physicality to the operator tensor setting.  Physicality now demands that we have $T$ positivity for operator tensors
\begin{equation}
0~\underset{T}{\leq}
\begin{Compose}{0}{0}\setdefaultfont{}\setsecondfont{\mathsfb}\setthirdfont{\mathtt}
\crectangledouble{B}{1.5}{1.5}{0,0} \csymbol{\hat{B}}
\thispoint{d}{0,-3} \csymbolalt[0,-20]{a}
\thispoint{u}{0,3}  \csymbolalt[0,20]{b}
\thispoint{l}{-3,0} \csymbolthird[-20,0]{x}
\thispoint{r}{3,0}  \csymbolthird[20,0]{y}
\jointbnoarrow{d}{0}{B}{0}
\jointbnoarrow{B}{0}{u}{0}
\joinrlnoarrowthick{l}{0}{B}{0}
\joinrlnoarrowthick{B}{0}{r}{0}
\end{Compose}
\end{equation}
and double causality
\begin{equation}
\Funitaldoublexyab{1}{1}{1}{1} ~~~~~~~~~~~~  \Bunitaldoubleyxba{1}{1}{1}{1}
\end{equation}
The $T$ positivity condition is the same as demanding that $\hat{B}$ is positive after we take the transpose on the input (or output) space.

Once we have the notions of correspondence and physicality we can state a mathematical axiom for Quantum Theory. This is as follows
\begin{quote}
\textbf{Quantum Axiom.}
Every physical operator has an operation corresponding to it.
\end{quote}
This provides a time symmetric operational formulation of Quantum Theory.  We provide a proof of this in Sec.\ \ref{sec:proofthatformulationistimesymmetric}.

In the standard formalism of Quantum Theory an important result is the Stinespring extension theorem.  This shows how any permissible operator corresponding to an operation can be modeled as a unitary acting on the original system and an ancilla prepared in some initial state and then traced out afterwards.  This theorem is not applicable here since the initial ancilla would have to be prepared in a totally mixed state to respect double causality.  We prove an extension theorem that works in this time symmetric setting.  This says that physical operators can be modeled as follows
\begin{equation}\label{extensiontheoremequationintro}
\begin{Compose}{0}{0}\setsecondfont{\mathtt} \setthirdfont{\mathsfb}
\crectangledouble{U}{2}{2}{0,0} \csymbol{B} \thispoint{AL}{-4,0} \csymbolalt[-20,0]{x} \thispoint{AR}{4,0} \csymbolalt[20,0]{y}
\joinlrnoarrowthick{U}{0}{AL}{0} \joinrlnoarrowthick{U}{0}{AR}{0}
\thispoint{a}{0,-4} \csymbolthird[0,-20]{a} \jointbnoarrow{a}{0}{U}{0}
\thispoint{d}{0,4} \csymbolthird[0,20]{b} \joinbtnoarrow{d}{0}{U}{0}
\end{Compose}
~~=~~
\begin{Compose}{0}{0} \setdefaultfont{}\setsecondfont{\mathsfb}\setthirdfont{\mathtt}
\crectangledouble{U}{3}{2}{0,0} \csymbol{\hat{U}_\mathsf{B}}
\relpoint{U}{-0.3,4.5}{Iupos}  \crectangledouble{Iu}{0.9}{0.9}{Iupos} \csymbol{\hat{I}}  \jointbnoarrow[left]{U}{-0.3}{Iu}{0}  \csymbolalt{z}
\relpoint{U}{0.3,-4.5}{Idpos}  \crectangledouble{Id}{0.9}{0.9}{Idpos} \csymbol{\hat{I}} \jointbnoarrow[left]{Id}{0}{U}{0.3} \csymbolalt[-3,0]{w}
\relpoint{U}{2.3,4.5}{Ypos}  \crectangledouble{Y}{0.9}{0.9}{Ypos} \csymbol{\hat{Y}}  \jointbnoarrow[left]{U}{2.3}{Y}{0}  \csymbolalt{y}
\relpoint{Y}{4,0}{r} \joinlrnoarrowthick[above]{r}{0}{Y}{0} \csymbolthird[0,3]{y}
\relpoint{U}{-2.3,-4.5}{Xpos}  \crectangledouble{X}{0.9}{0.9}{Xpos} \csymbol{\hat{X}}  \jointbnoarrow[left]{X}{0}{U}{-2.3} \csymbolalt{x} \relpoint{X}{-4,0}{l} \joinrlnoarrowthick[above]{l}{0}{X}{0} \csymbolthird{x}
\thispoint{down}{0,-7} \thispoint{up}{0,7}
\jointbnoarrow[right]{down}{2.3}{U}{2.3} \csymbolalt{a}
\jointbnoarrow[left]{U}{-2.3}{up}{-2.3} \csymbolalt{b}
\end{Compose}
\end{equation}
where $\hat{U}$ is unitary, and $\hat{X}$ and $\hat{Y}$ are maximal.  The proof of this uses the Hilbert space machinery adapted from Coecke and Kissinger.

Finally, we show that this formalism is equivalent to the standard (time asymmetric) operational formalism in Sec.\ \ref{sec:equivalencetostandardoperationalformulation}.

\part{Basic operational considerations}\label{part:operationalconsiderations}

\section{Time symmetric operations}\label{sec:timesymmetricoperations}

In operational theories the central object is the \emph{operation}.  An operation corresponds to a single use of an apparatus.  Operations can be wired together to form circuits.

\subsection{The standard notion of an operation}

The standard notion of an operation in operational theories can be depicted as follows.
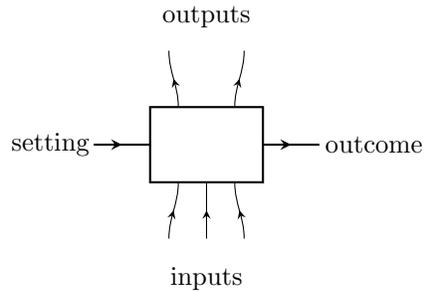
\[
\begin{Compose}{0}{0}
\crectangle{A}{3}{2}{0,0}
\thispoint{AL}{-6,0} \Thistexthere{setting}{-8.3,0} \thispoint{AR}{6,0} \Thistexthere{outcome}{8.9,0}
\thispoint{A1in}{-2,-5} 
\thispoint{A2in}{0,-5} 
\thispoint{A3in}{2,-5} 
\thispoint{A1out}{-2,5}
 \thispoint{A2out}{0,5}
 \thispoint{A3out}{2,5} 
\Thistexthere{inputs}{0,-7.1} \Thistexthere{outputs}{0,6.8}
\joinrlthick{AL}{0}{A}{0}  \joinrlthick{A}{0}{AR}{0}
\jointb{A1in}{0}{A}{-1.5} \jointb{A2in}{0}{A}{0} \jointb{A3in}{0}{A}{1.5}
\jointb{A}{-1.5}{A1out}{0} \jointb{A}{1.5}{A3out}{0}
\end{Compose}
\]
Settings are controlled by knobs, etc.   These are classical.  Outcomes correspond to pointer readings, detector clicks, etc.  These are also classical.
Inputs and outputs are for the physical systems (e.g.\ quantum systems like electrons, photons, \dots).

If we think about this a little more carefully, we should really depict an operation as follows.
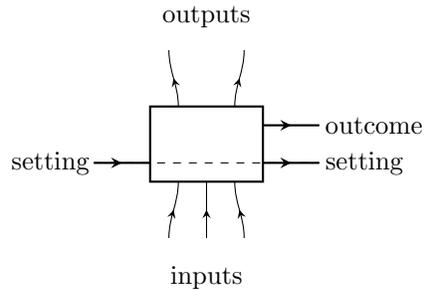
\[
\begin{Compose}{0}{0}
\crectangle{A}{3}{2}{0,0}
\thispoint{AL}{-6,0} \thispoint{AR}{6,0}
\thispoint{A1in}{-2,-5} 
\thispoint{A2in}{0,-5} 
\thispoint{A3in}{2,-5} 
\thispoint{A1out}{-2,5}
 \thispoint{A2out}{0,5}
 \thispoint{A3out}{2,5} 
\Thistexthere{inputs}{0,-7.1} \Thistexthere{outputs}{0,6.8}\relpoint{AL}{0,-1}{ALm} \relpoint{AR}{0,-1}{ARm}
\joinrlthick{AL}{-1}{A}{-1}  \joinrlthick{A}{-1}{AR}{-1}
\linebyhand[thin, dashed]{ALm}{ARm}  \Thistexthere{setting}{-8.3,-1} \Thistexthere{setting}{8.4,-1}
\joinrlthick{A}{1}{AR}{1}   \Thistexthere{outcome}{8.9,1}
\jointb{A1in}{0}{A}{-1.5} \jointb{A2in}{0}{A}{0} \jointb{A3in}{0}{A}{1.5}
\jointb{A}{-1.5}{A1out}{0} \jointb{A}{1.5}{A3out}{0}
\end{Compose}
\]
Settings are available before and after so we now show this in the diagram.  Outcomes are available afterwards but not before.   We can now see clearly that this basic idea of an operation is not time-symmetric. We have, right at the very beginning of our theoretical development of the subject, injected a basic time-asymmetry.

\subsection{Time symmetric operations} 

Fortunately, it is clear from the above picture what we have to do to make it time symmetric.
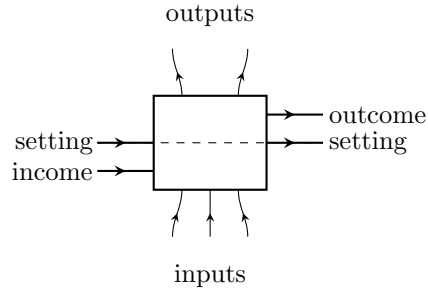
\[
\begin{Compose}{0}{0}
\crectangle{A}{3}{2.5}{0,0}
\thispoint{AL}{-6,0} \thispoint{AR}{6,0}
\thispoint{A1in}{-2,-5} 
\thispoint{A2in}{0,-5} 
\thispoint{A3in}{2,-5} 
\thispoint{A1out}{-2,5}
 \thispoint{A2out}{0,5}
 \thispoint{A3out}{2,5} 
\Thistexthere{inputs}{0,-7.1} \Thistexthere{outputs}{0,6.8}  
\joinrlthick{AL}{0}{A}{0}  \joinrlthick{A}{0}{AR}{0}
\linebyhand[thin, dashed]{AL}{AR}  \Thistexthere{setting}{-8.3,0} \Thistexthere{setting}{8.4,0}
\joinrlthick{AL}{-1.5}{A}{-1.5}   \Thistexthere{income}{-8.4,-1.5}
\joinrlthick{A}{1.5}{AR}{1.5}   \Thistexthere{outcome}{8.9,1.5}
\jointb{A1in}{0}{A}{-1.5} \jointb{A2in}{0}{A}{0} \jointb{A3in}{0}{A}{1.5}
\jointb{A}{-1.5}{A1out}{0} \jointb{A}{1.5}{A3out}{0}
\end{Compose}
\]
Now we have \emph{incomes} as well as outcomes.  Within the confines of an operation, an income is available before but not after (this is not to say that some communication structure extrinsic to the operation will not communicate this forward in time - this will be discussed later in Sec.\ \ref{sec:portinginformationarround}).  Like outcomes, incomes are classical.  Note that we use the English word \lq\lq income" in a way different from its usual modern usage (which has a financial meaning).  In fact, the original usage of the word is consistent with a broader definition (of something that \lq\lq comes in").

In this paper we will be primarily concerned with incomes and outcomes so we will suppress the settings
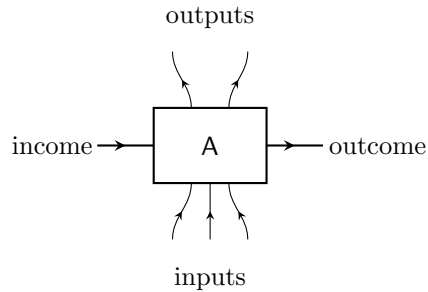
\[
\begin{Compose}{0}{0}
\Crectangle{A}{3}{2}{0,0}
\thispoint{AL}{-6,0} \thispoint{AR}{6,0}
\thispoint{A1in}{-2,-5} 
\thispoint{A2in}{0,-5} 
\thispoint{A3in}{2,-5} 
\thispoint{A1out}{-2,5}
 \thispoint{A2out}{0,5}
 \thispoint{A3out}{2,5} 
\Thistexthere{inputs}{0,-7.1} \Thistexthere{outputs}{0,6.8}  
\joinrlthick{AL}{0}{A}{0}   \Thistexthere{income}{-8.4,0}
\joinrlthick{A}{0}{AR}{0}   \Thistexthere{outcome}{8.9,0}
\jointb{A1in}{0}{A}{-1} \jointb{A2in}{0}{A}{0} \jointb{A3in}{0}{A}{1}
\jointb{A}{-1}{A1out}{0} \jointb{A}{1}{A3out}{0}
\end{Compose}
\]
The setting is now taken to be implicit in the symbol $\mathsf A$ (so if we have a different setting we would put a different symbol here).

We will associate symbols with the incomes, outcomes, inputs, and outputs as shown in this example
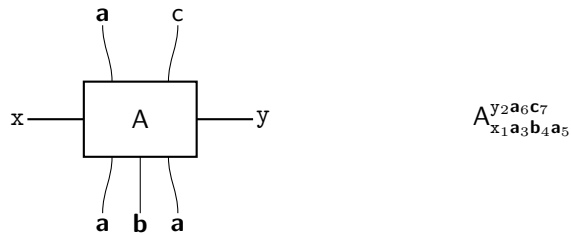
\[
\begin{Compose}{0}{-0.1}\setsecondfont{\mathtt}\setthirdfont{\mathsfb}
\Crectangle{A}{3}{2}{0,0}
\thispoint{AL}{-6,0} \csymbolalt[-15,0]{x} \thispoint{AR}{6,0} \csymbolalt[15,0]{y}
\thispoint{A1in}{-2,-5} \csymbolthird[-0,-18]{a} \thispoint{A2in}{0,-5} \csymbolthird[-0,-15]{b}\thispoint{A3in}{2,-5} \csymbolthird[-0,-18]{a}
\thispoint{A1out}{-2,5}\csymbolthird[0,15]{a} \thispoint{A2out}{0,5} \thispoint{A3out}{2,5} \csymbol[0,15]{c}
\joinrlnoarrowthick{AL}{0}{A}{0}  \joinrlnoarrowthick{A}{0}{AR}{0}
\jointbnoarrow{A1in}{0}{A}{-1.5} \jointbnoarrow{A2in}{0}{A}{0} \jointbnoarrow{A3in}{0}{A}{1.5}
\jointbnoarrow{A}{-1.5}{A1out}{0} \jointbnoarrow{A}{1.5}{A3out}{0}
\end{Compose}
~~~~~~~~~~~~~~~~~~~~~~
\mathsf{A}_{\mathtt{x}_1\mathsfb{a}_3\mathsfb{b}_4\mathsfb{a}_5}^{\mathtt{y}_2\mathsfb{a}_6\mathsfb{c}_7}
\]
We use $\mathtt x$, $\mathtt y$, etc.\ to denote income/outcome types (this is LaTeX \verb+\mathtt+ font).  For the sake of having a name, we will call these \emph{pointer types}.  They are classical and take values
\[
x=1, 2, \dots N_\mathtt{x}  ~~~~~~~~
y=1, 2, \dots N_\mathtt{y}
\]
We call the types associated with inputs and outputs \emph{system types}.
We use $\mathsfb{a}$, $\mathsfb{b}$, etc.\ to denote these (this is LaTeX \verb+\boldsymbol{\mathsf{a}}+ font).  These physical systems could be quantum systems.  We use LaTeX font \verb+\mathsf+ to denote the operation ($\mathsf A$ in the above example).   We give the diagrammatic notation for the operation on the left. On the right we give the symbolic notation wherein incomes and inputs are represented by subscripts and outcomes and outputs are represented by superscripts.  In the symbolic notation we also include integer labels on these subscripts and superscripts.  These label the wires as will be explained in Sec.\ \ref{sec:notation}.

A special type pointer type is the null pointer type, which we will denote by $\mathtt{0}$.  The null pointer type has $N_\mathtt{0}=1$ so it only has one value associated with it (namely $x=1$).   The null type can be useful.  For example, if we have an operation with no incomes, then we can append the null income without changing anything.  Then we can apply mathematical conditions that refer to incomes.

\subsection{Terminology for incomes and outcomes in standard Quantum Theory}

In the usual terminology of Quantum Theory we have examples of incomes and outcomes:
\begin{description}
\item[Incomes.] Here are examples of incomes taken from the usual way of thinking
\begin{enumerate}
  \item Initial pointer states (from maximal set)
  \item Initial ancilla state (from maximal set)
\end{enumerate}
\item[Outcomes.] Here are examples of outcomes taken from the usual way of thinking
\begin{enumerate}
  \item Final pointer state (from maximal set)
  \item Postselection on ancilla (onto element of a maximal set).
\end{enumerate}
\end{description}
Note that, usually, we take the initial pointer state to be in some fixed zero state.  However, we may more correctly need to take it to be one of a set of income states.   These examples are the reason for adopting the terminology \emph{pointer types}.  However, the notion of incomes and outcomes is a more general idea (or at least, it can represent other things not normally regarded as pointers).

\section{Joining operations together}

\subsection{Joining rules}

When we wire operations together there are two rules:
\begin{enumerate}
  \item We must match types.  Outputs must be matched with inputs of the same type. Outcomes must be matched with incomes of the same type.
  \item There must be no closed loops.  Thus, if we start on any particular operation, it must be impossible to get back to the same operation by following wires forward in time (whether they are pointer or system wires).
\end{enumerate}
The second rule means that we have a directed acyclic graph.

\subsection{Notation}\label{sec:notation}

If we wire together a bunch of operations we can notate it diagrammatically or symbolically as shown in this example.
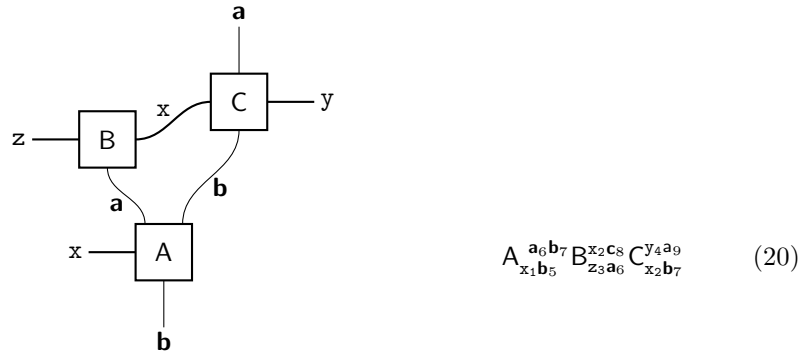
\begin{equation}\label{fragmentexample1}
\begin{Compose}{0}{0}\setsecondfont{\mathtt} \setthirdfont{\mathsfb}
\Crectangle{A}{1.5}{1.5}{0,0}  \relpoint{A}{-4,0}{AL}\csymbolalt[-20,0]{x} \relpoint{A}{0,-4}{AD}\csymbolthird[0,-20]{b}
\joinbtnoarrow{A}{0}{AD}{0} \joinlrnoarrowthick{A}{0}{AL}{0}
\Crectangle{B}{1.5}{1.5}{-3,6} \relpoint{B}{-4,0}{BL} \csymbolalt[-20,0]{z}
\joinlrnoarrowthick{B}{0}{BL}{0}
\Crectangle{C}{1.5}{1.5}{4,8}  \relpoint{C}{4,0}{CR} \csymbolalt[20,0]{y}  \relpoint{C}{0,4}{CU}\csymbolthird[0,20]{a}
\jointbnoarrow{C}{0}{CU}{0} \joinrlnoarrowthick{C}{0}{CR}{0}
\jointbnoarrow[below left]{A}{-1}{B}{0} \csymbolthird{a}
\jointbnoarrow[below right]{A}{1}{C}{0} \csymbolthird{b}
\joinrlnoarrowthick[above left]{B}{0}{C}{0}  \csymbolalt{x}
\end{Compose}
~~~~~~~~~~~~~~~~~~
\mathsf{A}_{\mathtt{x_1}\mathsfb{b}_5}^{~\mathsfb{a}_6\mathsfb{b}_7} \mathsf{B}_{\mathtt{z_3}\mathsfb{a}_6}^{\mathtt{x_2}\mathsfb{c}_8}
\mathsf{C}_{\mathtt{x_2}\mathsfb{b}_7}^{\mathtt{y_4}\mathsf{a}_9}
\end{equation}
In the symbolic notation on the left we have the same integer label for any given wire.  For example $\mathsfb{b}_7$ leaves $\mathsf A$ and enters $\mathsf{C}$ and so is repeated in the symbolic notation.  The integers are only labels. The diagram has the same meaning if we relabel them with different integers (as long as distinct labels remain distinct).

\subsection{Circuits and readout boxes}\label{sec:circuitsandreadoutboxes}

A \emph{circuit} is where we wire together operations and have no open wires left over.  For example,
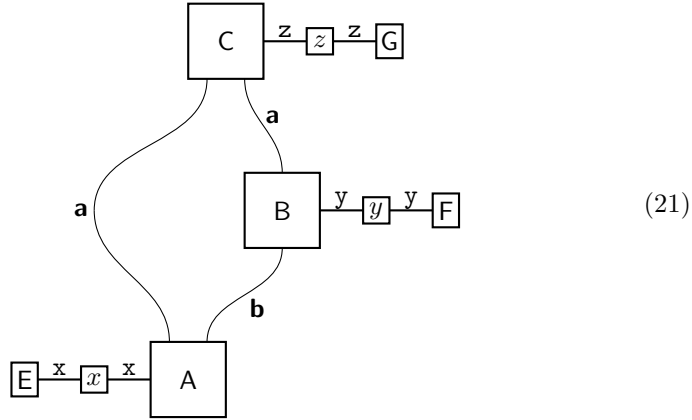
\begin{equation} 
\begin{Compose}{0}{-2.4} \setsecondfont{\mathsfb}
\crectangle{A}{2}{2}{0,0} \csymbol{A}  \crectangle{B}{2}{2}{5,9} \csymbol{B} \crectangle{C}{2}{2}{2,18} \csymbol{C}
\thispoint{amid}{-5,9}  \csymbolalt[-20,0]{a}
\jointbnoarrow[left]{A}{-1}{amid}{0}\jointbnoarrow{amid}{0}{C}{-1}
\jointbnoarrow[below right]{A}{1}{B}{0}  \csymbolalt[5,-5]{b}
\jointbnoarrow[above right]{B}{0}{C}{1}  \csymbolalt{a}
\GenxBoxincome{A}{2}{0}{x}{E}
\GenxBoxoutcome{B}{2}{0}{y}{F}
\GenxBoxoutcome{C}{2}{0}{z}{G}
\end{Compose}
\end{equation}
The small boxes with an $x$, $y$, and $z$ in this circuit are \emph{readout boxes}.  They can be thought of as reading the income/outcome they are attached to.  They have the property
\begin{equation}\label{xboxprop}
\begin{Compose}{0}{0}\setsecondfont{\mathtt}
\crectangle{AL}{0.7}{0.8}{-6,0} \csymbolthird{x} \relpoint{AL}{-2.5,0}{ALL} \csymbolalt[-18,0]{x}
\crectangle{AR}{0.7}{0.8}{-2,0} \csymbolthird{x'}
\relpoint{AR}{2.5,0}{ARR} \csymbolalt[15,0]{x}
\joinrlnoarrowthick{ALL}{0}{AL}{0}
\joinrlnoarrowthick[above]{AL}{0}{AR}{0}   \csymbolalt[0,4]{x}
\joinrlnoarrowthick{AR}{0}{ARR}{0}
\end{Compose}
~~~~
\equiv
~~~~
\left\{
\begin{array}{ll}
\begin{Compose}{0}{-0.1}\setsecondfont{\mathtt}
\crectangle{A}{0.6}{0.7}{0,0} \csymbolthird{x} \relpoint{A}{-2.5,0}{AL} \csymbolalt[-18,0]{x}
\relpoint{A}{2.5,0}{AR} \csymbolalt[15,0]{x}
\joinlrnoarrowthick{A}{0}{AL}{0}
\joinrlnoarrowthick{A}{0}{AR}{0}
\end{Compose}  &
\text{if} ~~x=x'  \\
\begin{Compose}{0}{-0.1}\setsecondfont{\mathtt}
\crectangle{A}{0.6}{0.7}{0,0} \csymbolthird{0} \relpoint{A}{-2.5,0}{AL} \csymbolalt[-18,0]{x}
\relpoint{A}{2.5,0}{AR} \csymbolalt[15,0]{x}
\joinlrnoarrowthick{A}{0}{AL}{0}
\joinrlnoarrowthick{A}{0}{AR}{0}
\end{Compose} &
\text{if} ~~x\not=x'
\end{array}
\right.
\end{equation}
since, if reading out the value twice is the same as reading it out once.  Here, the \emph{null box},
\begin{equation}
\begin{Compose}{0}{0}\setsecondfont{\mathtt}
\crectangle{A}{0.6}{0.7}{0,0} \csymbolthird{0} \relpoint{A}{-2.5,0}{AL} \csymbolalt[-18,0]{x}
\relpoint{A}{2.5,0}{AR} \csymbolalt[15,0]{x}
\joinlrnoarrowthick{A}{0}{AL}{0}
\joinrlnoarrowthick{A}{0}{AR}{0}
\end{Compose}
\end{equation}
has the property that, if it is added to any circuit, that circuit has probability equal to 0.

\subsection{Fragments and their causal structure}\label{sec:fragmentsandtheircausalstructure}

A \emph{fragment} is where we have some wires left open (\lq\lq fragment" here means fragment of a circuit). An example of a fragment is shown in \eqref{fragmentexample1}.

Any fragment, $\mathsf F$, has a certain \emph{causal structure} indicating the temporal relationships between the open wires which determines the set, $S_\mathsf{F}$,  of fragments, $\mathsf E$, with \emph{complementary causal structure} such that we can join $\mathsf F$ and any $\mathsf{E}\in S_\mathsf{F}$ such that we obtain a circuit.  This is a useful notion we will make use of later.    We say two fragments, $\mathsf A$ and $\mathsf B$, have \emph{compatible causal structure} if $S_\mathsf{A}\cap S_\mathsf{B}\not=\emptyset$.  This means there exist fragments that complete both $\mathsf A$ and $\mathsf B$ into circuits.

\subsection{The circuit probability assumption}\label{sec:circuitprobabilityassumption}

We make the following assumption
\begin{quote}
\textbf{Circuit probability assumption:}  Every circuit (wherein all inputs and outputs and all incomes and outcomes are closed) has a probability associated with it that depends only on the specification of that circuit.
\end{quote}
A similar assumption was used in earlier work \cite{hardy2011reformulating}.  When we have time symmetry, this assumption is more intriguing because we demand we get the same answer whether we calculate the probability forward in time or backwards in time.

We denote a circuit by $\mathsf{A}$, $\mathsf{B}$, \dots (since these are simply operations without any open wires).  Let us denote the readouts associated with circuit $\mathsf A$ by $x_\mathsf{A}$ and the specification of circuit $\mathsf A$ by $s_\mathsf{A}$.  The specification, $s_\mathsf{A}$, details the operations used and the arrangement of wires between these operations.  In this notation we denote the probability associated with circuit $\mathsf A$ by
\begin{equation}
\text{prob}(\mathsf{A}) = \text{prob}(x_\mathsf{A}|s_\mathsf{A})
\end{equation}
The circuit probability assumption can now be written as
\begin{equation}\label{extraneous}
\text{prob}(x_\mathsf{A}|s_\mathsf{A}, \text{extraneous conditions})= \text{prob}(x_\mathsf{A}|s_\mathsf{A})
\end{equation}
where $\text{extraneous conditions}$ are conditions that do not pertain to circuit $\mathsf A$.

\subsection{Probabilities for circuits factorize}\label{sec:circuitprobsfactorize}

If we have a circuit formed from two separate circuits then it is easy to show that the probabilities factorise
\begin{align*}
\text{prob}(\mathsf{AB}) & =  \text{prob}(x_\mathsf{A},x_\mathsf{B}|s_\mathsf{A},s_\mathsf{B})  \\
&= \text{prob}(x_\mathsf{A}|x_\mathsf{B}s_\mathsf{A},s_\mathsf{B})\text{prob}(x_\mathsf{B}|s_\mathsf{A},s_\mathsf{B}) \\
&= \text{prob}(x_\mathsf{A}|s_\mathsf{A})\text{prob}(x_\mathsf{B}|s_\mathsf{B}) \\
&= \text{prob}(\mathsf{A})\text{prob}(\mathsf{B})
\end{align*}
In going from the second to the third line of the proof we have used the circuit probability assumption - see \eqref{extraneous}.
This factorization property is very important when we come to set up the duotensor approach in Part \ref{part:duotensorformulation}

\subsection{Equivalence}\label{sec:equivalence}

We will say that two fragments, $\mathsf A$ and $\mathsf B$, are equivalent if they have compatible causal structure and give rise to the same probabilities when completed into a circuit by a fragment with complementary causal structure
\begin{equation}
\mathsf{A}\equiv \mathsf{B} ~~~\text{iff} ~~~ \text{prob}(\mathsf{AE}) = \text{prob}(\mathsf{BE}) ~~ \forall ~~ \mathsf{E} \in S_\mathsf{A}\cap S_\mathsf{B}
\end{equation}
We are suppressing subscripts and superscripts here for ease of expression.  Later, in Sec.\ \ref{sec:extendedequivalence}, we will introduce a more general notion of equivalence which linearly extends the present definition to weighted sums of fragments.

\subsection{Equivalence concerning readout boxes}

A useful property of readout boxes is the following equivalence
\begin{equation}\label{sumoverx}
\sum_x
\begin{Compose}{0}{-0.1}\setsecondfont{\mathtt}
\crectangle{A}{0.6}{0.7}{0,0} \csymbolthird{x} \relpoint{A}{-2.5,0}{AL} \csymbolalt[-18,0]{x}
\relpoint{A}{2.5,0}{AR} \csymbolalt[15,0]{x}
\joinlrnoarrowthick{A}{0}{AL}{0}
\joinrlnoarrowthick{A}{0}{AR}{0}
\end{Compose}
\equiv
\begin{Compose}{0}{0}\setsecondfont{\mathtt}
\thispoint{X}{-4,0}
\thispoint{Y}{4,0}
\joinrlnoarrowthick[above]{X}{0}{Y}{0}\csymbolalt[0,5]{x}
\end{Compose}
\end{equation}
This is true because, if we sum over the $x$ variables, we are taking the marginal of a joint probability.

\subsection{Definition of time symmetric operational theory}\label{sec:definitionoftimesymmetricoperationaltheory}

We will say an operational theory is time symmetric if,
\begin{enumerate}
\item Every allowed operation, $\mathsf A$, can be mapped to a \lq\lq time reversed" operation, $\utilde{\mathsf A}$, which is also allowed, wherein input and output types are swapped as are income and outcome types.    For example,
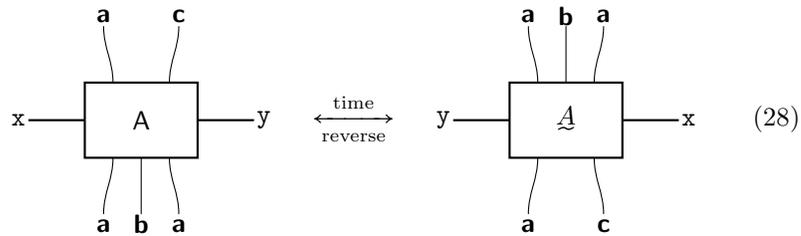
\begin{equation}
\begin{Compose}{0}{-0.1}\setsecondfont{\mathtt}\setthirdfont{\mathsfb}
\Crectangle{A}{3}{2}{0,0}
\thispoint{AL}{-6,0} \csymbolalt[-15,0]{x} \thispoint{AR}{6,0} \csymbolalt[15,0]{y}
\thispoint{A1in}{-2,-5} \csymbolthird[-0,-18]{a} \thispoint{A2in}{0,-5} \csymbolthird[-0,-15]{b}\thispoint{A3in}{2,-5} \csymbolthird[-0,-18]{a}
\thispoint{A1out}{-2,5}\csymbolthird[0,15]{a} \thispoint{A2out}{0,5} \thispoint{A3out}{2,5} \csymbolthird[0,15]{c}
\joinrlnoarrowthick{AL}{0}{A}{0}  \joinrlnoarrowthick{A}{0}{AR}{0}
\jointbnoarrow{A1in}{0}{A}{-1.5} \jointbnoarrow{A2in}{0}{A}{0} \jointbnoarrow{A3in}{0}{A}{1.5}
\jointbnoarrow{A}{-1.5}{A1out}{0} \jointbnoarrow{A}{1.5}{A3out}{0}
\end{Compose}
~~~ \extleftrightarrow[\text{reverse}]{\text{time}} ~~~
\begin{Compose}{0}{-0.1}\setsecondfont{\mathtt}\setthirdfont{\mathsfb}
\crectangle{A}{3}{2}{0,0} \csymbol{\utilde{A}}
\thispoint{AL}{-6,0} \csymbolalt[-15,0]{y} \thispoint{AR}{6,0} \csymbolalt[15,0]{x}
\thispoint{A1in}{-2,-5} \csymbolthird[-0,-18]{a} \thispoint{A2in}{0,5} \csymbolthird[0,15]{b}\thispoint{A3in}{2,-5} \csymbolthird[-0,-18]{c}
\thispoint{A1out}{-2,5}\csymbolthird[0,15]{a} \thispoint{A2out}{0,5} \thispoint{A3out}{2,5} \csymbolthird[0,15]{a}
\joinrlnoarrowthick{AL}{0}{A}{0}  \joinrlnoarrowthick{A}{0}{AR}{0}
\jointbnoarrow{A1in}{0}{A}{-1.5} \joinbtnoarrow{A2in}{0}{A}{0} \jointbnoarrow{A3in}{0}{A}{1.5}
\jointbnoarrow{A}{-1.5}{A1out}{0} \jointbnoarrow{A}{1.5}{A3out}{0}
\end{Compose}
\end{equation}
Note readout boxes are already time symmetric
\[
\begin{Compose}{0}{-0.1}\setsecondfont{\mathtt}
\crectangle{A}{0.6}{0.7}{0,0} \csymbolthird{x} \relpoint{A}{-2.5,0}{AL} \csymbolalt[-18,0]{x}
\relpoint{A}{2.5,0}{AR} \csymbolalt[15,0]{x}
\joinlrnoarrowthick{A}{0}{AL}{0}
\joinrlnoarrowthick{A}{0}{AR}{0}
\end{Compose}
~~~ \extleftrightarrow[\text{reverse}]{\text{time}} ~~~
\begin{Compose}{0}{-0.1}\setsecondfont{\mathtt}
\crectangle{A}{0.6}{0.7}{0,0} \csymbolthird{x} \relpoint{A}{-2.5,0}{AL} \csymbolalt[-18,0]{x}
\relpoint{A}{2.5,0}{AR} \csymbolalt[15,0]{x}
\joinlrnoarrowthick{A}{0}{AL}{0}
\joinrlnoarrowthick{A}{0}{AR}{0}
\end{Compose}
\]
This follows from their definition.
\item Given any circuit, $\mathsf E$, we can obtain the time reversed circuit, $\utilde{\mathsf E}$, by inverting the graph replacing operations with their time reversed counterparts. We require that $\text{prob}(\mathsf{E})= \text{prob}(\utilde{\mathsf{E}})$.  For example,
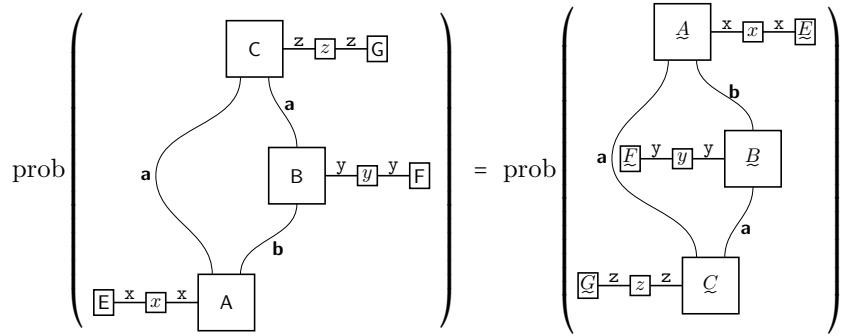
\[
\text{prob}\left(
\scalebox{0.75}{
\begin{Compose}{0}{-2.4} \setsecondfont{\mathsfb}
\crectangle{A}{2}{2}{0,0} \csymbol{A}  \crectangle{B}{2}{2}{5,9} \csymbol{B} \crectangle{C}{2}{2}{2,18} \csymbol{C}
\thispoint{amid}{-5,9}  \csymbolalt[-20,0]{a}
\jointbnoarrow[left]{A}{-1}{amid}{0}\jointbnoarrow{amid}{0}{C}{-1}
\jointbnoarrow[below right]{A}{1}{B}{0}  \csymbolalt[5,-5]{b}
\jointbnoarrow[above right]{B}{0}{C}{1}  \csymbolalt{a}
\GenxBoxincome{A}{2}{0}{x}{E}
\GenxBoxoutcome{B}{2}{0}{y}{F}
\GenxBoxoutcome{C}{2}{0}{z}{G}
\end{Compose}  }
\right)
~=~
\text{prob}\left( \hspace{-3mm}
\scalebox{0.75}{
\begin{Compose}{0}{2.4} \setsecondfont{\mathsfb}
\crectangle{A}{2}{2}{0,0} \csymbol{\utilde{A}}  \crectangle{B}{2}{2}{5,-9} \csymbol{\utilde{B}} \crectangle{C}{2}{2}{2,-18} \csymbol{\utilde{C}}
\thispoint{amid}{-5,-9}  \csymbolalt[-20,0]{a}
\joinbtnoarrow[left]{A}{-1}{amid}{0}\joinbtnoarrow{amid}{0}{C}{-1}
\joinbtnoarrow[above right]{A}{1}{B}{0}  \csymbolalt[5,-5]{b}
\joinbtnoarrow[below right]{B}{0}{C}{1}  \csymbolalt{a}
\GenxBoxoutcome{A}{2}{0}{x}{\utilde{E}}
\GenxBoxincome{B}{2}{0}{y}{\utilde{F}}
\GenxBoxincome{C}{2}{0}{z}{\utilde{G}}
\end{Compose} }
\right)
\]
The circuit on the right is the time reverse of the circuit on the left and has the same probability.
\end{enumerate}
This means that, for every process described in the forward direction, there is a corresponding process described in the backwards direction with the same predictions.

This definition is different from that of Leifer and Pusey in \cite{leifer2017time} which concerned conditional probabilities rather than joint probabilities.

\part{A simple classical situation}\label{sec:asimpleclassicalsituation}

\section{A circuit as a simple classical situation}\label{sec:sectionasimpleclassicalsituation}

Consider wiring together some operations so that there are no physical systems left open.  For example,
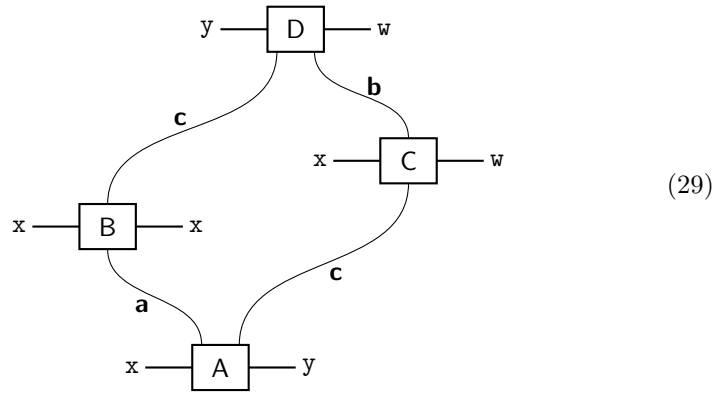
\begin{equation}\label{circuitexample}
\begin{Compose}{0}{-2.5}\setdefaultfont{\mathsf}\setsecondfont{\mathtt} \setthirdfont{\mathsfb}
\Crectangle{A}{1.5}{1.2}{0,0}
\relpoint{A}{-4,0}{AL}\csymbolalt[-20,0]{x}  \joinrlnoarrowthick[above]{AL}{0}{A}{0}
\relpoint{A}{4,0}{AR}  \csymbolalt[20,0]{y}\joinrlnoarrowthick[above]{A}{0}{AR}{0}
\Crectangle{D}{1.5}{1.2}{4,18}
\relpoint{D}{-4,0}{DL}  \csymbolalt[-20,0]{y}\joinrlnoarrowthick[above]{DL}{0}{D}{0}
\relpoint{D}{4,0}{DR} \csymbolalt[20,0]{w} \joinrlnoarrowthick[above]{D}{0}{DR}{0}
\Crectangle{B}{1.5}{1.2}{-6, 7.5}
\relpoint{B}{-4,0}{BL} \csymbolalt[-20,0]{x} \joinrlnoarrowthick[above]{BL}{0}{B}{0}
\relpoint{B}{4,0}{BR} \csymbolalt[20,0]{x} \joinrlnoarrowthick[above]{B}{0}{BR}{0}
\Crectangle{C}{1.5}{1.2}{10,11}
\relpoint{C}{-4,0}{CL} \csymbolalt[-20,0]{x} \joinrlnoarrowthick[above]{CL}{0}{C}{0}
\relpoint{C}{4,0}{CR} \csymbolalt[20,0]{w}\joinrlnoarrowthick[above]{C}{0}{CR}{0}
%
%
\jointbnoarrow[above left]{B}{0}{D}{-1} \csymbolthird[-4,0]{c} \jointbnoarrow[above right]{C}{0}{D}{1} \csymbolthird[5,0]{b}
\jointbnoarrow[below left]{A}{-1}{B}{0} \csymbolthird[-5,0]{a}
\jointbnoarrow[below right]{A}{1}{C}{0} \csymbolthird[4,0]{c}
\end{Compose}
\end{equation}
This is equivalent to
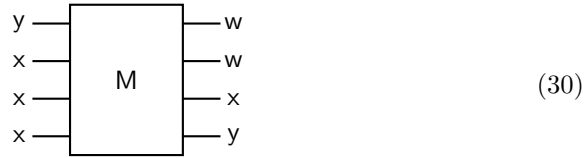
\begin{equation}\label{Mmanyports}
\begin{Compose}{0}{0}\setdefaultfont{\mathsf}\setsecondfont{\mathsf}
\crectangle{T}{3}{4}{0,0} \csymbol{M}
\thispoint{y7}{-5,3}   \csymbol[-20,0]{y} \joinrlnoarrowthick[above]{y7}{0}{T}{3}
\thispoint{x4}{-5,1}  \csymbol[-20,0]{x} \joinrlnoarrowthick[above]{x4}{0}{T}{1}
\thispoint{x3}{-5,-1}   \csymbol[-20,0]{x} \joinrlnoarrowthick[above]{x3}{0}{T}{-1}
\thispoint{x1}{-5,-3}   \csymbol[-20,0]{x} \joinrlnoarrowthick[above]{x1}{0}{T}{-3}
\thispoint{w8}{5,3}   \csymbol[20,0]{w} \joinrlnoarrowthick[above]{T}{3}{w8}{0}
\thispoint{w6}{5,1}   \csymbol[20,0]{w} \joinrlnoarrowthick[above]{T}{1}{w6}{0}
\thispoint{x5}{5,-1} \csymbol[20,0]{x} \joinrlnoarrowthick[above]{T}{-1}{x5}{0}
\thispoint{y2}{5,-3}   \csymbol[20,0]{y} \joinrlnoarrowthick[above]{T}{-3}{y2}{0}
\end{Compose}
\end{equation}
Here $\mathsf M$ is has classical incomes and classical outcomes. However, it has physics (possibly quantum) hidden inside.
We can simplify this situation further as
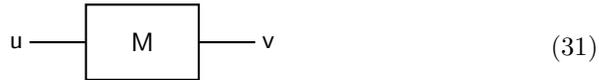
\begin{equation}\label{simpleclassicalsituation}
\begin{Compose}{0}{0}
\crectangle{M}{3}{2}{0,0} \csymbol{M}
\thispoint{u}{-6,0} \csymbol[-20,0]{u}
\thispoint{v}{6,0} \csymbol[20,0]{v}
\joinrlnoarrowthick[above]{u}{0}{M}{0}
\joinrlnoarrowthick[above]{M}{0}{v}{0}
\end{Compose}
\end{equation}
where we define the composite pointer types $\mathtt{u}=\mathtt{xxxy}$ and $\mathtt{v}=\mathtt{yxww}$.

The incredibly simple classical situation shown in \eqref{simpleclassicalsituation} is the key to understanding time-symmetry in operational theories (including Quantum Theory).  To this end we will, in this Part of the paper, describe a number of \lq\lq double" properties pertaining to this situation.
Included are the following five related double properties
\begin{description}
  \item[Double determinism.] Probabilities add up to one in both directions.
  \item[Double distribution uniqueness.] Income/outcome distribution pairs are unique.
  \item[Double pointer causality.] We cannot send information in either time direction.
  \item[Double flatness.] A flat distribution in gives flat distribution out in either time direction.
  \item[Double summation.]  This is a mathematical condition resulting from double determinism and double flatness.
\end{description}
We will explain these in more detail in what follows.  We use the word \lq\lq double" since, in each case, the double property is comprised of a \lq\lq forward" and a \lq\lq backward" property which are the time reverse of one another.

\section{Double determinism}\label{sec:doubledeterminism}

For the simple situation in \eqref{simpleclassicalsituation} we define
\begin{description}
\item[Forward determinism] is the condition that
\begin{equation} \sum_v p_\mathsf{M}(v|u) = 1  \end{equation}
This says that, for each $u$ on the income, the probabilities for the different outcomes add to one.
\item[Backward determinism] is the condition that
\begin{equation} \sum_u p_\mathsf{M}(u|v) = 1  \end{equation}
This says that, for each $v$ on the outcome, the probabilities for the different incomes add to one.
\end{description}
We say $\mathsf M$ is \emph{doubly deterministic} if it is both forward and backward deterministic.   Any operation, $\mathsf M$, that has no readout boxes (explicit or hidden), is taken to be doubly deterministic.

If $\mathsf M$ has only an outcome we call it a \emph{pointer preparation}. This case can be treated by making $\mathtt u$ the null type (discussed in Sec.\ \ref{sec:timesymmetricoperations}).  This means $N_\mathtt{u}=1$. Then we must necessarily have $u=1$ and so we can drop the sum over $u$ in the backwards causality condition.  Then the conditions for double determinism are $\sum_v p_\mathsf{M}(v|1) = 1$ and $p_\mathsf{M}(1|v)=1$.  We must necessarily have $p_\mathsf{M}(1|v)=1$ since there are no other incomes.  Hence, a pointer preparation is doubly deterministic if and only if it is forward deterministic.

If $\mathsf M$ has only an income we call it a \emph{pointer result} (we use the word \lq\lq result" in the time reversed sense to the use of the word \lq\lq preparation"). By similar reasoning,  a pointer result is doubly deterministic if and only if it is backwards deterministic.

We have defined the notion of double determinism for operations having only pointer types (but no system types).   We can extend the notion to general operations (that may have system types).  We will say that an operation which has no readouts (either explicit or implicit) is doubly deterministic.  We assume that, when we wire together such operations so that there are no system types left open, as illustrated in \eqref{circuitexample}, then we obtain a pointer type operation that is doubly deterministic in the sense defined above.  We will assume that operations are doubly deterministic unless otherwise stated.   Note that the readout box is not doubly deterministic.

\section{Pointer preparations and results}

\subsection{Pointer type closing operations}

For the operation $\mathsf M$ shown in \eqref{simpleclassicalsituation}, we can close the income, the outcome.   To do this we need pointer type closing operations having the form
\begin{equation}\label{pointerclosings}
\begin{Compose}{0}{0} \setsecondfont{\mathtt}
\crectangle{R}{0.7}{0.9}{0,0} \csymbol{E}\thispoint{p}{4,0} \csymbolalt[20,0]{u} \joinrlnoarrowthick{R}{0}{p}{0}
\end{Compose}
~~~~~~~~~~~~~~~~~~~~~~~~~~~~~
\begin{Compose}{0}{0}\setsecondfont{\mathtt}
\crectangle{R}{0.7}{0.9}{0,0}  \csymbol{F} \thispoint{p}{-4,0} \csymbolalt[-20,0]{v} \joinlrnoarrowthick{R}{0}{p}{0}
\end{Compose}
\end{equation}
The operation on the left is a pointer preparation.   The operation on the right is a pointer result.

Closing $\mathsf M$ with a pointer preparation gives
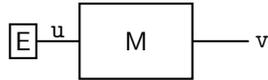
\[
\begin{Compose}{0}{0} \setsecondfont{\mathtt}
\crectangle{M}{3}{2}{0,0} \csymbol{M}
\crectangle{u}{0.7}{0.9}{-6,0} \csymbol{E}
\thispoint{v}{6,0} \csymbolalt[20,0]{v}
\joinrlnoarrowthick[above]{u}{0}{M}{0}\csymbolalt{u}
\joinrlnoarrowthick[above]{M}{0}{v}{0}
\end{Compose}
\]
which is another pointer preparation.  Closing $\mathsf M$ with a pointer result gives
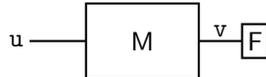
\[
\begin{Compose}{0}{0}\setsecondfont{\mathtt}
\crectangle{M}{3}{2}{0,0} \csymbol{M}
\thispoint{u}{-6,0} \csymbolalt[-20,0]{u}
\crectangle{v}{0.7}{0.9}{6,0} \csymbol{F}
\joinrlnoarrowthick[above]{u}{0}{M}{0}
\joinrlnoarrowthick[above]{M}{0}{v}{0}\csymbolalt{v}
\end{Compose}
\]
which is another pointer result.

As a consequence of assumptions we introduce, we will see that the only allowed pointer preparations correspond to the flat distribution over the outcome variables (when regarded forward in time).  Similarly, we will see that the only allowed pointer results correspond to the flat distribution over the income variables (when regarded backward in time).

\subsection{Double pointer tomography}\label{sec:doublepointer tomography}

We say that a set of probabilities, pertaining to some given operation, are \emph{tomographically complete} if all operations that are equivalent to the given operation have the same values for the probabilities in this set.  In this case, we will say that the set of probabilities fully specify the given operation.

Consider a circuit,
\begin{equation}
\begin{Compose}{0}{-0.06} \setsecondfont{\mathtt}
\crectangle{RR}{0.7}{0.9}{0,0}\csymbol{E}
\GenxBoxoutcome{RR}{0.9}{0}{x}{F}
\end{Compose}
\end{equation}
\emph{Forward pointer tomography} is the property that any pointer preparation, $\mathsf{E}$, is fully specified by the $N_\mathtt{x}$ probabilities
\begin{equation}\label{Nxprobabilities}
p(x) = \left(
\begin{Compose}{0}{-0.06} \setsecondfont{\mathtt}
\crectangle{RR}{0.7}{0.9}{0,0}\csymbol{E}
\GenxBoxoutcome{RR}{0.9}{0}{x}{F}
\end{Compose}
\right)
\end{equation}
if $\mathsf F$ is doubly deterministic (which, following the discussion in  Sec.\ \ref{sec:doubledeterminism} is the same as demanding that $\mathsf F$ is backwards deterministic since it has no outcomes).  \emph{Backward pointer tomography} is the property that any pointer result, $\mathsf F$, is fully specified by the $N_\mathtt{x}$ probabilities in \eqref{Nxprobabilities} if $\mathsf E$ is doubly deterministic (which is the same as demanding that $\mathsf E$ is forward deterministic).  Double pointer tomography is the property that both forward and backward pointer tomography hold.

It is noteworthy that, if both $\mathsf E$ and $\mathsf F$ are doubly deterministic, then they are both fully specified by the same probability distribution.  This issue will be explored in greater generality in Sec.\ \ref{sec:doubleuniqueness} below.

\subsection{Circuits}\label{sec:circuitagain}

We can close all incomes and outcomes in our example shown in \eqref{circuitexample} as follows.
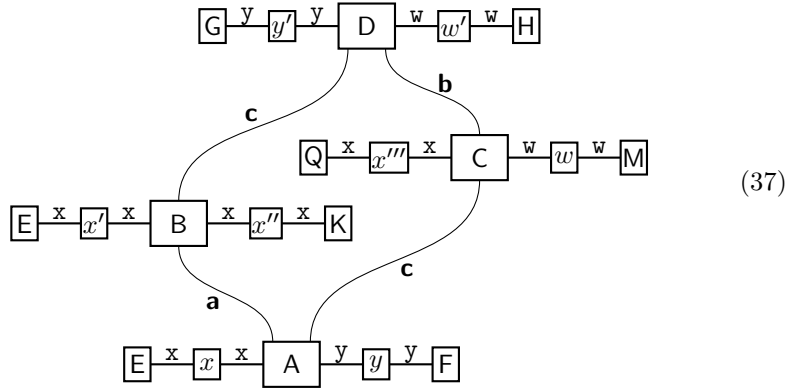
\begin{equation}\label{circuitexampleclosed}
\begin{Compose}{0}{-2.5}\setdefaultfont{\mathsf}\setsecondfont{\mathtt} \setthirdfont{\mathsfb}
\Crectangle{A}{1.5}{1.2}{0,0}
\GenxdifBoxincome{A}{1.5}{0}{x}{x}{E}   
\GenxdifBoxoutcome{A}{1.5}{0}{y}{y}{F}  
\Crectangle{D}{1.5}{1.2}{4,18}
\GenxdifBoxincome{D}{1.5}{0}{y}{y'}{G}  
\GenxdifBoxoutcome[1.2]{D}{1.5}{0}{w}{w'}{H}   
\Crectangle{B}{1.5}{1.2}{-6, 7.5}
\GenxdifBoxincome{B}{1.5}{0}{x}{x'}{E}  
\GenxdifBoxoutcome[1.2]{B}{1.5}{0}{x}{x''}{K}
\Crectangle{C}{1.5}{1.2}{10,11}
\GenxdifBoxincome[1.45]{C}{1.5}{0}{x}{x'''}{Q}
\GenxdifBoxoutcome{C}{1.5}{0}{w}{w}{M}
%
%
\jointbnoarrow[above left]{B}{0}{D}{-1} \csymbolthird[-4,0]{c} \jointbnoarrow[above right]{C}{0}{D}{1} \csymbolthird[5,0]{b}
\jointbnoarrow[below left]{A}{-1}{B}{0} \csymbolthird[-5,0]{a}
\jointbnoarrow[below right]{A}{1}{C}{0} \csymbolthird[4,0]{c}
\end{Compose}
\end{equation}
We can write this circuit as
\[
\begin{Compose}{0}{0} \setsecondfont{\mathtt}
\crectangle{T}{0.7}{1}{-7,0} \csymbol{E}
\crectangle{x}{0.7}{0.7}{-4,0} \csymbolthird{u}
\Crectangle{M}{1}{1}{0,0}
\joinrlnoarrowthick[above]{T}{0}{x}{0} \csymbolalt[0,5]{u}
\joinrlnoarrowthick[above]{x}{0}{M}{0} \csymbolalt[0,5]{u}
\crectangle{Tr}{0.7}{1}{7,0} \csymbol{F}
\crectangle{xr}{0.7}{0.7}{4,0} \csymbolthird{v}
\joinlrnoarrowthick[above]{Tr}{0}{xr}{0} \csymbolalt[0,5]{v}
\joinlrnoarrowthick[above]{xr}{0}{M}{0} \csymbolalt[0,5]{v}
\end{Compose}
\]
where we have followed steps similar to those in Sec.\ \ref{sec:asimpleclassicalsituation}.  The form of circuit shown in \eqref{circuitexampleclosed} is not the most general because we can allow an outcome from one operation to feed into the income of another operation.  It turns out that such examples can be modeled by circuits of the form in \eqref{circuitexampleclosed} if we include appropriate multiplicative factors. This is discussed in Sec.\ \ref{sec:midcomes}.

\subsection{Double uniqueness}\label{sec:doubleuniqueness}

Consider again the circuit
\[
\begin{Compose}{0}{0} \setsecondfont{\mathtt}
\crectangle{T}{0.7}{1}{-7,0} \csymbol{E}
\crectangle{x}{0.7}{0.7}{-4,0} \csymbolthird{u}
\Crectangle{M}{1}{1}{0,0}
\joinrlnoarrowthick[above]{T}{0}{x}{0} \csymbolalt[0,5]{u}
\joinrlnoarrowthick[above]{x}{0}{M}{0} \csymbolalt[0,5]{u}
\crectangle{Tr}{0.7}{1}{7,0} \csymbol{F}
\crectangle{xr}{0.7}{0.7}{4,0} \csymbolthird{v}
\joinlrnoarrowthick[above]{Tr}{0}{xr}{0} \csymbolalt[0,5]{v}
\joinlrnoarrowthick[above]{xr}{0}{M}{0} \csymbolalt[0,5]{v}
\end{Compose}
\]
We are interested in calculating
\begin{equation}\label{probuv}
\text{prob}(u, v)
\end{equation}
How do we actually calculate the probability associated with this circuit?    We will consider two ways of doing this. We can do the calculation \lq\lq forwards in time" by starting off with a distribution over the initial $u$ variables and evolve it forwards in time.  Or we can do the calculation \lq\lq backwards in time" by starting off with a distribution over the final variables and evolve this backwards in time.   By the circuit assumption we must get the same answer whether we do the calculation forward or backward in time. We will see that, if we are not careful, we run into problems with this.
Doing the calculation forwards we obtain
 \[p(u,v) = p_\mathsf{E}(u) p_\mathsf{M}(v|u)\]
Doing the calculation backwards we obtain
\[p(u,v) = p_\mathsf{F}(v) p_\mathsf{M}(u|v)\]
Hence
\begin{equation}\label{EMFequation}
p_\mathsf{E}(u) p_\mathsf{M}(v|u) = p_\mathsf{F}(v) p_\mathsf{M}(u|v)
\end{equation}
Consider, instead, the situation
\[
\begin{Compose}{0}{0} \setsecondfont{\mathtt}
\crectangle{T}{0.7}{1}{-7,0} \csymbol{E'}
\crectangle{x}{0.7}{0.7}{-4,0} \csymbolthird{u}
\Crectangle{M}{1}{1}{0,0}
\joinrlnoarrowthick[above]{T}{0}{x}{0} \csymbolalt[0,5]{u}
\joinrlnoarrowthick[above]{x}{0}{M}{0} \csymbolalt[0,5]{u}
\crectangle{Tr}{0.7}{1}{7,0} \csymbol{F'}
\crectangle{xr}{0.7}{0.7}{4,0} \csymbolthird{v}
\joinlrnoarrowthick[above]{Tr}{0}{xr}{0} \csymbolalt[0,5]{v}
\joinlrnoarrowthick[above]{xr}{0}{M}{0} \csymbolalt[0,5]{v}
\end{Compose}
\]
By similar reasoning we obtain
\begin{equation}\label{EpMFpequation}
p_\mathsf{E'}(u) p_\mathsf{M}(v|u) = p_\mathsf{F'}(v) p_\mathsf{M}(u|v)
\end{equation}
Dividing \eqref{EMFequation} by \eqref{EpMFpequation} gives
\[ \frac{p_\mathsf{E}(u)}{p_\mathsf{E'}(u)} = \frac{p_\mathsf{F}(v)}{p_\mathsf{F'}(v)}  \]
unless $p_\mathsf{M}(u|v)=0$ or $p_\mathsf{M}(v|u)=0$.
Using double determinism we obtain that
\begin{equation} \label{agreeuniqueness}
p_\mathsf{E}(u) = p_\mathsf{E'}(u) ~~~~\text{and}~~~~ p_\mathsf{F}(v) = p_\mathsf{F'}(v)
\end{equation}
Hence, for $u$ and $v$ having non-zero conditional probabilities, there is a unique distribution pair $p_\mathsf{E}(u)$ and $p_\mathsf{F}$ that \emph{agree} with $\mathsf{M}$ given that we demand we get the same joint probability $p(u,v)$ whether we do the calculation forwards or backwards in time.  We will call this property \emph{double uniqueness} since we have a unique distribution coming in from the past and a unique distribution coming in from the future.

There is another property related to double uniqueness.   To see this let us write
\[p_\text{forwards}(u,v) = p_\mathsf{E}(u) p_\mathsf{M}(v|u)\]
for the forwards calculation and
\[p_\text{backwards}(u,v) = p_\mathsf{F}(v) p_\mathsf{M}(u|v)\]
for the backwards calculation.  According to the circuit probability assumption we have
\begin{equation}
p_\text{forwards}(u,v) = p_\text{backwards}(u,v)
\end{equation}
If we sum this over $u$ and use backwards determinism, we obtain
\[ p_\mathsf{F}(v) = p(v|\mathsf{E,M}) ~~~~~\text{where}~~~~~~  p(v|\mathsf{E,M})= \sum_u p_\mathsf{E}(u) p_\mathsf{M}(v|u) \]
Similarly if we sum over $v$ and use forwards determinism, we obtain
\[ p_\mathsf{E}(u) = p(u|\mathsf{M, F}) ~~~~~\text{where}~~~~~~ p(u|\mathsf{M, F})= \sum_v p_\mathsf{F}(v) p_\mathsf{M}(u|v)   \]
In the first case we see that the forward evolved distribution agrees with the final distribution. In the second case we see that the backward evolved distribution agrees with the initial distribution.  We call this \emph{double agreement}.  This accords with the remarks at the end of Sec.\ \ref{sec:doublepointer tomography}.

\subsection{Disagreements and secrets}\label{sec:disagreementsandsecrets}

Double uniqueness and double agreement appear, on the face of it, to be obviously wrong. Assume, for the sake of argument, that $p_\mathsf{M}(u|v)$ and $p_\mathsf{M}(v|u)$ are strictly positive for all $u$ and $v$ so we have a unique matching pair. Now we can simply send in a distribution, $p_\mathsf{G}(u)$, that is not part of this matching pair.   We must, surely, obtain some distribution, $p_\mathsf{H}(v)$, on the other side.  Now, by construction, we have a pair of probability distributions where, at least, $p_\mathsf{G}(u)$ is not in the unique matching pair.  What has gone wrong?

To disentangle this, first we will sharpen the apparent contradiction. We have
\[p_\text{forwards}(u,v) = p_\mathsf{E}(u) p_\mathsf{M}(v|u)\]
for the forwards calculation and
\[p_\text{backwards}(u,v) = p_\mathsf{F}(v) p_\mathsf{M}(u|v)\]
for the backwards calculation.  According to the circuit probability assumption we have
\begin{equation}
p_\text{forwards}(u,v) = p_\text{backwards}(u,v)
\end{equation}
Consequently, we obtain
\[ p_\mathsf{F}(v) = p(v|\mathsf{E,M})  := \sum_u p_\mathsf{E}(u) p_\mathsf{M}(v|u) \]
and
\[ p_\mathsf{E}(u) = p(u|\mathsf{M, F}) := \sum_v p_\mathsf{F}(v) p_\mathsf{M}(u|v)   \]
Now we will look at an explicit example.   Consider the quantum circuit
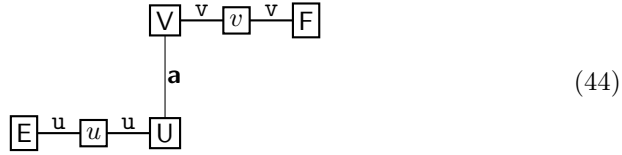
\begin{equation}
\begin{Compose}{0}{0} \setsecondfont{\mathsfb}\setthirdfont{\mathtt}
\Crectangle{U}{0.8}{0.8}{0,-3} \Crectangle{V}{0.8}{0.8}{0,3}
\jointbnoarrow[right]{U}{0}{V}{0} \csymbolalt{a}
\GenxBoxincome{U}{0.8}{0}{u}{E}
\GenxBoxoutcome{V}{0.8}{0}{v}{F}
\end{Compose}
\end{equation}
where $\mathsf U$ prepares quantum system $\mathsfb s$ in spin up state ($u=+1$) and spin down state($u=-1$) along the $\mathbf u$ direction and $\mathsf{V}$ prepares spin up state ($v=+1$) and spin down state ($v=-1$) along the $\mathbf v$ axis. Here $\mathbf u$ and $\mathbf v$ are taken to be unit vectors.   We can collapse this down to the circuit
\[
\begin{Compose}{0}{0} \setsecondfont{\mathtt}
\crectangle{T}{0.7}{1}{-7,0} \csymbol{E}
\crectangle{x}{0.7}{0.7}{-4,0} \csymbolthird{u}
\Crectangle{M}{1}{1}{0,0}
\joinrlnoarrowthick[above]{T}{0}{x}{0} \csymbolalt[0,5]{u}
\joinrlnoarrowthick[above]{x}{0}{M}{0} \csymbolalt[0,5]{u}
\crectangle{Tr}{0.7}{1}{7,0} \csymbol{F}
\crectangle{xr}{0.7}{0.7}{4,0} \csymbolthird{v}
\joinlrnoarrowthick[above]{Tr}{0}{xr}{0} \csymbolalt[0,5]{v}
\joinlrnoarrowthick[above]{xr}{0}{M}{0} \csymbolalt[0,5]{v}
\end{Compose}
\]
We can now easily calculate that
\begin{equation}
p_\mathsf{M}(v|u) = \frac{1}{2}\left(1+uv\mathbf{u}\cdot \mathbf{v}\right) = p_\mathsf{M}(u|v)
\end{equation}
Consider the case where $\mathbf{u}\cdot \mathbf{v}=0.75$.  We have
\begin{equation} \arraycolsep=1.4pt\def\arraystretch{1.3}
p_\mathsf{M}(v|u) = \left( \begin{array}{cc}
                             \frac{7}{8} & \frac{1}{8} \\
                             \frac{1}{8} & \frac{7}{8}
                           \end{array}  \right)
                           = p_\mathsf{M}(u|v)
\end{equation}
Note there are no zero entries in $p_\mathsf{M}(v|u)$ and $p_\mathsf{M}(u|v)$ so there exists a unique matching pair.

First consider the special case where $p_\mathsf{E}(+1)=1$ and $p_\mathsf{E}(-1)=0$ which we write as
\begin{equation}\label{pEuexample}
p_\mathsf{E}(u) =  \left(\begin{array}{c}
                            1 \\
                            0
                          \end{array}  \right)
\end{equation}
Then, evolving forward, we obtain
\begin{equation} \arraycolsep=1.4pt\def\arraystretch{1.3}
p(v|\mathsf{E,M}) =\sum_u p_\mathsf{M}(v|u) p_\mathsf{E}(u) =\left(\begin{array}{c}
                            \frac{7}{8} \\
                            \frac{1}{8}
                          \end{array}  \right)
\end{equation}
where we can use matrix multiplication to obtain this result.  As noted above, we require $p(v|\mathsf{E,M}) = p_\mathsf{F}(v)$
Hence, we can put
\begin{equation} \arraycolsep=1.4pt\def\arraystretch{1.3}
p_\mathsf{F}(v) =\sum_u p_\mathsf{M}(v|u) p_\mathsf{E}(u) =\left(\begin{array}{c}
                            \frac{7}{8} \\
                            \frac{1}{8}
                          \end{array}  \right)
\end{equation}
We can now evolve this backwards in time and we obtain
\begin{equation} \arraycolsep=1.4pt\def\arraystretch{1.3}
p(u|\mathsf{M,F}) =\sum_v p_\mathsf{M}(u|v) p_\mathsf{F}(v) =\left(\begin{array}{c}
                            \frac{50}{64} \\
                            \frac{14}{64}
                          \end{array}  \right)
\end{equation}
But, this contradicts the requirement that $p_\mathsf{E}(u)=p(u|\mathsf{E,M})$.  This is the disagreement.

One way to resolve this disagreement by allowing the preparation, $\mathsf E$, to have a secret income denoted by $\mathtt s$:
\begin{equation}
\begin{Compose}{0}{0} \setsecondfont{\mathsfb}\setthirdfont{\mathtt}
\Crectangle{E}{0.8}{0.8}{0,0}
\thispoint{l}{-3,0}\thispoint{r}{3,0}
\joinrlnoarrowthick[above]{l}{0}{E}{0} \csymbolthird{s}
\joinrlnoarrowthick[above]{E}{0}{r}{0} \csymbolthird{u}
\end{Compose}
\end{equation}
where we suppose that the income takes values $s=0$ and $s=1$.  Furthermore, we suppose that,
\begin{equation}\label{psEuexample}
p_\mathsf{E}(u|s=1) =  \left(\begin{array}{c}
                            1 \\
                            0
                          \end{array}  \right)
\end{equation}
so that the preparation we are interested in is actually only implemented when we have $s=1$.   In this case we are interested in $p(u,v|s=1)$. We can calculate if forward in time as
\begin{equation}
p_\text{forwards}(u,v|s=1) = p_\mathsf{E}(u|s=1)p_\mathsf{M}(v|u)
\end{equation}
and backward in time as
\begin{equation}
p_\text{backwards}(u,v|s=1) = \frac{p_\text{backwards}(u,v,s=1)}{p_\mathsf{E}(s=1)}
                            = \frac{p_\mathsf{F}(v)p_\mathsf{M}(u|v) p_\mathsf{E}{s=1|u} }{p_\mathsf{E}(s=1)}
\end{equation}
If we now equate $p_\text{forwards}(u,v|s=1)=p_\text{backwards}(u,v|s=1)$ we do not obtain a double uniqueness theorem as before.  It is worth thinking a little bit about the physics of what is happening here.  We could think of $s=1$ as corresponding to pressing the button that prepares $u=1$ thereby preparing the spin up state along the $\mathbf u$ direction.  If we time reverse this but send in from the future a spin down state then the dynamics (now evolved in reverse) will not necessarily undo the button press and so we have a different possible income, namely $s=0$.  This discussion illustrates that it is important we correctly characterize boxes in terms of their incomes and outcomes in representing physical situations within this theoretical framework.

\section{Double pointer causality and double flatness}\label{sec:doublepointercausalityanddoubleflatness}

\subsection{Flat pointer operations}

We define two special pointer operations, the \emph{flat pointer operations}, that are taken to be the time reverse of one another. These are
\begin{equation}
\begin{Compose}{0}{0} \setsecondfont{\mathtt}
\Crectangle{R}{0.7}{0.9}{0,0} \thispoint{p}{4,0} \csymbolalt[20,0]{x} \joinrlnoarrowthick{R}{0}{p}{0}
\end{Compose}
~~~ \extleftrightarrow[\text{reverse}]{\text{time}} ~~~
\begin{Compose}{0}{0}\setsecondfont{\mathtt}
\Crectangle{R}{0.7}{0.9}{0,0}  \thispoint{p}{-4,0} \csymbolalt[-20,0]{x} \joinlrnoarrowthick{R}{0}{p}{0}
\end{Compose}
\end{equation}
Strictly, we should use the notation $\utilde{\mathsf R}$ for one of these two operations (as per the notation introduced in Sec. \ref{sec:definitionoftimesymmetricoperationaltheory}). However, it is clear from the position of the wire that they are distinct so we omit the tilde.  We preserve the symbol $\mathsf R$ to refer to the flat pointer operations only.  We define the flat pointer operations by means of the following properties
\begin{description}
\item[Flat]
\begin{equation}\label{Rflatcond}
\text{prob}\left(
\begin{Compose}{0}{-0.06} \setsecondfont{\mathtt}
\crectangle{RR}{0.7}{0.9}{0,0}\csymbol{R}
\RxBoxoutcome{RR}{0.9}{0}{x}
\end{Compose}\right)
= \frac{1}{N_\mathtt{x}}
\end{equation}
\item[Deterministic]
\begin{equation}\label{Rdeterministiccond}
\text{prob}\left(
\begin{Compose}{0}{-0.06} \setsecondfont{\mathtt}
\crectangle{R}{0.7}{0.9}{0,0} \csymbol{R}
\crectangle{RR}{0.7}{0.9}{4,0} \csymbol{R}
\joinrlnoarrowthick{R}{0}{RR}{0} \csymbolalt{x}
\end{Compose}\right)
=1
\end{equation}
\end{description}
Using double pointer tomography and the property that probabilities over disjoint circuits factorize, we can prove the factorization properties
\begin{equation}\label{factorization}
\begin{Compose}{0}{0} \setsecondfont{\mathtt}
\Crectangle{R}{0.7}{0.9}{0,0} \thispoint{p}{4,0} \csymbolalt[25,0]{xy} \joinrlnoarrowthick{R}{0}{p}{0}
\end{Compose}
~~~\equiv~~~
\begin{Compose}{0}{0} \setsecondfont{\mathtt}
\Crectangle{R}{0.7}{0.9}{0,-1.3} \thispoint{p}{4,-1.3} \csymbolalt[20,0]{x} \joinrlnoarrowthick{R}{0}{p}{0}
\Crectangle{R}{0.7}{0.9}{0,1.3} \thispoint{p}{4,1.3} \csymbolalt[20,0]{y} \joinrlnoarrowthick{R}{0}{p}{0}
\end{Compose}
~~~~~~~~~~~~~~~
\begin{Compose}{0}{0}\setsecondfont{\mathtt}
\Crectangle{R}{0.7}{0.9}{0,0}  \thispoint{p}{-4,0} \csymbolalt[-25,0]{xy} \joinlrnoarrowthick{R}{0}{p}{0}
\end{Compose}
~~~\equiv~~~
\begin{Compose}{0}{0}\setsecondfont{\mathtt}
\Crectangle{R}{0.7}{0.9}{0,-1.3}  \thispoint{p}{-4,-1.3} \csymbolalt[-20,0]{x} \joinlrnoarrowthick{R}{0}{p}{0}
\Crectangle{R}{0.7}{0.9}{0,1.3}  \thispoint{p}{-4,1.3} \csymbolalt[-20,0]{x} \joinlrnoarrowthick{R}{0}{p}{0}
\end{Compose}
\end{equation}
This is clear since, if we form the operations,
\begin{equation}
\begin{Compose}{0}{0} \setsecondfont{\mathtt}
\crectangle{R}{0.7}{0.9}{0,0} \csymbol{Q} \thispoint{p}{4,0} \csymbolalt[25,0]{xy} \joinrlnoarrowthick{R}{0}{p}{0}
\end{Compose}
~~~:=~~~
\begin{Compose}{0}{0} \setsecondfont{\mathtt}
\Crectangle{R}{0.7}{0.9}{0,-1.3} \thispoint{p}{4,-1.3} \csymbolalt[20,0]{x} \joinrlnoarrowthick{R}{0}{p}{0}
\Crectangle{R}{0.7}{0.9}{0,1.3} \thispoint{p}{4,1.3} \csymbolalt[20,0]{y} \joinrlnoarrowthick{R}{0}{p}{0}
\end{Compose}
~~~~~~~~~~~~~~~
\begin{Compose}{0}{0}\setsecondfont{\mathtt}
\crectangle{R}{0.7}{0.9}{0,0} \csymbol{Q} \thispoint{p}{-4,0} \csymbolalt[-25,0]{xy} \joinlrnoarrowthick{R}{0}{p}{0}
\end{Compose}
~~~:=~~~
\begin{Compose}{0}{0}\setsecondfont{\mathtt}
\Crectangle{R}{0.7}{0.9}{0,-1.3}  \thispoint{p}{-4,-1.3} \csymbolalt[-20,0]{x} \joinlrnoarrowthick{R}{0}{p}{0}
\Crectangle{R}{0.7}{0.9}{0,1.3}  \thispoint{p}{-4,1.3} \csymbolalt[-20,0]{x} \joinlrnoarrowthick{R}{0}{p}{0}
\end{Compose}
\end{equation}
then using the property discussed in Sec.\ \ref{sec:circuitprobsfactorize} that, when circuits are disjoint, the probabilities multiply, we have
\begin{equation}\label{Qflatxy}
\text{prob}\left(
\begin{Compose}{0}{-0.06} \setsecondfont{\mathtt}
\crectangle{RR}{0.7}{0.9}{0,0}\csymbol{Q}  \crectangle{R}{0.7}{0.9}{7,0}\csymbol{Q}
\GenxdifBoxincome[1.3]{RR}{0.7}{0}{xy}{xy}{Q}
\end{Compose}\right)
= \frac{1}{N_\mathtt{xy}}
~~~~~~~~~~~~~~~~~~
\text{prob}\left(
\begin{Compose}{0}{-0.06} \setsecondfont{\mathtt}
\crectangle{R}{0.7}{0.9}{0,0} \csymbol{Q}
\crectangle{RR}{0.7}{0.9}{4,0} \csymbol{Q}
\joinrlnoarrow{R}{0}{RR}{0} \csymbolalt{xy}
\end{Compose}\right)
=1
\end{equation}
where $N_\mathtt{xy}=N_\mathtt{x} N_\mathtt{y}$.
Hence we see that the flat and deterministic properties are automatically satisfied.  We see that $\mathsf{Q}^{\mathtt{x}_1\mathtt{y}_2}$ and $\mathsf{Q}_{\mathtt{x}_1\mathtt{y}_2}$ are doubly deterministic since they satisfy the conditions given in Sec.\ \ref{sec:doubledeterminism}.  Hence, according to the assumption of double pointer tomography, $\mathsf{Q}^{\mathtt{x}_1\mathtt{y}_2}$, is fully specified by the probabilities in the equation on the left of \eqref{Qflatxy} and so we must have $\mathsf{Q}^{\mathtt{x}_1\mathtt{y}_2}=\mathsf{R}^{\mathtt{x}_1\mathtt{y}_2}$.  Similarly, we obtain $\mathsf{Q}_{\mathtt{x}_1\mathtt{y}_2}=\mathsf{R}_{\mathtt{x}_1\mathtt{y}_2}$.  This proves the factorisation properties in \eqref{factorization}.

Given these properties, we can interpret these operations.  The operation
\begin{equation}\label{flatforwardop}
\begin{Compose}{0}{0} \setsecondfont{\mathtt}
\Crectangle{R}{0.7}{0.9}{0,0} \thispoint{p}{4,0} \csymbolalt[20,0]{x} \joinrlnoarrowthick{R}{0}{p}{0}
\end{Compose}
\end{equation}
gives the flat distribution
\begin{equation}
p_R(x) = \frac{1}{N_\mathtt{x}}
\end{equation}
if we think in the forward in time point of view.
The operation
\begin{equation}
\begin{Compose}{0}{0}\setsecondfont{\mathtt}
\Crectangle{R}{0.7}{0.9}{0,0}  \thispoint{p}{-4,0} \csymbolalt[-20,0]{x} \joinlrnoarrowthick{R}{0}{p}{0}
\end{Compose}
\end{equation}
gives the flat probability distribution,
\begin{equation}
p_R(x) = \frac{1}{N_\mathtt{x}}
\end{equation}
if we think in the backwards in time point of view.

\subsection{Double maximality}\label{sec:doublemaximality}

Some physical theories (including Quantum Theory) we consider will have the following property
\begin{quote}
\emph{Double maximality} is the property that, for each pointer type, $\mathtt x$, we can associate a physical type $\mathsfb x$ such that there exist operations
\begin{equation}\label{maximaloperations}
\begin{Compose}{0}{0}\setthirdfont{\mathsfb}\setsecondfont{\mathtt}
\Crectangle{X}{0.8}{0.8}{0,0}
\thispoint{p}{0,3} \csymbolthird[0,20]{x} \jointbnoarrow[left]{X}{0}{p}{0}
\thispoint{pp}{-3,0} \csymbolalt[-20,0]{x} \joinlrnoarrowthick{X}{0}{pp}{0}
\end{Compose}
~~~~~~~~~~~~~~~~~~~~~~~~
\begin{Compose}{0}{0}\setsecondfont{\mathtt}\setthirdfont{\mathsfb}
\Crectangle{X}{0.8}{0.8}{0,3}
\thispoint{p}{0,0} \csymbolthird[0,-20]{x} \jointbnoarrow[left]{p}{0}{X}{0}
\thispoint{pp}{3,3} \csymbolalt[20,0]{x} \joinlrnoarrowthick[left]{pp}{0}{X}{0}
\end{Compose}
\end{equation}
for which we have the equivalence
\begin{equation}\label{doublemaximality}
\begin{Compose}{0}{0}\setsecondfont{\mathtt} \setthirdfont{\mathsfb}
\Crectangle{X}{0.8}{0.8}{0,-2} \relpoint{X}{-4,0}{XL} \csymbolalt[-20,0]{x} \joinrlnoarrowthick{XL}{0}{X}{0}
\crectangle{Y}{0.8}{0.8}{0,2}\csymbol{X} \relpoint{Y}{4,0}{YR} \csymbolalt[20,0]{x}\joinlrnoarrowthick{YR}{0}{Y}{0}
\jointbnoarrow[left]{X}{0}{Y}{0}\csymbolthird[0,5]{x}
\end{Compose}
~~~~\equiv ~~~~~~~
\begin{Compose}{0}{0}\setsecondfont{\mathtt}
\thispoint{X}{-4,0}
\thispoint{Y}{4,0}
\joinrlnoarrowthick[above]{X}{0}{Y}{0}\csymbolalt[0,5]{x}
\end{Compose}
\end{equation}
Furthermore, this is maximal in that there does not exist a similar equivalence for the same physical type, $\mathsfb x$, but for another pointer type, $\mathtt z$, having $N_\mathtt{z} > N_\mathtt{x}$.  The right hand side of \eqref{doublemaximality} is just the identity for the type $\mathtt x$.
\end{quote}
This means that if we have the value $x$ on the income (in the diagram on the left hand side above) then we will have the same value $x$ on the outcome (forward maximality) and if we have the value $x$ on the outcome then we will have the same value $x$ on the income (backward maximality).  We will call the operations in \eqref{maximaloperations} \emph{maximal operations}.  The maximal operation on the left in \eqref{maximaloperations} is called a \emph{maximal preparation} and the maximal operation on the right is called a \emph{maximal measurement}. Note that forward maximality does not imply backwards maximality.

Note that the operations shown in \eqref{maximaloperations} are the time reverse of one another (in the sense discussed in Sec.\ \ref{sec:definitionoftimesymmetricoperationaltheory}).  We should, strictly indicate this by putting a tilde under the $\mathsf X$ for one of the two operations. However, we omit this since it is clear whether we have a preparation or measurement from the positions of the wires.

\subsection{Double pointer causality}

Causality plays an important role in physics in general.  In the standard operational formalism it is a time asymmetric notion - namely that choices in the future should not influence probabilities in the present (this is the Pavia causality condition \cite{chiribella2010informational}).  We need a time symmetric notion of causality.  In Sec.\ \ref{sec:doublebisummationfromdoublecausality} we will present such a condition which we call \emph{double causality}.  In the meantime, in this section, we present an intermediate notion which we call \emph{double pointer causality}.  We will see in Sec.\ \ref{sec:doublebisummationfromdoublecausality} that double pointer causality is a consequence of double causality.

We define the following property:
\begin{quote}
\textbf{Double pointer causality.} Consider an operation, $\mathsf M$, having incomes and outcomes but no open inputs or outputs (for systems).
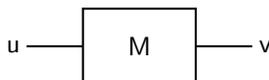
\[
\begin{Compose}{0}{0}
\crectangle{M}{3}{2}{0,0} \csymbol{M}
\thispoint{u}{-6,0} \csymbol[-20,0]{u}
\thispoint{v}{6,0} \csymbol[20,0]{v}
\joinrlnoarrowthick[above]{u}{0}{M}{0}
\joinrlnoarrowthick[above]{M}{0}{v}{0}
\end{Compose}
\]
The property of double pointer causality imposes that we have both
\begin{description}
  \item[{\it Forwards pointer causality}] is the property that there is no access to information about $\mathsf{M}$ from the outcome without some knowledge of the income.
  \item[{\it Backward pointer causality}] is the property that there is no access to information about $\mathsf{M}$ from the income without some knowledge of the outcome.
\end{description}
No knowledge means the flat distribution
\end{quote}
We can spell this out in a little more detail.   Forward pointer causality is the property that the probability distribution over $v$ associated with
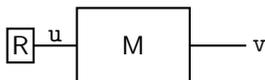
\[
\begin{Compose}{0}{0}\setsecondfont{\mathtt}
\crectangle{M}{3}{2}{0,0} \csymbol{M}
\crectangle{u}{0.7}{0.9}{-6,0} \csymbol{R}
\thispoint{v}{6,0} \csymbolalt[20,0]{v}
\joinrlnoarrowthick[above]{u}{0}{M}{0}\csymbolalt{u}
\joinrlnoarrowthick[above]{M}{0}{v}{0}
\end{Compose}
\]
is independent of $\mathsf M$.  Backward pointer causality is the property that the probability distribution over $u$ associated with
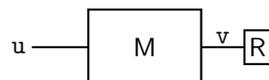
\[
\begin{Compose}{0}{0} \setsecondfont{\mathtt}
\crectangle{M}{3}{2}{0,0} \csymbol{M}
\thispoint{u}{-6,0} \csymbolalt[-20,0]{u}
\crectangle{v}{0.7}{0.9}{6,0} \csymbol{R}
\joinrlnoarrowthick[above]{u}{0}{M}{0}
\joinrlnoarrowthick[above]{M}{0}{v}{0}\csymbolalt{v}
\end{Compose}
\]
is independent of $\mathsf M$.

\subsection{Double flatness}

We are now in a position to define the following property.
\begin{quote}
  {\bf Double flatness.}  Consider an operation, $\mathsf M$, having incomes and outcomes but no open inputs or outputs (for systems). The property of double flatness imposes we have both
  \begin{description}
    \item[{\it Forward flatness:}] if we have a flat probability distribution over the incomes and evolve this forward in time through $\mathsf M$, then we obtain a flat probability distribution over the outcomes, and
    \item[{\it Backwards flatness:}] if we have a flat probability distribution over outcomes and evolve this backwards in time through $\mathsf M$, then we obtain a flat probability over the incomes.
  \end{description}
\end{quote}
In pictures the property of forward flatness says that
\begin{equation}\label{forwardflatnesspic}
\begin{Compose}{0}{0} \setsecondfont{\mathtt}
\crectangle{M}{3}{2}{0,0} \csymbol{M}
\crectangle{u}{0.7}{0.9}{-6,0} \csymbol{R}
\thispoint{v}{6,0} \csymbolalt[20,0]{v}
\joinrlnoarrowthick[above]{u}{0}{M}{0}\csymbolalt{u}
\joinrlnoarrowthick[above]{M}{0}{v}{0}
\end{Compose}
~~~~~\equiv ~~~~~
\begin{Compose}{0}{0}\setsecondfont{\mathtt}
\Crectangle{R}{0.7}{0.9}{0,0} \thispoint{p}{4,0} \csymbolalt[20,0]{v} \joinrlnoarrowthick{R}{0}{p}{0}
\end{Compose}
\end{equation}
If we have a flat distribution on the incomes $p(u)=\frac{1}{N_\mathsf{u}}$ then according to forward flatness
\begin{equation}
p(v) = \sum_u \frac{1}{N_\mathsf{u}} p(v|u) = \frac{1}{N_\mathsf{v}}
\end{equation}
and hence
\begin{equation}
\sum_u p(v|u) = \frac{N_\mathsf{u}}{N_\mathsf{v}}
\end{equation}
Similarly, the property of backward flatness says
\begin{equation}\label{backwardflatnesspic}
\begin{Compose}{0}{0}\setsecondfont{\mathtt}
\crectangle{M}{3}{2}{0,0} \csymbol{M}
\thispoint{u}{-6,0} \csymbolalt[-20,0]{u}
\crectangle{v}{0.7}{0.9}{6,0} \csymbol{R}
\joinrlnoarrowthick[above]{u}{0}{M}{0}
\joinrlnoarrowthick[above]{M}{0}{v}{0}\csymbolalt{v}
\end{Compose}
~~~~~ \equiv ~~~~~
\begin{Compose}{0}{0} \setsecondfont{\mathtt}
\Crectangle{R}{0.7}{0.9}{0,0} \thispoint{p}{-4,0} \csymbolalt[-20,0]{u} \joinlrnoarrowthick{R}{0}{p}{0}
\end{Compose}
\end{equation}
If we have $p(v)=\frac{1}{N_\mathsf{v}}$ then according to the principle
\begin{equation}
p(u) = \sum_v \frac{1}{N_\mathsf{v}} p(u|v) = \frac{1}{N_\mathsf{u}}
\end{equation}
and hence
\begin{equation}
\sum_v p(u|v) = \frac{N_\mathsf{v}}{N_\mathsf{u}}
\end{equation}

\subsection{Double pointer causality and double flatness}

It is clear that
\begin{equation}\text{double flatness} \Rightarrow \text{double pointer causality} \end{equation}
We can also argue for the converse
\begin{equation}\label{causalityimpliesflattness} \text{double pointer causality} \Rightarrow \text{double flatness} \end{equation}
under some reasonableness assumptions.  By causality we know that the flat distribution on the input, $p_\mathsf{E}(u)=\frac{1}{N_\mathtt{u}}$, evolves to some distribution, $p_\mathsf{F}(v)$, and that this is independent of $\mathsf M$.  Thus, it is reasonable to assume $p_\mathsf{E}(u)=\frac{1}{N_\mathtt{u}}$ and $p_\mathsf{F}(v)$ are a matching pair.  Second, we assume that there exists some $\mathsf{M}$ such that both $p_\mathsf{M}(u|v)$ and $p_\mathsf{M}(v|u)$ are nonzero for all $u$ and $v$ (in the simple example discussed in Sec.\ \ref{sec:disagreementsandsecrets} we saw that this was the case).  Hence, this matching pair is unique.  We can also think backwards in time.  Thus, we similarly argue that some $p_\mathsf{E}(u)$ and $p_\mathsf{F}(v)=\frac{1}{N_\mathtt{v}}$ are the unique matching pair.  It follows from uniqueness that the matching pair is $p_\mathsf{E}(u)=\frac{1}{N_\mathtt{u}}$ and $p_\mathsf{F}(v)=\frac{1}{N_\mathtt{v}}$.  In other words, we have double flatness.

The reason we have been able to argue for  \eqref{causalityimpliesflattness} in this way is because, when we defined double pointer causality, we defined the zero knowledge situation as corresponding to the flat distribution.  We could imagine changing this so that the zero knowledge distribution corresponds to some other distribution.

\subsection{Example of double pointer causality/double flatness}

It is illustrative to consider a simple example of double pointer causality/double flatness.

Backwards flatness implies
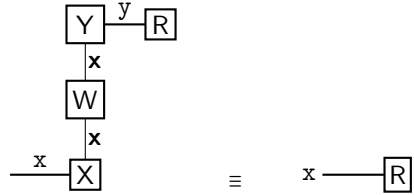
\[
\begin{Compose}{0}{0} \setsecondfont{\mathtt} \setthirdfont{\mathsfb}
\Crectangle{X}{0.8}{0.8}{0,0} \Crectangle{W}{1}{1}{0,4} \Crectangle{Y}{1}{1}{0,8}
\jointbnoarrow[right]{X}{0}{W}{0} \csymbolthird{x}
\jointbnoarrow[right]{W}{0}{Y}{0} \csymbolthird{x}
\thispoint{XL}{-4,0}
\crectangle{YR}{0.8}{0.8}{4,8} \csymbol{R}
\joinrlnoarrowthick{XL}{0}{X}{0} \csymbolalt[0,5]{x}
\joinrlnoarrowthick{Y}{0}{YR}{0} \csymbolalt[0,5]{y}
\end{Compose}
~~~~~\equiv~~~~~
\begin{Compose}{0}{0} \setsecondfont{\mathtt} \setthirdfont{\mathsfb}
\Crectangle{R}{0.7}{0.9}{0,0} \thispoint{p}{-4,0} \csymbolalt[-20,0]{x} \joinlrnoarrowthick{R}{0}{p}{0}
\end{Compose}
\]
We have no information about the choice of $\mathsf W$ as demanded by backward pointer causality.

Forwards flatness implies
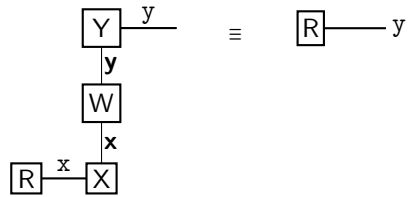
\[
\begin{Compose}{0}{-2} \setsecondfont{\mathtt} \setthirdfont{\mathsfb}
\Crectangle{X}{0.8}{0.8}{0,0} \Crectangle{W}{1}{1}{0,4} \Crectangle{Y}{1}{1}{0,8}
\jointbnoarrow[right]{X}{0}{W}{0} \csymbolthird{x}
\jointbnoarrow[right]{W}{0}{Y}{0} \csymbolthird{y}
\crectangle{XL}{0.8}{0.8}{-4,0}\csymbol{R}
\thispoint{YR}{4,8}
\joinrlnoarrowthick{XL}{0}{X}{0} \csymbolalt[0,5]{x}
\joinrlnoarrowthick{Y}{0}{YR}{0} \csymbolalt[0,5]{y}
\end{Compose}
~~~~~\equiv~~~~~
\begin{Compose}{0}{0} \setsecondfont{\mathtt} \setthirdfont{\mathsfb}
\Crectangle{R}{0.7}{0.9}{0,0} \thispoint{p}{4,0} \csymbolalt[20,0]{y} \joinrlnoarrowthick{R}{0}{p}{0}
\end{Compose}
\]
We have no information about $\mathsf W$ as demanded by forward pointer causality.

\subsection{Upshot - only flat distributions}

Given these assumptions, the upshot of all this is that we can only consider the flat distribution over incomes and outcomes (otherwise we get different answers for the probability of a circuit depending on whether we do the calculation forwards or backwards in time).

That is, we are interested in circuits that reduce to
\[
\begin{Compose}{0}{0} \setsecondfont{\mathtt}
\crectangle{T}{0.7}{1}{-7,0} \csymbol{R}
\crectangle{x}{0.7}{0.7}{-4,0} \csymbolthird{u}
\Crectangle{M}{1}{1}{0,0}
\joinrlnoarrowthick[above]{T}{0}{x}{0} \csymbolalt[0,5]{u}
\joinrlnoarrowthick[above]{x}{0}{M}{0} \csymbolalt[0,5]{u}
\crectangle{Tr}{0.7}{1}{7,0} \csymbol{R}
\crectangle{xr}{0.7}{0.7}{4,0} \csymbolthird{v}
\joinlrnoarrowthick[above]{Tr}{0}{xr}{0} \csymbolalt[0,5]{v}
\joinlrnoarrowthick[above]{xr}{0}{M}{0} \csymbolalt[0,5]{v}
\end{Compose}
\]
We can omit either or both of the $u$ box and the $v$ box.  We can omit some of the readout boxes. To elaborate this point further,  recall that $\mathtt u$ and $\mathtt v$ are composites as explained in Sec.\ \ref{sec:asimpleclassicalsituation} and so we can omit the readout boxes on some of the components on either side of $\mathsf M$.

\subsection{Double summation property}\label{sec:doublesummationproperty}

An alternative way of stating the double flatness property is what we will call the double summation property for certain matrices.
By forward flatness \eqref{forwardflatnesspic} and the defining property in \eqref{Rflatcond} for flat pointer operations (also applying \eqref{sumoverx}), we obtain the marginal probability
\begin{equation}\label{RMvR}
\sum_u \text{prob}\left(
\begin{Compose}{0}{0} \setsecondfont{\mathtt}
\crectangle{T}{0.7}{1}{-7,0} \csymbol{R}
\crectangle{x}{0.7}{0.7}{-4,0} \csymbolthird{u}
\Crectangle{M}{1}{1}{0,0}
\joinrlnoarrowthick[above]{T}{0}{x}{0} \csymbolalt[0,5]{u}
\joinrlnoarrowthick[above]{x}{0}{M}{0} \csymbolalt[0,5]{u}
\crectangle{Tr}{0.7}{1}{7,0} \csymbol{R}
\crectangle{xr}{0.7}{0.7}{4,0} \csymbolthird{v}
\joinlrnoarrowthick[above]{Tr}{0}{xr}{0} \csymbolalt[0,5]{v}
\joinlrnoarrowthick[above]{xr}{0}{M}{0} \csymbolalt[0,5]{v}
\end{Compose} \right)
=
\text{prob} \left(
\begin{Compose}{0}{0} \setsecondfont{\mathtt}
\crectangle{T}{0.7}{1}{-4,0} \csymbol{R}
\Crectangle{M}{1}{1}{0,0}
\joinrlnoarrowthick[above]{T}{0}{M}{0} \csymbolalt[0,5]{u}
\crectangle{Tr}{0.7}{1}{7,0} \csymbol{R}
\crectangle{xr}{0.7}{0.7}{4,0} \csymbolthird{v}
\joinlrnoarrowthick[above]{Tr}{0}{xr}{0} \csymbolalt[0,5]{v}
\joinlrnoarrowthick[above]{xr}{0}{M}{0} \csymbolalt[0,5]{v}
\end{Compose}  \right)  = \frac{1}{N_\mathsf{v}}
\end{equation}
We call this property \emph{forward summation}.

Similarly, by backwards flatness (using \eqref{backwardflatnesspic} and \eqref{Rflatcond}) we can obtain the marginal probability
\begin{equation}\label{RuMR}
\sum_v \text{prob}\left(
\begin{Compose}{0}{0} \setsecondfont{\mathtt}
\crectangle{T}{0.7}{1}{-7,0} \csymbol{R}
\crectangle{x}{0.7}{0.7}{-4,0} \csymbolthird{u}
\Crectangle{M}{1}{1}{0,0}
\joinrlnoarrowthick[above]{T}{0}{x}{0} \csymbolalt[0,5]{u}
\joinrlnoarrowthick[above]{x}{0}{M}{0} \csymbolalt[0,5]{u}
\crectangle{Tr}{0.7}{1}{7,0} \csymbol{R}
\crectangle{xr}{0.7}{0.7}{4,0} \csymbolthird{v}
\joinlrnoarrowthick[above]{Tr}{0}{xr}{0} \csymbolalt[0,5]{v}
\joinlrnoarrowthick[above]{xr}{0}{M}{0} \csymbolalt[0,5]{v}
\end{Compose} \right)
=
\text{prob} \left(
\begin{Compose}{0}{0} \setsecondfont{\mathtt}
\crectangle{T}{0.7}{1}{-7,0} \csymbol{R}
\crectangle{x}{0.7}{0.7}{-4,0} \csymbolthird{u}
\Crectangle{M}{1}{1}{0,0}
\joinrlnoarrowthick[above]{T}{0}{x}{0} \csymbolalt[0,5]{u}
\joinrlnoarrowthick[above]{x}{0}{M}{0} \csymbolalt[0,5]{u}
\crectangle{Tr}{0.7}{1}{4,0} \csymbol{R}
\joinlrnoarrowthick[above]{Tr}{0}{M}{0} \csymbolalt[0,5]{v}
\end{Compose} \right)  = \frac{1}{N_\mathsf{u}}
\end{equation}
We call this property \emph{backward summation}.

Double summation is when we have both forward summation and backward summation.  If we form the matrix with entries equal to
\begin{equation}
p(u,v)=
\text{prob}\left(
\begin{Compose}{0}{0} \setsecondfont{\mathtt}
\crectangle{T}{0.7}{1}{-7,0} \csymbol{R}
\crectangle{x}{0.7}{0.7}{-4,0} \csymbolthird{u}
\Crectangle{M}{1}{1}{0,0}
\joinrlnoarrowthick[above]{T}{0}{x}{0} \csymbolalt[0,5]{u}
\joinrlnoarrowthick[above]{x}{0}{M}{0} \csymbolalt[0,5]{u}
\crectangle{Tr}{0.7}{1}{7,0} \csymbol{R}
\crectangle{xr}{0.7}{0.7}{4,0} \csymbolthird{v}
\joinlrnoarrowthick[above]{Tr}{0}{xr}{0} \csymbolalt[0,5]{v}
\joinlrnoarrowthick[above]{xr}{0}{M}{0} \csymbolalt[0,5]{v}
\end{Compose}\right)
\end{equation}
then this matrix is an example what we will call a doubly summing matrix.  They have been studied a little in the literature as a generalisation of doubly stochastic matrices \cite{caron1996nonsquare}.  Doubly stochastic matrices (sometimes called bistochastic matrices) are matrices whose entries are nonnegative and whose columns add to 1 and whose rows add to 1.  They are necessarily square.  We define \emph{doubly summing matrices} to be matrices having the property that the sum, $s_\text{row}$,  of the entries in a row is the same for every row, and the sum, $s_\text{column}$, of the entries in a column is the same for every column (we omit the nonnegativity constraint here choosing, instead, to regard that separately - see Sec.\ \ref{sec:physicalityconstraints}).  Furthermore, we impose the normalisation condition that the columns sum to 1 (so $s_\text{column}=1$).  We note the following
\begin{enumerate}
\item For doubly summing matrices we have
\begin{equation}
\frac{s_\text{row}}{s_\text{column}} = \frac{N_\text{row}}{N_\text{column}}
\end{equation}
where $N_\text{row}$ is the number of entries in each row (i.e.\ the number of columns) and $N_\text{column}$ is the number of entries in each column (i.e.\ the number of rows). This equation follows from the fact that the sum of all the entries in the matrix can be written as $s_\text{column}N_\text{row}$ or as $s_\text{row}N_\text{column}$.
\item The case where $N_\text{row}=N_\text{column}$ is a square matrix.  Imposing  normalisation (that $s_\text{column}=1$ gives us a bistochastic matrix in the usual sense if we also impose that the entries are non-negative.
\item It can be proven that doubly summing matrices compose (in parallel and in sequence) to give doubly summing matrices.
\end{enumerate}
It is convenient, pedagogically, for us to impose non-negativity as a separate requirement to the doubly summing property since, for our application, they correspond to different constraints on operations (see Sec.\ \ref{sec:physicalityconstraints}).

We prove in Sec.\ \ref{sec:proofthatquantumtheoryisdoublysumming}that quantum theory is doubly summing.  This proof makes use of the extension theorem whereby the operator tensors associated with quantum operations can be written in terms of unitaries.

\subsection{Midcomes} \label{sec:midcomes}

Till now we have been considering circuits of the form
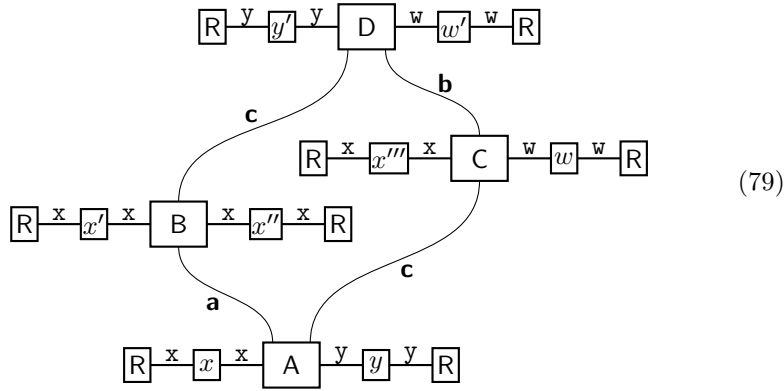
\begin{equation}
\begin{Compose}{0}{-2.5}\setdefaultfont{\mathsf}\setsecondfont{\mathtt} \setthirdfont{\mathsfb}
\Crectangle{A}{1.5}{1.2}{0,0}
\GenxdifBoxincome{A}{1.5}{0}{x}{x}{R}   
\GenxdifBoxoutcome{A}{1.5}{0}{y}{y}{R}  
\Crectangle{D}{1.5}{1.2}{4,18}
\GenxdifBoxincome{D}{1.5}{0}{y}{y'}{R}  
\GenxdifBoxoutcome[1.2]{D}{1.5}{0}{w}{w'}{R}   
\Crectangle{B}{1.5}{1.2}{-6, 7.5}
\GenxdifBoxincome{B}{1.5}{0}{x}{x'}{R}  
\GenxdifBoxoutcome[1.2]{B}{1.5}{0}{x}{x''}{R}
\Crectangle{C}{1.5}{1.2}{10,11}
\GenxdifBoxincome[1.45]{C}{1.5}{0}{x}{x'''}{R}
\GenxdifBoxoutcome{C}{1.5}{0}{w}{w}{R}
%
%
\jointbnoarrow[above left]{B}{0}{D}{-1} \csymbolthird[-4,0]{c} \jointbnoarrow[above right]{C}{0}{D}{1} \csymbolthird[5,0]{b}
\jointbnoarrow[below left]{A}{-1}{B}{0} \csymbolthird[-5,0]{a}
\jointbnoarrow[below right]{A}{1}{C}{0} \csymbolthird[4,0]{c}
\end{Compose}
\end{equation}
Though this form is quite general, it misses out an important possibility.  We can also feed outcomes into incomes as in the following example
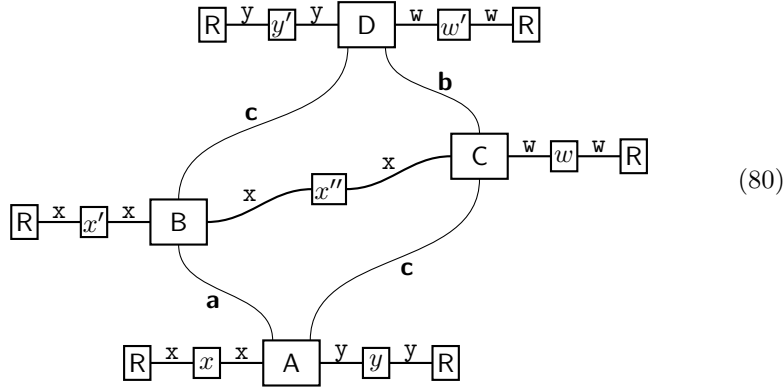
\begin{equation}\label{circuitexamplemidcomes}
\begin{Compose}{0}{-2.5}\setdefaultfont{\mathsf}\setsecondfont{\mathtt} \setthirdfont{\mathsfb}
\Crectangle{A}{1.5}{1.2}{0,0}
\GenxdifBoxincome{A}{1.5}{0}{x}{x}{R}   
\GenxdifBoxoutcome{A}{1.5}{0}{y}{y}{R}  
\Crectangle{D}{1.5}{1.2}{4,18}
\GenxdifBoxincome{D}{1.5}{0}{y}{y'}{R}  
\GenxdifBoxoutcome[1.2]{D}{1.5}{0}{w}{w'}{R}   
\Crectangle{B}{1.5}{1.2}{-6, 7.5}
\GenxdifBoxincome{B}{1.5}{0}{x}{x'}{R}  
\Crectangle{C}{1.5}{1.2}{10,11}
\GenxdifBoxoutcome{C}{1.5}{0}{w}{w}{R}
\crectangle{xm}{0.9}{0.8}{2, 9.25} \csymbolfourth{x''}
\joinrlnoarrowthick[above left]{B}{0}{xm}{0} \csymbolalt{x} \joinrlnoarrowthick[above left]{xm}{0}{C}{0} \csymbolalt{x}
%
%
%
\jointbnoarrow[above left]{B}{0}{D}{-1} \csymbolthird[-4,0]{c} \jointbnoarrow[above right]{C}{0}{D}{1} \csymbolthird[5,0]{b}
\jointbnoarrow[below left]{A}{-1}{B}{0} \csymbolthird[-5,0]{a}
\jointbnoarrow[below right]{A}{1}{C}{0} \csymbolthird[4,0]{c}
\end{Compose}
\end{equation}
We will call the $x''$ readout a \emph{midcome}.   Circuits with midcomes can be reduced to circuits without midcomes employing the following equivalence
\begin{equation}\label{midcomeequivalence}
\begin{Compose}{0}{-0.1} \setdefaultfont{} \setfourthfont{\mathtt}
\thispoint{l}{-4,0} \thispoint{r}{4,0}
\Crectangle{x}{0.7}{0.8}{0,0}
\joinrlnoarrowthick[above]{l}{0}{x}{0} \csymbolfourth{x}
\joinrlnoarrowthick[above]{x}{0}{r}{0} \csymbolfourth{x}
\end{Compose}
~~ \equiv ~~
\begin{Compose}{0}{-0.1}
\thispoint{l}{-7.6,0} \thispoint{r}{7.6,0}
\RxBoxoutcome{l}{0}{0}{x}\RxBoxincome{r}{0}{0}{x}
\crectangle[thin]{number}{1}{1}{0,2.3} \csymbolthird{N_\mathtt{x}}
\end{Compose}
\end{equation}
The $N_\mathtt{x}$ inside the rectangular box is an over all factor.  Including such factors goes a little beyond the definition of equivalence introduced in Sec.\ \ref{sec:equivalence}.  It is, however, a simple application of the notion of the linearly extended notion of equivalence introduced in Sec.\ \ref{sec:extendedequivalence}.    It is clear \eqref{midcomeequivalence} is true.  If we think forwards in time then there is a $\frac{1}{N_\mathtt{x}}$ probability that the $x$ in the second readout box will equal the $x$ in the first so this accounts for the $N_\mathtt{x}$ factor (similar remarks apply if we think in the backwards time direction).   This equivalence means that we can always reduce circuits with midcomes to ones without midcomes by including the appropriate $N_\mathtt{x}$ factors.

\subsection{Calculating some probabilities}

We can use these result to calculate conditional probabilites.  Here is a simple example:
\begin{equation}
p_\mathtt{M}(u|v) ~~=~~
\frac{\text{prob} \left(
\begin{Compose}{0}{0} \setsecondfont{\mathtt}
\crectangle{T}{0.7}{1}{-7,0} \csymbol{R}
\crectangle{x}{0.7}{0.7}{-4,0} \csymbolthird{u}
\Crectangle{M}{1}{1}{0,0}
\joinrlnoarrowthick[above]{T}{0}{x}{0} \csymbolalt[0,5]{u}
\joinrlnoarrowthick[above]{x}{0}{M}{0} \csymbolalt[0,5]{u}
\crectangle{Tr}{0.7}{1}{7,0} \csymbol{R}
\crectangle{xr}{0.7}{0.7}{4,0} \csymbolthird{v}
\joinlrnoarrowthick[above]{Tr}{0}{xr}{0} \csymbolalt[0,5]{v}
\joinlrnoarrowthick[above]{xr}{0}{M}{0} \csymbolalt[0,5]{v}
\end{Compose} \right)}
{ \text{prob} \left(
\begin{Compose}{0}{0} \setsecondfont{\mathtt}
\crectangle{T}{0.7}{1}{-7,0} \csymbol{R}
\crectangle{x}{0.7}{0.7}{-4,0} \csymbolthird{u}
\Crectangle{M}{1}{1}{0,0}
\joinrlnoarrowthick[above]{T}{0}{x}{0} \csymbolalt[0,5]{u}
\joinrlnoarrowthick[above]{x}{0}{M}{0} \csymbolalt[0,5]{u}
\crectangle{Tr}{0.7}{1}{7,0} \csymbol{R}
\joinlrnoarrowthick[above]{Tr}{0}{M}{0} \csymbolalt[0,5]{v}
\end{Compose} \right) }
\end{equation}
Using \eqref{RuMR} we obtain
\begin{equation}\label{pugvpic}
p_\mathtt{M}(u|v) ~~=~~
N_\mathtt{u} ~
\text{prob} \left(
\begin{Compose}{0}{0} \setsecondfont{\mathtt}
\crectangle{T}{0.7}{1}{-7,0} \csymbol{R}
\crectangle{x}{0.7}{0.7}{-4,0} \csymbolthird{u}
\Crectangle{M}{1}{1}{0,0}
\joinrlnoarrowthick[above]{T}{0}{x}{0} \csymbolalt[0,5]{u}
\joinrlnoarrowthick[above]{x}{0}{M}{0} \csymbolalt[0,5]{u}
\crectangle{Tr}{0.7}{1}{7,0} \csymbol{R}
\crectangle{xr}{0.7}{0.7}{4,0} \csymbolthird{v}
\joinlrnoarrowthick[above]{Tr}{0}{xr}{0} \csymbolalt[0,5]{v}
\joinlrnoarrowthick[above]{xr}{0}{M}{0} \csymbolalt[0,5]{v}
\end{Compose} \right)
\end{equation}
Similarly,
\begin{equation}\label{pvgupic}
p_\mathtt{M}(v|u) ~~=~~
N_\mathtt{v} ~
\text{prob} \left(
\begin{Compose}{0}{0} \setsecondfont{\mathtt}
\crectangle{T}{0.7}{1}{-7,0} \csymbol{R}
\crectangle{x}{0.7}{0.7}{-4,0} \csymbolthird{u}
\Crectangle{M}{1}{1}{0,0}
\joinrlnoarrowthick[above]{T}{0}{x}{0} \csymbolalt[0,5]{u}
\joinrlnoarrowthick[above]{x}{0}{M}{0} \csymbolalt[0,5]{u}
\crectangle{Tr}{0.7}{1}{7,0} \csymbol{R}
\crectangle{xr}{0.7}{0.7}{4,0} \csymbolthird{v}
\joinlrnoarrowthick[above]{Tr}{0}{xr}{0} \csymbolalt[0,5]{v}
\joinlrnoarrowthick[above]{xr}{0}{M}{0} \csymbolalt[0,5]{v}
\end{Compose} \right)
\end{equation}
We see here that
\begin{equation}\label{bayesinversion}
\frac{p_\mathtt{M}(u|v) }{N_\mathtt{u}} =  \frac{p_\mathtt{M}(v|u) }{N_\mathtt{v}}
\end{equation}
This is an example of Bayesian inversion.

\part{Equivalence, Linearity, and Physicality}\label{part:equivalencelinearityandphysicality}

\section{Extended equivalence}\label{sec:extendedequivalence}

\subsection{The $p(\cdot)$ function}

We wish to set up a notion of equivalence between different fragments of a circuit which linearly extends the simple notion introduced in Sec.\ \ref{sec:equivalence}.  To do this we define the $p(\cdot)$ function.  Consider various circuits, $\mathsf A$, $\mathsf B$, \dots. We know, by the circuit probability assumption (see Sec.\ \ref{sec:circuitprobabilityassumption}) that the probabilities $\text{prob}(\mathsf{A})$, $\text{prob}(\mathsf{B})$, \dots are defined.  Hence we can define
\begin{equation}
p(\alpha \mathsf{A} + \beta\mathsf{B} + \dots ) = \alpha \text{prob}(\mathsf{A}) + \beta \text{prob}(\mathsf{B}) + \dots
\end{equation}
where $\alpha$, $\beta$, \dots are real numbers (they can be positive, zero, or negative).
Here we have done something a little odd - we have formed a weighted linear sum of circuits.  The weights are mathematical while the circuits live in the physical world.

\subsection{Definition of equivalence}

A fragment of a circuit (which we will simply called a \emph{fragment} - see Sec.\ \ref{sec:fragmentsandtheircausalstructure}) is a collection of operations wired together that could be completed into a circuit by other operations.  Consider \emph{expressions} given by
\begin{equation}
\text{expression}=\alpha + \beta \mathsf{C} + \gamma \mathsf{D}+ \dots
\end{equation}
where $\mathsf{C}$, $\mathsf{D}$, \dots are fragments of circuits (we have suppressed the subscripts and superscripts which would normally appear for the sake of generality).   We will say that
\begin{equation}
\text{expression}_1 \equiv \text{expression}_2
\end{equation}
if all the terms in both expressions can be completed into a circuit by a fragment and, furthermore,
\begin{equation}
p(\text{expression}_1 \mathsf{E}) = p(\text{expression}_2 \mathsf{E})
\end{equation}
for all such fragments, $\mathsf{E}$, which have the property that they convert the expressions into a linear sum of circuits.  In order that fragments $\mathsf E$ exist, it is necessary that all the terms in both $\text{expression}_1$ and $\text{expression}_2$ have compatible causal structure (see Sec.\ \ref{sec:fragmentsandtheircausalstructure}).

\subsection{Examples of equivalence}

Here are two examples of equivalence.  First, we have
\begin{equation}\label{equivexample1}
\alpha\mathsf{A}_{\mathtt{x}_1}^{\mathsfb{x}} + \beta\mathsf{B}_{\mathtt{x}_1}^{\mathsfb{x}} \equiv
\gamma \mathsf{C}_{\mathtt{x}_1}^{\mathsfb{x}} + \delta \mathsf{D}_{\mathtt{x}_1}^{\mathsfb{x}}
\end{equation}
if
\begin{equation}
p(\alpha\mathsf{A}_{\mathtt{x}_1}^{\mathsfb{x}_2} \mathsf{E}^{\mathtt{x}_1}_{\mathsfb{x}_2} + \beta\mathsf{B}_{\mathtt{x}_1}^{\mathsfb{x}_2}\mathsf{E}^{\mathtt{x}_1}_{\mathsfb{x}_2} )
\equiv
p(\gamma \mathsf{C}_{\mathtt{x}_1}^{\mathsfb{x}_2}\mathsf{E}^{\mathtt{x}_1}_{\mathsfb{x}_2} +
\delta\mathsf{D}_{\mathtt{x}_1}^{\mathsfb{x}_2}\mathsf{E}^{\mathtt{x}_1}_{\mathsfb{x}_2})  ~~~~~ \forall ~~ \mathsf{E}^{\mathtt{x}_1}_{\mathsfb{x}_2}
\end{equation}
Second, for a circuit, $\mathsf F$
\begin{equation}\label{circuitequivprob}
\mathsf{F} \equiv \text{prob}(\mathsf{F})
\end{equation}
That is, a circuit is equivalent to its own probability.  The proof of this is simple.  The only fragments that can complete $\mathsf{F}$  into a circuit is another circuit.  We have
\begin{equation}
p(\mathsf{F}\mathsf{E})= p(\mathsf{F})p(\mathsf{E}) = \text{prob}(\mathsf{F})p(\mathsf{E})  ~~~\forall ~\mathsf{E}
\end{equation}
where we use the fact that probabilities factorise for circuits (see Sec.\ \ref{sec:circuitprobsfactorize}) and the linearity of the $p(\cdot)$ function on the right.  This proves \eqref{circuitequivprob}.  This second example illustrates that equivalence is a different notion from equality.  Clearly a circuit (which lives in the physical world) is not equal to a number (which is a mathematical object) but the two are equivalent under the definition used here.

\subsection{Spanning sets and equivalence}\label{sec:spanningsetsandequivalence}

Consider a set, $S$, of fragments. A subset, $S_\text{spanning}\subseteq S$, is a \emph{spanning set} for $S$ if every fragment in $S$ can be written as being equivalent to a weighted linear sum of the fragments in $S_\text{spanning}$.

Here we will provide a useful test for whether two fragments $\mathsf A$ and $\mathsf B$ are equivalent (here we omit the type indices for the sake of providing a general discussion).  First, recall from Sec.\ \ref{sec:equivalence} that a necessary condition for two fragments to be equivalent is that they have compatible causal structures (as defined in Sec.\ \ref{sec:fragmentsandtheircausalstructure}). This means  $S_\mathsf{A}\cap S_\mathsf{B}\not= \emptyset$ where $S_\mathsf{A}$ is the set of fragments that can complete $\mathsf A$ into a circuit (and similarly for $S_\mathsf{B}$).

From the definition of a spanning set above, it is clear that, if $S_\mathsf{A}\cap S_\mathsf{B}\not= \emptyset$, then
\begin{equation}
\mathsf{A}\equiv \mathsf{B} ~~\text{iff}~~ \text{prob}(\mathsf{AE})= \text{prob}(\mathsf{BE})~~ \text{for all}~~ \mathsf{E}\in [S_\mathsf{A}\cap S_\mathsf{B}]_\text{spanning}
\end{equation}
where $\mathsf{AE}$ and $\mathsf{BE}$ are circuits (the wiring is not shown explicitly for generality) and $[S_\mathsf{A}\cap S_\mathsf{B}]_\text{spanning}$ is any spanning set for $S_\mathsf{A}\cap S_\mathsf{B}$

\section{Spanning properties and identity operations}

\subsection{System preparations and results}

A general system preparation is a fragment having only a output left (or, possibly, multiple outputs) open.  If there are no midcomes (see Sec.\ \ref{sec:midcomes}) then a general system preparation can always be put in the form
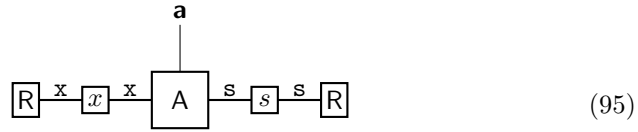
\begin{equation}\label{generalpreparation}
\begin{Compose}{0}{0}\setsecondfont{\mathsfb}
\Crectangle{A}{1.5}{1.5}{0,0}
\RxBoxincome{A}{1.5}{0}{x} \RxBoxoutcome{A}{1.5}{0}{s}
\thispoint{p}{0,4} \csymbolalt[0,20]{a} \jointbnoarrow{A}{0}{p}{0}
\end{Compose}
\end{equation}
If there are midcomes we can still put the fragment in this form though we accumulate appropriate $N_\mathtt{x}$ factors through use of the equivalence in \eqref{midcomeequivalence} as described in Sec.\ \ref{sec:midcomes}.
We can also consider \emph{income only system preparations}. These are preparations of the form
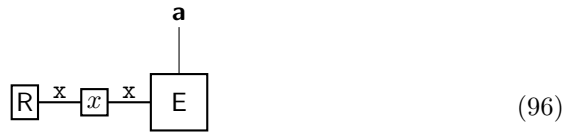
\begin{equation}\label{incomeonlyprep}
\begin{Compose}{0}{0}\setsecondfont{\mathsfb}
\crectangle{A}{1.5}{1.5}{0,0} \csymbol{E}
\RxBoxincome{A}{1.5}{0}{x} 
\thispoint{p}{0,4} \csymbolalt[0,20]{a} \jointbnoarrow{A}{0}{p}{0}
\end{Compose}
\end{equation}
where $\mathsf E$ is doubly deterministic.  This is more restricted than the preparation shown in \eqref{generalpreparation}.

Similarly, a \emph{general system result} is a fragment having only an input (or inputs) left open.  A general system result can be put in the form \begin{equation}\label{generalresult}
\begin{Compose}{0}{0}\setsecondfont{\mathsfb}
\crectangle{A}{1.5}{1.5}{0,0} \csymbol{B}
\RxBoxincome{A}{1.5}{0}{w} \RxBoxoutcome{A}{1.5}{0}{y}
\thispoint{p}{0,-4} \csymbolalt[0,-20]{a} \joinbtnoarrow{A}{0}{p}{0}
\end{Compose}
\end{equation}
(where there maybe additional $N_\mathtt{x}$ factors if there are midcomes in the fragment).  An \emph{outcome only results} is of the form
\begin{equation}\label{outcomeonlyresult}
\begin{Compose}{0}{0}\setsecondfont{\mathsfb}
\crectangle{A}{1.5}{1.5}{0,0} \csymbol{F}
\RxBoxoutcome{A}{1.5}{0}{y}
\thispoint{p}{0,-4} \csymbolalt[0,-20]{a} \joinbtnoarrow{A}{0}{p}{0}
\end{Compose}
\end{equation}
where $\mathsf F$ is doubly deterministic.

\subsection{Pure and homogeneous system preparations and results}\label{sec:pureandhomogeneous}

\begin{figure}
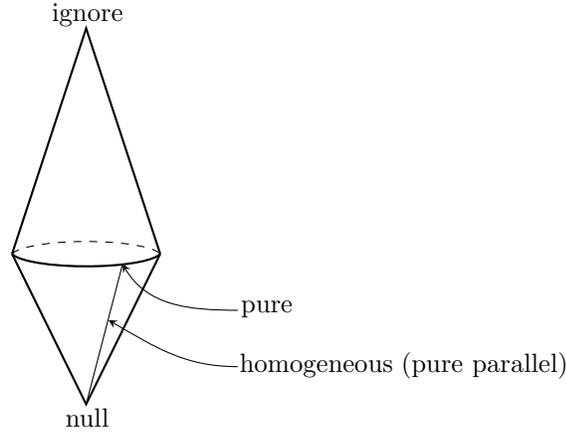

\begin{center}
\newcommand{\joinlccustom}[5][above]{\draw[->] ($(#2)-(#2 right)+(0,#3)$)to [out=180, in=#5] node [#1=0.07] (fontlocation){} ($(#4)+ cos(#5)*(#4 right) + sin(#5)*(#4 top)$);}
\begin{Compose}[2]{0}{0}
    \draw[dashed] (0,0) arc (170:10:2cm and 0.4cm)coordinate[pos=0] (a);
    \draw[thick] (0,0) arc (-170:-10:2cm and 0.4cm)coordinate (b);
    \draw[thick] (a) -- ([yshift=-4cm]$(a)!0.5!(b)$) -- (b);
    \draw[thick] (a) -- ([yshift=6cm]$(a)!0.5!(b)$) -- (b);
    \draw ([yshift=-4cm]$(a)!0.5!(b)$) -- ($(b)+(-1,-0.27)$);
    \thispoint{p}{$(b)+(-1,-0.27)$} 
    \thispoint{h}{$0.6*(b)+0.6*(-1,-0.27)+0.4*(2,-4)$} 
    \thispoint{n}{2,-4} \csymbol[0,-10]{\text{null}}
    \thispoint{I}{2,6} \csymbol[0,10]{\text{ignore}}
    \thispoint{pp}{6,-1.5} \csymbol[22,0.5]{\text{pure}} \joinlccustom{pp}{0}{p}{-60}
    \thispoint{hh}{6,-3} \csymbol[125,1]{\text{homogeneous (pure parallel)}}  \joinlccustom{hh}{0}{h}{-30}
    \thispoint{l}{-4,6}
\end{Compose}
\end{center}
\caption{Different types of preparation (or result).  The notions of pure and pure parallel (or homogeneous) are illustrated.}\label{fig:convexspace}
\end{figure}

It is useful to introduce some distinctions concerning types of system preparations and results.  First, we will say two system preparations (or results) are parallel if one is equivalent to the other times a positive number.  Now we define the following cases:
\begin{description}
  \item[Mixed.] A system preparation (or result) is mixed if it can be written as being equivalent to a convex sum of distinct preparations (or results).
  \item[Extremal.] A system preparation (or result) is extremal if it is not mixed.
  \item[Heterogeneous.] A system preparation (or result) is heterogeneous if it can be written as being equivalent to a sum, with non-negative weights, of non-parallel preparations (or results).
  \item[Homogeneous (pure parallel).] A system preparation (or result) is homogeneous if it is not heterogeneous.  This means that if it is written as being equivalent to a sum of other preparations (or results), all these preparations(or results) must be parallel.
  \item[Pure.]  A system preparation (or result) is pure if it is homogeneous and extremal.
\end{description}
The pure and homogenous cases are illustrated by the example shown in Fig.\ \ref{fig:convexspace}.   We will sometimes call homogeneous preparations or results \emph{pure parallel} since the concept of purity is more familiar. In this example we have a unique deterministic preparation (result) which we label \lq\lq ignore".  In this example, the ignore preparation (result) is extremal. Thus, the ignore preparation (or result) is not mixed - however it is heterogeneous (as is clear from looking at the figure).   This example illustrates the distinction between the concepts of being mixed and being heterogenous.  The null preparation (result) is equivalent to any preparation (result) weighted by zero.  If it is included in a circuit, the circuit will have probability equal to zero.

The classification shown here is subtlety different from that standardly used.  In particular, the pure case does not cover all extremal non-null cases.  The figure is just an example (though illustrative of what happens in time-symmetric Quantum Theory).  We could have a case where the top is cut off so there is not a unique ignore preparation or result (though, in the approach taken in this paper, the ignore operation will be seen to be unique both for preparations and results - see Sec.\ \ref{sec:identityoperations}).

A little more nomenclature.  For linguistic brevity, we will say that operations of the following sort
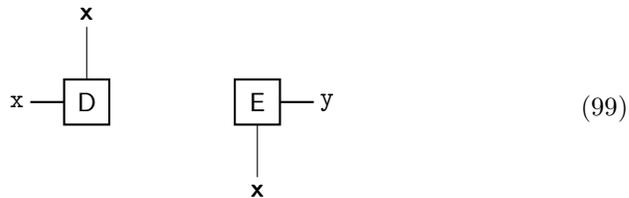
\begin{equation}\label{DEininoutout}
\begin{Compose}{0}{0}\setsecondfont{\mathsfb}
\crectangle{A}{1.2}{1.2}{0,0} \csymbol{D}
\incomewire{A}{0}{x} 
\thispoint{p}{0,4} \csymbolalt[0,20]{x} \jointbnoarrow{A}{0}{p}{0}
\end{Compose}
~~~~~~~~~~~~~~
\begin{Compose}{0}{0}\setsecondfont{\mathsfb}
\crectangle{A}{1.2}{1.2}{0,0} \csymbol{E}
\outcomewire{A}{0}{y}
\thispoint{p}{0,-4} \csymbolalt[0,-20]{x} \joinbtnoarrow{A}{0}{p}{0}
\end{Compose}
\end{equation}
are homogeneous (pure parallel) if the corresponding operations
\begin{equation}\label{DEinout}
\begin{Compose}{0}{0}\setsecondfont{\mathsfb}
\crectangle{A}{1.2}{1.2}{0,0} \csymbol{D}
\RxBoxincome{A}{1.2}{0}{x} 
\thispoint{p}{0,4} \csymbolalt[0,20]{x} \jointbnoarrow{A}{0}{p}{0}
\end{Compose}
~~~~~~~~~~~~~~
\begin{Compose}{0}{0}\setsecondfont{\mathsfb}
\crectangle{A}{1.2}{1.2}{0,0} \csymbol{E}
\RxBoxoutcome{A}{1.5}{0}{y}
\thispoint{p}{0,-4} \csymbolalt[0,-20]{x} \joinbtnoarrow{A}{0}{p}{0}
\end{Compose}
\end{equation}
are homogeneous (pure parallel) for all $x$, $y$.

\subsection{Double spanning property}

We define the following property
\begin{quote}
\emph{Forward spanning.}   A theory is \emph{forward spanning} if any general system preparation is equivalent to a weighted linear sum of income only preparations where the weights are real numbers (they can be positive, zero or negative).
\end{quote}
Once we go into the language of states (as we will in Part \ref{part:duotensorformulation}) then the forward spanning property means we can write the state associated with a general system preparation as weighted sum of states associated with income only system preparations.

The property of backwards spanning is defined as
\begin{quote}
\emph{Backwards spanning.}  A theory is \emph{backwards spanning} if any general system result is equivalent to a linear sum of outcome only results where the weights are real numbers (they can be positive, zero or negative).
\end{quote}
This is, of course, the time reverse of the forward spanning property.

A theory is said to have the \emph{double spanning} property if it is both forward spanning and backwards spanning.

\subsection{Double purity property}\label{sec:doublepurity}

We can also provide a more restrictive notion.   We define the following properties
\begin{quote}
\emph{Forward purity.}   A theory has the property of \emph{forward purity} if any general system preparation is equivalent to a weighted linear sum of income only preparations where the weights are non-negative (they can be positive or zero).
\end{quote}
and
\begin{quote}
\emph{Backwards purity.}  A theory has the property of \emph{backwards purity} if any general system result is equivalent to a linear sum of outcome only results where the weights are non-negative (they can be positive or zero).
\end{quote}
Naturally, we say a theory has the property of \emph{double purity} if it has both the property of forward purity and the property of backwards purity.

It is clear that
\[ \text{double purity property} \Longrightarrow \text{double spanning property}    \]
since, for double purity, we add the extra requirement that the weights are non-negative.

Given the definitions in Sec.\ \ref{sec:pureandhomogeneous}, it is clear that the property of forward purity is equivalent to the statement that there is a income only system preparation parallel to any given pure preparation.  Similarly, the property of backwards purity is equivalent to the statement that there is an outcome only system result parallel to any given pure system result.

\subsection{Double maximal purity}

We can impose an even stronger condition that is true in Quantum Theory (we include this for interest only since we not use this property in any subsequent proofs).  Recall maximal operations as discussed in Sec.\ \ref{sec:doublemaximality}. Adding an $\mathsf R$ operation and a readout box, we obtain
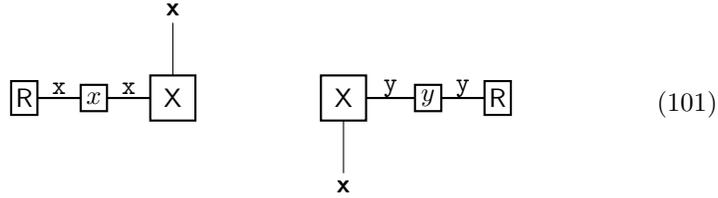
\begin{equation}
\begin{Compose}{0}{0}\setsecondfont{\mathsfb}
\crectangle{A}{1.2}{1.2}{0,0} \csymbol{X}
\RxBoxincome{A}{1.2}{0}{x} 
\thispoint{p}{0,4} \csymbolalt[0,20]{x} \jointbnoarrow{A}{0}{p}{0}
\end{Compose}
~~~~~~~~~~~~~~
\begin{Compose}{0}{0}\setsecondfont{\mathsfb}
\crectangle{A}{1.2}{1.2}{0,0} \csymbol{X}
\RxBoxoutcome{A}{1.5}{0}{y}
\thispoint{p}{0,-4} \csymbolalt[0,-20]{x} \joinbtnoarrow{A}{0}{p}{0}
\end{Compose}
\end{equation}
We call the operation on the left a \emph{maximal system preparation} and the object on the right a {maximal system result}.  They are special cases of the income only preparation shown in \eqref{incomeonlyprep}and the outcome only result illustrated in \eqref{outcomeonlyresult}.
We define
\begin{quote}
\emph{Forward maximal purity.}   A theory has the property of \emph{forward maximal purity} if any general system preparation is equivalent to a weighted linear sum of maximal preparations where the weights are non-negative (they can be positive or zero).
\end{quote}
and
\begin{quote}
\emph{Backwards maximal purity.}  A theory has the property of \emph{backwards maximal purity} if any general system result is equivalent to a linear sum of maximal results where the weights are non-negative (they can be positive or zero).
\end{quote}
Double maximal purity is when we have both properties.  This means that maximal system preparations and maximal system results are homogeneous (pure parallel).
Note
\[ \text{double maximal purity} \Longrightarrow \text{double purity} \Longrightarrow \text{double spanning}  \]
To prove the complete positivity condition below we need double purity. To prove the double causality condition below we need only the double spanning property.  We mention the more specific property of double maximal purity since it is a property that Quantum Theory has.

\subsection{Ignore operations}\label{sec:identityoperations}

Consider a doubly deterministic operation having only an input (no output, no incomes, no outcomes).
\begin{equation}
\begin{Compose}{0}{0}\setsecondfont{\mathsfb}
\Crectangle{I}{1}{1}{0,0} \thispoint{p}{0,-3} \csymbolalt[0,-20]{a} \joinbtnoarrow{I}{0}{p}{0}
\end{Compose}
\end{equation}
We call this an ignore result. Similarly, a doubly determistic operation having only an output
\begin{equation}
\begin{Compose}{0}{0}\setsecondfont{\mathsfb}
\Crectangle{I}{1}{1}{0,0} \thispoint{p}{0,3} \csymbolalt[0,20]{a} \jointbnoarrow{I}{0}{p}{0}
\end{Compose}
\end{equation}
is called an ignore preparation.

The most general circuit containing the ignore result is where we send in a general system preparation.  In the case that there are no midcomes any circuit containing the ignore result on $\mathsfb a$ can be put in the form
\begin{equation}
\begin{Compose}{0}{0}\setsecondfont{\mathsfb}
\Crectangle{A}{1.5}{1.5}{0,0}
\RxBoxincome{A}{1.5}{0}{x} \RxBoxoutcome{A}{1.5}{0}{s}
\Crectangle{I}{0.8}{1}{0,5} \jointbnoarrow[left]{A}{0}{I}{0}\csymbolalt{a}
\end{Compose}
\end{equation}
If there are midcomes we can still put it in this form where we accumulate appropriate $N_\mathtt{x}$ factors through use of the equivalence in \eqref{midcomeequivalence} as described in Sec.\ \ref{sec:midcomes}.

We can prove, using forward spanning, that all ignore results on a given system type are equivalent (we will refer to this as the uniqueness of the ignore preparation). To this end, consider two ignore results, $\mathsf{I}_{\mathsf{a}_1}$ and $\mathsf{I'}_{\mathsf{a}_1}$. By forward spanning we know that any general system preparation is spanned by income only preparations.   Hence, applying the ideas in Sec.\ \ref{sec:spanningsetsandequivalence}, we have equivalence iff
\begin{equation}\label{equivalenceIIprimeresults}
\text{prob}\left(
\begin{Compose}{0}{-0.65}\setsecondfont{\mathsfb}
\Crectangle{A}{1.5}{1.5}{0,0}
\RxBoxincome{A}{1.5}{0}{x} 
\Crectangle{I}{0.8}{1}{0,5} \jointbnoarrow[left]{A}{0}{I}{0}\csymbolalt{a}
\end{Compose}\right)
=
\text{prob}\left(
\begin{Compose}{0}{-0.65}\setsecondfont{\mathsfb}
\Crectangle{A}{1.5}{1.5}{0,0}
\RxBoxincome{A}{1.5}{0}{x} 
\crectangle{I}{0.8}{1}{0,5} \csymbol{I'}\jointbnoarrow[left]{A}{0}{I}{0}\csymbolalt{a}
\end{Compose}\right)
\end{equation}
for all $x$, $s$, and $\mathsf A$.  In fact, \eqref{equivalenceIIprimeresults} clearly holds since both sides are equal to $\frac{1}{N_\mathtt{x}}$ by backward flatness.   Hence, all ignore results for a given system type are equivalent.

Similarly, we can prove that all ignore preparations are equivalent employing backwards spanning and the fact that
\begin{equation}\label{equivalenceIIprimepreps}
\text{prob}\left(
\begin{Compose}{0}{0.65}\setsecondfont{\mathsfb}
\crectangle{A}{1.5}{1.5}{0,0} \csymbol{B}
\RxBoxoutcome{A}{1.5}{0}{y}
\Crectangle{I}{0.8}{1}{0,-5} \joinbtnoarrow[left]{A}{0}{I}{0}\csymbolalt{a}
\end{Compose}\right)
=
\text{prob}\left(
\begin{Compose}{0}{0.65}\setsecondfont{\mathsfb}
\crectangle{A}{1.5}{1.5}{0,0} \csymbol{B}
\RxBoxoutcome{A}{1.5}{0}{y}
\crectangle{I}{0.8}{1}{0,-5} \csymbol{I'}\joinbtnoarrow[left]{A}{0}{I}{0}\csymbolalt{a}
\end{Compose}\right)
=\frac{1}{N_\mathtt{y}}
\end{equation}
by forwards flatness.

The following results are easily proven by uniqueness.
\begin{equation}\label{IequalsII}
\begin{Compose}{0}{0}\setsecondfont{\mathtt}\setthirdfont{\mathsfb}
\crectangle{X}{0.8}{0.8}{0,2} \csymbol{I}
\thispoint{Y}{0,-1.5}\csymbolthird[0,-20]{ab} \joinbtnoarrow[right]{X}{0}{Y}{0}
\end{Compose}
~~\equiv~~
\begin{Compose}{0}{0}\setsecondfont{\mathtt}\setthirdfont{\mathsfb}
\crectangle{X}{0.8}{0.8}{0,2} \csymbol{I}
\thispoint{Y}{0,-1.5}\csymbolthird[0,-20]{a} \joinbtnoarrow[right]{X}{0}{Y}{0}
\end{Compose}
~~\begin{Compose}{0}{0}\setsecondfont{\mathtt}\setthirdfont{\mathsfb}
\crectangle{X}{0.8}{0.8}{0,2} \csymbol{I}
\thispoint{Y}{0,-1.5}\csymbolthird[0,-20]{b} \joinbtnoarrow[right]{X}{0}{Y}{0}
\end{Compose}
~~~~~~~~~~~~~~~~~~~~
\begin{Compose}{0}{0}\setsecondfont{\mathtt}\setthirdfont{\mathsfb}
\crectangle{X}{0.8}{0.8}{0,-2} \csymbol{I}
\thispoint{Y}{0,1.5}\csymbolthird[0,20]{ab} \jointbnoarrow[right]{X}{0}{Y}{0}
\end{Compose}
~~\equiv~~
\begin{Compose}{0}{0}\setsecondfont{\mathtt}\setthirdfont{\mathsfb}
\crectangle{X}{0.8}{0.8}{0,-2} \csymbol{I}
\thispoint{Y}{0,1.5}\csymbolthird[0,20]{a} \jointbnoarrow[right]{X}{0}{Y}{0}
\end{Compose}
~~\begin{Compose}{0}{0}\setsecondfont{\mathtt}\setthirdfont{\mathsfb}
\crectangle{X}{0.8}{0.8}{0,-2} \csymbol{I}
\thispoint{Y}{0,1.5}\csymbolthird[0,20]{b} \jointbnoarrow[right]{X}{0}{Y}{0}
\end{Compose}
\end{equation}
The equation on the left hand side follows since two separate ignore results constitutes a doubly deterministic operation having only inputs - i.e.\ it is an ignore result on $\mathtt{ab}$, and therefore by uniqueness it is the ignore result.  The equation on the right hand side follows by similar reasoning.

Note that
\begin{equation}
\text{prob}\left(
\begin{Compose}{0}{-0.05} \setsecondfont{\mathsfb}
\crectangle{Iu}{0.8}{1}{0,-2} \csymbol{I}  \crectangle{Id}{0.8}{1}{0,2} \csymbol{I} \joinbtnoarrow[left]{Id}{0}{Iu}{0} \csymbolalt{a}
\end{Compose} ~\right)
=1
\end{equation}
since ignore operations are doubly deterministic.

\section{Physicality}

\subsection{Physicality conditions}\label{sec:physicalityconstraints}

We will consider circuits without midcomes (see Sec.\ \ref{sec:midcomes}) in this section.  Every circuit has a probability associated with it. We require that these probabilities are nonnegative and satisfy the double summation property discussed in Sec.\ \ref{sec:doublesummationproperty}.   We proceed by imposing conditions that must be satisfied by operations (from which we can build circuits).  We call these \emph{physicality conditions},
\begin{description}
\item[Complete positivity] is a constraint imposed on operations which is such that any circuit built from those operations has nonnegative probability.
\item[Double causality] is a constraint imposed on operations which is such that any circuit built from those operations satisfies the double summation property.
\end{description}
In this section we will show how to implement physicality conditions on operations.

To obtain the complete positivity condition we employ double purity (see Sec.\ \ref{sec:doublepurity}).  To obtain the double causality condition we employ the double spanning condition (see Sec.\ \ref{sec:doublesummationproperty}).

We can model circuits that have midcomes with circuits that do not by including some multiplicative factors using the midcome equivalence equation \eqref{midcomeequivalence} as discussed in Sec.\ \ref{sec:midcomes}.  Since these multiplicative factors are positive, complete positivity is still satisfied when there are midcomes.  Further, the multiplicative factors introduced by \eqref{midcomeequivalence} ensure we have the correct properties for circuits with midcomes given that we impose the double causality condition on operations.

\subsection{Complete positivity}\label{sec:completepositivity}

We consider an arbitrary circuit wherein all incomes and outcomes are closed by a readout box and an $\mathsf R$ box.

Consider an arbitrary circuit containing some given operation, $\mathsf B$. We can write any such circuit as
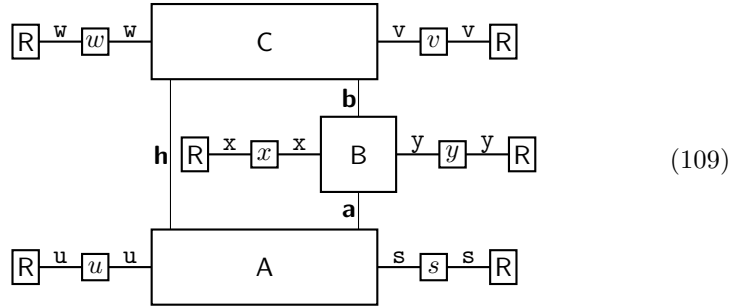
\begin{equation}\label{GeneralcircuitwithB}
\begin{Compose}{0}{0} \setsecondfont{\mathsfb}
\crectangle{A}{6}{2}{0,-6}  \csymbol{A}
\crectangle{C}{6}{2}{0,6}  \csymbol{C}
\Crectangle{B}{2}{2}{5,0}
\jointbnoarrow[left]{A}{-5}{C}{-5} \csymbolalt{h}
\jointbnoarrow[left]{A}{5}{B}{0}  \csymbolalt{a}
\jointbnoarrow[left]{B}{0}{C}{5}  \csymbolalt{b}
\RxBoxincome{B}{2}{0}{x} \RxBoxoutcome{B}{2}{0}{y}
\RxBoxincome{A}{6}{0}{u} \RxBoxoutcome{A}{6}{0}{s}
\RxBoxincome{C}{6}{0}{w} \RxBoxoutcome{C}{6}{0}{v}
\end{Compose}
\end{equation}
We impose the condition that probability associated with any such circuit is greater than or equal to zero. Under the assumption of double purity, a necessary and sufficient condition for this condition is that
\begin{equation}\label{GeneralcircuitwithBwithpureEF}
\text{prob}\left(
\begin{Compose}{0}{0} \setsecondfont{\mathsfb}
\crectangle{A}{6}{2}{0,-6}  \csymbol{D}
\crectangle{C}{6}{2}{0,6}  \csymbol{E}
\Crectangle{B}{2}{2}{5,0}
\jointbnoarrow[left]{A}{-5}{C}{-5} \csymbolalt{h}
\jointbnoarrow[left]{A}{5}{B}{0}  \csymbolalt{a}
\jointbnoarrow[left]{B}{0}{C}{5}  \csymbolalt{b}
\RxBoxincome{B}{2}{0}{x} \RxBoxoutcome{B}{2}{0}{y}
\RxBoxincome{A}{6}{0}{u} 
\RxBoxoutcome{C}{6}{0}{v}
\end{Compose}
\right)
~~\geq~ 0
\end{equation}
for all \emph{pure parallel} (i.e.\ homogeneous) $\mathsf D$ and $\mathsf E$. Note that the concept of pure parallel (homogeneous) was defined in  Sec.\ \ref{sec:pureandhomogeneous} (see also the notes at the end of that subsection on nomenclature).  Note that it is actually only necessary to check this condition for one representative of each parallel direction. This condition is necessary since $\mathsf D$ and $\mathsf E$ are allowed operations with respect to which we must have non-negative probabilities.  This condition is sufficient for the probability of the circuit in \eqref{GeneralcircuitwithB} to be non-negative because, by double purity, any operations $\mathsf A$ and $\mathsf B$ (as in \eqref{GeneralcircuitwithB}) can be written as being equivalent to a positive weighted sum of the pure parallel operations $\mathsf D$ and $\mathsf E$. Consequently, the circuit in \eqref{GeneralcircuitwithB} can be written as being equivalent to a positive weighted sum of circuits of the type in  circuit \eqref{GeneralcircuitwithBwithpureEF}.   We write the condition in \eqref{GeneralcircuitwithBwithpureEF} as
\begin{equation}\label{positivity}
0~\underset{T}{\leq} ~
\begin{Compose}{0}{0}\setsecondfont{\mathsfb}\setthirdfont{\mathtt}
\Crectangle{B}{1.5}{1.5}{0,0}
\thispoint{d}{0,-3} \csymbolalt[0,-20]{a}
\thispoint{u}{0,3}  \csymbolalt[0,20]{b}
\thispoint{l}{-3,0} \csymbolthird[-20,0]{x}
\thispoint{r}{3,0}  \csymbolthird[20,0]{y}
\jointbnoarrow{d}{0}{B}{0}
\jointbnoarrow{B}{0}{u}{0}
\joinrlnoarrowthick{l}{0}{B}{0}
\joinrlnoarrowthick{B}{0}{r}{0}
\end{Compose}
\end{equation}
where this means that $\mathsf B$ is positive with respect to any application of objects of the form
\begin{equation}\label{tester}
\begin{Compose}{0}{0} \setsecondfont{\mathsfb}
\crectangle{A}{6}{2}{0,-6}  \csymbol{D}
\crectangle{C}{6}{2}{0,6}  \csymbol{E}
\Crectangle[white]{B}{2}{2}{5,0}
\jointbnoarrow[left]{A}{-5}{C}{-5} \csymbolalt{h}
\jointbnoarrow[left]{A}{5}{B}{0}  \csymbolalt{a}
\jointbnoarrow[left]{B}{0}{C}{5}  \csymbolalt{b}
\RxBoxincome{B}{2}{0}{x} \RxBoxoutcome{B}{2}{0}{y}
\RxBoxincome{A}{6}{0}{u} 
\RxBoxoutcome{C}{6}{0}{v}
\end{Compose}
\end{equation}
with pure parallel $\mathsf D$ and $\mathsf E$.  We will call the object in \eqref{tester} an \emph{operation tester}. We place a $T$ under $\leq$ symbol to denote that the positivity is with respect to these testers.   We call the condition in \eqref{positivity} \emph{tester positivity}.   In the quantum case this condition ends up corresponding to the condition that corresponding operators are positive after we take the partial transpose with respect to the input (or output) part of the Hilbert space.  If we assume double maximal purity then it is sufficient to check for positivity just for the case where $\mathsf D$ and $\mathsf E$ are maximal.

It is interesting to consider the special cases where $\mathsf B$ is a system preparation (so $\mathsfb a$ is null) or a system result (so $\mathsfb b$ is null).  Consider the case where it is a system result so $\mathsfb b$ is null.  Then we require
\begin{equation}\label{GeneralcircuitwithBwithpureEFnullb}
\text{prob}\left(
\begin{Compose}{0}{0} \setsecondfont{\mathsfb}
\crectangle{A}{6}{2}{0,-6}  \csymbol{D}
\crectangle{C}{6}{2}{0,6}  \csymbol{E}
\Crectangle{B}{2}{2}{5,0}
\jointbnoarrow[left]{A}{-5}{C}{-5} \csymbolalt{h}
\jointbnoarrow[left]{A}{5}{B}{0}  \csymbolalt{a}
\RxBoxincome{B}{2}{0}{x} \RxBoxoutcome{B}{2}{0}{y}
\RxBoxincome{A}{6}{0}{u} 
\RxBoxoutcome{C}{6}{0}{v}
\end{Compose}
\right)
~~\geq~ 0
\end{equation}
The fragment consisting of pure parallel $\mathsf D$ and $\mathsf E$ constitutes a system preparation for system $\mathsfb a$.  By forward purity this is equivalent to a sum over pure system preparations with non-negative weights.  Hence, the condition \eqref{GeneralcircuitwithBwithpureEFnullb} is the same as the condition
\begin{equation}\label{GeneralcircuitwithBwithpureG}
\text{prob}\left(
\begin{Compose}{0}{0} \setsecondfont{\mathsfb}
\crectangle{A}{2}{2}{0,-3}  \csymbol{F}
\Crectangle{B}{2}{2}{0,3}
\jointbnoarrow[left]{A}{0}{B}{0}  \csymbolalt{a}
\RxBoxincome{B}{2}{0}{x} \RxBoxoutcome{B}{2}{0}{y}
\RxBoxincome{A}{2}{0}{u} 
\end{Compose}
\right)
~~\geq~ 0
\end{equation}
where $\mathsf F$ is pure parallel. The tester positivity condition for system preparations is that \eqref{GeneralcircuitwithBwithpureG} must hold for all pure parallel $\mathsf F$.   The tester positivity condition for system results, obtained by similar reasoning using backwards purity, is that
\begin{equation}\label{GeneralcircuitwithBwithpureH}
\text{prob}\left(
\begin{Compose}{0}{0} \setsecondfont{\mathsfb}
\crectangle{C}{2}{2}{0,3}  \csymbol{G}
\Crectangle{B}{2}{2}{0,-3}
\jointbnoarrow[left]{B}{0}{C}{0}  \csymbolalt{b}
\RxBoxincome{B}{2}{0}{x} \RxBoxoutcome{B}{2}{0}{y}
\RxBoxoutcome{C}{2}{0}{v}
\end{Compose}
\right)
~~\geq~ 0
\end{equation}
must hold for all pure parallel $\mathsf G$.

Having defined tester positivity, we are in a position to state and prove the following important theorem.
\begin{quote}
{\bf Complete positivity theorem:}  If the property of double purity is satisfied then sufficient a condition for any given circuit to have non-negative probability is that (i) all operations in the circuit satisfy tester positivity and (ii) all pure system preparations and pure system results satisfy tester positivity.
\end{quote}
To prove this, consider the following circuit
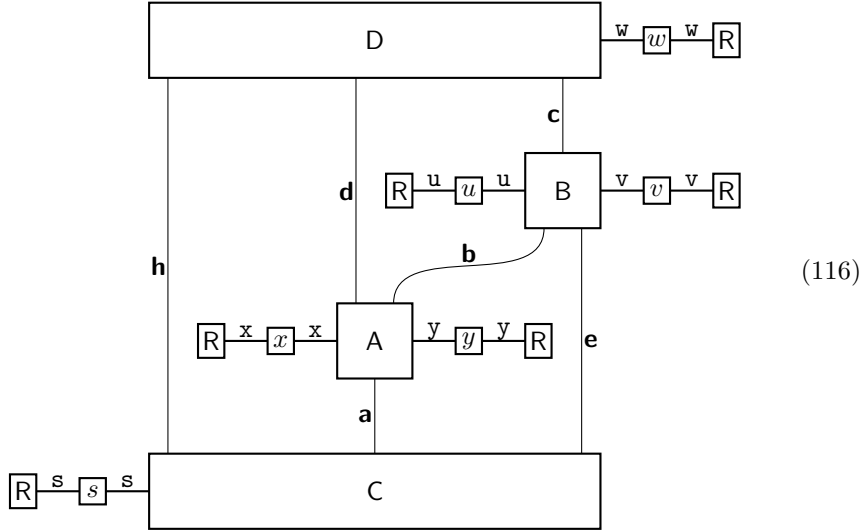
\begin{equation}\label{CABDcircuit}
\begin{Compose}{0}{0} \setsecondfont{\mathsfb}
\Crectangle{C}{12}{2}{0,-12} \RxBoxincome{C}{12}{0}{s}
\Crectangle{D}{12}{2}{0,12} \RxBoxoutcome{D}{12}{0}{w}
\Crectangle{A}{2}{2}{0,-4} \RxBoxincome{A}{2}{0}{x} \RxBoxoutcome{A}{2}{0}{y}
\Crectangle{B}{2}{2}{10,4} \RxBoxincome{B}{2}{0}{u}  \RxBoxoutcome{B}{2}{0}{v}
\jointbnoarrow[left]{C}{0}{A}{0} \csymbolalt{a}
\jointbnoarrow[right]{C}{11}{B}{1} \csymbolalt{e}
\jointbnoarrow[left]{A}{-1}{D}{-1} \csymbolalt{d}
\jointbnoarrow[left]{C}{-11}{D}{-11} \csymbolalt{h}
\jointbnoarrow[above]{A}{1}{B}{-1} \csymbolalt[0,5]{b}
\jointbnoarrow[left]{B}{0}{D}{10} \csymbolalt{c}
\end{Compose}
\end{equation}
where $\mathsf C$ and $\mathsf D$ are pure parallel and are, according to condition (ii) in the theorem statement, taken to satisfy tester positivity (if pure system preparations (or results) satisfy tester positivity then so must pure parallel system preparations (results)).  Furthermore, we assume that $\mathsf A$ and $\mathsf B$ satisfy tester positivity (this is condition (i) in the theorem statement).  The condition for tester positivity for a system preparation is given in \eqref{GeneralcircuitwithBwithpureH} (and the text below this equation). It is clear that this condition is satisfied for the system preparation
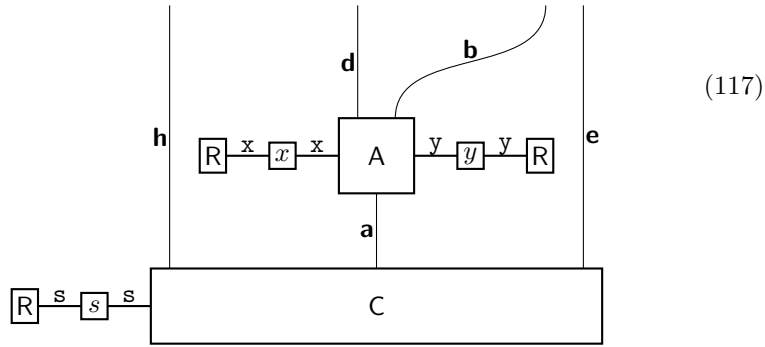
\begin{equation}\label{CAsystemprep}
\begin{Compose}{0}{0} \setsecondfont{\mathsfb}
\Crectangle{C}{12}{2}{0,-12} \RxBoxincome{C}{12}{0}{s}
\thispoint{D}{0,4} 
\Crectangle{A}{2}{2}{0,-4}  \RxBoxincome{A}{2}{0}{x} \RxBoxoutcome{A}{2}{0}{y}
\thispoint{B}{10,4}   
\jointbnoarrow[left]{C}{0}{A}{0} \csymbolalt{a}
\jointbnoarrow[right]{C}{11}{B}{1} \csymbolalt{e}
\jointbnoarrow[left]{A}{-1}{D}{-1} \csymbolalt{d}
\jointbnoarrow[left]{C}{-11}{D}{-11} \csymbolalt{h}
\jointbnoarrow[above]{A}{1}{B}{-1} \csymbolalt[0,5]{b}
\end{Compose}
\end{equation}
simply by considering the definition of tester positivity for $\mathsf A$.   Similarly,  it is clear that the system result
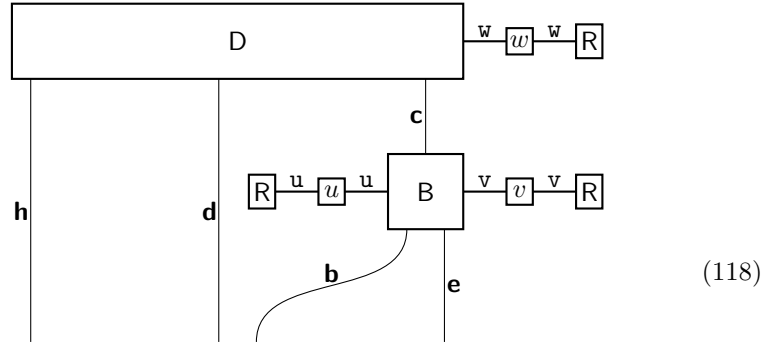
\begin{equation}\label{DBsystemresult}
\begin{Compose}{0}{0} \setsecondfont{\mathsfb}
\thispoint{C}{0,-4} 
\Crectangle{D}{12}{2}{0,12} \RxBoxoutcome{D}{12}{0}{w}
\thispoint{A}{0,-4}  
\Crectangle{B}{2}{2}{10,4} \RxBoxincome{B}{2}{0}{u}  \RxBoxoutcome{B}{2}{0}{v}
\jointbnoarrow[right]{C}{11}{B}{1} \csymbolalt{e}
\jointbnoarrow[left]{A}{-1}{D}{-1} \csymbolalt{d}
\jointbnoarrow[left]{C}{-11}{D}{-11} \csymbolalt{h}
\jointbnoarrow[above]{A}{1}{B}{-1} \csymbolalt[0,5]{b}
\jointbnoarrow[left]{B}{0}{D}{10} \csymbolalt{c}
\end{Compose}
\end{equation}
satisfies tester positivity by considering the definition of tester positivity for $\mathsf B$.   Now, by forward purity it is possible to write the system preparation in \eqref{CAsystemprep} as being equivalent to a weighted sum of pure paralle system preparations where the weights are non-negative.  Similarly, by backwards purity it is possible to write the system result in \eqref{DBsystemresult} as a weighted sum of pure parallel system results where the weights are non-negative. Consequently, it is possible to write the circuit in \eqref{CABDcircuit} as a non-negative coefficient weighted sum of circuits where each of these circuits is comprised of a pure parallel system preparation followed by a pure parallel system result.  Since these pureparallel system preparations and pure parallel system results satisfy tester positivity (by (ii)) the probability for the circuit in \eqref{CABDcircuit} is non-negative.   This means that
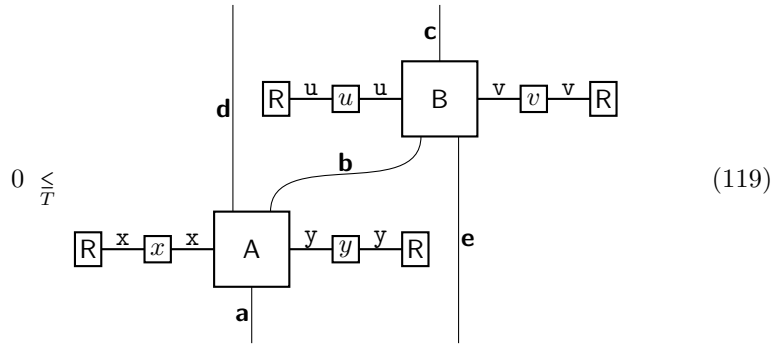
\begin{equation}\label{ABfragment}
0~\underset{T}{\leq} ~
\begin{Compose}{0}{0} \setsecondfont{\mathsfb}
\thispoint{C}{0,-9} 
\thispoint{D}{0,9} 
\Crectangle{A}{2}{2}{0,-4} \RxBoxincome{A}{2}{0}{x} \RxBoxoutcome{A}{2}{0}{y}
\Crectangle{B}{2}{2}{10,4} \RxBoxincome{B}{2}{0}{u}  \RxBoxoutcome{B}{2}{0}{v}
\jointbnoarrow[left]{C}{0}{A}{0} \csymbolalt{a}
\jointbnoarrow[right]{C}{11}{B}{1} \csymbolalt{e}
\jointbnoarrow[left]{A}{-1}{D}{-1} \csymbolalt{d}
\jointbnoarrow[above]{A}{1}{B}{-1} \csymbolalt[0,5]{b}
\jointbnoarrow[left]{B}{0}{D}{10} \csymbolalt{c}
\end{Compose}
\end{equation}
That is, the composite operation formed from $\mathsf A$ and $\mathsf B$ satisfies tester positivity.   We can build up any circuit by adding operations.  According to this result, at each step of the way we satisfy tester positivity.  Consequently, any circuit has non-negative probability. This proves the theorem.

\subsection{Double summation from double causality}\label{sec:doublebisummationfromdoublecausality}

We will prove the following theorem
\begin{quote}
\textbf{Double causality theorem.}  It follows from double flatness, double spanning, and double pointer tomography that a general operation,
\begin{equation}
\begin{Compose}{0}{0}\setsecondfont{\mathtt} \setthirdfont{\mathsfb}
\crectangle{U}{2}{2}{0,0} \csymbol{B} \thispoint{AL}{-4,0} \csymbolalt[-20,0]{x} \thispoint{AR}{4,0} \csymbolalt[20,0]{y}
\joinlrnoarrowthick{U}{0}{AL}{0} \joinrlnoarrowthick{U}{0}{AR}{0}
\thispoint{a}{0,-4} \csymbolthird[0,-20]{a} \jointbnoarrow{a}{0}{U}{0}
\thispoint{d}{0,4} \csymbolthird[0,20]{b} \joinbtnoarrow{d}{0}{U}{0}
\end{Compose}
\end{equation}
satisfies the following two conditions
\begin{description}
\item[\textit{Forward causality.}]
\begin{equation}\label{forwardunital}
\begin{Compose}{0}{0}\setsecondfont{\mathtt}\setthirdfont{\mathsfb}
\crectangle{C}{2}{2}{0,0} \csymbol{B}
\crectangle{CL}{0.8}{0.8}{-4.5,0}\csymbol{R} \joinlrnoarrowthick{C}{0}{CL}{0} \csymbolalt{x}
\thispoint{CR}{4,0} \csymbolalt[20,0]{y} \joinrlnoarrow{C}{0}{CR}{0}
\crectangle{A}{0.8}{0.8}{0,-5} \csymbol{I} \jointbnoarrow[left]{A}{0}{C}{0} \csymbolthird{a} 
\thispoint{B}{0,4}\csymbolthird[0,20]{b} \jointbnoarrow[left]{C}{0}{B}{0}
\end{Compose}
~\equiv ~
\begin{Compose}{0}{0}\setsecondfont{\mathtt}\setthirdfont{\mathsfb}
\crectangle{B}{0.8}{0.8}{0,0} \csymbol{I} \thispoint{BR}{0,4} \csymbolthird[20,0]{b} \jointbnoarrow{B}{0}{BR}{0}
\crectangle{BD}{0.8}{0.8}{2,0} \csymbol{R} \thispoint{BRD}{4,0} \csymbolalt[20,0]{y} \joinrlnoarrowthick{BD}{0}{BRD}{0}
\end{Compose}
\end{equation}
\item[\textit{Backwards causality.}]
\begin{equation}\label{backwardunital}
\begin{Compose}{0}{0}\setsecondfont{\mathtt}\setthirdfont{\mathsfb}
\crectangle{C}{2}{2}{0,0} \csymbol{B}
\thispoint{CL}{-4,0} \csymbolalt[-20,0]{x} \joinlrnoarrowthick{C}{0}{CL}{0}
\crectangle{CR}{0.8}{0.8}{4.5,0} \csymbol{R} \joinrlnoarrow{C}{0}{CR}{0} \csymbolalt[0,5]{y}
\thispoint{A}{0,-4} \csymbolthird[0,-20]{a} \jointbnoarrow{A}{0}{C}{0}
\crectangle{B}{0.8}{0.8}{0,5} \csymbol{I} \jointbnoarrow[left]{C}{0}{B}{0}\csymbolthird{b}
\end{Compose}
~\equiv ~
\begin{Compose}{0}{0}\setsecondfont{\mathtt}\setthirdfont{\mathsfb}
\crectangle{B}{0.8}{0.8}{0,0} \csymbol{I} \thispoint{BR}{0,-4} \csymbolthird[0,-20]{a} \joinbtnoarrow{B}{0}{BR}{0}
\crectangle{BD}{0.8}{0.8}{-2,0} \csymbol{R} \thispoint{BRD}{-4,0} \csymbolalt[-20,0]{x} \joinlrnoarrowthick{BD}{0}{BRD}{0}
\end{Compose}
\end{equation}
\end{description}
and, furthermore, if these conditions hold for all the operations comprising a circuit, then the double summation conditions hold for that circuit.
\end{quote}
It operators satisfy both forward and backwards causality then they are said to satisfy the property of \emph{double causality}. We will prove the double causality  theorem below.  First a few comments.  We can treat any of the types, $\mathtt x$, $\mathtt y$, $\mathsfb a$, and $\mathsfb b$ as null in which case we can omit them.  This gives rise to multiple special cases of the forward causality and backward causality.  These are given in Table \ref{table:causality}.  It is worth commenting on some special cases.  In row 7 we have the double flatness conditions (so double flatness is a special case of double causality).  In row 8 we see that the doubly deterministic preparation for systems is unique (up to equivalence) as is the doubly deterministic result. In row 9 we see that the doubly deterministic pointer preparation is unique (up to equivalence) as is the doubly deterministic result.
\begin{table}
\begin{center}
{\tabulinesep=1mm
\begin{tabu}{|c|c|c|}
  \hline
  1 & $\Funitalxyab[0.2]{1}{1}{1}{1}$ &  $\Bunitalyxba[-0.2]{1}{1}{1}{1}$   \\ \hline
  2 & $\Funitalxyab[-0.4]{1}{1}{0}{1}$ &  $\Bunitalyxba[0.4]{1}{1}{0}{1}$   \\ \hline
  3 & $\Funitalxyab[0.4]{1}{1}{1}{0}$ &  $\Bunitalyxba[-0.4]{1}{1}{1}{0}$   \\ \hline
  4 & $\Funitalxyab{0}{1}{1}{1}$ &  $\Bunitalyxba{0}{1}{1}{1}$   \\ \hline
  5 & $\Funitalxyab[0.2]{1}{0}{1}{1}$ &  $\Bunitalyxba[-0.2]{1}{0}{1}{1}$   \\ \hline
  6 & $\Funitalxyab[0.2]{0}{0}{1}{1}$ &  $\Bunitalyxba[-0.2]{0}{0}{1}{1}$   \\ \hline
  7 & $\Funitalxyab{1}{1}{0}{0}$ &  $\Bunitalyxba{1}{1}{0}{0}$   \\ \hline
  8 & $\Funitalxyab[-0.4]{0}{0}{0}{1}$ &  $\Bunitalyxba[0.4]{0}{0}{0}{1}$   \\ \hline
  9 & $\Funitalxyab{0}{1}{0}{0}$ &  $\Bunitalyxba{0}{1}{0}{0}$   \\ \hline
\end{tabu} }
\end{center}
\caption{This tabulates many of the different versions of forward causality (left column) and backward causality (right column). }
\label{table:causality}
\end{table}
Furthermore, note that if we have multiple inputs, outputs, incomes, and outcomes, we can employ the results that flat pointer operations, $\mathsf R$, and ignore operations, $\mathsf I$, factorize for composite systems (see \eqref{Qflatxy} and \eqref{IequalsII}).  For example,
\begin{equation}\label{backwardunitalcompositeexample}
\begin{Compose}{0}{0}\setsecondfont{\mathtt}\setthirdfont{\mathsfb}
\crectangle{C}{2}{2}{0,0} \csymbol{B}
\thispoint{CL}{-4,1} \csymbolalt[-20,0]{x} \joinlrnoarrowthick{C}{1}{CL}{0}
\thispoint{CL}{-4,-1} \csymbolalt[-20,0]{s} \joinlrnoarrowthick{C}{-1}{CL}{0}
\crectangle{CR}{0.8}{0.8}{4.5,1} \csymbol{R} \joinrlnoarrow{C}{1}{CR}{0} \csymbolalt[0,5]{y}
\crectangle{CR}{0.8}{0.8}{4.5,-1} \csymbol{R} \joinrlnoarrow{C}{-1}{CR}{0} \csymbolalt[0,5]{v}
\thispoint{A}{-1,-4} \csymbolthird[0,-20]{a} \jointbnoarrow{A}{0}{C}{-1}
\thispoint{A}{1,-4} \csymbolthird[0,-20]{d} \jointbnoarrow{A}{0}{C}{1}
\crectangle{B}{0.8}{0.8}{1,5} \csymbol{I} \jointbnoarrow[left]{C}{1}{B}{0}\csymbolthird{b}
\crectangle{B}{0.8}{0.8}{-1,5} \csymbol{I} \jointbnoarrow[left]{C}{-1}{B}{0}\csymbolthird{c}
\end{Compose}
~\equiv ~
\begin{Compose}{0}{0}\setsecondfont{\mathtt}\setthirdfont{\mathsfb}
\crectangle{B}{0.8}{0.8}{0,0} \csymbol{I} \thispoint{BR}{0,-4} \csymbolthird[0,-20]{a} \joinbtnoarrow{B}{0}{BR}{0}
\crectangle{B}{0.8}{0.8}{2,0} \csymbol{I} \thispoint{BR}{2,-4} \csymbolthird[0,-20]{d} \joinbtnoarrow{B}{0}{BR}{0}
\crectangle{BD}{0.8}{0.8}{-2,1} \csymbol{R} \thispoint{BRD}{-4,1} \csymbolalt[-20,0]{x} \joinlrnoarrowthick{BD}{0}{BRD}{0}
\crectangle{BD}{0.8}{0.8}{-2,-1} \csymbol{R} \thispoint{BRD}{-4,-1} \csymbolalt[-20,0]{s} \joinlrnoarrowthick{BD}{0}{BRD}{0}
\end{Compose}
\end{equation}

We will now prove backwards causality.  The proof for forward causality follows similarly.  Using backwards flatness we have
\begin{equation}
\begin{Compose}{0}{0}\setsecondfont{\mathtt}\setthirdfont{\mathsfb}
\crectangle{C}{2}{2}{0,0} \csymbol{B}
\thispoint{x}{-4,0} \csymbolalt[-20,0]{x}  \joinrlnoarrowthick{x}{0}{C}{0}
\crectangle{CR}{0.7}{0.9}{4.5,0} \csymbol{R} \joinrlnoarrowthick{C}{0}{CR}{0} \csymbolalt[0,5]{y}
\Crectangle{A}{1}{1}{0,-5} \jointbnoarrow[left]{A}{0}{C}{0} \csymbolthird{a}
\thispoint{AL}{-4,-5} \csymbolalt[-20,0]{s} \joinlrnoarrowthick{A}{0}{AL}{0}
\crectangle{I}{0.7}{0.9}{0,5} \csymbol{I} \jointbnoarrow[left]{C}{0}{I}{0}\csymbolthird{b}
\end{Compose}
~~\equiv ~~
\begin{Compose}{0}{0}\setsecondfont{\mathtt}\setthirdfont{\mathsfb}
\thispoint{RU}{0,0} \csymbolalt[-20,0]{x} \Routcome{RU}{0}{}
\thispoint{RD}{0,-5} \csymbolalt[-20,0]{s} \Routcome{RD}{0}{}
\end{Compose}
~~\equiv ~~
\begin{Compose}{0}{0}\setsecondfont{\mathtt}\setthirdfont{\mathsfb}
\thispoint{RU}{-5,0} \csymbolalt[-20,0]{x} \Routcome{RU}{0}{}
\crectangle{RD}{0.8}{0.9}{1,-5} \csymbol{A}
\thispoint{AL}{-5,-5} \csymbolalt[-20,0]{s} \joinrlnoarrowthick{AL}{0}{RD}{0} 
\Crectangle{I}{0.7}{0.9}{1,0} \joinbtnoarrow[left]{I}{0}{RD}{0} \csymbolthird{a}
\end{Compose}
\end{equation}
for all (doubly deterministic) $\mathsf A$. Note that backward flatness tells us that the fragment on the left and the fragment on the right are both equivalent to the fragment in the middle (and therefore equivalent to each other). Hence
\begin{equation}
\begin{Compose}{0}{0}\setsecondfont{\mathtt}\setthirdfont{\mathsfb}
\crectangle{C}{2}{2}{0,0} \csymbol{B}
\RxBoxincome{C}{2}{0}{x}
\crectangle{CR}{0.7}{0.9}{4.5,0} \csymbol{R} \joinrlnoarrow{C}{0}{CR}{0} \csymbolalt[0,5]{y}
\Crectangle{A}{1}{1}{0,-5} \jointbnoarrow[left]{A}{0}{C}{0} \csymbolthird{a}
\RxBoxincome{A}{1}{0}{s}
\crectangle{I}{0.8}{0.8}{0,5} \csymbol{I} \jointbnoarrow[left]{C}{0}{I}{0}\csymbolthird{b}
\end{Compose}
~\equiv ~
\begin{Compose}{0}{0}\setsecondfont{\mathtt}\setthirdfont{\mathsfb}
\crectangle{RU}{0.7}{0.9}{0,0} \csymbol{R} \RxBoxincome{RU}{0.7}{0}{x}
\crectangle{RD}{0.8}{0.9}{2,-5} \csymbol{A}   \RxBoxincome{RD}{0.7}{0}{s}
\Crectangle{I}{0.8}{0.8}{2,0} \joinbtnoarrow[left]{I}{0}{RD}{0} \csymbolthird{a}
\end{Compose}
\end{equation}
Equivalence means the circuit on the left has the same probability as the circuit on the right (the circuit on the right consists of two disjoint parts).
Now using the forward spanning property, we have that the income only system preparations
\begin{equation}
\begin{Compose}{0}{0}\setsecondfont{\mathtt}\setthirdfont{\mathsfb}
\crectangle{RD}{0.8}{0.9}{2,-4} \csymbol{A}   \RxBoxincome{RD}{0.7}{0}{s}
\thispoint{I}{2,0} \csymbolthird[0,20] {a}\joinbtnoarrow[left]{I}{0}{RD}{0}
\end{Compose}
\end{equation}
are spanning and therefore
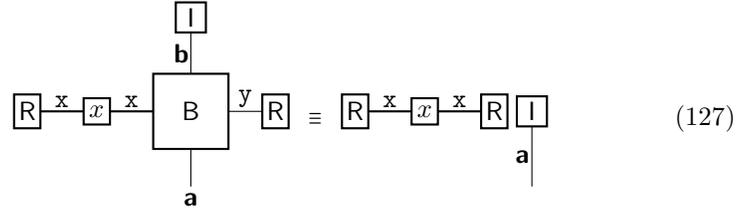
\begin{equation}
\begin{Compose}{0}{0}\setsecondfont{\mathtt}\setthirdfont{\mathsfb}
\crectangle{C}{2}{2}{0,0} \csymbol{B} \RxBoxincome{C}{2}{0}{x}
\crectangle{CR}{0.7}{0.9}{4.5,0} \csymbol{R} \joinrlnoarrow{C}{0}{CR}{0} \csymbolalt[0,5]{y}
\thispoint{A}{0,-4} \csymbolthird[0,-20]{a} \jointbnoarrow[left]{A}{0}{C}{0}
\crectangle{I}{0.8}{0.8}{0,5} \csymbol{I} \jointbnoarrow[left]{C}{0}{I}{0}\csymbolthird{b}
\end{Compose}
~ \equiv ~
\begin{Compose}{0}{0}\setsecondfont{\mathtt}\setthirdfont{\mathsfb}
\crectangle{RU}{0.7}{0.9}{0,0} \csymbol{R} \RxBoxincome{RU}{0.7}{0}{x}
\thispoint{RD}{2,-4} 
\Crectangle{I}{0.8}{0.8}{2,0} \joinbtnoarrow[left]{I}{0}{RD}{0} \csymbolthird{a}
\end{Compose}
\end{equation}
Finally, using backward pointer tomography (see Sec.\ \ref{sec:doublepointer tomography}) and the fact that $\mathsf R$ is doubly deterministic, the backwards  causality condition in \eqref{backwardunital} follows.  The forwards causality condition can be proven by similar means.

It is easy to prove that double bisummation follows for any circuit built out of operations satisfying double causality.  If, starting at the top of a circuit, we use backward causality to successively replace each operation then we can prove we have backwards flatness.  This is best illustrated by example. We can prove
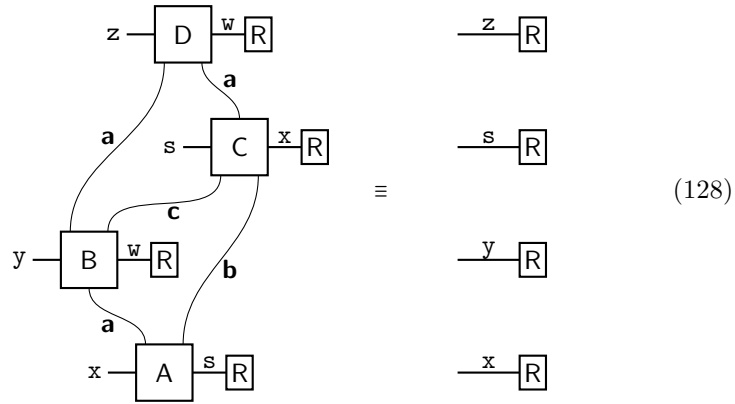
\begin{equation}\label{goodcircuitexample}
\begin{Compose}{0}{-2.5}\setsecondfont{\mathtt}\setthirdfont{\mathsfb}
\Crectangle{A}{1.5}{1.5}{0,0}  \incomewire{A}{0}{x} \Routcome{A}{0}{s}
\Crectangle{B}{1.5}{1.5}{-4,6} \incomewire{B}{0}{y} \Routcome{B}{0}{w}
\Crectangle{C}{1.5}{1.5}{4,12} \incomewire{C}{0}{s} \Routcome{C}{0}{x}
\Crectangle{D}{1.5}{1.5}{1,18} \incomewire{D}{0}{z} \Routcome{D}{0}{w}
\jointbnoarrow[below left]{A}{-1}{B}{0} \csymbolthird{a}
\jointbnoarrow[below right]{A}{1}{C}{1} \csymbolthird{b}
\jointbnoarrow[above left]{B}{-1}{D}{-1} \csymbolthird{a}
\jointbnoarrow[below right]{B}{1}{C}{-1} \csymbolthird{c}
\jointbnoarrow[above right]{C}{0}{D}{1} \csymbolthird{a}
\end{Compose}
~~~~\equiv ~~~~~~~
\begin{Compose}{0}{-2.5} \setsecondfont{\mathtt}\setthirdfont{\mathsfb}
\thispoint{A}{0,0} \Routcome{A}{0}{x} \Routcome{A}{6}{y} \Routcome{A}{12}{s} \Routcome{A}{18}{z}
\end{Compose}
\end{equation}
This is equivalent to backward summation.  Here are the steps of the proof (the dotted boxes are to help see where the replacements are to be made or have been made).
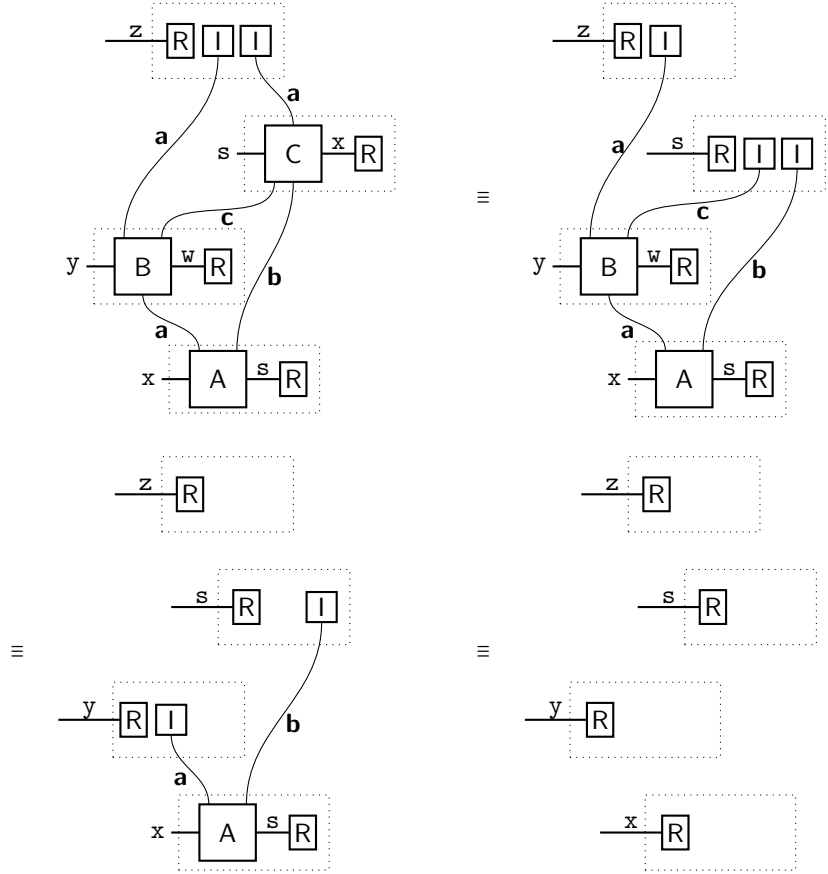
\begin{align*}\label{goodcircuitexample}
&
\begin{Compose}{0}{-2.5}\setsecondfont{\mathtt}\setthirdfont{\mathsfb} 
\crectangle[dotted]{A}{4}{1.8}{1.4,0}
\crectangle[dotted]{B}{4}{2}{-2.6,6}
\crectangle[dotted]{C}{4}{2}{5.4,12}
\crectangle[dotted]{D}{3.5}{2}{0,18}
\Crectangle{A}{1.5}{1.5}{0,0}  \incomewire{A}{0}{x} \Routcome{A}{0}{s}
\Crectangle{B}{1.5}{1.5}{-4,6} \incomewire{B}{0}{y} \Routcome{B}{0}{w}
\Crectangle{C}{1.5}{1.5}{4,12} \incomewire{C}{0}{s} \Routcome{C}{0}{x}
\thispoint{DR}{-6,18} \Routcome{DR}{0}{z}
\crectangle{DIL}{0.8}{0.8}{0,18} \csymbol{I}
\crectangle{DIR}{0.8}{0.8}{2,18} \csymbol{I}
\jointbnoarrow[below left]{A}{-1}{B}{0} \csymbolthird{a}
\jointbnoarrow[below right]{A}{1}{C}{0} \csymbolthird{b}
\jointbnoarrow[above left]{B}{-1}{DIL}{0} \csymbolthird{a}
\jointbnoarrow[below right]{B}{1}{C}{-1} \csymbolthird{c}
\jointbnoarrow[above right]{C}{0}{DIR}{0} \csymbolthird[15,-20]{a}
\end{Compose}
&  ~~~\equiv ~~~ &
\begin{Compose}{0}{-2.5}\setsecondfont{\mathtt}\setthirdfont{\mathsfb} 
\crectangle[dotted]{A}{4}{2}{1.4,0}
\crectangle[dotted]{B}{4}{2}{-2.6,6}
\crectangle[dotted]{C}{3.5}{2}{4,12}
\crectangle[dotted]{D}{3.5}{2}{-0.8,18}
\Crectangle{A}{1.5}{1.5}{0,0}  \incomewire{A}{0}{x} \Routcome{A}{0}{s}
\Crectangle{B}{1.5}{1.5}{-4,6} \incomewire{B}{0}{y} \Routcome{B}{0}{w}
\crectangle{CIL}{0.8}{0.8}{4,12} \csymbol{I}
\crectangle{CIR}{0.8}{0.8}{6,12} \csymbol{I}
\thispoint{CR}{-2,12} \Routcome{CR}{0}{s}
\thispoint{DR}{-7,18} \Routcome{DR}{0}{z}
\crectangle{DIL}{0.8}{0.8}{-1,18} \csymbol{I}
\jointbnoarrow[below left]{A}{-1}{B}{0} \csymbolthird{a}
\jointbnoarrow[below right]{A}{1}{CIR}{0} \csymbolthird{b}
\jointbnoarrow[left]{B}{-1}{DIL}{0} \csymbolthird{a}
\jointbnoarrow[below right]{B}{1}{CIL}{0} \csymbolthird{c}
\end{Compose}
\\[2.5ex]
 \equiv ~~~ &
\begin{Compose}{0}{-2.5}\setsecondfont{\mathtt}\setthirdfont{\mathsfb} 
\crectangle[dotted]{B}{3.5}{2}{-2.6,6}%
\crectangle[dotted]{A}{4}{2}{1.4,0}
\crectangle[dotted]{C}{3.5}{2}{3,12}
\crectangle[dotted]{D}{3.5}{2}{0,18}
\Crectangle{A}{1.5}{1.5}{0,0}  \incomewire{A}{0}{x} \Routcome{A}{0}{s}
\thispoint{BR}{-9,6} \Routcome{BR}{0}{y}
\crectangle{BI}{0.8}{0.8}{-3,6} \csymbol{I}
\crectangle{CIR}{0.8}{0.8}{5,12} \csymbol{I}
\thispoint{CR}{-3,12} \Routcome{CR}{0}{s}
\thispoint{DR}{-6,18} \Routcome{DR}{0}{z}
\jointbnoarrow[below left]{A}{-1}{BI}{0} \csymbolthird{a}
\jointbnoarrow[below right]{A}{1}{CIR}{0} \csymbolthird{b}
\end{Compose}
& ~~~ \equiv ~~~ &
\begin{Compose}{0}{-2.5}\setsecondfont{\mathtt}\setthirdfont{\mathsfb} 
\crectangle[dotted]{A}{4}{2}{1.4,0}
\crectangle[dotted]{B}{4}{2}{-2.6,6}
\crectangle[dotted]{C}{3.5}{2}{3,12}
\crectangle[dotted]{D}{3.5}{2}{0,18}
\thispoint{AR}{-5,0} \Routcome{AR}{0}{x}
\thispoint{BR}{-9,6} \Routcome{BR}{0}{y}
\thispoint{CR}{-3,12} \Routcome{CR}{0}{s}
\thispoint{DR}{-6,18} \Routcome{DR}{0}{z}
\end{Compose}
\end{align*}
We can prove forward flatness (and, thereby, forward summation) in the same way by replacing operations with their forward unital version, this time starting at the bottom of the circuit.  Following through with these steps, we obtain
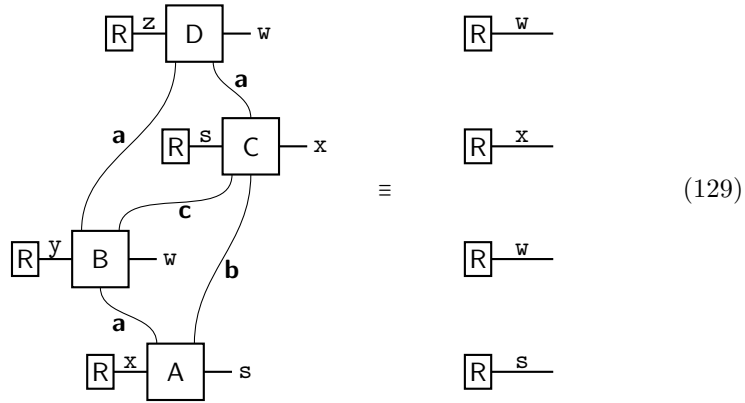
\begin{equation}\label{goodcircuitexampleforwardsflatness}
\begin{Compose}{0}{-2.5}\setsecondfont{\mathtt}\setthirdfont{\mathsfb}
\Crectangle{A}{1.5}{1.5}{0,0}  \outcomewire{A}{0}{s} \Rincome{A}{0}{x}
\Crectangle{B}{1.5}{1.5}{-4,6} \outcomewire{B}{0}{w} \Rincome{B}{0}{y}
\Crectangle{C}{1.5}{1.5}{4,12} \outcomewire{C}{0}{x} \Rincome{C}{0}{s}
\Crectangle{D}{1.5}{1.5}{1,18} \outcomewire{D}{0}{w} \Rincome{D}{0}{z}
\jointbnoarrow[below left]{A}{-1}{B}{0} \csymbolthird{a}
\jointbnoarrow[below right]{A}{1}{C}{0} \csymbolthird{b}
\jointbnoarrow[above left]{B}{-1}{D}{-1} \csymbolthird{a}
\jointbnoarrow[below right]{B}{1}{C}{-1} \csymbolthird{c}
\jointbnoarrow[above right]{C}{0}{D}{1} \csymbolthird{a}
\end{Compose}
~~~~\equiv ~~~~~~~
\begin{Compose}{0}{-2.5} \setsecondfont{\mathtt}\setthirdfont{\mathsfb}
\thispoint{A}{0,0} \Rincome{A}{0}{s} \Rincome{A}{6}{w} \Rincome{A}{12}{x} \Rincome{A}{18}{w}
\end{Compose}
\end{equation}
Clearly this technique will work with any circuit.  One way to see this is to foliate the circuit with \lq\lq space-like" foliation lines such that each operation is sandwiched between a pair of these foliation lines.  Then starting at the top (if we are proving backwards flatness) or bottom (if we are proving forward flatness) replace operations with their backwards or forwards unital form as appropriate until all replacements have been made.

\subsection{Contrast with the standard time asymmetric framework}

The standard operational picture is time asymmetric (see for example \cite{chiribella2010informational, hardy2011reformulating}).  Chiribella, D'Ariano, and Perinotti show that the principle that choices in the future cannot influence corresponds to the mathematical statement that \lq\lq the deterministic effect is unique".  In our present context, this is the application of backwards causality shown on the right row 8 of Table \ref{table:causality}, namely
\[   \Bunitalyxba{0}{0}{0}{1}   \]
On the other hand, in the usual framework, choices in the past can influence probabilities in the future.  However, here we have the result on the left of row 8 that
\[  \Funitalxyab{0}{0}{0}{1}   \]
Hence, choices in the past cannot influence the future.  If we understand \lq\lq choices" to mean settings as discussed in Sec.\ \ref{sec:timesymmetricoperations} then it is true that, in the time symmetric framework, we cannot signal into the past.  However, we should also note that
double causality is consistent with the following
\begin{equation}
\begin{Compose}{0}{0}\setsecondfont{\mathtt}\setthirdfont{\mathsfb}
\crectangle{C}{1.3}{1.3}{0,0} \csymbol{B}
\thispoint{CL}{-4,0} \joinlrnoarrowthick{C}{0}{CL}{0} \csymbolalt{x}
\thispoint{B}{0,4} \jointbnoarrow[left]{C}{0}{B}{0} \csymbolthird{b}
\end{Compose}
\not\equiv
\begin{Compose}{0}{0}\setsecondfont{\mathtt}\setthirdfont{\mathsfb}
\crectangle{B}{0.8}{0.8}{0,0} \csymbol{I} \thispoint{BR}{0, 3.5} \jointbnoarrow[left]{B}{0}{BR}{0} \csymbolthird{a}
\crectangle{BD}{0.8}{0.8}{-2,0} \csymbol{R} \thispoint{BRD}{-5.5,0} \joinlrnoarrowthick[above]{BD}{0}{BRD}{0}\csymbolalt{x}
\end{Compose}
\end{equation}
This means that, if we condition on an income in the past, then the future can be influenced by the past.  Of course, the time reverse statement is also true.  Double causality is consistent with
\begin{equation}
\begin{Compose}{0}{0}\setsecondfont{\mathtt}\setthirdfont{\mathsfb}
\crectangle{C}{1.3}{1.3}{0,0} \csymbol{B}
\thispoint{CL}{4,0} \joinrlnoarrowthick{C}{0}{CL}{0} \csymbolalt{x}
\thispoint{B}{0,-4} \joinbtnoarrow[left]{C}{0}{B}{0} \csymbolthird{b}
\end{Compose}
\not\equiv
\begin{Compose}{0}{0}\setsecondfont{\mathtt}\setthirdfont{\mathsfb}
\crectangle{B}{0.8}{0.8}{-2,0} \csymbol{I} \thispoint{BR}{-2, -3.5} \joinbtnoarrow[left]{B}{0}{BR}{0} \csymbolthird{a}
\crectangle{BD}{0.8}{0.8}{0,0} \csymbol{R} \thispoint{BRD}{3.5,0} \joinrlnoarrowthick[above]{BD}{0}{BRD}{0}\csymbolalt{x}
\end{Compose}
\end{equation}
This means that, if we condition on an outcome in the future, then the future can be influenced by the past.   With this change in perspective, the time symmetric framework can accommodate the features of the standard time asymmetric framework.

In general, it is interesting what influences are, and are not, possible.  We will discuss this in the next two subsections.

\subsection{Z flatness}\label{sec:doubleZflatness}

We have seen that we can obtain double flatness (or, equivalently, double summation) from double causality (this is row 7 of Table \ref{table:causality}).  It turns out that we can prove an additional important and very basic flatness property.  We will call this \emph{Z flatness} (pronounced \lq\lq zed flatness").  

Consider the following where we replace $\mathsf E$ with an equivalent expression using forward causality and $\mathsf F$ with an equivalent expression using backwards causality
\begin{equation}\label{doubleZflatness}
\begin{Compose}{0}{0}\setsecondfont{\mathtt}\setthirdfont{\mathsfb}
\Crectangle{E}{1.5}{1.5}{0,-3} \Crectangle{F}{1.5}{1.5}{0,3}
\incomewire[2]{F}{0}{x} \outcomewire[2]{E}{0}{y}
\jointbnoarrow[left]{E}{0}{F}{0} \csymbol{a}
\end{Compose}
~~~\equiv~~~
\begin{Compose}{0}{0}\setsecondfont{\mathtt}\setthirdfont{\mathsfb}
\thispoint{o}{-6.8,3} \Routcome[1.2]{o}{0}{x}
\thispoint{o}{6.8,-3}\Rincome[1.2]{o}{0}{y}
\crectangle{Iup}{0.8}{0.8}{0,3} \csymbol{I} \crectangle{Ido}{0.8}{0.8}{0,-3} \csymbol{I} \jointbnoarrow[left]{Ido}{0}{Iup}{0} \csymbol{a}
\end{Compose}
~~~\equiv~~~
\begin{Compose}{0}{0}\setsecondfont{\mathtt}\setthirdfont{\mathsfb}
\thispoint{o}{-4.8,3} \Routcome[1.2]{o}{0}{x}
\thispoint{o}{4.8,-3}\Rincome[1.2]{o}{0}{y}
\end{Compose}
\end{equation}
We take $\mathsf E$ and $\mathsf F$ to be doubly deterministic.  The Z flatness property is that the $Z$ shaped object on the left reduces to the flat pointer objects on the right.   Many situations will reduce to the form shown on the left in \eqref{doubleZflatness}.  For example, we have
\begin{equation}\label{goodcircuitexampleZflatness}
\begin{Compose}{0}{-2.5}\setsecondfont{\mathtt}\setthirdfont{\mathsfb}
\Crectangle{A}{1.5}{1.5}{0,0}  \outcomewire{A}{0}{s} \Rincome{A}{0}{x}
\Crectangle{B}{1.5}{1.5}{-4,6} \outcomewire{B}{0}{w} \Rincome{B}{0}{y}
\Crectangle{C}{1.5}{1.5}{4,12} \Routcome{C}{0}{x} \incomewire{C}{0}{s}
\Crectangle{D}{1.5}{1.5}{1,18} \Routcome{D}{0}{w} \incomewire{D}{0}{z}
\jointbnoarrow[below left]{A}{-1}{B}{0} \csymbolthird{a}
\jointbnoarrow[below right]{A}{1}{C}{0} \csymbolthird{b}
\jointbnoarrow[above left]{B}{-1}{D}{-1} \csymbolthird{a}
\jointbnoarrow[below right]{B}{1}{C}{-1} \csymbolthird{c}
\jointbnoarrow[above right]{C}{0}{D}{1} \csymbolthird{a}
\end{Compose}
~~\equiv~~
\begin{Compose}{0}{-2.5}\setsecondfont{\mathtt}\setthirdfont{\mathsfb}
\Crectangle{R}{0.8}{0.8}{0,0} \relpoint{R}{4,0}{p}\csymbol[20,0]{s} \joinrlnoarrowthick{R}{0}{p}{0}
\Crectangle{R}{0.8}{0.8}{0,6} \relpoint{R}{4,0}{p}\csymbol[20,0]{w} \joinrlnoarrowthick{R}{0}{p}{0}
\Crectangle{R}{0.8}{0.8}{0,12} \relpoint{R}{-4,0}{p}\csymbol[-20,0]{s} \joinlrnoarrowthick{R}{0}{p}{0}
\Crectangle{R}{0.8}{0.8}{0,18} \relpoint{R}{-4,0}{p}\csymbol[-20,0]{z} \joinlrnoarrowthick{R}{0}{p}{0}
\end{Compose}
\end{equation}
since the object on the left hand side can be put in Z form.

\subsection{Simple dependence and independence configurations}

Z flatness contrasts with double maximality whereby we can have dependence either way between an earlier input and a later outcome.  The reason we obtain Z flatness is that we assumed the double spanning property.  From a mathematical point of view, this was a choice breaking the symmetry between Z shaped objects and "reversed Z" shaped objects.  Physically, the double spanning property helps enforce the appropriate dependence channels between past and future via incomes and outcomes.  It is noteworthy that we have also ruled out dependence between past and future incomes (by backward flatness), and between past and future outcomes (by forwards flatness).  For reference we tabulate these simple situations (and two others) in Table \ref{ZCTtable}.
\begin{table}
\begin{center}
{\tabulinesep=1.2mm
\begin{tabu}{|c|c|p{0.41\linewidth}|}
  \hline
  \Ccase & no & By application of forward flatness. \\ \hline
  \revCcase & no & By application of backward flatness. \\ \hline
  \Zcase & no & By application of Z flatness. \\ \hline
  \revZcase & yes & Double maximality is example. \\ \hline
  \Tcase & yes & Can signal between income and outcome of a given operation. \\ \hline
  \flippedTcase & yes & Can signal between income and outcome of a given operation. \\ \hline
\end{tabu} }
\end{center}
\caption{This tabulates whether there is a dependence (\lq \lq yes" in second column) or not (\lq\lq no" in second column) between the open pointer wires. }
\label{ZCTtable}
\end{table}
This does note exhaust the set of interesting cases that can be uncovered by application of double causality.

\subsection{Simplifying circuits and fragments}

Double causality can often be used to simplify circuits.  Here is an example to illustrate this
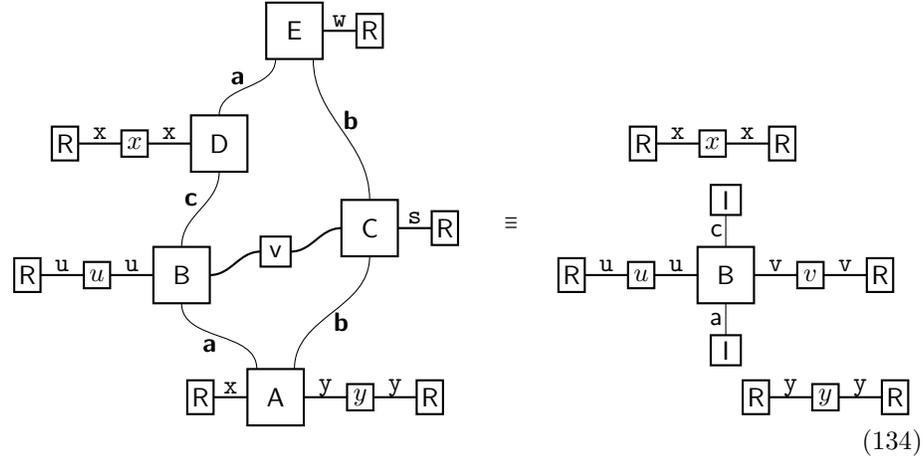
\begin{equation}
\begin{Compose}{0}{-2.4}\setsecondfont{\mathtt}\setthirdfont{\mathsfb}
\Crectangle{A}{1.5}{1.5}{0,0}\Rincome{A}{0}{x} \RxBoxoutcome{A}{1.5}{0}{y}
\Crectangle{B}{1.5}{1.5}{-5,6.5} \RxBoxincome{B}{1.5}{0}{u}
\Crectangle{C}{1.5}{1.5}{5,9} \Routcome{C}{0}{s}
\Crectangle{D}{1.5}{1.5}{-3,13.5}\RxBoxincome{D}{1.5}{0}{x}
\Crectangle{E}{1.5}{1.5}{1,19.5} \Routcome{E}{0}{w}
\Crectangle{v}{0.8}{0.8}{0,7.75} \joinrlnoarrowthick{B}{0}{v}{0} \joinrlnoarrowthick{v}{0}{C}{0}
\jointbnoarrow[below left]{A}{-1}{B}{0}   \csymbolthird{a}
\jointbnoarrow[below right]{A}{1}{C}{0}   \csymbolthird{b}
\jointbnoarrow[above left]{B}{0}{D}{0}   \csymbolthird{c}
\jointbnoarrow[above left]{D}{0}{E}{-1}   \csymbolthird{a}
\jointbnoarrow[above right]{C}{0}{E}{1}   \csymbolthird{b}
\end{Compose}
~~~~~~\equiv~~~~~
\begin{Compose}{0}{-2.4}\setsecondfont{\mathtt}\setthirdfont{\mathsfb}
\crectangle{RR}{0.7}{0.9}{-2,13.5} \csymbol{R}\RxBoxincome{RR}{0.7}{0}{x}
\Crectangle{B}{1.5}{1.5}{-5,6.5} \RxBoxincome{B}{1.5}{0}{u}  \RxBoxoutcome{B}{1.5}{0}{v}
\Crectangle{I}{0.8}{0.8}{-5,10.5} \jointbnoarrow[left]{B}{0}{I}{0} \csymbol{c}
\Crectangle{I}{0.8}{0.8}{-5,2.5} \joinbtnoarrow[left]{B}{0}{I}{0} \csymbol{a}
\crectangle{RR}{0.7}{0.9}{4,0} \csymbol{R}\RxBoxincome{RR}{0.7}{0}{y}
\end{Compose}
\end{equation}
Consequently, the probability associated with this circuit is
\begin{equation}
\frac{1}{N_\mathtt{x}N_\mathtt{y}}\text{prob}\left(
\begin{Compose}{0}{-1.8}\setsecondfont{\mathtt}\setthirdfont{\mathsfb}
\Crectangle{B}{1.5}{1.5}{0,6.5} \RxBoxincome{B}{1.5}{0}{u}  \RxBoxoutcome{B}{1.5}{0}{v}
\Crectangle{I}{0.8}{0.8}{0,10.5} \jointbnoarrow[left]{B}{0}{I}{0} \csymbol{c}
\Crectangle{I}{0.8}{0.8}{0,2.5} \joinbtnoarrow[left]{B}{0}{I}{0} \csymbol{a}
\end{Compose}
\right)
\end{equation}
It is noteworthy that the probability for the readout boxes that are connected by a \lq\lq Z" shape (namely the $x$ and $y$ boxes) factorise, while the probabilities for readout boxes for which we can trace forward in time from incomes to outcomes do not factorize.  Thus, we can see the different causal structures illustrated in Table \ref{ZCTtable} at work in the same circuit.

\part{Duotensor formalism}\label{part:duotensorformulation}

In this part of the paper we will introduce a formulation for general probabilistic theories that is general enough to formulate Quantum Theory.  We will use an elaborated duotensor notation (as used previously in\cite{hardy2010formalism}).   The elaboration is that we need to treat the classical pointer wires as well as the system wires by duotensor methods.  As we will see, this is very natural for a situation in which calculations are done by means of the conditional probability chain rule.  Furthermore, the duotensor formulation presented here allows us very naturally to provide an \emph{operator tensor} formulation of time symmetric operational Quantum Theory as presented in Part \ref{TSOQT}.

\section{A precursor - notating the chain rule}\label{sec:precursor}

As a precursor to this, note that we can calculate
\[
\text{prob}\left(
\begin{Compose}{0}{0} \setsecondfont{\mathtt}
\crectangle{T}{0.7}{1}{-7,0} \csymbol{R}
\crectangle{x}{0.7}{0.7}{-4,0} \csymbolthird{u}
\Crectangle{M}{1}{1}{0,0}
\joinrlnoarrowthick[above]{T}{0}{x}{0} \csymbolalt[0,5]{u}
\joinrlnoarrowthick[above]{x}{0}{M}{0} \csymbolalt[0,5]{u}
\crectangle{Tr}{0.7}{1}{7,0} \csymbol{R}
\crectangle{xr}{0.7}{0.7}{4,0} \csymbolthird{v}
\joinlrnoarrowthick[above]{Tr}{0}{xr}{0} \csymbolalt[0,5]{v}
\joinlrnoarrowthick[above]{xr}{0}{M}{0} \csymbolalt[0,5]{v}
\end{Compose}\right)
\]
in the following way
\begin{equation}
p(u,v) = \sum_{u_1,u_2,v_3,v_4}p_\mathsf{R}(u_1) p_u(u_2|u_1) p_\mathsf{M}(v_3|u_2) p_v(v_4|v_3) p_\mathsf{R}(-|v_4)
\end{equation}
by applying the chain rule for conditional probabilities.  Here we have $p_\mathsf{R}(u_1)=\frac{1}{N_\mathtt{u}}$.  Further $p_u(u_2|u_1)$ is equal to 1 if $u_1=u$ and $u_2=u$, otherwise it is equal to 0. And $p_v(v_2|v_1)$ is defined similarly. Also, $p_\mathsf{R}(-|v_2)$ is equal to 1 for all $v_2$.
We can notate this calculation in more suggestive terms as follows
\begin{equation}\label{RMRcalcsymbolic}
p(u,v) = R^{u_1} B_{u_1}^{u_2}[u] M_{u_2}^{v_3} B_{v_3}^{v_4}[v] R_{v_4}
\end{equation}
where we use Einstein summation convention when there is a raised and lowered index.  Here we define $R^{u_1}=p_\mathsf{R}(u_1)$, $B_{u_1}^{u_2}[u]=p_u(u_2|u_1)$, $M_{u_2}^{v_3}=p_\mathsf{M}(v_3|u_2)$, $B_{v_3}^{v_4}[v]=p_v(v_4|v_3)$, and $R_{v_4}=p_\mathsf{R}(-|v_4)$.  We can also notate this calculation diagrammatically as follows
\begin{equation}\label{RMRcalcdiagram}
p(u,v) ~~=~~~
\begin{Compose}{0}{0}\setdefaultfont{\mathnormal} \setsecondfont{\mathnormal}
\crectangle{T}{0.8}{0.9}{-11,0} \csymbol{R}
\crectangle{x}{0.7}{0.7}{-5.5,0} \csymbolthird{u}
\Crectangle{M}{1}{1}{0,0}
\joinrlnoarrowbw[above]{T}{0}{x}{0} \csymbolalt[0,5]{u}
\joinrlnoarrowbw[above]{x}{0}{M}{0} \csymbolalt[0,5]{u}
\crectangle{Tr}{0.8}{0.9}{11,0} \csymbol{R}
\crectangle{xr}{0.7}{0.7}{5.5,0} \csymbolthird{v}
\joinlrnoarrowwb[above]{Tr}{0}{xr}{0} \csymbolalt[0,5]{v}
\joinlrnoarrowwb[above]{xr}{0}{M}{0} \csymbolalt[0,5]{v}
\end{Compose}
\end{equation}
Unlike previous diagrams, this diagram actually represents a calculation.  We will go into much more detail as to where this diagram comes from. However, a word on notation is warranted at this point.  We have summation when a small black and white square are matched (by being placed just next to each other). The dummy variable for the summation is indicated just above these matchings.  Each variable that lives on a different wire is different. We could indicate this by using integer subscripts as in \eqref{RMRcalcsymbolic}.  However these integer subscripts are unnecessary since we can see from the diagram that the summations happen in different places. On the other hand, the variables that appear inside the boxes are not dummy variables.    The calculation in \eqref{RMRcalcdiagram} is exactly the same as the calculation in \eqref{RMRcalcsymbolic} and, reading from left to right, the terms have the same meaning.

The calculation for the probability can be done backwards in time instead.  Then the calculation, represented diagrammatically, is
\begin{equation}\label{RMRcalcdiagrambackwards}
p(u,v) ~~=~~~
\begin{Compose}{0}{0}\setdefaultfont{\mathnormal} \setsecondfont{\mathnormal}
\crectangle{T}{0.8}{0.9}{-11,0} \csymbol{R}
\crectangle{x}{0.7}{0.7}{-5.5,0} \csymbolthird{u}
\Crectangle{M}{1}{1}{0,0}
\joinrlnoarrowwb[above]{T}{0}{x}{0} \csymbolalt[0,5]{u}
\joinrlnoarrowwb[above]{x}{0}{M}{0} \csymbolalt[0,5]{u}
\crectangle{Tr}{0.8}{0.9}{11,0} \csymbol{R}
\crectangle{xr}{0.7}{0.7}{5.5,0} \csymbolthird{v}
\joinlrnoarrowbw[above]{Tr}{0}{xr}{0} \csymbolalt[0,5]{v}
\joinlrnoarrowbw[above]{xr}{0}{M}{0} \csymbolalt[0,5]{v}
\end{Compose}
\end{equation}
Note that the black and white squares are now in the reverse order.  This calculation represents application of the conditional probability chain rule in the backwards time direction.

\section{Square dots and pointer types}

\subsection{Pointer fiducials}

First we define fiducial elements for the pointer types.  Consider
\begin{equation}
\begin{Compose}{0}{0}
\Crectangle{R}{0.7}{0.7}{-3,0}
\crectangle{x}{0.6}{0.6}{0,0} \csymbolthird{x}
\relpoint{x}{3,0}{xR}
\joinrlnoarrowthick[above]{R}{0}{x}{0} \csymbol[0,5]{x}
\joinrlnoarrowthick[above]{x}{0}{xR}{0} \csymbol[0,5]{x}
\end{Compose}
\end{equation}
We will represent this by different notation, namely
\begin{equation}
\begin{Compose}{0}{-0.09}
\Crectangle{R}{0.7}{0.7}{-4,0}
\blackdotsq{S}{-6,-2}  \csymbolthird[-20,0]{x}
\crectangle{x}{0.65}{2}{0,-1} 
\relpoint{x}{3,1}{xR}  \csymbol[20,0]{x}
\joinrlnoarrowthick[above]{R}{0}{x}{1}   \csymbol[0,5]{x}
\joinrlnoarrow[above]{S}{0}{x}{-1}
\joinrlnoarrowthick[above]{x}{1}{xR}{0}
\end{Compose}
\end{equation}
where the $x$ on the black dot is the variable that \lq\lq belongs" inside the empty box.  This will be the outcome fiducial element (that closes incomes) and we write
\begin{equation}
\begin{Compose}{0}{0} \setsecondfont{\mathtt}
\sleftdatafid{F}{0,0} \blackdotsq{FL}{-2.5,0} \csymbolthird[-24,0]{x}
\relpoint{F}{2.5,0}{FR} \csymbolalt[15,0]{x}
\joinrlnoarrowthick[above]{F}{0}{FR}{0}  \joinrlnoarrow[above]{FL}{0}{F}{0}
\end{Compose}
~~~=~~~
\begin{Compose}{0}{0.2}
\Crectangle{R}{0.7}{0.7}{-4,0}
\blackdotsq{S}{-6,-2}  \csymbolthird[-20,0]{x}
\crectangle{x}{0.65}{2}{0,-1} 
\relpoint{x}{3,1}{xR}  \csymbol[20,0]{x}
\joinrlnoarrowthick[above]{R}{0}{x}{1}   \csymbol[0,5]{x}
\joinrlnoarrow[above]{S}{0}{x}{-1}
\joinrlnoarrowthick[above]{x}{1}{xR}{0}
\end{Compose}
\end{equation}
Similarly, we write
\begin{equation}
\begin{Compose}{0}{0} \setsecondfont{\mathtt}
\srightdatafid{F}{0,0} \relpoint{F}{-2.5,0}{FL} \csymbolalt[-18,0]{x}
\blackdotsq{FR}{2.5,0} \csymbolthird[21,0]{x}
\joinrlnoarrow[above]{F}{0}{FR}{0}  \joinrlnoarrowthick{FL}{0}{F}{0}
\end{Compose}
~~~=~~~
\begin{Compose}{0}{-0.45}
\Crectangle{R}{0.7}{0.7}{4,0}
\blackdotsq{S}{6,2}  \csymbolthird[20,0]{x}
\crectangle{x}{0.65}{2}{0,1} 
\relpoint{x}{-3,-1}{xR}\csymbol[-20,0]{x}
\joinlrnoarrowthick[above]{R}{0}{x}{-1}   \csymbol[0,5]{x}
\joinlrnoarrow[above]{S}{0}{x}{1}
\joinlrnoarrowthick[above]{x}{-1}{xR}{0}
\end{Compose}
\end{equation}
for the income fiducial element.

\subsection{The pointer hopping matrix}\label{sec:thepointerhoppingmatrix}

Now we have defined the pointer fiducial elements, we can define the \emph{pointer hopping matrix}.  This is
\begin{equation}
\begin{Compose}{0}{0}
\bbmatrixnoarrowsq{h}{0,0} \csymbolthird{x}
\end{Compose}
~~=~~
\text{prob}\left(
\begin{Compose}{0}{0}
\blackdotsq{FL}{-2.5,0} \csymbolthird[-24,0]{x}
\sleftdatafid{F}{0,0}
\srightdatafid{rF}{3,0}
\blackdotsq{rFR}{5.5,0} \csymbolthird[24,0]{x}
 \joinrlnoarrow[above]{FL}{0}{F}{0}
\joinrlnoarrowthick[above]{F}{0}{rF}{0} \csymbol{x}
\joinrlnoarrow[above]{rF}{0}{rFR}{0}
\end{Compose}
\right)
\end{equation}
I explain the reason for the terminology \lq\lq hopping matrix" in Sec.\ \ref{sec:symbolicduotensornotation}.
We indicate, by placing the $x$ over the centre, the variable associated with this hopping matrix. Since this is a matrix, it is associated with two copies of this variable. Since circuits are equivalent to their own probabilities (see \eqref{circuitequivprob}) we have
\begin{equation}\label{hoppingequivfidcircuitpointer}
\begin{Compose}{0}{0}
\bbmatrixnoarrowsq{h}{0,0} \csymbolthird{x}
\end{Compose}
~~\equiv~~
\begin{Compose}{0}{0}
\blackdotsq{FL}{-2.5,0} \csymbolthird[-24,0]{x}
\sleftdatafid{F}{0,0}
\srightdatafid{rF}{3,0}
\blackdotsq{rFR}{5.5,0} \csymbolthird[24,0]{x}
 \joinrlnoarrow[above]{FL}{0}{F}{0}
\joinrlnoarrowthick[above]{F}{0}{rF}{0} \csymbol{x}
\joinrlnoarrow[above]{rF}{0}{rFR}{0}
\end{Compose}
\end{equation}
This is very useful in manipulating diagrams between equivalent forms as we will see in Sec.\ \ref{sec:calculatingprobabilitiesforcircuits}.

It is clear that
\begin{equation}
\begin{Compose}{0}{0}
\bbmatrixnoarrowsq{h}{0,0} \csymbolthird{x}
\end{Compose}
~~=~~
\frac{1}{N_\mathtt{x}}
\left(
\begin{array}{cccc}
  1 &   &   &   \\
    & 1  &   &   \\
    &    & \ddots &  \\
    &   &    & 1
\end{array}
\right)
\end{equation}
since
\begin{equation}
\begin{Compose}{0}{0}
\blackdotsq{FL}{-2.5,0} \csymbolthird[-24,0]{x}
\sleftdatafid{F}{0,0}
\srightdatafid{rF}{3,0}
\blackdotsq{rFR}{5.5,0} \csymbolthird[24,0]{x}
 \joinrlnoarrow[above]{FL}{0}{F}{0}
\joinrlnoarrowthick[above]{F}{0}{rF}{0} \csymbol{x}
\joinrlnoarrow[above]{rF}{0}{rFR}{0}
\end{Compose}
\end{equation}
is the same as
\begin{equation}
\begin{Compose}{0}{0} \setsecondfont{\mathtt}
\crectangle{RL}{0.7}{0.9}{0,0} \csymbol{R} \crectangle{xL}{0.7}{0.7}{3.5,0}\csymbolthird{x}
\crectangle{xR}{0.7}{0.7}{7,0}\csymbolthird{x'} \crectangle{RR}{0.7}{0.9}{10.5,0} \csymbol{R}
\joinrlnoarrowthick[above]{RL}{0}{xL}{0} \csymbol{x}
\joinrlnoarrowthick[above]{xL}{0}{xR}{0} \csymbol{x}
\joinrlnoarrowthick[above]{xR}{0}{RR}{0} \csymbol{x}
\end{Compose}
\end{equation}
where we have now distinguished the two instances of the variable $x$ (as $x$ and $x'$) since they are not on wires.  The hopping matrix can be used to change white dots into black dots as we will see shortly.

The inverse of the hopping matrix is
 \begin{equation}
\begin{Compose}{0}{0}
\wwmatrixnoarrowsq{h}{0,0} \csymbolthird{x}
\end{Compose}
~~=~~
N_\mathtt{x}
\left(
\begin{array}{cccc}
  1 &   &   &   \\
    & 1  &   &   \\
    &    & \ddots &  \\
    &   &    & 1
\end{array}
\right)
\end{equation}
This can be used to change black dots into white dots as we will see below.

We have the relationships
\begin{equation}\label{wbidentity}
\begin{Compose}{0}{-0.08}
\wwmatrixnoarrowsq{h}{0,0} 
\bbmatrixnoarrowsq{g}{4,0} 
\end{Compose}
~=~
\begin{Compose}{0}{-0.08}
\wbmatrixnoarrowsq{h}{0,0} 
\end{Compose}
\end{equation}
and
\begin{equation}
\begin{Compose}{0}{-0.08}
\bbmatrixnoarrowsq{h}{0,0} 
\wwmatrixnoarrowsq{g}{4,0} 
\end{Compose}
~=~
\begin{Compose}{0}{-0.08}
\bwmatrixnoarrowsq{h}{0,0} 
\end{Compose}
\end{equation}
where $\wbdotsnoarrowsq$ and $\bwdotsnoarrowsq$ are the identity matrix.

We can change the colour of the square dots by applying the hopping matrix or its inverse as follows
\begin{align*}
\nbdotsnoarrowsq \hspace{2pt} \wwdotsnoarrowsq = \nwdotsnoarrowsq
~~~~ \nwdotsnoarrowsq \hspace{2pt} \bbdotsnoarrowsq = \nbdotsnoarrowsq  \\
\wwdotsnoarrowsq \hspace{2pt} \bndotsnoarrowsq = \wndotsnoarrowsq
~~~~ \bbdotsnoarrowsq \hspace{2pt} \wndotsnoarrowsq = \bndotsnoarrowsq
\end{align*}
This is clear since we can apply the hopping matrix or its inverse again to undo the effect.  Additionally, we see that
\begin{equation}
\nbdotsnoarrowsq \hspace{2pt} \wndotsnoarrowsq ~ = ~ \nbdotsnoarrowsq \hspace{2pt} \wwdotsnoarrowsq \hspace{2pt} \bbdotsnoarrowsq \hspace{2pt} \wndotsnoarrowsq
~ = ~ \nwdotsnoarrowsq \hspace{2pt} \bndotsnoarrowsq
\end{equation}
by insertion of the identity in \eqref{wbidentity}.  This means we can write
\begin{equation}\label{bwtowb}
\nbdotsnoarrowsq \hspace{2pt} \wndotsnoarrowsq ~ = ~ \customdotsnoarrowvar[]{9mm}
~=~ \nwdotsnoarrowsq \hspace{2pt} \bndotsnoarrowsq
\end{equation}
with the understanding that, when we do a calculation, we can insert a black and white dot in either order.

\subsection{Duotensors with square dots}

We can introduce duotensors with square dots in two ways.  We can expand an arbitrary operation only having open pointer wires as being equivalent to a weighted sum of fiducials as follows
\begin{equation}\label{Lexpansion}
\begin{Compose}{0}{0}\setdefaultfont{\mathtt}
\crectangle{T}{3}{3}{0,0} \csymbol{\mathsf{L}}
\thispoint{y7}{-7,1.5}   \csymbol[-24,0]{y} \joinrlnoarrowthick[above]{y7}{0}{T}{1.5}
\thispoint{x1}{-7,-1.5}   \csymbol[-24,0]{x} \joinrlnoarrowthick[above]{x1}{0}{T}{-1.5}
\thispoint{w8}{7,2.5}   \csymbol[24,0]{x} \joinrlnoarrowthick[above]{T}{2.5}{w8}{0}
\thispoint{w6}{7,0}   \csymbol[24,0]{w} \joinrlnoarrowthick[above]{T}{0}{w6}{0}
\thispoint{y2}{7,-2.5}   \csymbol[24,0]{y} \joinrlnoarrowthick[above]{T}{-2.5}{y2}{0}
\end{Compose}
~~~\equiv~~~
\begin{Compose}{0}{0}\setdefaultfont{\mathtt}\setsecondfont{\mathsf}
\crectangle{T}{3}{3}{0,0} \csymbolthird{L}
\srightdatafid{y7}{-7,1.5}   \joinrlnoarrowbw[above]{y7}{0}{T}{1.5}\csymbolthird[0,10]{y}
\thispoint{yL7}{-9.5,1.5}   \csymbolalt[-24,0]{y}  \joinrlnoarrowthick{yL7}{0}{y7}{0}
\srightdatafid{x1}{-7,-1.5}  \joinrlnoarrowbw[above]{x1}{0}{T}{-1.5} \csymbolthird[0,10]{x}
\thispoint{xL1}{-9.5,-1.5} \csymbolalt[-24,0]{x} \joinrlnoarrowthick[above]{xL1}{0}{x1}{0}
\sleftdatafid{w8}{7,2.5}  \joinrlnoarrowwb[above]{T}{2.5}{w8}{0} \csymbolthird[0,10]{x}
\thispoint{wR8}{9.5,2.5} \csymbolalt[24,0]{x}  \joinrlnoarrowthick[above]{w8}{0}{wR8}{0}
\sleftdatafid{w6}{7,0}   \joinrlnoarrowwb[above]{T}{0}{w6}{0} \csymbolthird[0,10]{w}
\thispoint{wR6}{9.5,0} \csymbolalt[24,0]{w}  \joinrlnoarrowthick[above]{w6}{0}{wR6}{0}
\sleftdatafid{y2}{7,-2.5} \joinrlnoarrowwb[above]{T}{-2.5}{y2}{0}  \csymbolthird[0,10]{y}
\thispoint{yR2}{9.5,-2.5} \csymbolalt[24,0]{y}  \joinrlnoarrowthick[above]{y2}{0}{yR2}{0}
\end{Compose}
\end{equation}
The weights are given by a duotensor with all white dots.
Or we can take the probability of the circuit formed when we close the operation with fiducial elements as follows
\begin{equation}\label{Lprob}
\begin{Compose}{0}{0}\setdefaultfont{\mathnormal}\setsecondfont{\mathnormal}
\crectangle{T}{3}{3}{0,0} \csymbol{L}
\blackdotsq{y7}{-7,1.5}   \csymbol[-24,0]{y} \joinrlnoarrowthick[above]{y7}{0}{T}{1.5}
\blackdotsq{x1}{-7,-1.5}   \csymbol[-24,0]{x} \joinrlnoarrowthick[above]{x1}{0}{T}{-1.5}
\blackdotsq{w8}{7,2.5}   \csymbol[24,0]{x} \joinrlnoarrowthick[above]{T}{2.5}{w8}{0}
\blackdotsq{w6}{7,0}   \csymbol[24,0]{w} \joinrlnoarrowthick[above]{T}{0}{w6}{0}
\blackdotsq{y2}{7,-2.5}   \csymbol[24,0]{y} \joinrlnoarrowthick[above]{T}{-2.5}{y2}{0}
\end{Compose}
~~~=~~~
\text{prob}\left(
\begin{Compose}{0}{0}\setdefaultfont{\mathsf}\setsecondfont{\mathsf}
\crectangle{T}{3}{3}{0,0} \csymbolthird{\mathsf{L}}
\sleftdatafid{y7}{-6,1.5}   \joinrlnoarrowthick[above]{y7}{0}{T}{1.5}\csymbol[0,4]{y}
\blackdotsq{yL7}{-8.5,1.5}   \csymbolthird[-24,0]{y}  \joinrlnoarrowthick{yL7}{0}{y7}{0}
\sleftdatafid{x1}{-6,-1.5}  \joinrlnoarrowthick[above]{x1}{0}{T}{-1.5} \csymbol[0,4]{x}
\blackdotsq{xL1}{-8.5,-1.5} \csymbolthird[-24,0]{x} \joinrlnoarrowthick[above]{xL1}{0}{x1}{0}
\srightdatafid{w8}{6,2.5}  \joinrlnoarrowthick[above]{T}{2.5}{w8}{0} \csymbol{x}
\blackdotsq{wR8}{8.5,2.5} \csymbolthird[24,0]{x}  \joinrlnoarrowthick[above]{w8}{0}{wR8}{0}
\srightdatafid{w6}{6,0}   \joinrlnoarrowthick[above]{T}{0}{w6}{0} \csymbol{w}
\blackdotsq{wR6}{8.5,0} \csymbolthird[24,0]{w}  \joinrlnoarrowthick[above]{w6}{0}{wR6}{0}
\srightdatafid{y2}{6,-2.5} \joinrlnoarrowthick[above]{T}{-2.5}{y2}{0}  \csymbol[0,4]{y}
\blackdotsq{yR2}{8.5,-2.5} \csymbolthird[24,0]{y}  \joinrlnoarrowthick[above]{y2}{0}{yR2}{0}
\end{Compose}
\right)
\end{equation}
Now we have a duotensor with all black dots.  We can change black dots to white dots and white to black by appropriate application of the hopping matrix and its inverse.  It is a simple matter to prove that \eqref{Lexpansion} and \eqref{Lprob} are consistent with one another under such transformations.

\subsection{Duotensors for flat pointer operations}

We can write
\begin{equation}\label{Requivalence}
\begin{Compose}{0}{0} \setsecondfont{\mathtt}
\Crectangle{R}{0.7}{0.9}{0,0} \thispoint{p}{4,0} \csymbolalt[20,0]{x} \joinrlnoarrowthick{R}{0}{p}{0}
\end{Compose} ~~~\equiv~~~
\begin{Compose}{0}{0} \setsecondfont{\mathtt}
\sleftdatafid{F}{0,0} \crectangle{FL}{0.7}{0.9}{-5,0} \csymbolthird{R}
\relpoint{F}{2.5,0}{FR} \csymbolalt[15,0]{x}
\joinrlnoarrowthick[above]{F}{0}{FR}{0}  \joinrlnoarrowwb[above]{FL}{0}{F}{0} \csymbolthird[0,10]{x}
\end{Compose}
\end{equation}
where
\begin{equation}
\begin{Compose}{0}{0} \setsecondfont{\mathtt}
\crectangle{R}{0.7}{0.9}{0,0} \csymbolthird{R} \whitedotsq{p}{4,0} \csymbolthird[25,0]{x} \joinrlnoarrow{R}{0}{p}{0}
\end{Compose}
~~=~~
\left(
\begin{array}{cccc}
  1 & 1 & \dots & 1
\end{array}
\right)
\end{equation}
(i.e. it is equal to 1 for each value of $x$). The equivalence in \eqref{Requivalence} says that $\mathsf R$ (which prepares the flat distribution) is equivalent to a sum of the fiducial preparations.  Similarly, we can write
\begin{equation}
\begin{Compose}{0}{0}\setsecondfont{\mathtt}
\Crectangle{R}{0.7}{0.9}{0,0}  \thispoint{p}{-4,0} \csymbolalt[-20,0]{x} \joinlrnoarrowthick{R}{0}{p}{0}
\end{Compose}
~~~\equiv~~~
\begin{Compose}{0}{0} \setsecondfont{\mathtt}
\srightdatafid{F}{0,0} \relpoint{F}{-2.5,0}{FL} \csymbolalt[-18,0]{x}
\crectangle{FR}{0.7}{0.9}{5,0} \csymbolthird{R}
\joinrlnoarrowbw[above]{F}{0}{FR}{0} \csymbolthird[0,10]{x} \joinrlnoarrowthick{FL}{0}{F}{0}
\end{Compose}
\end{equation}
where
\begin{equation}
\begin{Compose}{0}{0}\setsecondfont{\mathtt}
\crectangle{R}{0.7}{0.9}{0,0} \csymbolthird{R}  \whitedotsq{p}{-4,0} \csymbolthird[-25,0]{x} \joinlrnoarrow{R}{0}{p}{0}
\end{Compose}
~~=~~
\left(
\begin{array}{cccc}
  1 & 1 & \dots & 1
\end{array}
\right)
\end{equation}
(i.e. it is equal to 1 for each value of $x$).

We can use the hopping matrix to change the colour of the dot.  We obtain
\begin{equation}
\begin{Compose}{0}{0} \setsecondfont{\mathtt}
\crectangle{R}{0.7}{0.9}{0,0} \csymbolthird{R} \blackdotsq{p}{4,0} \csymbolthird[25,0]{x} \joinrlnoarrow{R}{0}{p}{0}
\end{Compose}
~~=~~
\frac{1}{N_\mathtt{x}}
\left(
\begin{array}{c}
  1 \\
  1 \\
  \vdots \\
  1
\end{array}
\right)
~~~~~~~~~~~~
\begin{Compose}{0}{0}\setsecondfont{\mathtt}
\crectangle{R}{0.7}{0.9}{0,0} \csymbolthird{R}  \blackdotsq{p}{-4,0} \csymbolthird[-25,0]{x} \joinlrnoarrow{R}{0}{p}{0}
\end{Compose}
~~=~~
\frac{1}{N_\mathtt{x}}
\left(
\begin{array}{c}
  1 \\
  1 \\
  \vdots \\
  1
\end{array}
\right)
\end{equation}
confirming that $R$ prepares a flat probability distribution.

We can verify the deterministic property \eqref{Rdeterministiccond} for flat pointer operations
\begin{equation}
\text{prob}\left(
\begin{Compose}{0}{-0.06} \setsecondfont{\mathtt}
\crectangle{R}{0.7}{0.9}{0,0} \csymbolthird{R}
\crectangle{RR}{0.7}{0.9}{4,0} \csymbolthird{R}
\joinrlnoarrowthick{R}{0}{RR}{0} \csymbolalt{x}
\end{Compose}\right)
=
\begin{Compose}{0}{-0.06} \setsecondfont{\mathtt}
\crectangle{R}{0.7}{0.9}{0,0} \csymbol{R}
\crectangle{RR}{0.7}{0.9}{4,0} \csymbol{R}
\joinrlnoarrowbw{R}{0}{RR}{0} \csymbolthird[0,10]{x}
\end{Compose}
=\begin{Compose}{0}{-0.06} \setsecondfont{\mathtt}
\crectangle{R}{0.7}{0.9}{0,0} \csymbolthird{R}
\crectangle{RR}{0.7}{0.9}{4,0} \csymbolthird{R}
\joinrlnoarrowwb{R}{0}{RR}{0} \csymbolthird[0,10]{x}
\end{Compose}
=1
\end{equation}

\subsection{Duotensors for readout boxes}\label{sec:duotensorsforreadoutboxes}

Readout boxes were defined Sec.\ \ref{sec:circuitsandreadoutboxes} as having the property
\begin{equation}\label{xboxpropertyinproof}
\begin{Compose}{0}{0}\setsecondfont{\mathtt}
\crectangle{AL}{0.7}{0.8}{-6,0} \csymbolthird{x} \relpoint{AL}{-2.5,0}{ALL} \csymbolalt[-18,0]{x}
\crectangle{AR}{0.7}{0.8}{-2,0} \csymbolthird{x'}
\relpoint{AR}{2.5,0}{ARR} \csymbolalt[15,0]{x}
\joinrlnoarrowthick{ALL}{0}{AL}{0}
\joinrlnoarrowthick[above]{AL}{0}{AR}{0}   \csymbolalt[0,4]{x}
\joinrlnoarrowthick{AR}{0}{ARR}{0}
\end{Compose}
~~~~
\equiv
~~~~
\arraycolsep=1.4pt\def\arraystretch{1.5}
\left\{
\begin{array}{ll}
\begin{Compose}{0}{-0.1}\setsecondfont{\mathtt}
\crectangle{A}{0.6}{0.7}{0,0} \csymbolthird{x} \relpoint{A}{-2.5,0}{AL} \csymbolalt[-18,0]{x}
\relpoint{A}{2.5,0}{AR} \csymbolalt[15,0]{x}
\joinlrnoarrowthick{A}{0}{AL}{0}
\joinrlnoarrowthick{A}{0}{AR}{0}
\end{Compose}  &
\text{if} ~~x=x'  \\
\begin{Compose}{0}{-0.1}\setsecondfont{\mathtt}
\crectangle{A}{0.6}{0.7}{0,0} \csymbolthird{0} \relpoint{A}{-2.5,0}{AL} \csymbolalt[-18,0]{x}
\relpoint{A}{2.5,0}{AR} \csymbolalt[15,0]{x}
\joinlrnoarrowthick{A}{0}{AL}{0}
\joinrlnoarrowthick{A}{0}{AR}{0}
\end{Compose} &
\text{if} ~~x\not=x'
\end{array}
\right.
\end{equation}
If we expand each of the result boxes in terms of fiducial elements, we obtain the equivalent expression
\begin{equation}\label{xboxpropexpansion}
\begin{Compose}{0}{0}\setsecondfont{\mathtt}
\crectangle{AL}{0.7}{0.8}{-6,0} \csymbolthird{x}
\crectangle{AR}{0.7}{0.8}{6,0} \csymbolthird{x'}
\fiddyleftwire{AL}{0}{x}
\joinwithpointerfidsx{AL}{0}{AR}{0}{x} 
\fiddyrightwire{AR}{0}{x}
\end{Compose}
\end{equation}
We can substitute in the hopping matrix to obtain the equivalent expression
\begin{equation}\label{xboxpropexpansionwithhopping}
\begin{Compose}{0}{0}\setsecondfont{\mathtt}
\crectangle{AL}{0.7}{0.8}{-4,0} \csymbolthird{x}
\crectangle{AR}{0.7}{0.8}{4,0} \csymbolthird{x'}
\fiddyleftwire{AL}{0}{x}
\joinrlnoarrowsqwbbw[above]{AL}{0}{AR}{0}  \csymbolfourth[0,10]{x}
\fiddyrightwire{AR}{0}{x}
\end{Compose}
\end{equation}
We can use \eqref{bwtowb} to obtain
\begin{equation}\label{xboxpropexpansionbw}
\begin{Compose}{0}{0}\setsecondfont{\mathtt}
\crectangle{AL}{0.7}{0.8}{-2.5,0} \csymbolthird{x}
\crectangle{AR}{0.7}{0.8}{2.5,0} \csymbolthird{x'}
\fiddyleftwire{AL}{0}{x}
\joinrlnoarrowbw[above]{AL}{0}{AR}{0}  \csymbolfourth[0,10]{x}
\fiddyrightwirebw{AR}{0}{x}
\end{Compose}
\end{equation}
Meanwhile, a single result box is equivalent to the following expansion
\begin{equation}\label{xboxexpansionbw}
\begin{Compose}{0}{0}\setsecondfont{\mathtt}
\crectangle{AL}{0.7}{0.8}{-6,0} \csymbolthird{x}
\fiddyleftwire{AL}{0}{x}
\fiddyrightwirebw{AL}{0}{x}
\end{Compose}
\end{equation}
Hence, in order for \eqref{xboxpropertyinproof} to be satisfied, we require that
\begin{equation}\label{xboxbw}
\begin{Compose}{0}{0}\setsecondfont{\mathtt}
\crectangle{AL}{0.7}{0.8}{-6,0} \csymbolthird{x}
\blacksquarerightwire{AL}{0}{x}
\whitesquareleftwire{AL}{0}{x}
\end{Compose}
~=~
\left(
\begin{array}{ccccc}
0 &   & & &  \\
  & 0 & & &  \\
  &   & \ddots &   &  \\
  &   &        & 1 &  \\
  &   &        &   & \ddots
\end{array}
\right)
\end{equation}
where the only non-zero entry is a 1 in the $x$ position along the diagonal (where $x$ is the entry in the box).

With this we can easily confirm the flat property given in \eqref{Rflatcond}, namely that
\begin{equation}
\text{prob}\left(
\begin{Compose}{0}{-0.06} \setsecondfont{\mathtt}
\crectangle{RR}{0.7}{0.9}{0,0}\csymbol{R}
\RxBoxoutcome{RR}{0.9}{0}{x}
\end{Compose}\right)
~=~
\begin{Compose}{0}{-0.06} \setdefaultfont{} \setsecondfont{\mathtt}
\Crectangle{R}{0.7}{0.9}{-5,0}
\crectangle{RR}{0.7}{0.9}{5,0}\csymbol{R}
\Crectangle{x}{0.7}{0.7}{0,0}
\joinrlnoarrowbw[above]{R}{0}{x}{0} \csymbol[0,4]{x}
\joinrlnoarrowwb[above]{x}{0}{RR}{0} \csymbol[0,4]{x}
\end{Compose}
~=~
\frac{1}{N_\mathtt{x}}
\end{equation}
by inserting in the expressions for these duotensors we obtained above.

Another interesting exercise is to prove the equivalence for midcomes \eqref{midcomeequivalence} discussed in Sec.\ \ref{sec:midcomes}.

\subsection{Interpretation of duotensors with square dots}\label{sec:interpretationofduotensorswithsquaredots}

Duotensors with black and white square dots can be interpreted as conditional probabilities of the form
\begin{equation}\label{squaredotinterpretation}
\text{prob}(\text{variables having black square dots}|\text{variables having white square dots})
\end{equation}
For example
\begin{equation}
\begin{Compose}{0}{0}
\crectangle{M}{3}{2}{0,0} \csymbol{M}
\whitedotsq{u}{-5,0} \csymbolthird[-20,0]{u}
\blackdotsq{v}{5,0} \csymbolthird[20,0]{v}
\joinrlnoarrowthick[above]{u}{0}{M}{0}
\joinrlnoarrowthick[above]{M}{0}{v}{0}
\end{Compose}
~=~ \text{prob}(v|u)
\end{equation}
Given this interpretation we can now understand that the diagram
\begin{equation}
p(u,v) ~~=~~~
\begin{Compose}{0}{0}\setdefaultfont{\mathnormal} \setsecondfont{\mathnormal}
\crectangle{T}{0.8}{0.9}{-11,0} \csymbol{R}
\crectangle{x}{0.7}{0.7}{-5.5,0} \csymbolthird{u}
\Crectangle{M}{1}{1}{0,0}
\joinrlnoarrowbw[above]{T}{0}{x}{0} \csymbolalt[0,5]{u}
\joinrlnoarrowbw[above]{x}{0}{M}{0} \csymbolalt[0,5]{u}
\crectangle{Tr}{0.8}{0.9}{11,0} \csymbol{R}
\crectangle{xr}{0.7}{0.7}{5.5,0} \csymbolthird{v}
\joinlrnoarrowwb[above]{Tr}{0}{xr}{0} \csymbolalt[0,5]{v}
\joinlrnoarrowwb[above]{xr}{0}{M}{0} \csymbolalt[0,5]{v}
\end{Compose}
\end{equation}
as one way of writing down the calculation
\begin{equation}
p(u,v) = \sum_{u_1,u_2,v_3,v_4}p_\mathsf{R}(u_1) p_u(u_2|u_1) p_\mathsf{M}(v_3|u_2) p_v(v_4|v_3) p_\mathsf{R}(-|v_4)
\end{equation}
as discussed in Sec.\ \ref{sec:precursor}. This corresponds to doing the calculation forwards in time.   We can, alternately, reverse the order of the black and white dots as permitted by \eqref{bwtowb} and write this calculation down as
\begin{equation}
p(u,v) ~~=~~~
\begin{Compose}{0}{0}\setdefaultfont{\mathnormal} \setsecondfont{\mathnormal}
\crectangle{T}{0.8}{0.9}{-11,0} \csymbol{R}
\crectangle{x}{0.7}{0.7}{-5.5,0} \csymbolthird{u}
\Crectangle{M}{1}{1}{0,0}
\joinrlnoarrowwb[above]{T}{0}{x}{0} \csymbolalt[0,5]{u}
\joinrlnoarrowwb[above]{x}{0}{M}{0} \csymbolalt[0,5]{u}
\crectangle{Tr}{0.8}{0.9}{11,0} \csymbol{R}
\crectangle{xr}{0.7}{0.7}{5.5,0} \csymbolthird{v}
\joinlrnoarrowbw[above]{Tr}{0}{xr}{0} \csymbolalt[0,5]{v}
\joinlrnoarrowbw[above]{xr}{0}{M}{0} \csymbolalt[0,5]{v}
\end{Compose}
\end{equation}
This corresponds to doing the calculation backwards in time.  We can also invert some of the black and white dot pairs but not all of them. For example,
\begin{equation}
p(u,v) ~~=~~~
\begin{Compose}{0}{0}\setdefaultfont{\mathnormal} \setsecondfont{\mathnormal}
\crectangle{T}{0.8}{0.9}{-11,0} \csymbol{R}
\crectangle{x}{0.7}{0.7}{-5.5,0} \csymbolthird{u}
\Crectangle{M}{1}{1}{0,0}
\joinrlnoarrowwb[above]{T}{0}{x}{0} \csymbolalt[0,5]{u}
\joinrlnoarrowwb[above]{x}{0}{M}{0} \csymbolalt[0,5]{u}
\crectangle{Tr}{0.8}{0.9}{11,0} \csymbol{R}
\crectangle{xr}{0.7}{0.7}{5.5,0} \csymbolthird{v}
\joinlrnoarrowwb[above]{Tr}{0}{xr}{0} \csymbolalt[0,5]{v}
\joinlrnoarrowwb[above]{xr}{0}{M}{0} \csymbolalt[0,5]{v}
\end{Compose}
\end{equation}
This is also a calculation permitted by the chain rule of probability - though one which is not interpretable as doing the calculation either forwards or backwards in time.  Given this freedom, we can safely write the calculation down as
\begin{equation}
p(u,v) ~~=~~~
\begin{Compose}{0}{0}\setdefaultfont{\mathnormal} \setsecondfont{\mathnormal}
\crectangle{T}{0.8}{0.9}{-9,0} \csymbol{R}
\crectangle{x}{0.7}{0.7}{-4.5,0} \csymbolthird{u}
\Crectangle{M}{1}{1}{0,0}
\joinrlnoarrow[above]{T}{0}{x}{0} \csymbolalt[0,5]{u}
\joinrlnoarrow[above]{x}{0}{M}{0} \csymbolalt[0,5]{u}
\crectangle{Tr}{0.8}{0.9}{9,0} \csymbol{R}
\crectangle{xr}{0.7}{0.7}{4.5,0} \csymbolthird{v}
\joinlrnoarrow[above]{Tr}{0}{xr}{0} \csymbolalt[0,5]{v}
\joinlrnoarrow[above]{xr}{0}{M}{0} \csymbolalt[0,5]{v}
\end{Compose}
\end{equation}
meaning that we are free to insert black and white dot pairs in whichever order we wish.

\section{Round dots and system types}

\subsection{Fiducial elements for system types}

Consider a preparation for a system type such as
\begin{equation}
\begin{Compose}{0}{0}\setsecondfont{\mathsfb}
\Crectangle{A}{1.2}{1.2}{0,0}  \thispoint{p}{0,4} \csymbolalt[0,20]{x} \jointbnoarrow{A}{0}{p}{0}
\end{Compose}
\end{equation}
We have absorbed any pointer wires and readout boxes into the $\mathsf A$ box (this means that, against our usual convention, $\mathsf A$ may not be doubly deterministic).  There are many such preparations we can consider.  We introduce a fiducial set of preparations for system $\mathsfb{x}$ labeled by $\mathpzc{x}=1, 2, \dots K_\mathsfb{x}$ (we are using the \verb+\mathpzc+ alphabet defined in \cite{pakin2009comprehensive}).  We represent these fiducial preparations by
\begin{equation}
\begin{Compose}{0}{0}\setsecondfont{\mathsfb} \setthirdfont{\mathpzc}
\supsysfid{A}{0,0}  \thispoint{p}{0,3} \csymbolalt[0,20]{x} \jointbnoarrow{A}{0}{p}{0} \blackdot{d}{0,-2} \csymbolthird[0,-23]{x}\joinbtnoarrow{A}{0}{d}{0}
\end{Compose}
\end{equation}
This fiducial set must be such that we can write an arbitrary preparation as being equivalent to a weighted sum of fiducials
\begin{equation}
\begin{Compose}{0}{-0.5}\setsecondfont{\mathsfb}
\Crectangle{A}{1.2}{1.2}{0,0}  \thispoint{p}{0,4} \csymbolalt[0,20]{x} \jointbnoarrow{A}{0}{p}{0}
\end{Compose}
~~\equiv~~
\begin{Compose}{0}{0}\setsecondfont{\mathsfb} \setthirdfont{\mathpzc}
\supsysfid{A}{0,1}  \thispoint{p}{0,3} \csymbolalt[0,20]{x} \jointbnoarrow{A}{0}{p}{0} \crectangle{d}{1.2}{1.2}{0,-4}  \csymbolfourth{A} \joinbtnoarrowdotsbw[right]{A}{0}{d}{0}\csymbolthird[10,0]{x}
\end{Compose}
\end{equation}
where the black and white (round) dots placed next to each other indicate we sum over the associated variable ($\mathpzc{x}$ in this case).
We assume that the fiducial set is complete but not over-complete (so it does not have more elements than required to write down an expression equivalent to an arbitrary preparation for the given system type).

We can, in the same way, define fiducial results so we have
\begin{equation}
\begin{Compose}{0}{-0.5}\setsecondfont{\mathsfb}
\Crectangle{B}{1.2}{1.2}{0,4}  \thispoint{p}{0,0} \csymbolalt[0,-20]{x} \joinbtnoarrow{B}{0}{p}{0}
\end{Compose}
~~\equiv~~
\begin{Compose}{0}{0}\setsecondfont{\mathsfb} \setthirdfont{\mathpzc}
\Crectangle{B}{1.2}{1.2}{0,4}  \fiddydownwire{B}{0}{x}
\end{Compose}
\end{equation}
where, again, the black and white dot placed next to each like this indicate summation.

\subsection{Hopping matrix for system types}

We define the hopping matrix for system types as follows
\begin{equation}\label{operationhoppingmatrix}
\begin{Compose}{0}{0} \setfourthfont{\mathpzc}
\blackdot{u}{0,1.5}\blackdot{d}{0,-1.5} \jointbnoarrow[left]{d}{0}{u}{0}\csymbolfourth{x}
\end{Compose}
~~=~~
\text{prob}
\left(
\begin{Compose}{0}{0} \setsecondfont{\mathsfb} \setfourthfont{\mathpzc}
\blackdot{u}{0,3.5} \csymbolfourth[0,20]{x}\sdownsysfid{up}{0,1.5} \supsysfid{do}{0,-1.5} \blackdot{d}{0,-3.5}\csymbolfourth[0,-20]{x}
\jointbnoarrow[left]{d}{0}{do}{0} \jointbnoarrow[left]{do}{0}{up}{0} \csymbolalt{x} \jointbnoarrow[left]{up}{0}{u}{0}
\end{Compose}
\right)
\end{equation}
Thus, the entries in the hopping matrix are equal to the probabilities formed from a fiducial preparation followed by a fiducial measurement.

Since a circuit is equivalent to the probabilities for that circuit (by \eqref{circuitequivprob}), we have
\begin{equation}\label{hoppingequivfidcircuitsystem}
\begin{Compose}{0}{0} \setfourthfont{\mathpzc}
\blackdot{u}{0,1.5}\blackdot{d}{0,-1.5} \jointbnoarrow[left]{d}{0}{u}{0}\csymbolfourth{x}
\end{Compose}
~~\equiv~~
\begin{Compose}{0}{0} \setsecondfont{\mathsfb} \setfourthfont{\mathpzc}
\blackdot{u}{0,3.5} \csymbolfourth[0,20]{x}\sdownsysfid{up}{0,1.5} \supsysfid{do}{0,-1.5} \blackdot{d}{0,-3.5}\csymbolfourth[0,-20]{x}
\jointbnoarrow[left]{d}{0}{do}{0} \jointbnoarrow[left]{do}{0}{up}{0} \csymbolalt{x} \jointbnoarrow[left]{up}{0}{u}{0}
\end{Compose}
\end{equation}
This will be useful in Sec.\ \ref{sec:calculatingprobabilitiesforcircuits}.

We define the inverse of the hopping matrix as
\begin{equation}
\begin{Compose}{0}{0} \setfourthfont{\mathpzc}
\whitedot{u}{0,1.5}\whitedot{d}{0,-1.5} \jointbnoarrow[left]{d}{0}{u}{0}\csymbolfourth{x}
\end{Compose}
\end{equation}
so that we have
\begin{equation}\label{rounddotsidentity}
\begin{Compose}{0}{0} \setfourthfont{\mathpzc}
\blackdot{u}{0,2.5}\whitedot{d}{0,-2.5} \jointbnoarrowdotswb[left]{d}{0}{u}{0}
\end{Compose}
~=~
\begin{Compose}{0}{0} \setfourthfont{\mathpzc}
\blackdot{u}{0,1.5}\whitedot{d}{0,-1.5} \jointbnoarrow[left]{d}{0}{u}{0}
\end{Compose}
~~~~~~ \text{and}~~~~~
\begin{Compose}{0}{0} \setfourthfont{\mathpzc}
\whitedot{u}{0,2.5}\blackdot{d}{0,-2.5} \jointbnoarrowdotsbw[left]{d}{0}{u}{0}
\end{Compose}
~=~
\begin{Compose}{0}{0} \setfourthfont{\mathpzc}
\whitedot{u}{0,1.5}\blackdot{d}{0,-1.5} \jointbnoarrow[left]{d}{0}{u}{0}
\end{Compose}
\end{equation}
where the left hand side of each equation above is equal to the identity.

As with square dots, we can change the colour of a dot by applying the hopping matrix or its inverse.  For example
\begin{equation}
\begin{Compose}{0}{0} \setfourthfont{\mathpzc}
\blackdot{u}{0,2.5}\thispoint{d}{0,-2.5} \jointbnoarrowdotswb[left]{d}{0}{u}{0}
\end{Compose}
~=~
\begin{Compose}{0}{0} \setfourthfont{\mathpzc}
\blackdot{u}{0,1.5}\thispoint{d}{0,-1.5} \jointbnoarrow[left]{d}{0}{u}{0}
\end{Compose}
\end{equation}
Also, as with square dots, we can write
\begin{equation}\label{bwtowbround}
\begin{Compose}{0}{0} \setfourthfont{\mathpzc}
\thispoint{u}{0,2.5}\thispoint{d}{0,-2.5} \jointbnoarrowdotswb[left]{d}{0}{u}{0}
\end{Compose}
~=~
\begin{Compose}{0}{0} \setfourthfont{\mathpzc}
\thispoint{u}{0,2.5}\thispoint{d}{0,-2.5} \jointbnoarrow[left]{d}{0}{u}{0}
\end{Compose}
~=~
\begin{Compose}{0}{0} \setfourthfont{\mathpzc}
\thispoint{u}{0,2.5}\thispoint{d}{0,-2.5} \jointbnoarrowdotsbw[left]{d}{0}{u}{0}
\end{Compose}
\end{equation}
This is clear since we can insert the identity and then use \eqref{rounddotsidentity}.

\subsection{Duotensors with round dots}\label{sec:duotensorswithrounddots}

We assume we expand a general operation as being equivalent to a weighted sum of fiducials as follows
\begin{equation}
\begin{Compose}{0}{0}\setsecondfont{\mathsfb} \setthirdfont{\mathpzc}
\Crectangle{A}{4}{2}{0,0}
\upwire{A}{-3}{x} \upwire{A}{0}{w} \upwire{A}{3}{y}
\downwire{A}{-2}{y} \downwire{A}{2}{z}
\end{Compose}
~~~\equiv~~~
\begin{Compose}{0}{0}\setsecondfont{\mathsfb} \setthirdfont{\mathpzc}
\crectangle{A}{4}{2}{0,0} \csymbolfourth{A}
\fiddyupwire{A}{-3}{x} \fiddyupwire{A}{0}{w} \fiddyupwire{A}{3}{y}
\fiddydownwire{A}{-2}{y} \fiddydownwire{A}{2}{z}
\end{Compose}
\end{equation}
In this case the duotensor providing the weights has all white dots. The assumption that we can expand an arbitrary operation as being equivalent to a such a weighted sum is equivalent to the assumption of tomographic locality (see \cite{hardy2011reformulating} for proof of this).  In Quantum Theory this assumption holds because we have underlying complex Hilbert spaces (rather than real or quaternionic for example).

We can also consider the probability of the circuit formed by appending fiducial elements to every input and output.
\begin{equation}
\begin{Compose}{0}{0}\setsecondfont{\mathsfb} \setthirdfont{\mathpzc}
\crectangle{A}{4}{2}{0,0} \csymbolfourth{A}
\dotupwire{A}{-3}{x} \dotupwire{A}{0}{w} \dotupwire{A}{3}{y}
\dotdownwire{A}{-2}{y} \dotdownwire{A}{2}{z}
\end{Compose}
~~~=~~~
\text{prob}\left(
\begin{Compose}{0}{0}\setsecondfont{\mathsfb} \setthirdfont{\mathpzc}
\Crectangle{A}{4}{2}{0,0}
\dotfidupwire{A}{-3}{x} \dotfidupwire{A}{0}{w} \dotfidupwire{A}{3}{y}
\dotfiddownwire{A}{-2}{y} \dotfiddownwire{A}{2}{z}
\end{Compose}
\right)
\end{equation}
In this case the duotensor has all black dots.  We can change the colour of dots using the hopping matrix.  It is a simple matter to prove consistency of the above two ways of introducing duotensors under these colour changes (see also \cite{hardy2011reformulating} for proof of this).

A duotensor that has only black round dots can be interpreted as a list of probabilities.  If, however, it has some white dots then it may contain negative numbers since the inverse of the hopping matrix (with round dots) can contain negative numbers.  Hence, we cannot adopt an interpretation of duotensors with black and white round dots as conditional probabilities as we did in \eqref{squaredotinterpretation} for the square dot case.

\section{General duotensor framework}

In this section we will describe how to put all this together so we have round and square dots at the same time.  This will enable us to do calculations for arbitrary circuits.

\subsection{General duotensors with round and square dots}\label{sec:generalduotensorswithroundandsquaredots}

We assume we can expand an arbitrary operation as follows
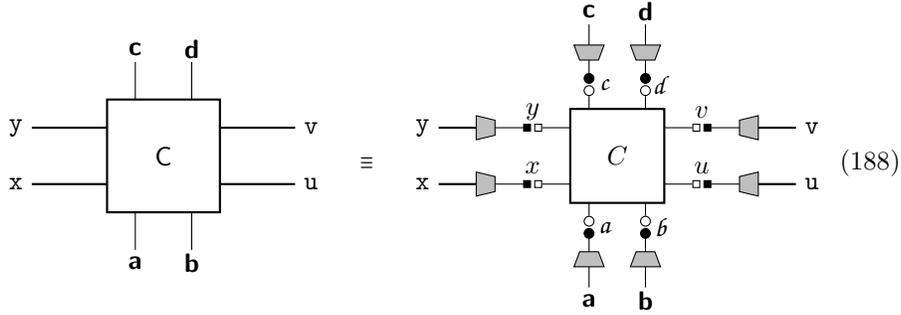
\begin{equation}\label{Lexpansiongeneral}
\begin{Compose}{0}{0}\setsecondfont{\mathsfb} \setthirdfont{\mathpzc}  \setdefaultfont{\mathtt}
\crectangle{T}{3}{3}{0,0} \csymbol{\mathsf{C}}
\thispoint{y7}{-7,1.5}   \csymbol[-24,0]{y} \joinrlnoarrowthick[above]{y7}{0}{T}{1.5}
\thispoint{x1}{-7,-1.5}   \csymbol[-24,0]{x} \joinrlnoarrowthick[above]{x1}{0}{T}{-1.5}
\thispoint{w8}{7,1.5}   \csymbol[24,0]{v} \joinrlnoarrowthick[above]{T}{1.5}{w8}{0}
\thispoint{y2}{7,-1.5}   \csymbol[24,0]{u} \joinrlnoarrowthick[above]{T}{-1.5}{y2}{0}
\upwire{T}{-1.5}{c}  \upwire{T}{1.5}{d}
\downwire{T}{-1.5}{a} \downwire{T}{1.5}{b}
\end{Compose}
~~~\equiv~~~
\begin{Compose}{0}{0}\setsecondfont{\mathsfb} \setthirdfont{\mathpzc} \setdefaultfont{\mathtt}
\crectangle{T}{2.5}{2.5}{0,0} \csymbolfourth{C}
\srightdatafid{y7}{-7,1.5}   \joinrlnoarrowbw[above]{y7}{0}{T}{1.5}\csymbolfourth[0,10]{y}
\thispoint{yL7}{-9.5,1.5}   \csymbol[-24,0]{y}  \joinrlnoarrowthick{yL7}{0}{y7}{0}
\srightdatafid{x1}{-7,-1.5}  \joinrlnoarrowbw[above]{x1}{0}{T}{-1.5} \csymbolfourth[0,10]{x}
\thispoint{xL1}{-9.5,-1.5} \csymbol[-24,0]{x} \joinrlnoarrowthick[above]{xL1}{0}{x1}{0}
\sleftdatafid{w8}{7,1.5}  \joinrlnoarrowwb[above]{T}{1.5}{w8}{0} \csymbolfourth[0,10]{v}
\thispoint{wR8}{9.5,1.5} \csymbol[24,0]{v}  \joinrlnoarrowthick[above]{w8}{0}{wR8}{0}
%
%
\sleftdatafid{y2}{7,-1.5} \joinrlnoarrowwb[above]{T}{-1.5}{y2}{0}  \csymbolfourth[0,10]{u}
\thispoint{yR2}{9.5,-1.5} \csymbol[24,0]{u}  \joinrlnoarrowthick[above]{y2}{0}{yR2}{0}
\fiddyupwire{T}{-1.5}{c}  \fiddyupwire{T}{1.5}{d}
\fiddydownwire{T}{-1.5}{a} \fiddydownwire{T}{1.5}{b}
\end{Compose}
\end{equation}
We call this assumption \emph{full decomposability}.  It is equivalent to the assumption of tomographic locality \cite{hardy2011reformulating}.
In this expansion the duotensor with white square and round dots provides the weights.

We can also write
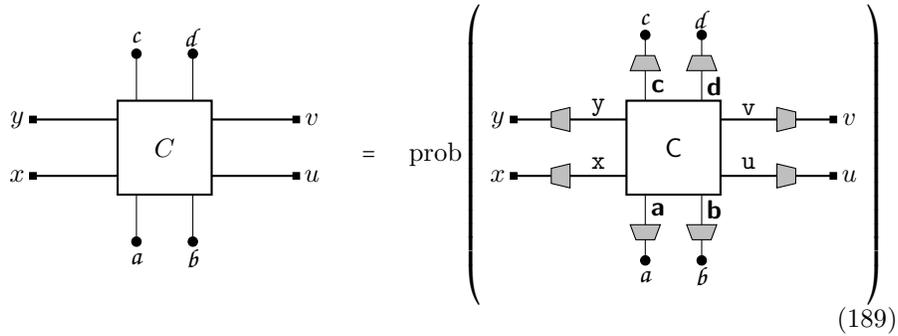
\begin{equation}\label{Lfidprobs}
\begin{Compose}{0}{0}\setsecondfont{\mathsfb} \setthirdfont{\mathpzc}  \setdefaultfont{\mathtt}
\crectangle{T}{2.5}{2.5}{0,0} \csymbolfourth{C}
\blackdotsq{y7}{-7,1.5}   \csymbolfourth[-24,0]{y} \joinrlnoarrowthick[above]{y7}{0}{T}{1.5}
\blackdotsq{x1}{-7,-1.5}   \csymbolfourth[-24,0]{x} \joinrlnoarrowthick[above]{x1}{0}{T}{-1.5}
\blackdotsq{w8}{7,1.5}   \csymbolfourth[24,0]{v} \joinrlnoarrowthick[above]{T}{1.5}{w8}{0}
\blackdotsq{y2}{7,-1.5}   \csymbolfourth[24,0]{u} \joinrlnoarrowthick[above]{T}{-1.5}{y2}{0}
\dotupwire{T}{-1.5}{c}  \dotupwire{T}{1.5}{d}
\dotdownwire{T}{-1.5}{a} \dotdownwire{T}{1.5}{b}
\end{Compose}
~~~=~~~
\text{prob}\left(
\begin{Compose}{0}{0} \setdefaultfont{\mathtt}\setsecondfont{\mathsfb} \setthirdfont{\mathpzc}
\crectangle{T}{2.5}{2.5}{0,0} \csymbolthird{\mathsf{C}}
\sleftdatafid{y7}{-6,1.5}   \joinrlnoarrowthick[above]{y7}{0}{T}{1.5}\csymbol[0,4]{y}
\blackdotsq{yL7}{-8.5,1.5}   \csymbolfourth[-24,0]{y}  \joinrlnoarrowthick{yL7}{0}{y7}{0}
\sleftdatafid{x1}{-6,-1.5}  \joinrlnoarrowthick[above]{x1}{0}{T}{-1.5} \csymbol[0,4]{x}
\blackdotsq{xL1}{-8.5,-1.5} \csymbolfourth[-24,0]{x} \joinrlnoarrowthick[above]{xL1}{0}{x1}{0}
\srightdatafid{w8}{6,1.5}  \joinrlnoarrowthick[above]{T}{1.5}{w8}{0} \csymbol{v}
\blackdotsq{wR8}{8.5,1.5} \csymbolfourth[24,0]{v}  \joinrlnoarrowthick[above]{w8}{0}{wR8}{0}
\srightdatafid{y2}{6,-1.5} \joinrlnoarrowthick[above]{T}{-1.5}{y2}{0}  \csymbol[0,4]{u}
\blackdotsq{yR2}{8.5,-1.5} \csymbolfourth[24,0]{u}  \joinrlnoarrowthick[above]{y2}{0}{yR2}{0}
\dotfidupwire{T}{-1.5}{c}  \dotfidupwire{T}{1.5}{d}
\dotfiddownwire{T}{-1.5}{a} \dotfiddownwire{T}{1.5}{b}
\end{Compose}
\right)
\end{equation}
where, now, the duotensor has black round and square dots.  We can change the colour of these dots by applying the appropriate hopping matrices.

\subsection{Expanding the wire}\label{sec:expandingthewire}

We can think of a wire as an operation and expand it in terms of fiducial elements.
\begin{equation}
\begin{Compose}{0}{-0.09} \setsecondfont{\mathtt}
\linebyhand[thick]{-3,0}{3,0}  \csymbolalt[0,20]{x}
\end{Compose}
~\equiv~
\begin{Compose}{0}{-0.09}\setsecondfont{\mathtt}
\Crectangle{W}{1}{1}{0,0}
\thispoint{l}{-3.5,0}  \joinrlnoarrowthick[above]{l}{0}{W}{0} \csymbolalt{x}
\thispoint{r}{3.5,0}   \joinrlnoarrowthick[above]{W}{0}{r}{0} \csymbolalt{x}
\end{Compose}
\end{equation}
We can expand the right hand side in terms of fiducials
\begin{equation}\label{wireexpandedwithfids}
\begin{Compose}{0}{-0.09} \setsecondfont{\mathtt}
\linebyhand[thick]{-3,0}{3,0}  \csymbolalt[0,20]{x}
\end{Compose}
~\equiv~
\begin{Compose}{0}{-0.09}\setsecondfont{\mathtt}
\Crectangle{W}{1}{1}{0,0}
\fiddyleftwire{W}{0}{x}
\fiddyrightwire{W}{0}{x}
\end{Compose}
\end{equation}
We can place a fiducial on each end to obtain
\begin{equation}
\begin{Compose}{0}{-0.09} \setsecondfont{\mathtt}
\sleftdatafid{l}{-2,0} \blacksquareleftwire{l}{0}{x}
\srightdatafid{r}{2,0} \blacksquarerightwire{r}{0}{x}
\joinrlnoarrowthick[above]{l}{0}{r}{0}   \csymbolalt{x}
\end{Compose}
~\equiv~
\begin{Compose}{0}{-0.09}\setsecondfont{\mathtt}
\Crectangle{W}{1}{1}{0,0}
\fiddyleftwire{W}{0}{x}
\fiddyrightwire{W}{0}{x}
\sleftdatafid{l}{-7.55,0} \blacksquareleftwire{l}{0}{x}
\srightdatafid{r}{7.55,0} \blacksquarerightwire{r}{0}{x}
\end{Compose}
\end{equation}
Using \eqref{hoppingequivfidcircuitpointer} we obtain
\begin{equation}
\begin{Compose}{0}{-0.09}\setsecondfont{\mathtt}
\Crectangle{W}{1}{1}{0,0}
\whitesquareleftwire{W}{0}{x}
\whitesquarerightwire{W}{0}{x}
\end{Compose}
~\equiv~
\begin{Compose}{0}{-0.09}\setsecondfont{\mathtt}
\thispoint{W}{0,0} \csymbolthird[0,20]{x}
\whitesquareleftwire{W}{0}{}
\whitesquarerightwire{W}{0}{}
\end{Compose}
\end{equation}
Inserting this back into \eqref{wireexpandedwithfids} gives the important result
\begin{equation}\label{wirepointertomloc}
\begin{Compose}{0}{-0.09} \setsecondfont{\mathtt}
\linebyhand[thick]{-3,0}{3,0}  \csymbolalt[0,20]{x}
\end{Compose}
~\equiv~
\begin{Compose}{0}{-0.09}\setsecondfont{\mathtt}
\srightdatafid{l}{-1.5,0} \thispoint{ll}{-4,0} \csymbolalt[-20,0]{x} \joinlrnoarrowthick{l}{0}{ll}{0}
\sleftdatafid{r}{1.5,0}   \thispoint{rr}{4,0} \csymbolalt[20,0]{x} \joinlrnoarrowthick{rr}{0}{r}{0}
\joinrlnoarrow[above]{l}{0}{r}{0} \csymbolthird{x}
\end{Compose}
\end{equation}
where we have cancelled over pairs of black and white dots. We call \eqref{wirepointertomloc} the \emph{pointer wire expansion property}.  We can reinsert pairs of black and white dots on the $x$ wire as we wish
\begin{equation}
\begin{Compose}{0}{-0.09} \setsecondfont{\mathtt}
\linebyhand[thick]{-3,0}{3,0}  \csymbolalt[0,20]{x}
\end{Compose}
~\equiv~
\begin{Compose}{0}{-0.09}\setsecondfont{\mathtt}
\srightdatafid{l}{-2,0} \thispoint{ll}{-4,0} \csymbolalt[-20,0]{x} \joinlrnoarrowthick{l}{0}{ll}{0}
\sleftdatafid{r}{2,0}   \thispoint{rr}{4,0} \csymbolalt[20,0]{x} \joinlrnoarrowthick{rr}{0}{r}{0}
\joinrlnoarrowbw[above]{l}{0}{r}{0} \csymbolthird[0,5]{x}
\end{Compose}
~\equiv~
\begin{Compose}{0}{-0.09}\setsecondfont{\mathtt}
\srightdatafid{l}{-2,0} \thispoint{ll}{-4,0} \csymbolalt[-20,0]{x} \joinlrnoarrowthick{l}{0}{ll}{0}
\sleftdatafid{r}{2,0}   \thispoint{rr}{4,0} \csymbolalt[20,0]{x} \joinlrnoarrowthick{rr}{0}{r}{0}
\joinrlnoarrowwb[above]{l}{0}{r}{0} \csymbolthird[0,5]{x}
\end{Compose}
\end{equation}
as this is useful when doing calculations.

By exactly analogous reasoning, we can prove the \emph{system wire expansion property}
\begin{equation} \label{wiresystomloc}
\begin{Compose}{0}{-0.09} \setsecondfont{\mathsfb} \setthirdfont{\mathpzc}
\linebyhand[thick]{0,-3}{0,3}  \csymbolalt[-20,0]{x}
\end{Compose}
~\equiv~
\begin{Compose}{0}{-0.09}\setsecondfont{\mathsfb} \setthirdfont{\mathpzc}
\supsysfid{u}{0, 1.5} \thispoint{uu}{0,4} \csymbolalt[0,20]{x} \jointbnoarrowthick{u}{0}{uu}{0}
\sdownsysfid{d}{0,-1.5}   \thispoint{dd}{0,-4} \csymbolalt[0,-20]{x} \jointbnoarrowthick{dd}{0}{d}{0}
\jointbnoarrow[right]{d}{0}{u}{0} \csymbolthird{x}
\end{Compose}
\end{equation}
We can insert black and white dot in either order
\begin{equation}
\begin{Compose}{0}{-0.09} \setsecondfont{\mathsfb} \setthirdfont{\mathpzc}
\linebyhand[thick]{0,-3}{0,3}  \csymbolalt[-20,0]{x}
\end{Compose}
~\equiv~
\begin{Compose}{0}{-0.09}\setsecondfont{\mathsfb} \setthirdfont{\mathpzc}
\supsysfid{u}{0, 2} \thispoint{uu}{0,4} \csymbolalt[0,20]{x} \jointbnoarrowthick{u}{0}{uu}{0}
\sdownsysfid{d}{0,-2}   \thispoint{dd}{0,-4} \csymbolalt[0,-20]{x} \jointbnoarrowthick{dd}{0}{d}{0}
\jointbnoarrowdotswb[right]{d}{0}{u}{0} \csymbolthird[8,0]{x}
\end{Compose}
~\equiv~
\begin{Compose}{0}{-0.09}\setsecondfont{\mathsfb} \setthirdfont{\mathpzc}
\supsysfid{u}{0, 2} \thispoint{uu}{0,4} \csymbolalt[0,20]{x} \jointbnoarrowthick{u}{0}{uu}{0}
\sdownsysfid{d}{0,-2}   \thispoint{dd}{0,-4} \csymbolalt[0,-20]{x} \jointbnoarrowthick{dd}{0}{d}{0}
\jointbnoarrowdotsbw[right]{d}{0}{u}{0} \csymbolthird[8,0]{x}
\end{Compose}
\end{equation}
for doing calculations.

The wire expansion equations \eqref{wirepointertomloc} and \eqref{wiresystomloc} taken together are equivalent to the assumption of full decomposability (and, therefore, tomographic locality) in \eqref{Lexpansiongeneral}.  The proof of equivalence is as follows. We have already seen that full decomposability implies the wire expansion properties.  To see that the wire expansion properties implies full decomposability consider replacing each of the wires on the operator in \eqref{Lexpansiongeneral} with their wire expansions such that the black dot is on the fiducial element next to the operation, then apply the $p(\cdot)$ function.  This gives \eqref{Lexpansiongeneral} (using \eqref{Lfidprobs} as a definition).

\subsection{Symbolic duotensor notation}\label{sec:symbolicduotensornotation}

We can also notate duotensors symbolically. For the most part, we will not use this notation since the diagrammatic notation is more immediate.  Naturally incomes and inputs will be associated with subscripts, while outcomes and outputs will be associated with superscripts.  But we need a little more since we also have black and white dots.  To notate this we use prescripts and postscripts.  For example,
\begin{equation}
\begin{Compose}{0}{0}\setsecondfont{\mathsfb} \setthirdfont{\mathpzc}  \setdefaultfont{\mathtt}
\crectangle{T}{2.5}{2.5}{0,0} \csymbolfourth{C}
\whitedotsq{y7}{-7,1.5}   \csymbolfourth[-24,0]{y} \joinrlnoarrowthick[above]{y7}{0}{T}{1.5}
\blackdotsq{x1}{-7,-1.5}   \csymbolfourth[-24,0]{x} \joinrlnoarrowthick[above]{x1}{0}{T}{-1.5}
\blackdotsq{w8}{7,1.5}   \csymbolfourth[24,0]{x} \joinrlnoarrowthick[above]{T}{1.5}{w8}{0}
\whitedotsq{y2}{7,-1.5}   \csymbolfourth[24,0]{y} \joinrlnoarrowthick[above]{T}{-1.5}{y2}{0}
\whitedotupwire{T}{-1.5}{x}  \dotupwire{T}{1.5}{y}
\whitedotdownwire{T}{-1.5}{y} \dotdownwire{T}{1.5}{z}
\end{Compose}
~~~~~~~~~~~~~~~~~~~~ \prescript{\mathpzc{z}_4}{\mathpzc{z}_6} \qcwavyl \prescript{x_1}{y_2}C^{x_4}_{y_2}\qcwavyr_{\mathpzc{y}_5}^{\mathpzc{y}_8}
\end{equation}
We have included the wavy line $\qcwavyl$ to demarcate the classical/non-classical divide (the font is also indicative of this but it is harder to tell apart the fonts by themselves).  The classical indices are closer to the $C$ symbol.
Here we are placing the indices according to the key
\begin{equation}
\tensor*[^\smallwhitecircle_\smallblackcircle]{\qcwavyl \hspace{1.5pt}}{^\smallwhitesquare_\smallblacksquare}
C
\tensor*[^\smallblacksquare_\smallwhitesquare]{\hspace{2pt}\qcwavyr}{_\smallwhitecircle^\smallblackcircle}
\end{equation}
This ensures that we can match black dots/squares on the prescript side as raised and lowered indices, or we can match them on the postscript side.  We have already given an example of a duotensor calculation in symbolic notation in \eqref{RMRcalcsymbolic} where postscripts were matched (the wavy lines are absent since we only have pointer types).

The term \lq\lq hopping matrix" (I have also called this the \lq\lq hopping metric" \cite{hardy2010formalism}) comes from the fact that, when we apply this or its inverse, the indices hop over the symbol while maintaining their level as raised or lowered indices.  For example, pre-superscripts will become post-superscripts on application of the appropriate (square dot or round dot) hopping matrix.

\subsection{Calculating probabilities for circuits}\label{sec:calculatingprobabilitiesforcircuits}

We now have sufficient formalism to write down an expression for the calculation of the probability associated with a general circuit.  We will illustrate this with the following circuit
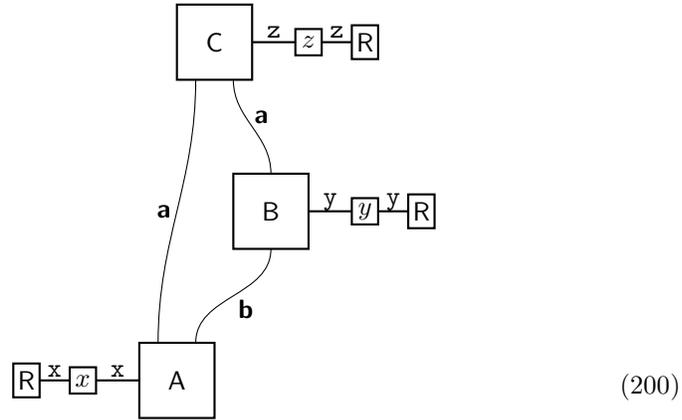
\begin{equation}\label{ABCcircuit}
\begin{Compose}{0}{0} \setsecondfont{\mathsfb}
\crectangle{A}{2}{2}{0,0} \csymbol{A}  \crectangle{B}{2}{2}{5,9} \csymbol{B} \crectangle{C}{2}{2}{2,18} \csymbol{C}
\jointbnoarrow[left]{A}{-1}{C}{-1} \csymbolalt[-5,0]{a}
\jointbnoarrow[below right]{A}{1}{B}{0}  \csymbolalt[5,-5]{b}
\jointbnoarrow[above right]{B}{0}{C}{1}  \csymbolalt{a}
\Rxboxincome{A}{0}{x}
\Rxboxoutcome{B}{0}{y}
\Rxboxoutcome{C}{0}{z}
\end{Compose}
\end{equation}
We now replace each operation with an equivalent sum over fiducials (of the form shown in \eqref{Lexpansiongeneral}) and we obtain the equivalent diagram
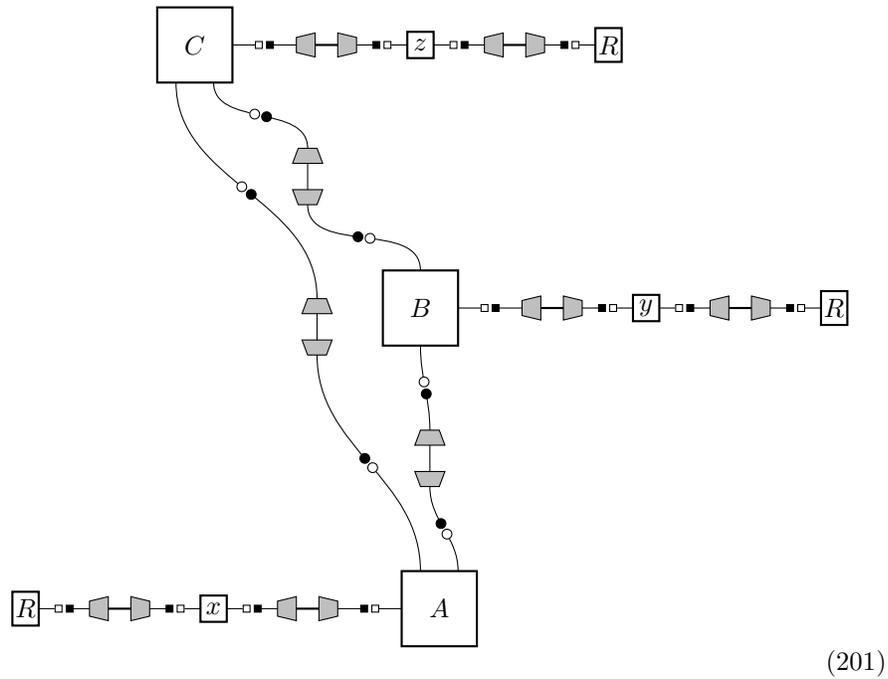
\begin{equation}\label{ABCcircuitwithfids}
\begin{Compose}{0}{0} \setsecondfont{\mathsfb}
\crectangle{A}{2}{2}{0,0} \csymbolfourth{A}  \crectangle{B}{2}{2}{-1,16} \csymbolfourth{B} \crectangle{C}{2}{2}{-13,30} \csymbolfourth{C}
\joinwithsystemfids{A}{-1}{C}{-1} 
\joinwithsystemfids{A}{1}{B}{0} 
\joinwithsystemfids{B}{0}{C}{1} 
\relpoint{A}{-2,0}{AL} \Rxboxincomefids{AL}{0}{x}
\relpoint{B}{2,0}{BR}\Rxboxoutcomefids{BR}{0}{y}
\relpoint{C}{2,0}{CR}\Rxboxoutcomefids{CR}{0}{z}
\end{Compose}
\end{equation}
(we have moved the boxes a little horizontally to fit the diagram across the page and suppressed the wire labels so the diagram doesn't get overcrowded).  Each of the matched fiducial pairs is equivalent to the corresponding hopping matrix by \eqref{hoppingequivfidcircuitpointer} and \eqref{hoppingequivfidcircuitsystem}. Hence, we obtain the equivalent diagram
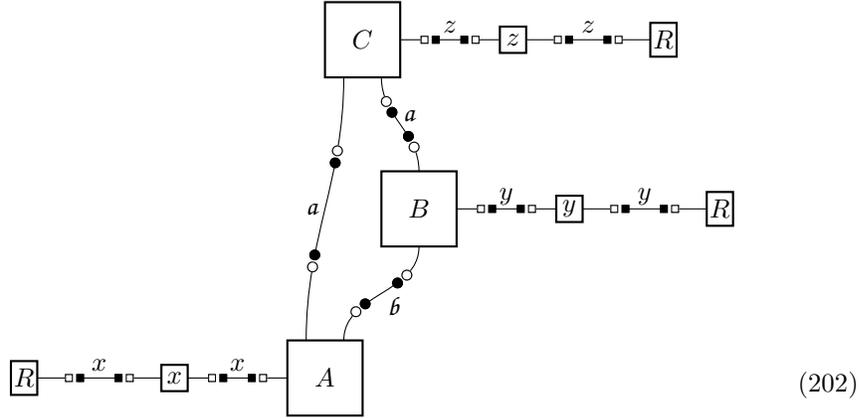
\begin{equation}\label{ABCduotensor}
\begin{Compose}{0}{0} \setsecondfont{\mathpzc}
\crectangle{A}{2}{2}{0,0} \csymbolfourth{A}  \crectangle{B}{2}{2}{5,9} \csymbolfourth{B} \crectangle{C}{2}{2}{2,18} \csymbolfourth{C}
\jointbnoarrowdotswbbw[left]{A}{-1}{C}{-1} \csymbolalt[-5,0]{a}
\jointbnoarrowdotswbbw[below right]{A}{1}{B}{0}  \csymbolalt[5,-5]{b}
\jointbnoarrowdotswbbw[above right]{B}{0}{C}{1}  \csymbolalt{a}
\Rxboxincomewbbw{A}{0}{x}
\Rxboxoutcomewbbw{B}{0}{y}
\Rxboxoutcomewbbw{C}{0}{z}
\end{Compose}
\end{equation}
This diagram corresponds to a calculation for the probability so we have gone from a circuit in \eqref{ABCcircuit} through the intermediate step in \eqref{ABCcircuitwithfids} to the calculation in \eqref{ABCduotensor}.  As a matter of notation we note that, using \eqref{bwtowb} and \eqref{bwtowbround}, we can replace black and white dots (round or square) with a straight line to obtain
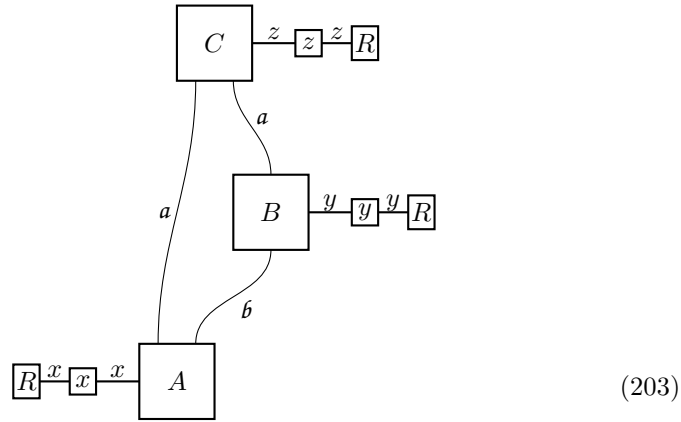
\begin{equation}\label{ABCduotensorsimple}
\begin{Compose}{0}{0} \setsecondfont{\mathpzc}
\crectangle{A}{2}{2}{0,0} \csymbolfourth{A}  \crectangle{B}{2}{2}{5,9} \csymbolfourth{B} \crectangle{C}{2}{2}{2,18} \csymbolfourth{C}
\jointbnoarrow[left]{A}{-1}{C}{-1} \csymbolalt[-5,0]{a}
\jointbnoarrow[below right]{A}{1}{B}{0}  \csymbolalt[5,-5]{b}
\jointbnoarrow[above right]{B}{0}{C}{1}  \csymbolalt{a}
\Rxboxincomecalc{A}{0}{x}
\Rxboxoutcomecalc{B}{0}{y}
\Rxboxoutcomecalc{C}{0}{z}
\end{Compose}
\end{equation}
We can reinsert the black and white dots in any order we like to do the calculation.  The final diagram for the calculation looks the same as the original diagram for the circuit it is a calculation for (in \cite{hardy2013theory} I called this the composition principle). This will work for any circuit and so we have a means to calculate probabilities for arbitrary circuits.

Note that we could have got this result by application of the wire expansion properties (that is, by replacing each of the wires in \eqref{ABCcircuit} with their expanded forms according to \eqref{wirepointertomloc} and \eqref{wiresystomloc}.

\section{Calculating conditional probabilities}

Usually, when people calculate probabilities for circuits, they are interested in the probability of some given outcome conditioned on some given income (the incomes are normally taken to be part of the description of the operation in the first place).  More generally, we might be interested in the probability of seeing some particular readouts (be they incomes or outcomes) conditioned on other readouts. One way to calculate such a probability is to read it off from the duotensor by arranging the colour of the dots appropriately as described in Sec.\ \ref{sec:interpretationofduotensorswithsquaredots}.  Here I want to describe a way to do determine this probability directly in terms of calculations involving circuits.  Consider, again, the circuit
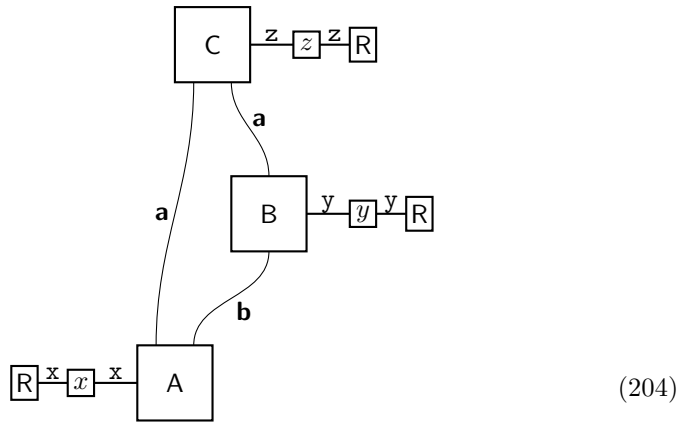
\begin{equation}
\begin{Compose}{0}{0} \setsecondfont{\mathsfb}
\crectangle{A}{2}{2}{0,0} \csymbol{A}  \crectangle{B}{2}{2}{5,9} \csymbol{B} \crectangle{C}{2}{2}{2,18} \csymbol{C}
\jointbnoarrow[left]{A}{-1}{C}{-1} \csymbolalt[-5,0]{a}
\jointbnoarrow[below right]{A}{1}{B}{0}  \csymbolalt[5,-5]{b}
\jointbnoarrow[above right]{B}{0}{C}{1}  \csymbolalt{a}
\Rxboxincome{A}{0}{x}
\Rxboxoutcome{B}{0}{y}
\Rxboxoutcome{C}{0}{z}
\end{Compose}
\end{equation}
We can calculate the probability
\begin{equation}
\text{prob}(y|xz) =
\frac{ \left(
\begin{Compose}{0}{-1.5} \setsecondfont{\mathpzc}
\crectangle{A}{2}{2}{0,0} \csymbolfourth{A}  \crectangle{B}{2}{2}{5,6} \csymbolfourth{B} \crectangle{C}{2}{2}{2,12} \csymbolfourth{C}
\jointbnoarrow[left]{A}{-1}{C}{-1} \csymbolalt[-5,0]{a}
\jointbnoarrow[below right]{A}{1}{B}{0}  \csymbolalt[5,-5]{b}
\jointbnoarrow[above right]{B}{0}{C}{1}  \csymbolalt{a}
\Rxboxincomecalc{A}{0}{x}
\Rxboxoutcomecalc{B}{0}{y}
\Rxboxoutcomecalc{C}{0}{z}
\end{Compose} \right)
}{ \left(~~
\begin{Compose}{0}{-1.5} \setsecondfont{\mathpzc}
\crectangle{A}{2}{2}{0,0} \csymbolfourth{A}  \crectangle{B}{2}{2}{5,6} \csymbolfourth{B} \crectangle{C}{2}{2}{2,12} \csymbolfourth{C}
\jointbnoarrow[left]{A}{-1}{C}{-1} \csymbolalt[-5,0]{a}
\jointbnoarrow[below right]{A}{1}{B}{0}  \csymbolalt[5,-5]{b}
\jointbnoarrow[above right]{B}{0}{C}{1}  \csymbolalt{a}
%
\crectangle{RA}{0.7}{0.9}{-6,0}\csymbolfourth{R} \joinrlnoarrow[above]{RA}{0}{A}{0} \csymbolfourth{x}
\Rxboxoutcomecalc{B}{0}{y}
\crectangle{RC}{0.7}{0.9}{8,12}\csymbolfourth{R} \joinrlnoarrow[above]{C}{0}{RC}{0} \csymbolfourth{z}
\end{Compose} ~~ \right)
}
\end{equation}
The advantage of this way of doing the calculation is that we are working entirely directly with the circuits.  The calculation for each of these circuits is gives the same answer whether we do it forwards or backwards in time.  To do the calculation forwards in time we insert
\begin{equation}\label{forwardbwdots}
\nbdotsnoarrowsq \hspace{2pt} \wndotsnoarrowsq  ~~~~~~~~\text{and}~~~~~~~~
\begin{Compose}{0}{0} \setfourthfont{\mathpzc}
\thispoint{u}{0,2.5}\thispoint{d}{0,-2.5} \jointbnoarrowdotsbw[left]{d}{0}{u}{0}
\end{Compose}
\end{equation}
on each wire as appropriate.  To do it backwards in time we insert
\begin{equation}\label{backwardswbdots}
\nwdotsnoarrowsq \hspace{2pt} \bndotsnoarrowsq  ~~~~~~~~\text{and}~~~~~~~~
\begin{Compose}{0}{0} \setfourthfont{\mathpzc}
\thispoint{u}{0,2.5}\thispoint{d}{0,-2.5} \jointbnoarrowdotswb[left]{d}{0}{u}{0}
\end{Compose}
\end{equation}
Indeed, we could also insert a mixture of white/black and black/white dot pairs and still get the same answer.  The standard way of thinking about quantum calculations is in terms of a quantum state evolving forwards in time.  We do not need to think in this way here since we simply do the calculation for a circuit by using the corresponding duotensor diagram then contract (by doing the summation) over the wires in any order we wish.  However, we can do the calculation explicitly in terms of the evolution of a state forwards or, indeed, backwards in time.

To do the calculation by explicitly evolving forward in time we insert black and white dots according \eqref{forwardbwdots} then we foliate the circuit into a sequence of spacelike hypersurfaces \cite{hardy2010formalism} that pass through the small gaps between the black and white dot pairs.  We start at the earliest hypersurface.  The part of the duotensor circuit prior to this hypersurface provides a state at this point in time.  Next we progress to the next hypersurface and contract over all the black/white pairs in between this and the earlier hypersurface.  This provides us with a new state.  In the case that the hypersurface cuts through only square dots our state is a classical probability distribution over pointer variables.  In the case that the hypersurface cuts only though round black/white dot pairs we have a quantum state.  More generally, we have a hybrid.

We can do the calculation by explicitly evolving backwards in time instead. To do this we insert dot pairs according to \eqref{backwardswbdots}, foliate, and work backwards from the last hypersurface.

\part{Operator Tensor formalism}\label{part:operatortensorformalism}

\section{Hilbert spaces, operators, and diagrammatics}\label{sec:hilbertspacesoperatorsanddiagrammatics}

In the operator tensor formulation of Quantum Theory there is a correspondence from operations to what we will call \emph{operator tensors}.  Before setting up operator tensors we need to dive \lq\lq under the hood" and look at the machinery that makes them work.  We need some finite dimensional complex vector spaces with equipped with input/output structure and having appropriate scalar products defined on them.  We will refer to these spaces as Hilbert spaces.

The operator tensor framework was introduced in \cite{hardy2011reformulating} with further developments in \cite{hardy2012operator, hardy2015quantum, hardy2016operational}.  In this work calculations are represented either by tensor-type symbolic expressions or by tensor-type diagrams.   Operator tensors are associated with operations and the duotensor formalism mediates between the two. In this way, the operator tensor approach can be construed as the natural way of formulating Quantum Theory in accord with an operational approach based around fiducial elements.  It is also possible to view operator tensors as arising from doubling up objects that pertain to Hilbert space.  This is the approach that is taken in the book by Coecke and Kissinger \cite{coecke2017picturing}.  In this section we will present this Hilbert space picture along with the doubling up picture in so far as is useful to provide a few proofs that come later (pertaining to physicality).  Pretty much all the diagrammatics in this section can be found in Coecke and Kissinger \cite{coecke2017picturing} (which distills much of the literature on this subject as well as adding new results). Key works in the diagrammatic story are Coecke \cite{quant-ph/0402014, coecke2007linearizing}, Selinger \cite{selinger2007dagger}.  A very useful summary of the main points is provided by \v Skori\'c and Wolffs \cite{skoric2021diagrammatic}.  For a more in depth literature review, see \cite{skoric2021diagrammatic}.  In the present work, I have modified the diagrammatic notation to connect with operator tensor notation I have previously adopted.  In particular, operator tensors are represented by double border boxes (to distinguish them from operations) as in my previous papers.  We split these in half (so we have a box with a single-border on just one side) to represent objects in Hilbert space.  We will also adopt a different notation on the arrows for reasons to be explained.

In the symbolic notation for operator tensors we emphasize the tensorial-style nature of the objects which mirrors the tensorial nature of the operator tensors rather than adopting the usual symbolic notation used in category theory in this context (which makes heavy use of $\otimes$ and $\circ$ denoting the tensor product and composition respectively).  The indices in the tensor-style notation allows operations to be carried out on non-adjacent objects as you read along a mathematical expression and, in this way, we can faithfully represent the diagrammatic calculations.  From a typographic point of view, this tensor-style notation is aided by using lower case letters for the systems rather than upper case letters as is typical in the category approach.

\subsection{Input and output Hilbert spaces}\label{sec:inputandoutputhilbertspaces}

In this subsection we will develop symbolic and diagrammatic notation for these input and output Hilbert spaces. For a quantum system $\mathsf{a}$ we have the Hilbert spaces.
\begin{equation}
\begin{array}{ll}
  \mathcal{H}_{\mathsfb{a}_1} ~~~~~ & ~~~~~\bar{\mathcal{H}}_{\mathsfb{a}_1} \\[5mm]
  \mathcal{H}^{\mathsfb{a}_1} ~~~~~& ~~~~~\bar{\mathcal{H}}^{\mathsfb{a}_1}
\end{array}
\end{equation}
We use the superscript when it is associated with an output and we use the subscript when it is associated with a input.  The dimension of these Hilbert spaces is $N_\mathsfb{a}$.   We write
\begin{equation}
\begin{array}{cc}
  \presub{\mathsfb{a}_1}\langle B|\in \mathcal{H}_{\mathsfb{a}_1} ~~~~ & ~~~~ |B\rangle_{\mathsfb{a_1}} \in \bar{\mathcal{H}}^{\mathsfb{a}_1} \\[5mm]
  |A\rangle^{\mathsfb{a_1}} \in \mathcal{H}^{\mathsfb{a}_1} ~~~~ & ~~~~ \presup{\mathsfb{a}_1}\langle A|\in \bar{\mathcal{H}}^{\mathsfb{a}_1}
\end{array}
\end{equation}
This symbolic notation is antiquated but relatively familiar to researchers in Quantum Theory.  An alternative possible symbolic notation is the following
\begin{equation}
\begin{array}{cc}
  \flatacute{B}_{\mathsfb{a}_1} \in \mathcal{H}_{\mathsfb{a}_1} ~~~~ & ~~~~ \flatgrave{B}_{\mathsfb{a_1}} \in \bar{\mathcal{H}}^{\mathsfb{a}_1} \\[5mm]
  \flatacute{A}^{\mathsfb{a_1}} \in \mathcal{H}^{\mathsfb{a}_1} ~~~~ & ~~~~ \flatgrave{A}^{\mathsfb{a}_1} \in \bar{\mathcal{H}}^{\mathsfb{a}_1}
\end{array}
\end{equation}
where we have placed an acute accent on the element of the unbarred Hilbert space and a grave accent on the element of the barred Hilbert space.  The motivation for this choice is that an acute and a grave accent placed next to one another looks like the hat (as in $\hat{A}$) associated with an operator.  In what follows we will stick with the standard bra and ket notation because of its familiarity.   We will, however, adopt a diagrammatic notation for the same objects as follows.
\begin{equation}
\begin{array}{cc}
\begin{Compose}{0}{0}\setdefaultfont{} \setsecondfont{\mathsfb}
\Crectangledleft{B}{1}{1.5}{0,0.3}  \relpoint{B}{0,-3.5}{Bd} \jointbleft[right]{Bd}{0}{B}{0} \csymbolalt{a}
\end{Compose}
~~\in \mathcal{H}_{\mathsfb{a}_1}
~~~~~ &  ~~~~~
\begin{Compose}{0}{0}\setdefaultfont{} \setsecondfont{\mathsfb}
\Crectangledright{B}{1}{1.5}{0,0.3}  \relpoint{B}{0,-3.5}{Bd} \jointbright[right]{Bd}{0}{B}{0} \csymbolalt{a}
\end{Compose}
~~\in \bar{\mathcal{H}}_{\mathsfb{a}_1}
\\[10mm]
\begin{Compose}{0}{0}\setdefaultfont{} \setsecondfont{\mathsfb}
\Crectangledleft{A}{1}{1.5}{0,-0.3}  \relpoint{A}{0,3.5}{Ad} \jointbleft[right]{A}{0}{Ad}{0} \csymbolalt{a}
\end{Compose}
~~\in \mathcal{H}^{\mathsfb{a}_1}
 ~~~~~ & ~~~~~
\begin{Compose}{0}{0}\setdefaultfont{} \setsecondfont{\mathsfb}
\Crectangledright{A}{1}{1.5}{0,-0.3}  \relpoint{A}{0,3.5}{Ad} \jointbright[right]{A}{0}{Ad}{0} \csymbolalt{a}
\end{Compose}
~~\in \bar{\mathcal{H}}^{\mathsfb{a}_1}
\end{array}
\end{equation}
Note we use a small left-box marking on the wire for elements of the unbarred Hilbert spaces (shown on the left) and a small right-box marking on the wire for elements of the barred Hilbert spaces (shown on the right).  We do not adopt the (sometimes used) convention of using a forward pointing arrow for the unbarred case and a backwards pointing arrow for the barred case because this would suggest that the unbarred case pertains to the forward direction in time and the barred case pertains to the backwards direction in time. In a paper on time-symmetry, this could be misleading.
Note also that one side of the box is not double lined (this will allow us to join such boxes together to create a double lined box corresponding to an operator tensor).

Since inputs feed into outputs it is natural to equip these spaces with scalar products
\begin{equation}\label{scalarproductsinout}
\presub{\mathsfb{a}_1}\langle B|A\rangle^{\mathsfb{a}_1}
~~~~~~~~~~~~~
\presup{\mathsfb{a}_1}\langle A|B\rangle_{\mathsfb{a}_1}
\end{equation}
These scalar products can also be represented diagrammatically as
\begin{equation}
\begin{Compose}{0}{0}\setdefaultfont{} \setsecondfont{\mathsfb}
\Crectangledleft{B}{1}{1.5}{0,3}
\Crectangledleft{A}{1}{1.5}{0,-3}
\jointbleft[right]{A}{0}{B}{0} \csymbolalt{a}
\end{Compose}
~~~~~~~~~~
\begin{Compose}{0}{0}\setdefaultfont{} \setsecondfont{\mathsfb}
\Crectangledright{B}{1}{1.5}{0,3}
\Crectangledright{A}{1}{1.5}{0,-3}
\jointbright[right]{A}{0}{B}{0} \csymbolalt{a}
\end{Compose}
\end{equation}
We impose that the scalar product on the left is the complex conjugate of the scalar product on the right so
\begin{equation}\label{compleximposition}
\presub{\mathsfb{a}_1}\langle B|A\rangle^{\mathsfb{a}_1}
= \left(
\presup{\mathsfb{a}_1}\langle A|B\rangle_{\mathsfb{a}_1}
\right)^*
\end{equation}
This means we convert vectors between the barred and unbarred Hilbert spaces by taking the complex conjugate of coefficients.  Symbolically we have
\begin{equation}
\begin{array}{ccc}
\presub{\mathsfb{a}_1}\langle B|=\sum_k \alpha_k\presub{\mathsfb{a}_1}\langle D[k]|
~~~~ & \Leftrightarrow & ~~~~
|B\rangle_{\mathsfb{a_1}}=\sum_k \alpha_k^*|D[k]\rangle_{\mathsfb{a_1}}
\\[5mm]
|A\rangle^{\mathsfb{a_1}}=\sum_l  \beta_l|E[l]\rangle^{\mathsfb{a_1}}
~~~~ & \Leftrightarrow & ~~~~
\presup{\mathsfb{a}_1}\langle A|=\sum_l \beta_l\presup{\mathsfb{a}_1}\langle E[l]|
\end{array}
\end{equation}
This situation can be better represented diagrammatically
\begin{equation}
\begin{array}{ccc}
\begin{Compose}{0}{0}\setdefaultfont{} \setsecondfont{\mathsfb}
\Crectangledleft{B}{1}{1.5}{0,0.3}  \relpoint{B}{0,-3.5}{Bd} \jointbleft[right]{Bd}{0}{B}{0} \csymbolalt{a}
\end{Compose}
~~=~~
\begin{Compose}{0}{0}\setdefaultfont{} \setsecondfont{\mathsfb}
\crectangledleft{B}{1}{1.5}{0,0.3} \csymbol{D}  \relpoint{B}{0,-3.5}{Bd} \jointbleft[right]{Bd}{0}{B}{0} \csymbolalt{a}
\crectangle{alpha}{1}{1}{4,0.3} \csymbol{\alpha} \joinrlnoarrow{B}{0}{alpha}{0} \csymbol[0,10]{k}
\end{Compose}
~~~~~ & \Leftrightarrow & ~~~~~
\begin{Compose}{0}{0}\setdefaultfont{} \setsecondfont{\mathsfb}
\Crectangledright{B}{1}{1.5}{0,0.3}  \relpoint{B}{0,-3.5}{Bd} \jointbright[right]{Bd}{0}{B}{0} \csymbolalt{a}
\end{Compose}
~~=~~
\begin{Compose}{0}{0}\setdefaultfont{} \setsecondfont{\mathsfb}
\crectangledright{B}{1}{1.5}{0,0.3} \csymbol{D} \relpoint{B}{0,-3.5}{Bd} \jointbright[right]{Bd}{0}{B}{0} \csymbolalt{a}
\crectangle{alpha}{1}{1}{-4,0.3} \csymbol{\alpha^*} \joinlrnoarrow{B}{0}{alpha}{0} \csymbol[0,10]{k}
\end{Compose}
\\[10mm]
\begin{Compose}{0}{0}\setdefaultfont{} \setsecondfont{\mathsfb}
\Crectangledleft{A}{1}{1.5}{0,-0.3}  \relpoint{A}{0,3.5}{Ad} \jointbleft[right]{A}{0}{Ad}{0} \csymbolalt{a}
\end{Compose}
~~=~~
\begin{Compose}{0}{0}\setdefaultfont{} \setsecondfont{\mathsfb}
\Crectangledleft{A}{1}{1.5}{0,-0.3}  \relpoint{A}{0,3.5}{Ad} \jointbleft[right]{A}{0}{Ad}{0} \csymbolalt{a}
\crectangle{beta}{1}{1}{4,-0.3} \csymbol{\beta} \joinrlnoarrow{A}{0}{beta}{0} \csymbol[0,10]{l}
\end{Compose}
 ~~~~~ & \Leftrightarrow & ~~~~~
\begin{Compose}{0}{0}\setdefaultfont{} \setsecondfont{\mathsfb}
\Crectangledright{A}{1}{1.5}{0,-0.3}  \relpoint{A}{0,3.5}{Ad} \jointbright[right]{A}{0}{Ad}{0} \csymbolalt{a}
\end{Compose}
~~=~~
\begin{Compose}{0}{0}\setdefaultfont{} \setsecondfont{\mathsfb}
\Crectangledright{A}{1}{1.5}{0,-0.3}  \relpoint{A}{0,3.5}{Ad} \jointbright[right]{A}{0}{Ad}{0} \csymbolalt{a}
\crectangle{beta}{1}{1}{-4,-0.3} \csymbol{\beta^*} \joinlrnoarrow{A}{0}{beta}{0} \csymbol[0,10]{l}
\end{Compose}
\end{array}
\end{equation}
The process of flipping a diagram (in this notation) horizontally is called \emph{conjugation}.

\subsection{Orthonormal bases}

Next, we introduce an orthonormal basis for each vector space
\begin{equation}
\begin{array}{cc}
  \left\{  \presub{\mathsfb{a}_1}\langle a|: a=1 ~\text{to}~ N_\mathsfb{a} \right\} ~~~ & ~~~ \left\{ |a\rangle_{\mathsfb{a}_1} : a=1 ~\text{to}~ N_\mathsfb{a} \right\} \\[5mm]
  \left\{ |a\rangle^{\mathsfb{a}_1} : a=1 ~\text{to}~ N_\mathsfb{a} \right\}~~~ & ~~~ \left\{  \presup{\mathsfb{a}_1}\langle a|: a=1 ~\text{to}~ N_\mathsfb{a} \right\}
\end{array}
\end{equation}
Orthonomality is the condition that
\begin{equation}\label{symbolicorthog}
\presup{\mathsfb{a}_1}\langle a|a'\rangle_{\mathsfb{a}_1} = \delta_{aa'}
~~~~~~~~~~~~~
\presub{\mathsfb{a}_1}\langle a|a'\rangle^{\mathsfb{a}_1} = \delta_{aa'}
~~~~~
\text{for all} ~~ a, a'
\end{equation}
Diagrammatically the orthonormal basis elements can be represented by having a basis label associated by a wire sticking out the side of a small empty box
\begin{equation}
\begin{array}{cc}
\begin{Compose}{0}{0}\setdefaultfont{} \setsecondfont{\mathsfb}
\crectangledleft{B}{0.8}{0.8}{0,0.3}  \relpoint{B}{0,-3.5}{Bd} \jointbleft[right]{Bd}{0}{B}{0} \csymbolalt{a}
\relpoint{B}{2.5,0}{Bs} \csymbol[20,0]{a} \joinrlnoarrow{B}{0}{Bs}{0}
\end{Compose}
~~~~~ &  ~~~~~
\begin{Compose}{0}{0}\setdefaultfont{} \setsecondfont{\mathsfb}
\crectangledright{B}{0.8}{0.8}{0,0.3}  \relpoint{B}{0,-3.5}{Bd} \jointbright[right]{Bd}{0}{B}{0} \csymbolalt{a}
\relpoint{B}{-2.5,0}{Bs} \csymbol[-20,0]{a} \joinlrnoarrow{B}{0}{Bs}{0}
\end{Compose}
\\[10mm]
\begin{Compose}{0}{0}\setdefaultfont{} \setsecondfont{\mathsfb}
\crectangledleft{A}{0.8}{0.8}{0,-0.3}  \relpoint{A}{0,3.5}{Ad} \jointbleft[right]{A}{0}{Ad}{0} \csymbolalt{a}
\relpoint{A}{2.5,0}{As} \csymbol[20,0]{a} \joinrlnoarrow{A}{0}{As}{0}
\end{Compose}
 ~~~~~ & ~~~~~
\begin{Compose}{0}{0}\setdefaultfont{} \setsecondfont{\mathsfb}
\crectangledright{A}{0.8}{0.8}{0,-0.3}  \relpoint{A}{0,3.5}{Ad} \jointbright[right]{A}{0}{Ad}{0} \csymbolalt{a}
\relpoint{A}{-2.5,0}{As} \csymbol[-20,0]{a} \joinlrnoarrow{A}{0}{As}{0}
\end{Compose}
\end{array}
\end{equation}
where $a=1$ to $N_\mathsf{a}$.   Note that, while in this example the label sticking out the side labels elements of an orthonormal set - in other examples this label will refer to non-orthonormal elements.  The orthogonality relations \eqref{symbolicorthog} are
\begin{equation}\label{diagrammaticorthog}
\begin{Compose}{0}{0}\setdefaultfont{} \setsecondfont{\mathsfb}
\crectangledleft{B}{0.8}{0.8}{0,2}
\crectangledleft{A}{0.8}{0.8}{0,-2}
\jointbleft[left]{A}{0}{B}{0}  \csymbolalt{a}
\thispoint{d}{3,-2}\thispoint{u}{3,2}
\joinrlnoarrow[above]{A}{0}{d}{0}\csymbol{a}
\joinrlnoarrow[above]{B}{0}{u}{0}\csymbol{a}
\end{Compose}
~~ = ~~
\begin{Compose}{0}{0}\setdefaultfont{}\setsecondfont{\mathsfb}
\thispoint{d}{0,-2}\thispoint{u}{0,2}
\joinllnoarrow[left]{d}{0}{u}{0} \csymbol{a}
\end{Compose}
~~~~~~~~~~
\begin{Compose}{0}{0}\setdefaultfont{} \setsecondfont{\mathsfb}
\crectangledright{B}{0.8}{0.8}{0,2}
\crectangledright{A}{0.8}{0.8}{0,-2}
\jointbright[right]{A}{0}{B}{0}  \csymbolalt{a}
\thispoint{d}{-3,-2}\thispoint{u}{-3,2}
\joinlrnoarrow[above]{A}{0}{d}{0}\csymbol{a}
\joinlrnoarrow[above]{B}{0}{u}{0}\csymbol{a}
\end{Compose}
~~ = ~~
\begin{Compose}{0}{0}\setdefaultfont{}\setsecondfont{\mathsfb}
\thispoint{d}{0,-2}\thispoint{u}{0,2}
\joinrrnoarrow[right]{d}{0}{u}{0} \csymbol{a}
\end{Compose}
\end{equation}
in diagrammatic form.
Note that the wires labeled $a$ simply represent sums and they are free to join to any box with that label without regard for the direction the wire is pointing (correspondingly, we are not employing subscripts and superscripts in \eqref{symbolicorthog}).  Another way of viewing this is that the above equations provides left/right cups and caps so we can bend a wire in any direction we want.  Also, note the following
\begin{equation}
\begin{Compose}{0}{0}\setdefaultfont{} \setsecondfont{\mathsfb}
\crectangledleft{B}{0.8}{0.8}{0,1.3}
\crectangledleft{A}{0.8}{0.8}{0,-1.3}
\joinrrnoarrow[right]{A}{0}{B}{0}  \csymbol{a}
\thispoint{d}{0,-4}\thispoint{u}{0,4}
\jointbleft[left]{d}{0}{A}{0}\csymbolalt{a}
\jointbleft[left]{B}{0}{u}{0}\csymbolalt{a}
\end{Compose}
~~ = ~~
\begin{Compose}{0}{0}\setdefaultfont{}\setsecondfont{\mathsfb}
\thispoint{d}{0,-4}\thispoint{u}{0,4}
\jointbleft[left]{d}{0}{u}{0} \csymbolalt{a}
\end{Compose}
~~~~~~~~~~
\begin{Compose}{0}{0}\setdefaultfont{} \setsecondfont{\mathsfb}
\crectangledright{B}{0.8}{0.8}{0,1.3}
\crectangledright{A}{0.8}{0.8}{0,-1.3}
\joinllnoarrow[left]{A}{0}{B}{0}  \csymbol{a}
\thispoint{d}{0,-4}\thispoint{u}{0,4}
\jointbright[right]{d}{0}{A}{0}\csymbolalt{a}
\jointbright[right]{B}{0}{u}{0}\csymbolalt{a}
\end{Compose}
~~ = ~~
\begin{Compose}{0}{0}\setdefaultfont{}\setsecondfont{\mathsfb}
\thispoint{d}{0,-4}\thispoint{u}{0,4}
\jointbright[right]{d}{0}{u}{0} \csymbolalt{a}
\end{Compose}
\end{equation}
This is just the decomposition of the identity.

\subsection{Operators}

We can combine elements of Hilbert spaces in various ways to obtain operators.  We can take the tensor product of various elements. For example,
\begin{equation}
\presub{\mathsfb{a}_1}\langle A|\otimes |C\rangle^{\mathsfb{a}_2} \otimes \presup{\mathsfb{b}_3}\langle E|
\in \mathcal{H}_{\mathsfb{a}_1} \mathcal{H}^{\mathsfb{a}_2} \bar{\mathcal{H}}^{\mathsfb{b}_3}
~~~~~~~~~~~~~~
\begin{Compose}{0}{0}\setdefaultfont{} \setsecondfont{\mathsfb}
\crectangledleft{B}{1}{1.5}{0,0} \csymbol{A} \relpoint{B}{0,-3.5}{Bd} \jointbleft[right]{Bd}{0}{B}{0} \csymbolalt{a}
\crectangledleft{B}{1}{1.5}{3,0} \csymbol{C} \relpoint{B}{0,3.5}{Bd} \jointbleft[right]{B}{0}{Bd}{0} \csymbolalt{a}
\crectangledright{B}{1}{1.5}{6,0} \csymbol{E} \relpoint{B}{0,3.5}{Bd} \jointbright[right]{B}{0}{Bd}{0} \csymbolalt{b}
\end{Compose}
\in \mathcal{H}_{\mathsfb{a}_1} \mathcal{H}^{\mathsfb{a}_2} \bar{\mathcal{H}}^{\mathsfb{b}_3}
\end{equation}
Symbolic notation is on the left and diagrammatic on the right.
We have inserted tensor product symbols, $\otimes$, in the symbolic notation to make it clear that we are not actually taking the scalar product.  However, these tensor product symbols are not necessary (and sometimes we will omit them). We only take the scalar product if we have a repeated index as in \eqref{scalarproductsinout}.
We can sum over such elements.
\begin{equation}
\presub{\mathsfb{a}_1}\langle A|\otimes |C\rangle^{\mathsfb{a}_2} \otimes \presup{\mathsfb{b}_3}\langle E|
\in \mathcal{H}_{\mathsfb{a}_1} \mathcal{H}^{\mathsfb{a}_2} \bar{\mathcal{H}}^{\mathsfb{b}_3}
~~~~~~~~~~~~~~
\begin{Compose}{0}{0}\setdefaultfont{} \setsecondfont{\mathsfb}
\crectangledleft{A}{1}{1.5}{0,-2} \csymbol{A} \relpoint{A}{0,-3.5}{Ad} \jointbleft[right]{Ad}{0}{A}{0} \csymbolalt{a}
\crectangledleft{C}{1}{1.5}{0,2} \csymbol{C} \relpoint{C}{0,3.5}{Cd} \jointbleft[right]{C}{0}{Cd}{0} \csymbolalt{a}
\crectangledright{E}{1}{1.5}{10.5,0} \csymbol{E} \relpoint{E}{0,3.5}{Ed} \jointbright[right]{E}{0}{Ed}{0} \csymbolalt{b}
\crectangle[thin]{G}{1.5}{1.5}{5.5,0}  \csymbol{\beta}
\joinrlnoarrow[below]{A}{0}{G}{-1}  \csymbol[5,-7]{l}
\joinrlnoarrow[above]{C}{0}{G}{1} \csymbol[5,7]{k}
\joinrlnoarrow[above]{G}{0}{E}{0} \csymbol{n}
\end{Compose}
\in \mathcal{H}_{\mathsfb{a}_1} \mathcal{H}^{\mathsfb{a}_2} \bar{\mathcal{H}}^{\mathsfb{b}_3}
\end{equation}
There are some important special cases which we will discuss in the next couple of subsections.

\subsection{Only unbarred or only barred elements}

We have unbarred Hilbert spaces ($\mathcal{H}_{\mathsfb{a}_1}$ and $\mathcal{H}^{\mathsfb{a}_1}$) and barred Hilbert spaces ($\bar{\mathcal{H}}_{\mathsfb{a}_1}$ and $\bar{\mathcal{H}}^{\mathsfb{a}_1}$).  We can build object out of elements from only unbarred Hilbert spaces (example on left below) or out of elements from only barred Hilbert spaces (example on right).
\begin{equation}
\begin{Compose}{0}{0}\setdefaultfont{} \setsecondfont{\mathsfb}
\crectangledleft{A}{1.5}{2.3}{0,0} \csymbol{B}
\relpoint{A}{0,-4.5}{Ad} \jointbleft[right]{Ad}{0}{A}{0} \csymbolalt{a}
\relpoint{A}{1,4.5}{Cd} \jointbleft[right]{A}{1}{Cd}{0} \csymbolalt{a}
\relpoint{A}{-1,4.5}{Ed} \jointbleft[right]{A}{-1}{Ed}{0} \csymbolalt{b}
\end{Compose}
~=~
\begin{Compose}{0}{0}\setdefaultfont{} \setsecondfont{\mathsfb}
\crectangledleft{A}{1}{1.5}{0,-3} \csymbol{A} \relpoint{A}{0,-3.5}{Ad} \jointbleft[right]{Ad}{0}{A}{0} \csymbolalt{a}
\crectangledleft{C}{1}{1.5}{0,3} \csymbol{C} \relpoint{C}{0,3.5}{Cd} \jointbleft[right]{C}{0}{Cd}{0} \csymbolalt{a}
\crectangledleft{E}{1}{1.5}{-3,0} \csymbol{E} \relpoint{E}{0,3.5}{Ed} \jointbleft[right]{E}{0}{Ed}{0} \csymbolalt{b}
\crectangle[thin]{G}{1.5}{1.5}{5.5,0}  \csymbol{\beta}
\joinrlnoarrow[below]{A}{0}{G}{-1}  \csymbol[5,-7]{l}
\joinrlnoarrow[above]{C}{0}{G}{1} \csymbol[6,8]{k}
\joinlrnoarrow[above]{G}{0}{E}{0} \csymbol{n}
\end{Compose}
~~~~~~~~~~
\begin{Compose}{0}{0}\setdefaultfont{} \setsecondfont{\mathsfb}
\crectangledright{A}{1.5}{2.3}{0,0} \csymbol{B}
\relpoint{A}{0,-4.5}{Ad} \jointbright[right]{Ad}{0}{A}{0} \csymbolalt{a}
\relpoint{A}{1,4.5}{Cd} \jointbright[right]{A}{1}{Cd}{0} \csymbolalt{b}
\relpoint{A}{-1,4.5}{Ed} \jointbright[right]{A}{-1}{Ed}{0} \csymbolalt{a}
\end{Compose}
~=~
\begin{Compose}{0}{0}\setdefaultfont{} \setsecondfont{\mathsfb}
\crectangledright{A}{1}{1.5}{0,-3} \csymbol{A} \relpoint{A}{0,-3.5}{Ad} \jointbright[left]{Ad}{0}{A}{0} \csymbolalt{a}
\crectangledright{C}{1}{1.5}{0,3} \csymbol{C} \relpoint{C}{0,3.5}{Cd} \jointbright[left]{C}{0}{Cd}{0} \csymbolalt{a}
\crectangledright{E}{1}{1.5}{3,0} \csymbol{E} \relpoint{E}{0,3.5}{Ed} \jointbright[right]{E}{0}{Ed}{0} \csymbolalt{b}
\crectangle[thin]{G}{1.5}{1.5}{-5.5,0}  \csymbol{\beta^*}
\joinlrnoarrow[below]{A}{0}{G}{-1}  \csymbol[0,-7]{l}
\joinlrnoarrow[above]{C}{0}{G}{1} \csymbol[0,7]{k}
\joinrlnoarrow[above]{G}{0}{E}{0} \csymbol{n}
\end{Compose}
\end{equation}
The example on the right is conjugate to the example on the left.

\subsection{Operator tensors}

We can form \emph{operator tensors} as follows
\begin{equation}
\begin{Compose}{0}{0}\setdefaultfont{} \setsecondfont{\mathsfb}
\crectangledouble{A}{2.5}{2.3}{0,0} \csymbol{\hat{C}}
\relpoint{A}{0,-4.5}{Ad} \jointbnoarrow[right]{Ad}{0}{A}{0} \csymbolalt{a}
\relpoint{A}{1,4.5}{Cd} \jointbnoarrow[right]{A}{1}{Cd}{0} \csymbolalt{a}
\relpoint{A}{-1,4.5}{Ed} \jointbnoarrow[right]{A}{-1}{Ed}{0} \csymbolalt{b}
\end{Compose}
~~=~~
\begin{Compose}{0}{0}\setdefaultfont{} \setsecondfont{\mathsfb}
\crectangledleft{A}{1.5}{2.3}{-3.7,0} \csymbol{A}
\relpoint{A}{0,-4.5}{Ad} \jointbleft[right]{Ad}{0}{A}{0} \csymbolalt{a}
\relpoint{A}{1,4.5}{Cd} \jointbleft[right]{A}{1}{Cd}{0} \csymbolalt{a}
\relpoint{A}{-1,4.5}{Ed} \jointbleft[right]{A}{-1}{Ed}{0} \csymbolalt{b}
\crectangledright{AA}{1.5}{2.3}{3.7,0} \csymbol{B}
\relpoint{AA}{0,-4.5}{Ad} \jointbright[right]{Ad}{0}{AA}{0} \csymbolalt{a}
\relpoint{AA}{1,4.5}{Cd} \jointbright[right]{AA}{1}{Cd}{0} \csymbolalt{b}
\relpoint{AA}{-1,4.5}{Ed} \jointbright[right]{AA}{-1}{Ed}{0} \csymbolalt{a}
\crectangle[thin]{M}{1}{1}{0,0} \csymbol{\alpha}
\joinrlnoarrow[above]{A}{0}{M}{0} \csymbol[0,5]{l} \joinrlnoarrow[above]{M}{0}{AA}{0} \csymbol[0,5]{l}
\end{Compose}
\end{equation}
Here we sum over the tensor product of an unbarred and a barred object each having complementary input output structure.  Further,
\begin{equation}
\begin{Compose}{0}{0}\setdefaultfont{} \setsecondfont{\mathsfb}
\thispoint{t}{0,2} \thispoint{b}{0,-2}
\jointbnoarrow[left]{b}{0}{t}{0}  \csymbolalt{a}
\end{Compose}
~~ = ~~
\begin{Compose}{0}{0}\setdefaultfont{} \setsecondfont{\mathsfb}
\thispoint{t}{0,2.5} \thispoint{b}{0,-2.5}
\jointbleft[left]{b}{-0.3}{t}{-0.3}\csymbolalt{a}
\jointbright[right]{b}{0.3}{t}{0.3} \csymbolalt{a}
\end{Compose}
\end{equation}
So a left marked wire and a right marked wire form a wire for a operator tensor.

We can expand an arbitrary operator tensor as a sum over basis elements as follows
\begin{equation}\label{Cbasisexpansion}
\begin{Compose}{0}{0}\setdefaultfont{} \setsecondfont{\mathsfb}
\crectangledouble{C}{3}{3}{0,0} \csymbol{\hat{C}}
\thispoint{u}{0,6} \thispoint{d}{0,-6}
\jointbnoarrow[left]{C}{-1.5}{u}{-1.5} \csymbolalt{b}
\jointbnoarrow[right]{C}{1.5}{u}{1.5} \csymbolalt{a}
\joinbtnoarrow[left]{C}{0}{d}{0} \csymbolalt{a}
\end{Compose}
~~=~~
\begin{Compose}{0}{0}\setdefaultfont{} \setsecondfont{\mathsfb}
\Crectangle[thin]{C}{3}{3}{0,0}
\thispoint{u}{0,6} \thispoint{d}{0,-6}
\crectangledleft{bul}{0.8}{0.8}{-9,0} \crectangledleft{aul}{0.8}{0.8}{-6,2} \crectangledleft{adl}{0.8}{0.8}{-6,-2}
\joinlrnoarrow[above]{C}{0}{bul}{0} \csymbol[45,4]{b} \joinlrnoarrow[above]{C}{2}{aul}{0} \csymbol{a} \joinlrnoarrow[below]{C}{-2}{adl}{0} \csymbol{a}
\jointbleft[left]{bul}{0}{u}{-9} \csymbolalt{b}
\jointbleft[left]{aul}{0}{u}{-6} \csymbolalt{a}
\jointbleft[left]{d}{-6}{adl}{0} \csymbolalt{a}
\crectangledright{bur}{0.8}{0.8}{9,0} \crectangledright{aur}{0.8}{0.8}{6,2} \crectangledright{adr}{0.8}{0.8}{6,-2}
\joinrlnoarrow[above]{C}{0}{bur}{0} \csymbol[-45,4]{b} \joinrlnoarrow[above]{C}{2}{aur}{0} \csymbol{a} \joinrlnoarrow[below]{C}{-2}{adr}{0} \csymbol{a}
\jointbright[right]{bur}{0}{u}{9} \csymbolalt{b}
\jointbright[right]{aur}{0}{u}{6} \csymbolalt{a}
\jointbright[right]{d}{6}{adr}{0} \csymbolalt{a}
\end{Compose}
\end{equation}
Any operator tensor can be written in this form.  We will not, however, pursue the mathematical structure arising from this expansion in this work.  This is because we will be interested in the space of Hermitian operator tensors as they correspond to physical situations.  Hermitian operator tensors admit an expansion in terms of fiducial elements that mirrors the expansion of operations we discussed in Sec.\ \ref{sec:generalduotensorswithroundandsquaredots}.

\subsection{Hermitian operator tensors}\label{sec:hermitianoperatortensors}

If the matrix, $C$ in the expansion \eqref{Cbasisexpansion} above is Hermitian, it can be diagonalised using a unitary matrix and its adjoint.  Diagrammatically, this takes the following form
\begin{equation} \label{Ceigenvalues}
\begin{Compose}{0}{0} \setdefaultfont{}
\Crectangle[thin]{C}{3}{3}{0,0}
\thispoint{l}{-6,0} \thispoint{r}{6,0}
\joinrlnoarrow[above]{l}{2}{C}{2}  \csymbol{a} \joinrlnoarrow[above]{l}{0}{C}{0} \csymbol[0,3]{b} \joinrlnoarrow[above]{l}{-2}{C}{-2} \csymbol{a}
\joinrlnoarrow[above]{C}{2}{r}{2}  \csymbol{a} \joinrlnoarrow[above]{C}{0}{r}{0} \csymbol[0,3]{b} \joinrlnoarrow[above]{C}{-2}{r}{-2} \csymbol{a}
\end{Compose}
=
\begin{Compose}{0}{0} \setdefaultfont{}
\crectangle[thin]{BL}{1.5}{3}{-5,0} \csymbol{B} \crectangle[thin]{BR}{1.5}{3}{5,0} \csymbol{B^*}
\relpoint{BL}{-4.5,0}{l}
\joinrlnoarrow[above]{l}{2}{BL}{2}  \csymbol{a} \joinrlnoarrow[above]{l}{0}{BL}{0} \csymbol[0,3]{b} \joinrlnoarrow[above]{l}{-2}{BL}{-2} \csymbol{a}
\relpoint{BR}{4.5,0}{r}
\joinrlnoarrow[above]{BR}{2}{r}{2}  \csymbol{a} \joinrlnoarrow[above]{BR}{0}{r}{0} \csymbol[0,3]{b} \joinrlnoarrow[above]{BR}{-2}{r}{-2} \csymbol{a}
\cdiamond[thin]{d}{1.2}{1.2}{0,0} \csymbol{\lambda}
\joinrlnoarrow[above]{BL}{0}{d}{0} \csymbol{l} \joinrlnoarrow[above]{d}{0}{BR}{0} \csymbol{l}
\end{Compose}
\end{equation}
where the matrix represented by the diamond box with $\lambda$ is diagonal and the $\lambda_l$ are real (so this introduces a weight $\lambda_l$ in the sum).  The $C$ matrix is a $N_\mathsfb{a}N_\mathsfb{b}N_\mathsfb{a}$ by $N_\mathsfb{a}N_\mathsfb{b}N_\mathsfb{a}$ matrix.  Therefore, the index, $l$, runs from $1$ to $N_\mathsfb{a}N_\mathsfb{b}N_\mathsfb{a}$.  The matrix, $B$, on the left is unitary.  The entries of the matrix, $B^*$, on the right are the complex conjugates of $B$ on the left and also, by virtue of the indices being flipped, it is also the transpose. Hence, the $B^*$ matrix on the right is the adjoint of the matrix on the left (as we required above).  Operator tensors for which the expansion matrix is Hermitian will be called Hermitian operator tensors.  In general, we will only be interested in Hermitian operator tensors in this work.  Substituting \eqref{Ceigenvalues} into \eqref{Cbasisexpansion} we see these take the form
\begin{equation}
\begin{Compose}{0}{0}\setdefaultfont{} \setsecondfont{\mathsfb}
\crectangledouble{A}{2.5}{2.3}{0,0} \csymbol{\hat{C}}
\relpoint{A}{0,-4.5}{Ad} \jointbnoarrow[right]{Ad}{0}{A}{0} \csymbolalt{a}
\relpoint{A}{1,4.5}{Cd} \jointbnoarrow[right]{A}{1}{Cd}{0} \csymbolalt{a}
\relpoint{A}{-1,4.5}{Ed} \jointbnoarrow[right]{A}{-1}{Ed}{0} \csymbolalt{b}
\end{Compose}
~~=~~
\begin{Compose}{0}{0}\setdefaultfont{} \setsecondfont{\mathsfb}
\crectangledleft{A}{1.5}{2.3}{-3.7,0} \csymbol{B}
\relpoint{A}{0,-4.5}{Ad} \jointbleft[right]{Ad}{0}{A}{0} \csymbolalt{a}
\relpoint{A}{1,4.5}{Cd} \jointbleft[right]{A}{1}{Cd}{0} \csymbolalt{a}
\relpoint{A}{-1,4.5}{Ed} \jointbleft[right]{A}{-1}{Ed}{0} \csymbolalt{b}
\crectangledright{AA}{1.5}{2.3}{3.7,0} \csymbol{B}
\relpoint{AA}{0,-4.5}{Ad} \jointbright[right]{Ad}{0}{AA}{0} \csymbolalt{a}
\relpoint{AA}{1,4.5}{Cd} \jointbright[right]{AA}{1}{Cd}{0} \csymbolalt{b}
\relpoint{AA}{-1,4.5}{Ed} \jointbright[right]{AA}{-1}{Ed}{0} \csymbolalt{a}
\cdiamond[thin]{M}{1.2}{1.2}{0,0} \csymbol{\lambda}
\joinrlnoarrow[above]{A}{0}{M}{0} \csymbol[0,5]{l} \joinrlnoarrow[above]{M}{0}{AA}{0} \csymbol[0,5]{l}
\end{Compose}
\end{equation}
where the matrix represented by the diamond box with $\lambda$ is diagonal and the $\lambda_l$ are real (so this introduces a weight $\lambda_l$ in the sum).  Hermitian operators can be written in this form (since we can diagonalize a Hermitian matrix).   We denote the space of Hermitian operators with $\mathcal{V}$ having appropriate subscripts and superscripts.  So, for example,
\begin{equation}
\begin{Compose}{0}{0}\setdefaultfont{} \setsecondfont{\mathsfb}
\crectangledouble{A}{2.5}{2.3}{0,0} \csymbol{\hat{C}}
\relpoint{A}{0,-4.5}{Ad} \jointbnoarrow[right]{Ad}{0}{A}{0} \csymbolalt{a}
\relpoint{A}{1,4.5}{Cd} \jointbnoarrow[right]{A}{1}{Cd}{0} \csymbolalt{a}
\relpoint{A}{-1,4.5}{Ed} \jointbnoarrow[right]{A}{-1}{Ed}{0} \csymbolalt{b}
\end{Compose}
\in \mathcal{V}_{\mathsfb{a}_1}^{\mathsfb{b}_2\mathsfb{c}_3}
\end{equation}
All physically interesting operator tensors are Hermitian so all the tensor operators we will consider in what follows will be taken to be Hermitian. The space of Hermitian operators has the important property that
\begin{equation}
\mathcal{V}_{\mathsfb{a}_1}^{\mathsfb{b}_2\mathsfb{c}_3} = \mathcal{V}_{\mathsfb{a}_1}\otimes \mathcal{V}^{\mathsfb{b}_2} \otimes \mathcal{V}^{\mathsfb{b}_2\mathsfb{c}_3}
\end{equation}
it is appropriate to drop the tensor product symbol in this notation
\begin{equation}
\mathcal{V}_{\mathsfb{a}_1}^{\mathsfb{b}_2\mathsfb{c}_3} = \mathcal{V}_{\mathsfb{a}_1} \mathcal{V}^{\mathsfb{b}_2}  \mathcal{V}^{\mathsfb{b}_2\mathsfb{c}_3}
\end{equation}
since the indices provide all the pertinent information.    As we will discuss in Sec.\ \ref{sec:ageneraloperatortensor} this enables us to expand an operator tensor as a sum over fiducial operators mirroring the structure of the decomposition of operations discussed in Sec.\ \ref{sec:generalduotensorswithroundandsquaredots}.

\subsection{$T$-positive operator tensors}\label{sec:Tpositiveoperatortensors}

If all the $\lambda$'s are positive then we can absorb them as $\sqrt{\lambda}$ factor into the $B$ boxes on either side.  Then we have
\begin{equation}
\begin{Compose}{0}{0}\setdefaultfont{} \setsecondfont{\mathsfb}
\crectangledouble{A}{2.5}{2.3}{0,0} \csymbol{\hat{B}}
\relpoint{A}{0,-4.5}{Ad} \jointbnoarrow[right]{Ad}{0}{A}{0} \csymbolalt{a}
\relpoint{A}{1,4.5}{Cd} \jointbnoarrow[right]{A}{1}{Cd}{0} \csymbolalt{a}
\relpoint{A}{-1,4.5}{Ed} \jointbnoarrow[right]{A}{-1}{Ed}{0} \csymbolalt{b}
\end{Compose}
~~=~~
\begin{Compose}{0}{0}\setdefaultfont{} \setsecondfont{\mathsfb}
\crectangledleft{A}{1.5}{2.3}{-3,0} \csymbol{B}
\relpoint{A}{0,-4.5}{Ad} \jointbleft[right]{Ad}{0}{A}{0} \csymbolalt{a}
\relpoint{A}{1,4.5}{Cd} \jointbleft[right]{A}{1}{Cd}{0} \csymbolalt{a}
\relpoint{A}{-1,4.5}{Ed} \jointbleft[right]{A}{-1}{Ed}{0} \csymbolalt{b}
\crectangledright{AA}{1.5}{2.3}{3,0} \csymbol{B}
\relpoint{AA}{0,-4.5}{Ad} \jointbright[right]{Ad}{0}{AA}{0} \csymbolalt{a}
\relpoint{AA}{1,4.5}{Cd} \jointbright[right]{AA}{1}{Cd}{0} \csymbolalt{b}
\relpoint{AA}{-1,4.5}{Ed} \jointbright[right]{AA}{-1}{Ed}{0} \csymbolalt{a}
\joinrlnoarrow[above]{A}{0}{AA}{0}\csymbol{l}
\end{Compose}
\end{equation}
Since the index, $l$, runs from $1$ to $N_\mathsfb{a}N_\mathsfb{b}N_\mathsfb{a}$ (as we noted in Sec.\ \ref{sec:hermitianoperatortensors}) it can be replaced with the label $(a,b,a')$. In this case, we have
\begin{equation} \label{BBdecompabaform}
\begin{Compose}{0}{0}\setdefaultfont{} \setsecondfont{\mathsfb}
\crectangledouble{A}{2.5}{2.3}{0,0} \csymbol{\hat{B}}
\relpoint{A}{0,-4.5}{Ad} \jointbnoarrow[right]{Ad}{0}{A}{0} \csymbolalt{a}
\relpoint{A}{1,4.5}{Cd} \jointbnoarrow[right]{A}{1}{Cd}{0} \csymbolalt{a}
\relpoint{A}{-1,4.5}{Ed} \jointbnoarrow[right]{A}{-1}{Ed}{0} \csymbolalt{b}
\end{Compose}
~~=~~
\begin{Compose}{0}{0}\setdefaultfont{} \setsecondfont{\mathsfb}
\crectangledleft{A}{1.5}{2.3}{-3,0} \csymbol{B}
\relpoint{A}{0,-4.5}{Ad} \jointbleft[right]{Ad}{0}{A}{0} \csymbolalt{a}
\relpoint{A}{1,4.5}{Cd} \jointbleft[right]{A}{1}{Cd}{0} \csymbolalt{a}
\relpoint{A}{-1,4.5}{Ed} \jointbleft[right]{A}{-1}{Ed}{0} \csymbolalt{b}
\crectangledright{AA}{1.5}{2.3}{3,0} \csymbol{B}
\relpoint{AA}{0,-4.5}{Ad} \jointbright[right]{Ad}{0}{AA}{0} \csymbolalt{a}
\relpoint{AA}{1,4.5}{Cd} \jointbright[right]{AA}{1}{Cd}{0} \csymbolalt{b}
\relpoint{AA}{-1,4.5}{Ed} \jointbright[right]{AA}{-1}{Ed}{0} \csymbolalt{a}
\joinrlnoarrow[above]{A}{1.7}{AA}{1.7}\csymbol{a}
\joinrlnoarrow[above]{A}{0}{AA}{0}\csymbol[0,3]{b}
\joinrlnoarrow[above]{A}{-1.7}{AA}{-1.7}\csymbol{a}
\end{Compose}
\end{equation}
This form will be useful when we prove the extension theorem in Sec.\ \ref{sec:extensiontheorem}.  The operators $B$ on the right of \eqref{BBdecompabaform} are eigenvectors of the diagonalisation process. Thus, they can be made to be orthogonal.  Though, conversely, we can start with an operator tensor in the form above where the $B$'s are not orthogonal, and then go through the diagonalisation process to obtain new $B$'s which are.
Operators of this form are positive in a certain key sense (to be discussed in depth in Sec.\ \ref{sec:completeoperatorpositivity}).  We will say satisfy $T$-positivity written
\begin{equation}
0 \underset{T}{\leq} ~~
\begin{Compose}{0}{0}\setdefaultfont{} \setsecondfont{\mathsfb}
\crectangledouble{A}{2.5}{2.3}{0,0} \csymbol{\hat{B}}
\relpoint{A}{0,-4.5}{Ad} \jointbnoarrow[right]{Ad}{0}{A}{0} \csymbolalt{a}
\relpoint{A}{1,4.5}{Cd} \jointbnoarrow[right]{A}{1}{Cd}{0} \csymbolalt{a}
\relpoint{A}{-1,4.5}{Ed} \jointbnoarrow[right]{A}{-1}{Ed}{0} \csymbolalt{b}
\end{Compose}
\end{equation}
The $T$ can stand for \lq\lq tester", or \lq\lq transpose" (terminology that will be explained in Sec.\ \ref{sec:completeoperatorpositivity}).  The concept of $T$-positivity is equivalent to what Coecke and Kissinger \cite{coecke2017picturing} call  $\otimes$-positivity and they trace it back to Selinger's 2007 paper \cite{selinger2007dagger}.

We can also form objects of the form
\begin{equation}\label{homogeneousoptensor}
\begin{Compose}{0}{0}\setdefaultfont{} \setsecondfont{\mathsfb}
\crectangledouble{A}{2.5}{2.3}{0,0} \csymbol{\hat{A}}
\relpoint{A}{0,-4.5}{Ad} \jointbnoarrow[right]{Ad}{0}{A}{0} \csymbolalt{a}
\relpoint{A}{1,4.5}{Cd} \jointbnoarrow[right]{A}{1}{Cd}{0} \csymbolalt{a}
\relpoint{A}{-1,4.5}{Ed} \jointbnoarrow[right]{A}{-1}{Ed}{0} \csymbolalt{b}
\end{Compose}
~~=~~
\begin{Compose}{0}{0}\setdefaultfont{} \setsecondfont{\mathsfb}
\crectangledleft{A}{1.5}{2.3}{-2,0} \csymbol{A}
\relpoint{A}{0,-4.5}{Ad} \jointbleft[right]{Ad}{0}{A}{0} \csymbolalt{a}
\relpoint{A}{1,4.5}{Cd} \jointbleft[right]{A}{1}{Cd}{0} \csymbolalt{a}
\relpoint{A}{-1,4.5}{Ed} \jointbleft[right]{A}{-1}{Ed}{0} \csymbolalt{b}
\crectangledright{AA}{1.5}{2.3}{2,0} \csymbol{A}
\relpoint{AA}{0,-4.5}{Ad} \jointbright[right]{Ad}{0}{AA}{0} \csymbolalt{a}
\relpoint{AA}{1,4.5}{Cd} \jointbright[right]{AA}{1}{Cd}{0} \csymbolalt{b}
\relpoint{AA}{-1,4.5}{Ed} \jointbright[right]{AA}{-1}{Ed}{0} \csymbolalt{a}
\end{Compose}
\end{equation}
where there is no sum.  We will say such operator tensors are \emph{homogeneous} since the cannot be written as a sum of operator tensors that are distinct (and not proportional to one another).

These examples explain the choice of diagrammatic notation. The two boxes \lq\lq join together" at the portions of those boxes that are not double-lined.

\subsection{Identity operators}\label{sec:identityoperators}

We can define an identity operator with inputs or with outputs as follows
\begin{equation}
\hat{\mathds{1}}_{\mathsfb{a}_1} = \sum_{a} |a\rangle_{\mathsfb{a}_1} \negs\negs\negs\langle a|  ~~~~~~~~~~~
\hat{\mathds{1}}^{\mathsfb{a}_1} = \sum_{a} |a\rangle^{\mathsfb{a}_1} \negs\negs \langle a|
\end{equation}
in symbolic notation.  Note that we read the notation $|a\rangle_{\mathsfb{a}_1}\negs\negs\negs \langle a|$ as meaning the tensor product of $|a\rangle_{\mathsfb{a}_1}$ and $\presub{\mathsfb{a}_1}\langle a|$.

In diagrammatic notation the identity with an input is as
 \begin{equation}
\begin{Compose}{0}{0}\setdefaultfont{} \setsecondfont{\mathsfb}
\crectangledleft{A}{0.8}{1}{-1.5,0}
\relpoint{A}{0,-2.5}{Ad} \jointbleft[left]{Ad}{0}{A}{0} \csymbolalt{a}
\crectangledright{AA}{0.8}{1}{1.5,0}
\relpoint{AA}{0,-2.5}{Ad} \jointbright[right]{Ad}{0}{AA}{0} \csymbolalt{a}
\joinrlnoarrow[above]{A}{0}{AA}{0}
\end{Compose}
~~=~~
\begin{Compose}{0}{0}\setdefaultfont{} \setsecondfont{\mathsfb}
\crectangledouble{A}{1.4}{1}{0,0} \csymbol{\hat{\mathds{1}}}
\relpoint{A}{0,-2.5}{Ad} \jointbleft[left]{Ad}{-0.65}{A}{-0.65} \csymbolalt{a}
\relpoint{A}{0,-2.5}{Ad} \jointbright[right]{Ad}{0.65}{A}{0.65} \csymbolalt{a}
\end{Compose}
~~=~~
\begin{Compose}{0}{0}\setdefaultfont{} \setsecondfont{\mathsfb}
\crectangledouble{A}{1.4}{1}{0,0} \csymbol{\hat{\mathds{1}}}
\relpoint{A}{0,-2.5}{Ad} \jointbnoarrow[right]{Ad}{0}{A}{0} \csymbolalt{a}
\end{Compose}
~~=~~
\begin{Compose}{0}{0} \setsecondfont{\mathsfb}
\thispoint{a}{0,-1}
\ccap{t}{2}{1}{0,1}
\jointbleft[left]{a}{-2}{t}{-2}  \csymbolalt{a}
\jointbright[right]{a}{2}{t}{2}  \csymbolalt{a}
\end{Compose}
\end{equation}
On the left is the definition (in terms of basis elements) and the three diagrams on the right are alternative ways of representing this.
The corresponding diagrams for the identity with an output are
\begin{equation}
\begin{Compose}{0}{0}\setdefaultfont{} \setsecondfont{\mathsfb}
\crectangledleft{A}{0.8}{1}{-1.5,0}
\relpoint{A}{0,2.5}{Ad} \jointbleft[left]{A}{0}{Ad}{0} \csymbolalt{a}
\crectangledright{AA}{0.8}{1}{1.5,0}
\relpoint{AA}{0,2.5}{Ad} \jointbright[right]{AA}{0}{Ad}{0} \csymbolalt{a}
\joinrlnoarrow[above]{A}{0}{AA}{0}
\end{Compose}
~~=~~
\begin{Compose}{0}{0}\setdefaultfont{} \setsecondfont{\mathsfb}
\crectangledouble{A}{1.4}{1}{0,0} \csymbol{\hat{\mathds{1}}}
\relpoint{A}{0,2.5}{Ad} \jointbleft[left]{A}{-0.65}{Ad}{-0.65} \csymbolalt{a}
\relpoint{A}{0,2.5}{Ad} \jointbright[right]{A}{0.65}{Ad}{0.65} \csymbolalt{a}
\end{Compose}
~~=~~
\begin{Compose}{0}{0}\setdefaultfont{} \setsecondfont{\mathsfb}
\crectangledouble{A}{1.4}{1}{0,0} \csymbol{\hat{\mathds{1}}}
\relpoint{A}{0,2.5}{Ad} \joinbtnoarrow[right]{Ad}{0}{A}{0} \csymbolalt{a}
\end{Compose}
~~=~~
\begin{Compose}{0}{0} \setsecondfont{\mathsfb}
\thispoint{a}{0,1}
\ccup{t}{2}{1}{0,-1} \csymbol[0,-50]{a}
\jointbleft[left]{t}{-2}{a}{-2} \csymbolalt{a}
\jointbright[right]{t}{2}{a}{2} \csymbolalt{a}
\end{Compose}
\end{equation}
The cap and cup notation was introduced into the quantum literature by Coecke, Perdrix, and Paquette,\cite{coecke2008bases}.  It is justified because the identity operator, by virtue of its definition, acts like a $\delta_{aa'}$ between the unbarred and barred legs.

We can also define caps and cups for the operator tensor wires (i.e.\ the doubled up wires) as follows
\begin{equation}\label{optenscapandcup}
\begin{Compose}{0}{0} \setsecondfont{\mathsfb}
\thispoint{a}{0,-1}
\ccap{t}{2.25}{1}{0,1}
\jointbnoarrow[left]{a}{-2.25}{t}{-2.25}  \csymbolalt{a}
\jointbnoarrow[right]{a}{2.25}{t}{2.25}  \csymbolalt{a}
\end{Compose}
~~=~~
\begin{Compose}{0}{0} \setsecondfont{\mathsfb}
\thispoint{a}{0,-1}
\ccap{t}{2.5}{1}{0,1}
\jointbleft[left]{a}{-2.5}{t}{-2.5}  \csymbolalt{a}
\jointbright[right]{a}{2.5}{t}{2.5}  \csymbolalt{a}
\thispoint{a}{0,-1}
\ccap{t}{2}{1}{0,1}
\jointbright[right]{a}{-2}{t}{-2}  \csymbolalt{a}
\jointbleft[left]{a}{2}{t}{2}  \csymbolalt{a}
\end{Compose}
~~~~~~~~~~~~~~~
\begin{Compose}{0}{0} \setsecondfont{\mathsfb}
\thispoint{a}{0,1}
\ccup{t}{2.25}{1}{0,-1}
\jointbnoarrow[left]{t}{-2.25}{a}{-2.25}  \csymbolalt{a}
\jointbnoarrow[right]{t}{2.25}{a}{2.25}  \csymbolalt{a}
\end{Compose}
~~=~~
\begin{Compose}{0}{0} \setsecondfont{\mathsfb}
\thispoint{a}{0,1}
\ccup{t}{2.5}{1}{0,-1}
\jointbleft[left]{t}{-2.5}{a}{-2.5}  \csymbolalt{a}
\jointbright[right]{t}{2.5}{a}{2.5}  \csymbolalt{a}
\thispoint{a}{0,1}
\ccup{t}{2}{1}{0,-1}
\jointbright[right]{t}{-2}{a}{-2}  \csymbolalt{a}
\jointbleft[left]{t}{2}{a}{2}  \csymbolalt{a}
\end{Compose}
\end{equation}

\subsection{Adjoint} \label{sec:adjoint}

We define the adjoint as follows
\begin{equation}
\left(
\begin{Compose}{0}{0}\setdefaultfont{} \setsecondfont{\mathsfb}
\crectangledleft{B}{1}{1.5}{0,0} \csymbol{B}
\thispoint{d}{0,-4} \thispoint{u}{0,4}
\jointbleft[left]{d}{0}{B}{0} \csymbolalt{a} \jointbleft[left]{B}{0}{u}{0}\csymbolalt{b}
\end{Compose} \right)^\dagger
~~=~~
\begin{Compose}{0}{0}\setdefaultfont{} \setsecondfont{\mathsfb}
\crectangledleft{B}{1}{1.5}{0,0} \csymbol{B^\dagger}
\thispoint{d}{0,-4} \thispoint{u}{0,4}
\jointbleft[left]{d}{0}{B}{0} \csymbolalt{b} \jointbleft[left]{B}{0}{u}{0}\csymbolalt{a}
\end{Compose}
~~ := ~~
\begin{Compose}{0}{0}\setdefaultfont{} \setsecondfont{\mathsfb}
\crectangledright{B}{1}{1.5}{0,0} \csymbol{B}
\ccap{u}{2}{2}{-2,5}  \ccup{d}{2}{2}{2,-5}
\jointbright[right]{d}{-2}{B}{0} \csymbolalt{a} \jointbright[right]{B}{0}{u}{2} \csymbolalt{b}
\thispoint{dd}{-4,-4} \thispoint{uu}{4,4}
\jointbleft[left]{dd}{0}{u}{-2}  \csymbolalt{b}
\jointbleft[left]{d}{2}{uu}{0} \csymbolalt{a}
\end{Compose}
\end{equation}
This is because, on flipping horizontally, we take the complex conjugate of the coefficients and then the cup and cap interchange rows and columns.
Clearly the adjoint of the adjoint returns the original operator.  The diagrammatic notation can be enriched to take advantage of this by representing operations that are adjoints of each other by a small black square (following Selinger \cite{selinger2007dagger}) as follows
\begin{equation}
\left(
\begin{Compose}{0}{0}\setdefaultfont{} \setsecondfont{\mathsfb}
\crectangledleft[-1]{B}{1}{1.5}{0,0} \csymbol{B}
\thispoint{d}{0,-4} \thispoint{u}{0,4}
\jointbleft[left]{d}{0}{B}{0} \csymbolalt{a} \jointbleft[left]{B}{0}{u}{0}\csymbolalt{b}
\end{Compose} \right)^\dagger
~~=~~
\begin{Compose}{0}{0}\setdefaultfont{} \setsecondfont{\mathsfb}
\crectangledleft[+1]{B}{1}{1.5}{0,0} \csymbol{B}
\thispoint{d}{0,-4} \thispoint{u}{0,4}
\jointbleft[left]{d}{0}{B}{0} \csymbolalt{b} \jointbleft[left]{B}{0}{u}{0}\csymbolalt{a}
\end{Compose}
~~:=~~
\begin{Compose}{0}{0}\setdefaultfont{} \setsecondfont{\mathsfb}
\crectangledright[-1]{B}{1}{1.5}{0,0} \csymbol{B}
\ccap{u}{2}{2}{-2,5}  \ccup{d}{2}{2}{2,-5}
\jointbright[right]{d}{-2}{B}{0} \csymbolalt{a} \jointbright[right]{B}{0}{u}{2} \csymbolalt{b}
\thispoint{dd}{-4,-4} \thispoint{uu}{4,4}
\jointbleft[left]{dd}{0}{u}{-2}  \csymbolalt{b}
\jointbleft[left]{d}{2}{uu}{0} \csymbolalt{a}
\end{Compose}
\end{equation}
We can visualise this as \lq\lq sliding" the box along the wire (then straightening out the wire).  Note, though, that there are different types ($\mathsfb{a}$ and $\mathsfb b$) on the ends so as the box slides, it \lq\lq eats up" the wire of one type replacing it with wire of the other type.

We can use this to define the adjoint of an operator tensor by using the caps and cups defined in \eqref{optenscapandcup} as follows
\begin{equation}\label{optensoradjoint}
\left(
\begin{Compose}{0}{0}\setdefaultfont{} \setsecondfont{\mathsfb}
\crectangledmark[-1]{B}{2}{1.5}{0,0} \csymbol{B}
\thispoint{d}{0,-4} \thispoint{u}{0,4}
\jointbnoarrow[left]{d}{0}{B}{0} \csymbolalt{a} \jointbnoarrow[left]{B}{0}{u}{0}\csymbolalt{b}
\end{Compose} \right)^\dagger
~~=~~
\begin{Compose}{0}{0}\setdefaultfont{} \setsecondfont{\mathsfb}
\crectangledmark[+1]{B}{2}{1.5}{0,0} \csymbol{B}
\thispoint{d}{0,-4} \thispoint{u}{0,4}
\jointbnoarrow[left]{d}{0}{B}{0} \csymbolalt{b} \jointbnoarrow[left]{B}{0}{u}{0}\csymbolalt{a}
\end{Compose}
~~:=~~
\begin{Compose}{0}{0}\setdefaultfont{} \setsecondfont{\mathsfb}
\crectangledmark[-1]{B}{2}{1.5}{0,0} \csymbol{B}
\ccap{u}{2}{2}{-2,5}  \ccup{d}{2}{2}{2,-5}
\jointbnoarrow[right]{d}{-2}{B}{0} \csymbolalt{a} \jointbnoarrow[right]{B}{0}{u}{2} \csymbolalt{b}
\thispoint{dd}{-4,-4} \thispoint{uu}{4,4}
\jointbnoarrow[left]{dd}{0}{u}{-2}  \csymbolalt{b}
\jointbnoarrow[left]{d}{2}{uu}{0} \csymbolalt{a}
\end{Compose}
\end{equation}
We will see in Sec.\ \ref{sec:proofthatformulationistimesymmetric} that the adjoint is associated with the time reverse (up to an over all gauge factor).

\subsection{Unitary operators}\label{sec:unitaryoperators}

Now we have defined the adjoint operation, we can define unitary operators. A unitary operator, $U$, has the properties that
\begin{equation}\label{Udefiningproperties}
\begin{Compose}{0}{0} \setdefaultfont{}\setsecondfont{\mathsfb}
\crectangledleft[-1]{U}{1}{1.5}{0,-3}\csymbol{U} \crectangledleft[1]{UA}{1}{1.5}{0,3}\csymbol{U}
\thispoint{d}{0,-7}\thispoint{u}{0,7}
\jointbleft[left]{d}{0}{U}{0} \csymbolalt{a}
\jointbleft[left]{U}{0}{UA}{0} \csymbolalt{b}
\jointbleft[left]{UA}{0}{u}{0} \csymbolalt{a}
\end{Compose}
~~=~~
\begin{Compose}{0}{0}\setdefaultfont{}\setsecondfont{\mathsfb}
\thispoint{d}{0,-7}\thispoint{u}{0,7}
\jointbleft[left]{d}{0}{u}{0} \csymbolalt{a}
\end{Compose}
~~~~~~~~~~~~
\begin{Compose}{0}{0} \setdefaultfont{}\setsecondfont{\mathsfb}
\crectangledleft[1]{U}{1}{1.5}{0,-3}\csymbol{U} \crectangledleft[-1]{UA}{1}{1.5}{0,3}\csymbol{U}
\thispoint{d}{0,-7}\thispoint{u}{0,7}
\jointbleft[left]{d}{0}{U}{0} \csymbolalt{b}
\jointbleft[left]{U}{0}{UA}{0} \csymbolalt{a}
\jointbleft[left]{UA}{0}{u}{0} \csymbolalt{b}
\end{Compose}
~~=~~
\begin{Compose}{0}{0}\setdefaultfont{}\setsecondfont{\mathsfb}
\thispoint{d}{0,-7}\thispoint{u}{0,7}
\jointbleft[left]{d}{0}{u}{0} \csymbolalt{b}
\end{Compose}
\end{equation}
where $N_\mathsfb{a}=N_\mathsfb{b}$.

We can form a unitary operator tensor by doubling up
\begin{equation}
\begin{Compose}{0}{0} \setdefaultfont{\hat}\setsecondfont{\mathsfb}
\Crectangledmark[-1]{U}{2.5}{1.5}{0,0}
\thispoint{d}{0,-4}\thispoint{u}{0,4}
\jointbnoarrow[left]{d}{0}{U}{0} \csymbolalt{a}
\jointbnoarrow[left]{U}{0}{u}{0} \csymbolalt{b}
\end{Compose}
~~=~~
\begin{Compose}{0}{0} \setdefaultfont{}\setsecondfont{\mathsfb}
\crectangledleft[-1]{U}{1}{1.5}{-1.5,0}\csymbol{U} \crectangledright[-1]{Ur}{1}{1.5}{1.5,0}\csymbol{U}
\thispoint{d}{0,-4}\thispoint{u}{0,4}
\jointbleft[left]{d}{-1.5}{U}{0} \csymbolalt{a}
\jointbleft[left]{U}{0}{u}{-1.5} \csymbolalt{b}
\jointbright[right]{d}{1.5}{Ur}{0} \csymbolalt{a}
\jointbright[right]{Ur}{0}{u}{1.5} \csymbolalt{b}
\end{Compose}
\end{equation}
It can be verified that this has the properties
\begin{equation}
\begin{Compose}{0}{0} \setdefaultfont{\hat}\setsecondfont{\mathsfb}
\crectangledmark[-1]{U}{2.5}{1.5}{0,-3}\csymbol{U} \crectangledmark[1]{UA}{2.5}{1.5}{0,3}\csymbol{U}
\thispoint{d}{0,-7}\thispoint{u}{0,7}
\jointbnoarrow[left]{d}{0}{U}{0} \csymbolalt{a}
\jointbnoarrow[left]{U}{0}{UA}{0} \csymbolalt{b}
\jointbnoarrow[left]{UA}{0}{u}{0} \csymbolalt{a}
\end{Compose}
~~=~~
\begin{Compose}{0}{0}\setdefaultfont{}\setsecondfont{\mathsfb}
\thispoint{d}{0,-7}\thispoint{u}{0,7}
\jointbnoarrow[left]{d}{0}{u}{0} \csymbolalt{a}
\end{Compose}
~~~~~~~~~~~~
\begin{Compose}{0}{0} \setdefaultfont{\hat}\setsecondfont{\mathsfb}
\crectangledmark[1]{U}{2.5}{1.5}{0,-3}\csymbol{U} \crectangledmark[-1]{UA}{2.5}{1.5}{0,3}\csymbol{U}
\thispoint{d}{0,-7}\thispoint{u}{0,7}
\jointbnoarrow[left]{d}{0}{U}{0} \csymbolalt{b}
\jointbnoarrow[left]{U}{0}{UA}{0} \csymbolalt{a}
\jointbnoarrow[left]{UA}{0}{u}{0} \csymbolalt{b}
\end{Compose}
~~=~~
\begin{Compose}{0}{0}\setdefaultfont{}\setsecondfont{\mathsfb}
\thispoint{d}{0,-7}\thispoint{u}{0,7}
\jointbnoarrow[left]{d}{0}{u}{0} \csymbolalt{b}
\end{Compose}
\end{equation}
We are using the position of the small black square to indicate the adjoint.

The defining properties of a unitary given in \eqref{Udefiningproperties} are equivalent to the following properties
\begin{equation}\label{Udefinitioncupcapversion}
\begin{Compose}{0}{0} \setdefaultfont{}\setsecondfont{\mathsfb}
\crectangledleft[-1]{U}{1}{1.5}{-1.5,0}\csymbol{U} \crectangledright[-1]{Ur}{1}{1.5}{1.5,0}\csymbol{U}
\thispoint{d}{0,-4}\thispoint{u}{0,4}
\jointbleft[left]{d}{-1.5}{U}{0} \csymbolalt{a}
\jointbleft[left]{U}{0}{u}{-1.5} \csymbolalt{b}
\jointbright[right]{d}{1.5}{Ur}{0} \csymbolalt{a}
\jointbright[right]{Ur}{0}{u}{1.5} \csymbolalt{b}
\jointtnoarrow{u}{-1.5}{u}{1.5}
\end{Compose}
~~=~~
\begin{Compose}{0}{0} \setdefaultfont{}\setsecondfont{\mathsfb}
\thispoint{d}{0,-4}\thispoint{u}{0,2}
\jointbleft[left]{d}{-1.5}{u}{-1.5} \csymbolalt{a}
\jointbright[right]{d}{1.5}{u}{1.5} \csymbolalt{a}
\jointtnoarrow{u}{-1.5}{u}{1.5}
\end{Compose}
~~~~~~~~~~~~~~~
\begin{Compose}{0}{0} \setdefaultfont{}\setsecondfont{\mathsfb}
\crectangledleft[-1]{U}{1}{1.5}{-1.5,0}\csymbol{U} \crectangledright[-1]{Ur}{1}{1.5}{1.5,0}\csymbol{U}
\thispoint{d}{0,-4}\thispoint{u}{0,4}
\jointbleft[left]{d}{-1.5}{U}{0} \csymbolalt{a}
\jointbleft[left]{U}{0}{u}{-1.5} \csymbolalt{b}
\jointbright[right]{d}{1.5}{Ur}{0} \csymbolalt{a}
\jointbright[right]{Ur}{0}{u}{1.5} \csymbolalt{b}
\joinbbnoarrow{d}{-1.5}{d}{1.5}
\end{Compose}
~~=~~
\begin{Compose}{0}{0} \setdefaultfont{}\setsecondfont{\mathsfb}
\thispoint{d}{0,-2}\thispoint{u}{0,4}
\jointbleft[left]{d}{-1.5}{u}{-1.5} \csymbolalt{b}
\jointbright[right]{d}{1.5}{u}{1.5} \csymbolalt{b}
\joinbbnoarrow{d}{-1.5}{d}{1.5}
\end{Compose}
\end{equation}
This is clear since we can add a cup or cap as appropriate and employ the sliding manoeuvre to convert between these equivalent expressions.  In so doing, we also employ the much discussed \lq\lq yanking" equations \cite{coecke2017picturing}
\begin{equation}
\begin{Compose}{0}{0}\setdefaultfont{}\setsecondfont{\mathsfb}
\thispoint{d}{0,-4}\thispoint{u}{0,4}
\jointbleft[left]{d}{0}{u}{0} \csymbolalt{a}
\end{Compose}
~~=~~
\begin{Compose}{0}{0} \setdefaultfont{}\setsecondfont{\mathsfb}
\thispoint{d}{0,-4}\thispoint{u}{0,2} \thispoint{e}{0,-2}  \thispoint{uu}{0,4}
\jointbleft[left]{d}{-1.5}{u}{-1.5} \csymbolalt{a}
\jointbright[right]{e}{1.5}{u}{1.5} \csymbolalt{a}
\jointtnoarrow{u}{-1.5}{u}{1.5}
\joinbbnoarrow{e}{1.5}{e}{4.5}
\jointbleft[left]{e}{4.5}{uu}{4.5} \csymbolalt{a}
\end{Compose}
~~~~~~~~~~~~~~~
\begin{Compose}{0}{0}\setdefaultfont{}\setsecondfont{\mathsfb}
\thispoint{d}{0,-4}\thispoint{u}{0,4}
\jointbleft[left]{d}{0}{u}{0} \csymbolalt{b}
\end{Compose}
~~=~~
\begin{Compose}{0}{0} \setdefaultfont{}\setsecondfont{\mathsfb}
\thispoint{d}{0,-4}\thispoint{u}{0,2} \thispoint{e}{0,-2}  \thispoint{uu}{0,4}
\jointbleft[left]{d}{1.5}{u}{1.5} \csymbolalt{b}
\jointbright[right]{e}{-1.5}{u}{-1.5} \csymbolalt{b}
\jointtnoarrow{u}{1.5}{u}{-1.5}
\joinbbnoarrow{e}{-1.5}{e}{-4.5}
\jointbleft[left]{e}{-4.5}{uu}{-4.5} \csymbolalt{b}
\end{Compose}
\end{equation}
These are easily obtained from the identity relations we have given.   We will employ the definition \eqref{Udefinitioncupcapversion} in proving the extension theorem in Sec.\ \ref{sec:extensiontheorem}.   An alternative definition of unitarity for an operator tensor is to say that $\hat{U}$ is homogeneous (see \eqref{homogeneousoptensor}) and
\begin{equation}
\begin{Compose}{0}{0} \setdefaultfont{\hat}\setsecondfont{\mathsfb}
\Crectangledmark[-1]{U}{2.5}{1.5}{0,0} \crectangledouble{I}{1}{1}{0,4.5} \csymbol{\mathds{1}}
\thispoint{d}{0,-4}\thispoint{u}{0,4}
\jointbnoarrow[left]{d}{0}{U}{0} \csymbolalt{a}
\jointbnoarrow[left]{U}{0}{I}{0} \csymbolalt{b}
\end{Compose}
~~=~~
\begin{Compose}{0}{0}\setdefaultfont{\hat}\setsecondfont{\mathsfb}
\crectangledouble{I}{1}{1}{0,1} \csymbol{\mathds{1}}
\thispoint{d}{0,-4}
\jointbnoarrow[left]{d}{0}{I}{0} \csymbolalt{a}
\end{Compose}
~~~~~~~~~~~~
\begin{Compose}{0}{0} \setdefaultfont{\hat}\setsecondfont{\mathsfb}
\Crectangledmark[-1]{U}{2.5}{1.5}{0,0} \crectangledouble{I}{1}{1}{0,-4.5} \csymbol{\mathds{1}}
\thispoint{d}{0,-4}\thispoint{u}{0,4}
\jointbnoarrow[left]{I}{0}{U}{0} \csymbolalt{a}
\jointbnoarrow[left]{U}{0}{u}{0} \csymbolalt{b}
\end{Compose}
~~=~~
\begin{Compose}{0}{0}\setdefaultfont{\hat}\setsecondfont{\mathsfb}
\crectangledouble{I}{1}{1}{0,-1} \csymbol{\mathds{1}}
\thispoint{u}{0,4}
\jointbnoarrow[left]{I}{0}{u}{0} \csymbolalt{b}
\end{Compose}
\end{equation}
These equations are interesting, in the context of this paper, because they relate to the double causality condition when there are no incomes or outcomes, as we will see.

\subsection{More scalar products}

We can use the identity operators we have just defined to define a scalar product between barred and unbarred elements as follows
\begin{equation}
\begin{Compose}{0}{0}\setdefaultfont{} \setsecondfont{\mathsfb}
\crectangledleft[-1]{B}{1}{1.5}{-2,-2} \csymbol{B} 
\crectangledright[-1]{C}{1}{1.5}{2,-2} \csymbol{A} 
\ccap{i}{2}{2}{0,4}
\jointbleft[left]{B}{0}{i}{-2} \csymbolalt{a}
\jointbright[right]{C}{0}{i}{2} \csymbolalt{a}
\end{Compose}
~~~~~~~~~~
\begin{Compose}{0}{0}\setdefaultfont{} \setsecondfont{\mathsfb}
\crectangledleft[1]{B}{1}{1.5}{-2,2} \csymbol{A}
\crectangledright[1]{C}{1}{1.5}{2,2} \csymbol{B}
\ccup{i}{2}{2}{0,-4}
\jointbleft[left]{i}{-2}{B}{0} \csymbolalt{a}
\jointbright[right]{i}{2}{C}{0} \csymbolalt{a}
\end{Compose}
\end{equation}
We have included the small black squares since we see that we can slide the $A$ box along the wire on the left and slide the $B$ box along the wire on the right to get the same scalar product (we can visualise this as the $\mathsf a$ wire getting eaten up and being replaced by a null wire not shown).

\subsection{Joining operator tensors}\label{sec:joiningoperatortensors}

We can wire operator tensors together.  Consider, first, the simple example of wiring together two homogeneous operator tensors.
\begin{equation}
\begin{Compose}{0}{0} \setdefaultfont{} \setsecondfont{\mathsfb}
\crectangledouble{A}{2}{1.5}{0,0} \csymbol{\hat{A}}
\thispoint{p}{0,5} \jointbnoarrow[left]{A}{0}{p}{0} \csymbolalt{a}
\end{Compose}
=
\begin{Compose}{0}{0} \setdefaultfont{} \setsecondfont{\mathsfb}
\crectangledleft{A}{1.2}{1.5}{-1.5,0} \csymbol{A}
\thispoint{p}{-1.5,5} \jointbleft[left]{A}{0}{p}{0} \csymbolalt{a}
\crectangledright{A}{1.2}{1.5}{1.5,0} \csymbol{A}
\thispoint{p}{1.5,5} \jointbright[right]{A}{0}{p}{0} \csymbolalt{a}
\end{Compose}
~~~~~~~~~~~~~~~~
\begin{Compose}{0}{0} \setdefaultfont{} \setsecondfont{\mathsfb}
\crectangledouble{A}{2}{1.5}{0,0} \csymbol{\hat{B}}
\thispoint{p}{0,5} \jointbnoarrow[left]{A}{0}{p}{0} \csymbolalt{a}
\end{Compose}
=
\begin{Compose}{0}{0} \setdefaultfont{} \setsecondfont{\mathsfb}
\crectangledleft{A}{1.2}{1.5}{-1.5,0} \csymbol{B}
\thispoint{p}{-1.5,-5} \joinbtleft[left]{A}{0}{p}{0} \csymbolalt{a}
\crectangledright{A}{1.2}{1.5}{1.5,0} \csymbol{B}
\thispoint{p}{1.5,-5} \joinbtright[right]{A}{0}{p}{0} \csymbolalt{a}
\end{Compose}
\end{equation}
When we wire these together we get
\begin{equation}
\begin{Compose}{0}{0} \setdefaultfont{} \setsecondfont{\mathsfb}
\crectangledouble{A}{2}{1.5}{0,-4} \csymbol{\hat{A}}
\crectangledouble{B}{2}{1.5}{0,4} \csymbol{\hat{B}}
\jointbnoarrow[left]{A}{0}{B}{0} \csymbolalt{a}
\end{Compose}
~~=~~
\begin{Compose}{0}{0} \setdefaultfont{} \setsecondfont{\mathsfb}
\crectangledleft{A}{1.2}{1.5}{-1.5,-4} \csymbol{A}
\crectangledleft{B}{1.2}{1.5}{-1.5,4} \csymbol{B}
\jointbleft[left]{A}{0}{B}{0} \csymbolalt{a}
\crectangledright{A}{1.2}{1.5}{1.5,-4} \csymbol{A}
\crectangledright{B}{1.2}{1.5}{1.5,4} \csymbol{B}
\jointbright[right]{A}{0}{B}{0} \csymbolalt{a}
\end{Compose}
\end{equation}
Inserting in more familiar kets and bras from Sec.\ \ref{sec:inputandoutputhilbertspaces} we see this expression is equal to
\begin{equation}
\presub{\mathsfb{a}_1}\langle B|A\rangle^{\mathsfb{a}_1} ~ \presup{\mathsfb{a}_1}\langle A| B\rangle_{\mathsfb{a}_1}
= \text{trace}(\hat{A}\hat{B})
\end{equation}
The trace operation is linear. Hence even if $\hat{A}$ and $\hat{B}$ are not homogeneous, we still get the trace when we wire two operators together in this fashion. In symbolic operator tensor notation we notate this as
\begin{equation}
\hat{A}^{\mathsfb{a}_1} \hat{B}_{\mathsfb{a}_1}
\end{equation}
so, when we have a repeated index like this, we take the trace. In fact, we take the trace just over the associated part of the Hilbert space (i.e.\ the partial trace).   To understand this, first consider the example
\begin{equation}\label{ACBcase} \hat{A}^{\mathsfb{a}_1}\hat{C}_{\mathsf{b}_3}  \hat{B}_{\mathsf{a}_1}  \end{equation}
is equal to the trace of $\hat{A}^{\mathsfb{a}_1} \hat{B}_{\mathsf{a}_1}$ times the operator tensor $\hat{C}_{\mathsf{b}_3}$. In more traditional notation we would write
\begin{equation} \text{trace}(\hat{A}^{\mathsfb{a}_1}  \hat{B}_{\mathsf{a}_1})\hat{C}_{\mathsf{b}_3}   ~~~~~~~\text{traditional notation}  \end{equation}
The object
\begin{equation}
\hat{D}^{\mathsfb{a}_1}_{\mathsf{b}_3}  \hat{B}_{\mathsf{a}_1}
\end{equation}
is obtained by taking the partial trace over the part of the Hilbert space corresponding to $\mathsfb{a}_1$.  Diagrammatically, this example is represented as
\begin{equation}
\begin{Compose}{0}{0} \setdefaultfont{} \setsecondfont{\mathsfb}
\crectangledouble{A}{2}{1.5}{0,-4} \csymbol{\hat{D}}
\crectangledouble{B}{2}{1.5}{0,4} \csymbol{\hat{B}}
\jointbnoarrow[left]{A}{0}{B}{0} \csymbolalt{a}
\thispoint{p}{0,-9} \jointbnoarrow[left]{p}{0}{A}{0} \csymbolalt{b}
\end{Compose}
\end{equation}
The partial trace is a well defined operation. One way to understand it is to expand $\hat{D}^{\mathsfb{a}_1}_{\mathsf{b}_3}$ as a linear sum of tensor product operators then apply the reasoning in \eqref{ACBcase} to each term in the expansion.  In fact, we can use fiducials to perform such an expansion.   In the next section we will discuss how to do this.  These ideas are better explained in diagrammatic notation.

Note, in the symbolic operator tensor notation, the order of the operators is unimportant since all the salient information is in the indices.  Thus, for example,
\begin{equation} \hat{A}^{\mathsfb{a}_1}  \hat{B}_{\mathsf{a}_2} = \hat{B}_{\mathsf{a}_2} \hat{A}^{\mathsfb{a}_1} \end{equation}
(this example is just the tensor product of two operator tensors since there are no repeated indices).

\section{Fiducials}

In this section we will consider expanding Hermitian operator tensors (henceforth the \lq\lq Hermitian" will be implicit) in terms of fiducial operators. We do this to mirror the expansion of operations in terms of fiducials as discussed in Sec.\ \ref{sec:generalduotensorswithroundandsquaredots}.  We also add fiducial structure mirroring that associated with the pointer types so we have both square and round dots on the associated duotensors.

\subsection{Fiducial input and output elements}

We define a fiducial output set of operator tensors corresponding to each type, $\mathsfb{a}$, that span the corresponding space.  We represent these as (symbolic notation on left, diagrammatic on the right)
\begin{equation}
\presub{\mathpzc{a}_2}{\hat{Q}}^{\mathsfb{a}_1}
~~~~~~~~~~~~~~~
\begin{Compose}{0}{-0.1}\setdefaultfont{\hat} \setsecondfont{\mathsfb} \setthirdfont{\mathpzc}
\supopfid{Q}{0,0}
\thispoint{QU}{0,2.5}\csymbolalt[0,20]{a} \jointbnoarrow{Q}{0}{QU}{0}
\blackdot{QD}{0,-2.5}\csymbolthird[0,-20]{a} \jointbnoarrow{QD}{0}{Q}{0}
\end{Compose}
\end{equation}
where $\mathpzc{a}=1$ to $K_\mathsfb{a}= N_\mathsfb{a}^2$ since we require $N_\mathsfb{a}^2$ Hermitian operator to span the space $\mathcal{V}^{\mathsfb{a}_1}$.
Similarly, we define a fiducial input set of operators as (symbolic notation on left, diagrammatic on the right)
\begin{equation}
{\hat{Q}}_{\mathsfb{a}_1}^{\mathpzc{a}_2}
~~~~~~~~~~~~~~~
\begin{Compose}{0}{-0.1}\setdefaultfont{\hat} \setsecondfont{\mathsfb} \setthirdfont{\mathpzc}
\sdownopfid{Q}{0,0}
\thispoint{QU}{0,-2.5}\csymbolalt[0,-20]{a} \joinbtnoarrow{Q}{0}{QU}{0}
\blackdot{QD}{0,2.5}\csymbolthird[0,25]{a} \joinbtnoarrow{QD}{0}{Q}{0}
\end{Compose}
\end{equation}
where, similarly, $\mathpzc{a}=1$ to $K_\mathsfb{a}= N_\mathsfb{a}^2$.

\subsection{Hopping matrix for inputs and outputs}\label{sec:hoppingmatrixforinputsandoutputs}

We define the hopping matrix for inputs and outputs in terms of the associated operator tensor fiducial elements as follows
\begin{equation}\label{operatorhoppingmetric}
\begin{Compose}{0}{0} \setfourthfont{\mathpzc}
\blackdot{u}{0,1.5}\blackdot{d}{0,-1.5} \jointbnoarrow[left]{d}{0}{u}{0}\csymbolfourth{x}
\end{Compose}
~~=~~
\begin{Compose}{0}{0} \setsecondfont{\mathsfb} \setfourthfont{\mathpzc}
\blackdot{u}{0,3.5} \csymbolfourth[0,20]{x}\sdownopfid{up}{0,1.5} \supopfid{do}{0,-1.5} \blackdot{d}{0,-3.5}\csymbolfourth[0,-20]{x}
\jointbnoarrow[left]{d}{0}{do}{0} \jointbnoarrow[left]{do}{0}{up}{0} \csymbolalt{x} \jointbnoarrow[left]{up}{0}{u}{0}
\end{Compose}
\end{equation}
Note that we use the same symbol for the hopping matrix in the operator case here as we do in the case of the operation theory (see \eqref{operationhoppingmatrix}) even though these are distinct objects. This is for convenience since we will subsequently equate these objects to set up a correspondence between operations and operators (see Sec.\ \ref{sec:correspondencebetweenoperationsandoperators}
The wire connecting the two operators in \eqref{operatorhoppingmetric} indicates taking the trace as discussed in Sec.\ \ref{sec:joiningoperatortensors}.  In symbolic notation we can write \eqref{operatorhoppingmetric} as
\begin{equation}
\presub{\mathpzc{a}_2}h^{\mathpzc{a}_3} =  \presub{\mathpzc{a}_2}{\hat{Q}}^{\mathsfb{a}_1} {\hat{Q}}_{\mathsfb{a}_1}^{\mathpzc{a}_3}
\end{equation}
where $\presub{\mathpzc{a}_2}h^{\mathpzc{a}_3}$ is symbolic notation for the hopping matrix.

We define the inverse of the hopping matrix to be
\begin{equation}
\begin{Compose}{0}{0} \setfourthfont{\mathpzc}
\whitedot{u}{0,1.5}\whitedot{d}{0,-1.5} \jointbnoarrow[left]{d}{0}{u}{0}\csymbolfourth{x}
\end{Compose}
\end{equation}
We can show, using the same technique as in Sec.\ \ref{sec:hoppingmatrixforinputsandoutputs}, that
\begin{equation}\label{bwtowbroundop}
\begin{Compose}{0}{0} \setfourthfont{\mathpzc}
\thispoint{u}{0,2.5}\thispoint{d}{0,-2.5} \jointbnoarrowdotswb[left]{d}{0}{u}{0}
\end{Compose}
~=~
\begin{Compose}{0}{0} \setfourthfont{\mathpzc}
\thispoint{u}{0,2.5}\thispoint{d}{0,-2.5} \jointbnoarrow[left]{d}{0}{u}{0}
\end{Compose}
~=~
\begin{Compose}{0}{0} \setfourthfont{\mathpzc}
\thispoint{u}{0,2.5}\thispoint{d}{0,-2.5} \jointbnoarrowdotsbw[left]{d}{0}{u}{0}
\end{Compose}
\end{equation}
so we can insert and delete pairs of black and white dots as we like.

\subsection{Vector space for incomes and outcomes}\label{sec:vectorspaceforincomesandoutcomes}

We also need spaces to represent the incomes and outcomes.   We define
\begin{equation}
\mathcal{P}^{\mathtt{x}_1}  ~~~\text{and}~~~ \mathcal{P}_{\mathtt{x}_1}
\end{equation}
to be $N_\mathtt{x}$ dimensional real vector spaces.   Further, we define an orthogonal set of vectors for these spaces
\begin{equation}
\presub{x_2}{\vec{P}}^{\mathtt{x}_1} \in \mathcal{P}^{\mathtt{x}_1}  ~~~ \text{where}~~~ x=1, 2, \dots, N_\mathtt{x}
~~~~~~~~\text{and}~~~~~~~~
{\vec{P}}_{\mathtt{x}_1}^{x_2} \in \mathcal{P}_{\mathtt{x}_1}  ~~~ \text{where}~~~ x=1, 2, \dots, N_\mathtt{x}
\end{equation}
These vectors are not taken to be normalized. Rather, we choose them so that
\begin{equation}
\presub{x_2}{\vec{P}}^{\mathtt{x}_1} \cdot {\vec{P}}_{\mathtt{x}_1}^{x_3} = \frac{1}{N_\mathtt{x}}\presub{x_2}\delta^{x_3}
\end{equation}
where $\presub{x_2}\delta^{x_3}$ is equal to 1 if $x_2=x_3$, else it is equal to 0.
We will take the dot product to be implicit when we have a repeated index in the pointer type (the $\mathtt{x}_1$ index in this case) is repeated without explicitly putting writing a dot.  Thus we write
\begin{equation}\label{vechoppingsymbolic}
\presub{x_2}{\vec{P}}^{\mathtt{x}_1} {\vec{P}}_{\mathtt{x}_1}^{x_3} = \frac{1}{N_\mathtt{x}} \presub{x_2}\delta^{x_3}
\end{equation}
where the dot is taken as implicit.  We can represent this mathematics diagrammatically.  We put
\begin{equation}
\presub{x_2}{\vec{P}}^{\mathtt{x}_1} ~~~~~~~~~~~
\begin{Compose}{0}{0} \setsecondfont{\mathtt}
\blackdotsq{b}{-2.5,0}\csymbolfourth[-20,0]{x} \sleftvecfid{P}{0,0} \thispoint{p}{2.5,0} \csymbolalt[20,0]{x}
\joinrlnoarrow{b}{0}{P}{0} \joinrlnoarrowthick{P}{0}{p}{0}
\end{Compose}
\end{equation}
where the diagrammatic notation is given on the right.  These represent the outcome fiducial set.  Similarly, we have
\begin{equation}
{\vec{P}}_{\mathtt{x}_1}^{x_2} ~~~~~~~~~~~
\begin{Compose}{0}{0} \setsecondfont{\mathtt}
\blackdotsq{b}{2.5,0}\csymbolfourth[20,0]{x} \srightvecfid{P}{0,0} \thispoint{p}{-2.5,0} \csymbolalt[-20,0]{x}
\joinlrnoarrow{b}{0}{P}{0} \joinlrnoarrowthick{P}{0}{p}{0}
\end{Compose}
\end{equation}
for the income fiducial set.

\subsection{Income outcome hopping matrix}\label{sec:incomeoutcomehoppingmatrix}

With this diagrammatic notation in place we can write \eqref{vechoppingsymbolic} as
\begin{equation}\label{hoppingequivfidcircuitvec}
\begin{Compose}{0}{0}
\bbmatrixnoarrowsq{h}{0,0} \csymbolthird{x}
\end{Compose}
~~=~~
\begin{Compose}{0}{0} \setsecondfont{\mathtt}
\blackdotsq{FL}{-2.5,0} \csymbolthird[-24,0]{x}
\sleftvecfid{F}{0,0}
\srightvecfid{rF}{3,0}
\blackdotsq{rFR}{5.5,0} \csymbolthird[24,0]{x}
 \joinrlnoarrow[above]{FL}{0}{F}{0}
\joinrlnoarrowthick[above]{F}{0}{rF}{0} \csymbolalt{x}
\joinrlnoarrow[above]{rF}{0}{rFR}{0}
\end{Compose}
~~=~~ \frac{1}{N_\mathtt{x}}\left(\begin{array}{cccc}
                                    1 &   &        &  \\
                                      & 1 &        &  \\
                                      &   & \ddots &  \\
                                      &   &        & 1
                                  \end{array}\right)
\end{equation}
This is clearly the hopping matrix associated the real vector space $\mathcal{P}$.  We should compare \eqref{hoppingequivfidcircuitvec} with \eqref{hoppingequivfidcircuitpointer}.

We can define the inverse of this to be $\wwdotsnoarrowsq$.  Further, following the same steps as in Sec.\ \ref{sec:thepointerhoppingmatrix}, we can show
\begin{equation}\label{bwtowbvec}
\nbdotsnoarrowsq \hspace{2pt} \wndotsnoarrowsq ~ = ~ \customdotsnoarrowvar[]{9mm}
~=~ \nwdotsnoarrowsq \hspace{2pt} \bndotsnoarrowsq
\end{equation}
Hence, we can insert and delete pairs of black and white dots as we wish.

\subsection{A general operator tensor}\label{sec:ageneraloperatortensor}

We can expand an operator tensor in terms of the fiducial elements.  For example we can write
\begin{equation}\label{Lexpansiongeneraloperator}
\begin{Compose}{0}{0}\setsecondfont{\mathsfb} \setthirdfont{\mathpzc}  \setdefaultfont{\mathtt}
\crectangledouble{T}{3}{3}{0,0} \csymbolfourth{\hat{C}}
\thispoint{y7}{-7,1.5}   \csymbol[-24,0]{y} \joinrlnoarrowthick[above]{y7}{0}{T}{1.5}
\thispoint{x1}{-7,-1.5}   \csymbol[-24,0]{x} \joinrlnoarrowthick[above]{x1}{0}{T}{-1.5}
\thispoint{w8}{7,1.5}   \csymbol[24,0]{v} \joinrlnoarrowthick[above]{T}{1.5}{w8}{0}
\thispoint{y2}{7,-1.5}   \csymbol[24,0]{u} \joinrlnoarrowthick[above]{T}{-1.5}{y2}{0}
\upwire{T}{-1.5}{c}  \upwire{T}{1.5}{d}
\downwire{T}{-1.5}{a} \downwire{T}{1.5}{b}
\end{Compose}
~~~=~~~
\begin{Compose}{0}{0}\setsecondfont{\mathsfb} \setthirdfont{\mathpzc} \setdefaultfont{\mathtt}
\crectangle{T}{2.5}{2.5}{0,0} \csymbolfourth{C}
\srightvecfid{y7}{-7,1.5}   \joinrlnoarrowbw[above]{y7}{0}{T}{1.5}\csymbolfourth[0,10]{y}
\thispoint{yL7}{-9.5,1.5}   \csymbol[-24,0]{y}  \joinrlnoarrowthick{yL7}{0}{y7}{0}
\srightvecfid{x1}{-7,-1.5}  \joinrlnoarrowbw[above]{x1}{0}{T}{-1.5} \csymbolfourth[0,10]{x}
\thispoint{xL1}{-9.5,-1.5} \csymbol[-24,0]{x} \joinrlnoarrowthick[above]{xL1}{0}{x1}{0}
\sleftvecfid{w8}{7,1.5}  \joinrlnoarrowwb[above]{T}{1.5}{w8}{0} \csymbolfourth[0,10]{v}
\thispoint{wR8}{9.5,1.5} \csymbol[24,0]{v}  \joinrlnoarrowthick[above]{w8}{0}{wR8}{0}
\sleftvecfid{y2}{7,-1.5} \joinrlnoarrowwb[above]{T}{-1.5}{y2}{0}  \csymbolfourth[0,10]{u}
\thispoint{yR2}{9.5,-1.5} \csymbol[24,0]{u}  \joinrlnoarrowthick[above]{y2}{0}{yR2}{0}
\fidopupwire{T}{-1.5}{c}  \fidopupwire{T}{1.5}{d}
\fidopdownwire{T}{-1.5}{a} \fidopdownwire{T}{1.5}{b}
\end{Compose}
\end{equation}
This is possible because the tensor products of the fiducial elements form spanning sets for the space of operator tensors (this fact only works when we start with a base Hilbert space that is complex \cite{hardy2011reformulating}). This expansion should be compared with the expansion for an operation in \eqref{Lexpansiongeneral} where we have an $\equiv$ symbol because the two sides are equivalent.  In \eqref{Lexpansiongeneraloperator} we have an $=$ sign because the two sides are, simply, equal to one another.

\subsection{Wire expansion properties for operators}\label{sec:wireexpansionpropertiesforoperators}

Following analogous proofs to those in Sec.\ \ref{sec:expandingthewire} we obtain
\begin{equation}\label{wirepointertomlocoperator}
\begin{Compose}{0}{-0.09} \setsecondfont{\mathtt}
\linebyhand[thick]{-3,0}{3,0}  \csymbolalt[0,20]{x}
\end{Compose}
~=~
\begin{Compose}{0}{-0.09}\setsecondfont{\mathtt}
\srightvecfid{l}{-1.5,0} \thispoint{ll}{-4,0} \csymbolalt[-20,0]{x} \joinlrnoarrowthick{l}{0}{ll}{0}
\sleftvecfid{r}{1.5,0}   \thispoint{rr}{4,0} \csymbolalt[20,0]{x} \joinlrnoarrowthick{rr}{0}{r}{0}
\joinrlnoarrow[above]{l}{0}{r}{0} \csymbolthird{x}
\end{Compose}
\end{equation}

\begin{equation} \label{wiresystomlocoperator}
\begin{Compose}{0}{-0.09} \setsecondfont{\mathsfb} \setthirdfont{\mathpzc}
\linebyhand[thick]{0,-3}{0,3}  \csymbolalt[-20,0]{x}
\end{Compose}
~=~
\begin{Compose}{0}{-0.09}\setsecondfont{\mathsfb} \setthirdfont{\mathpzc}
\supopfid{u}{0, 1.5} \thispoint{uu}{0,4} \csymbolalt[0,20]{x} \jointbnoarrowthick{u}{0}{uu}{0}
\sdownopfid{d}{0,-1.5}   \thispoint{dd}{0,-4} \csymbolalt[0,-20]{x} \jointbnoarrowthick{dd}{0}{d}{0}
\jointbnoarrow[right]{d}{0}{u}{0} \csymbolthird{x}
\end{Compose}
\end{equation}
These equations are very useful.

\section{Calculations with operator tensors}

\subsection{Using fiducials}

If we have two operator tensor sets that are connected by a wire connecting an output to an input like this
\begin{equation}
\hat{A}_{\mathtt{x_1}\mathsfb{b_5}}^{~\mathsfb{a_6b_7}}
\hat{C}_{\mathsfb{b_7}}^{\mathtt{y_4}\mathsfb{a_9}}
~~~~~~~~~~~~~~~~~~~~~~~
\begin{Compose}{0}{0} \setdefaultfont{\mathsfb} \setthirdfont{\hat} \setsecondfont{\mathtt}
\crectangledouble{A}{1.5}{1.5}{0,0} \csymbolthird{A} \relpoint{A}{-4,0}{AL}\csymbolalt[-20,0]{x} \relpoint{A}{0,-4}{AD}\csymbol[0,-20]{b} \relpoint{A}{-2.1,4}{AUL}\csymbol[0,20]{a}
\joinbtnoarrow{A}{0}{AD}{0} \joinlrnoarrowthick{A}{0}{AL}{0} \jointbnoarrow{A}{-1}{AUL}{0}
\crectangledouble{C}{1.5}{1.5}{4,8} \csymbolthird{C}
\relpoint{C}{4,0}{CR} \csymbolalt[20,0]{y}  \relpoint{C}{0,4}{CU}\csymbol[0,20]{a}
\jointbnoarrow{C}{0}{CU}{0} 
\joinrlnoarrowthick{C}{0}{CR}{0}
\jointbnoarrow[below right]{A}{1}{C}{0} \csymbol{b}
\end{Compose}
\end{equation}
As discussed in Sec.\ \ref{sec:joiningoperatortensors}, this means we take the partial trace over the Hilbert space associated with the connecting wire (wire $\mathsfb{b_7}$ in this example).
We can substitute expansions of the type shown in \eqref{Lexpansiongeneraloperator} in here.  This gives
\begin{equation}
\begin{Compose}{0}{0} \setdefaultfont{\mathtt} \setthirdfont{\mathpzc} \setsecondfont{\mathsfb}
\crectangle{A}{1.5}{1.5}{0,0}\csymbolfourth{A} \crectangle{C}{1.5}{1.5}{6,10} \csymbolfourth{C}
\fidopupwire{A}{-1}{a} \fidopdownwire{A}{0}{b}
\fidopupwire{C}{0}{a}
\joinwithopfids{A}{1}{C}{0}
\srightvecfid{xLA}{-5,0}   \joinrlnoarrowbw[above]{xLA}{0}{A}{0}\csymbolfourth[0,10]{x}
\thispoint{pxLA}{-7,0} \csymbol[-24,0]{x}  \joinrlnoarrowthick{pxLA}{0}{xLA}{0}
\sleftvecfid{yRC}{11,10}  \joinrlnoarrowwb[above]{C}{0}{yRC}{0} \csymbolfourth[0,10]{y}
\thispoint{pyRC}{13,10} \csymbol[24,0]{y}  \joinrlnoarrowthick[above]{yRC}{0}{pyRC}{0}
\end{Compose}
\end{equation}
Here we see that the duotensors provide the weightings in the linear sum over fiducials.

\subsection{Operator circuits}\label{sec:operatorcircuits}

We can extend this to an operator circuit (an circuit composed of operator tensors).  Consider the following operator tensor circuit
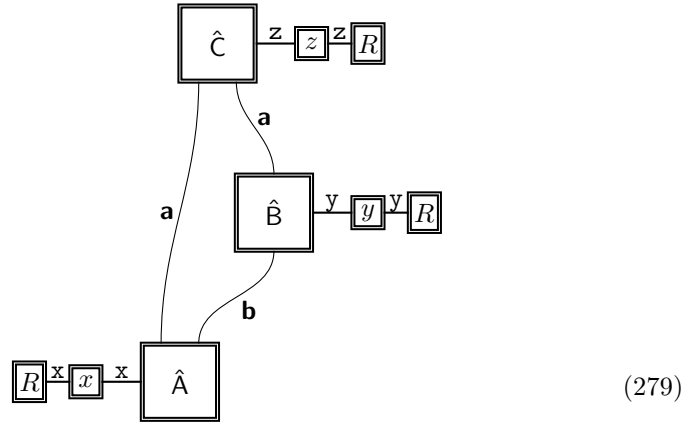
\begin{equation}\label{ABCcircuitop}
\begin{Compose}{0}{0}  \setsecondfont{\mathsfb}
\crectangledouble{A}{2}{2}{0,0} \csymbol{\hat{A}}  \crectangledouble{B}{2}{2}{5,9} \csymbol{\hat{B}} \crectangledouble{C}{2}{2}{2,18} \csymbol{\hat{C}}
\jointbnoarrow[left]{A}{-1}{C}{-1} \csymbolalt[-5,0]{a}
\jointbnoarrow[below right]{A}{1}{B}{0}  \csymbolalt[5,-5]{b}
\jointbnoarrow[above right]{B}{0}{C}{1}  \csymbolalt{a}
\Rxboxincomedouble{A}{0}{x}
\Rxboxoutcomedouble{B}{0}{y}
\Rxboxoutcomedouble{C}{0}{z}
\end{Compose}
\end{equation}
We can substitute each operator tensor for the expansion in terms of fiducials giving
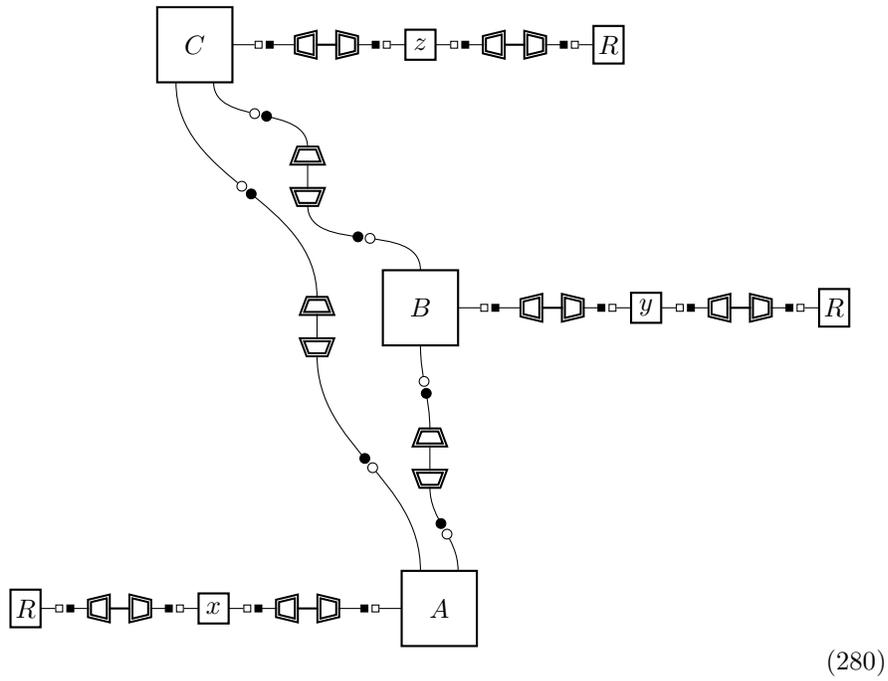
\begin{equation}\label{ABCcircuitwithfidsop}
\begin{Compose}{0}{0} \setsecondfont{\mathsfb}
\crectangle{A}{2}{2}{0,0} \csymbolfourth{A}  \crectangle{B}{2}{2}{-1,16} \csymbolfourth{B} \crectangle{C}{2}{2}{-13,30} \csymbolfourth{C}
\joinwithopfids{A}{-1}{C}{-1} 
\joinwithopfids{A}{1}{B}{0} 
\joinwithopfids{B}{0}{C}{1} 
\relpoint{A}{-2,0}{AL} \Rxboxincomefidsdouble{AL}{0}{x}
\relpoint{B}{2,0}{BR}\Rxboxoutcomefidsdouble{BR}{0}{y}
\relpoint{C}{2,0}{CR}\Rxboxoutcomefidsdouble{CR}{0}{z}
\end{Compose}
\end{equation}
We can now use \eqref{operatorhoppingmetric} and \eqref{hoppingequivfidcircuitvec} and to obtain
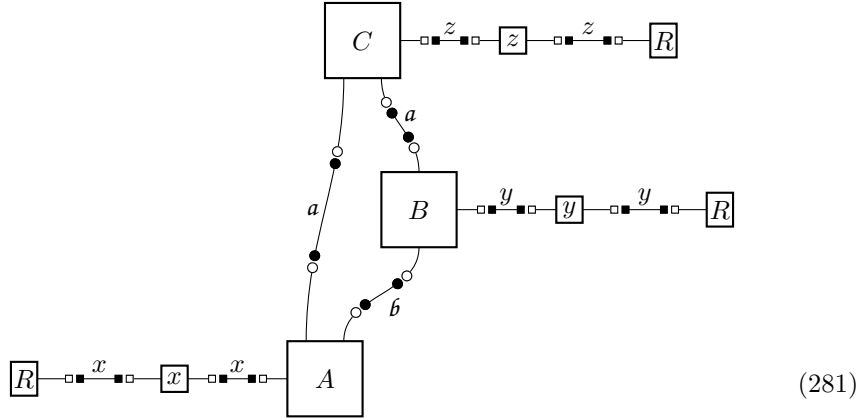
\begin{equation}\label{ABCduotensorop}
\begin{Compose}{0}{0} \setsecondfont{\mathpzc}
\crectangle{A}{2}{2}{0,0} \csymbolfourth{A}  \crectangle{B}{2}{2}{5,9} \csymbolfourth{B} \crectangle{C}{2}{2}{2,18} \csymbolfourth{C}
\jointbnoarrowdotswbbw[left]{A}{-1}{C}{-1} \csymbolalt[-5,0]{a}
\jointbnoarrowdotswbbw[below right]{A}{1}{B}{0}  \csymbolalt[5,-5]{b}
\jointbnoarrowdotswbbw[above right]{B}{0}{C}{1}  \csymbolalt{a}
\Rxboxincomewbbw{A}{0}{x}
\Rxboxoutcomewbbw{B}{0}{y}
\Rxboxoutcomewbbw{C}{0}{z}
\end{Compose}
\end{equation}
Finally, we can replace black and white square dots or black and white round dots by a straight line as in \eqref{bwtowbvec} and \eqref{bwtowbroundop} giving
\begin{equation}\label{ABCduotensorsimpleB}
\begin{Compose}{0}{0} \setsecondfont{\mathpzc}
\crectangle{A}{2}{2}{0,0} \csymbolfourth{A}  \crectangle{B}{2}{2}{5,9} \csymbolfourth{B} \crectangle{C}{2}{2}{2,18} \csymbolfourth{C}
\jointbnoarrow[left]{A}{-1}{C}{-1} \csymbolalt[-5,0]{a}
\jointbnoarrow[below right]{A}{1}{B}{0}  \csymbolalt[5,-5]{b}
\jointbnoarrow[above right]{B}{0}{C}{1}  \csymbolalt{a}
\Rxboxincomecalc{A}{0}{x}
\Rxboxoutcomecalc{B}{0}{y}
\Rxboxoutcomecalc{C}{0}{z}
\end{Compose}
\end{equation}
We can use the wire expansion properties in Sec.\ \ref{sec:wireexpansionpropertiesforoperators} to go directly from \eqref{ABCcircuitop} to \eqref{ABCduotensorsimpleB}.

\section{Special operator tensors}

We wish to define some operator tensors that are analogous to objects we have defined for operations.  Table \ref{table:operatortable} shows the operators we wish to consider.
\begin{table}
\begin{center}
{\tabulinesep=1.2mm
\begin{tabu}{|c|c|p{0.35\linewidth}|}
  \hline
  \RleftTSop ~ and  ~ \RrightTSop & ${R}^{\mathtt{x}_1}$ and ${R}_{\mathtt{x}_1}$ & Flat distribution operators \\ \hline
  \xboxTSop  & $B_{\mathtt{x}_1}^{\mathtt{x}_2}[x]$ & Readout box operator \\ \hline
  \XfidoutcomeTSop ~~ and ~~ \XfidincomeTSop & $\hat{X}_{\mathtt{x}_1}^{\mathsfb{x}_2}$ and $\hat{X}^{\mathtt{x}_1}_{\mathsfb{x}_2}$ & Maximal output and maximal input operators \\ \hline
  \IoutcomeTSop ~~ and ~~ \IincomeTSop & $\hat{I}^{\mathsfb{a}_1}$ and $\hat{I}_{\mathsfb{a}_1}$ & Ignore output and input operators \\ \hline
\end{tabu} }
\end{center}
\caption{Operators that play a special role.  We provide diagrammatic and symbolic notation for these operators.}
\label{table:operatortable}
\end{table}
We will now describe the properties we require of these mathematical objects.

\subsection{The flat distribution operators}\label{sec:flatdistributionoperators}

The flat distribution operators must satisfy
\begin{equation}\label{flatrequirementoperators}
\begin{Compose}{0}{0}\setsecondfont{\mathtt} \setthirdfont{\mathsfb}
\crectangledouble{RR}{0.7}{0.9}{3,0}\csymbolfourth{R}
\RxBoxincomedouble{RR}{0.7}{0}{x}
\end{Compose}
= \frac{1}{N_\mathtt{x}}
~~~~~~
\begin{Compose}{0}{0}\setsecondfont{\mathtt} \setthirdfont{\mathsfb}
\crectangledouble{RR}{0.9}{0.9}{3,0}\csymbolfourth{R}
\crectangledouble{RL}{0.9}{0.9}{-3,0}\csymbolfourth{R}
\joinlrnoarrowthick{RR}{0}{RL}{0} \csymbolalt{x}
\end{Compose}
= 1
\end{equation}
We can expand these operators in terms of fiducials.  Then we obtain
\begin{equation} \label{Requivalenceoperators}
\begin{Compose}{0}{0} \setsecondfont{\mathtt}
\crectangledouble{R}{0.8}{1}{0,0} \csymbolfourth{R} \thispoint{p}{4,0} \csymbolalt[20,0]{x} \joinrlnoarrowthick{R}{0}{p}{0}
\end{Compose} ~~~=~~~
\begin{Compose}{0}{0} \setsecondfont{\mathtt}
\sleftvecfid{F}{0,0} \crectangle{FL}{0.7}{0.9}{-5,0} \csymbolthird{R}
\relpoint{F}{2.5,0}{FR} \csymbolalt[15,0]{x}
\joinrlnoarrowthick[above]{F}{0}{FR}{0}  \joinrlnoarrowwb[above]{FL}{0}{F}{0} \csymbolthird[0,10]{x}
\end{Compose}
\end{equation}
where
\begin{equation}
\begin{Compose}{0}{0} \setsecondfont{\mathtt}
\crectangle{R}{0.7}{0.9}{0,0} \csymbolthird{R} \whitedotsq{p}{4,0} \csymbolthird[25,0]{x} \joinrlnoarrow{R}{0}{p}{0}
\end{Compose}
~~=~~
\left(
\begin{array}{cccc}
  1 & 1 & \dots & 1
\end{array}
\right)
\end{equation}
Using the hopping matrix we obtain
\begin{equation}\label{Rforwardsblackdot}
\begin{Compose}{0}{0} \setsecondfont{\mathtt}
\crectangle{R}{0.7}{0.9}{0,0} \csymbolthird{R} \blackdotsq{p}{4,0} \csymbolthird[25,0]{x} \joinrlnoarrow{R}{0}{p}{0}
\end{Compose}
~~=~~
\frac{1}{N_\mathtt{x}}
\left(
\begin{array}{c}
  1 \\
  1 \\
  \vdots \\
  1
\end{array}
\right)
\end{equation}
So we get a flat distribution for the duotensor ending with a black dot (in accord with the interpretation given in Sec.\ \ref{sec:interpretationofduotensorswithsquaredots}).

Similarly, we obtain
\begin{equation}
\begin{Compose}{0}{0} \setsecondfont{\mathtt}
\crectangle{R}{0.7}{0.9}{0,0} \csymbolthird{R} \whitedotsq{p}{-4,0} \csymbolthird[-25,0]{x} \joinlrnoarrow{R}{0}{p}{0}
\end{Compose}
~~=~~
\left(
\begin{array}{cccc}
  1 & 1 & \dots & 1
\end{array}
\right)
\end{equation}
and
\begin{equation} \label{Rbackwardsblackdot}
\begin{Compose}{0}{0} \setsecondfont{\mathtt}
\crectangle{R}{0.7}{0.9}{0,0} \csymbolthird{R} \blackdotsq{p}{-4,0} \csymbolthird[-25,0]{x} \joinlrnoarrow{R}{0}{p}{0}
\end{Compose}
~~=~~
\frac{1}{N_\mathtt{x}}
\left(
\begin{array}{c}
  1 \\
  1 \\
  \vdots \\
  1
\end{array}
\right)
\end{equation}
using the hopping matrix.

Given the flat nature of the duotensors associated with these $R$ boxes, we can easily prove the following factorization properties (analogous to \eqref{factorization})
\begin{equation}\label{factorizationoperators}
\begin{Compose}{0}{0} \setdefaultfont{\hat} \setsecondfont{\mathtt}
\Crectangledouble{R}{0.8}{1}{0,0} \thispoint{p}{4,0} \csymbolalt[25,0]{xy} \joinrlnoarrowthick{R}{0}{p}{0}
\end{Compose}
~~~\equiv~~~
\begin{Compose}{0}{0} \setdefaultfont{\hat} \setsecondfont{\mathtt}
\Crectangledouble{R}{0.8}{1}{0,-1.3} \thispoint{p}{4,-1.3} \csymbolalt[20,0]{x} \joinrlnoarrowthick{R}{0}{p}{0}
\Crectangledouble{R}{0.8}{1}{0,1.3} \thispoint{p}{4,1.3} \csymbolalt[20,0]{y} \joinrlnoarrowthick{R}{0}{p}{0}
\end{Compose}
~~~~~~~~~~~~~~~
\begin{Compose}{0}{0} \setdefaultfont{\hat}\setsecondfont{\mathtt}
\Crectangledouble{R}{0.8}{1}{0,0}  \thispoint{p}{-4,0} \csymbolalt[-25,0]{xy} \joinlrnoarrowthick{R}{0}{p}{0}
\end{Compose}
~~~\equiv~~~
\begin{Compose}{0}{0} \setdefaultfont{\hat} \setsecondfont{\mathtt}
\Crectangledouble{R}{0.8}{1}{0,-1.3}  \thispoint{p}{-4,-1.3} \csymbolalt[-20,0]{x} \joinlrnoarrowthick{R}{0}{p}{0}
\Crectangledouble{R}{0.8}{1}{0,1.3}  \thispoint{p}{-4,1.3} \csymbolalt[-20,0]{x} \joinlrnoarrowthick{R}{0}{p}{0}
\end{Compose}
\end{equation}
from the form of the associated duotensors \eqref{Rforwardsblackdot} and \eqref{Rbackwardsblackdot}.

We can substitute these expressions to verify the condition on the right hand side of \eqref{flatrequirementoperators} by inserting black and white pairs of dots to verify these conditions are satisfied.   To verify the condition on the left hand side of \eqref{flatrequirementoperators} we need to characterize the operator tensor version of the readout box.

\subsection{Readout operator boxes}

The readout boxes are defined by the property
\begin{equation}
\begin{Compose}{0}{0}\setsecondfont{\mathtt}
\crectangledouble{AL}{0.77}{0.87}{-6,0} \csymbolthird{x} \relpoint{AL}{-2.5,0}{ALL} \csymbolalt[-18,0]{x}
\crectangledouble{AR}{0.77}{0.87}{-2,0} \csymbolthird{x'}
\relpoint{AR}{2.5,0}{ARR} \csymbolalt[15,0]{x}
\joinrlnoarrowthick{ALL}{0}{AL}{0}
\joinrlnoarrowthick[above]{AL}{0}{AR}{0}   \csymbolalt[0,4]{x}
\joinrlnoarrowthick{AR}{0}{ARR}{0}
\end{Compose}
~~~~
\equiv
~~~~
\arraycolsep=1.4pt\def\arraystretch{1.5}
\left\{
\begin{array}{ll}
\begin{Compose}{0}{-0.1}\setsecondfont{\mathtt}
\crectangledouble{A}{0.6}{0.7}{0,0} \csymbolthird{x} \relpoint{A}{-2.5,0}{AL} \csymbolalt[-18,0]{x}
\relpoint{A}{2.5,0}{AR} \csymbolalt[15,0]{x}
\joinlrnoarrowthick{A}{0}{AL}{0}
\joinrlnoarrowthick{A}{0}{AR}{0}
\end{Compose}  &
\text{if} ~~x=x'  \\
\begin{Compose}{0}{-0.1}\setsecondfont{\mathtt}
\crectangledouble{A}{0.6}{0.7}{0,0} \csymbolthird{0} \relpoint{A}{-2.5,0}{AL} \csymbolalt[-18,0]{x}
\relpoint{A}{2.5,0}{AR} \csymbolalt[15,0]{x}
\joinlrnoarrowthick{A}{0}{AL}{0}
\joinrlnoarrowthick{A}{0}{AR}{0}
\end{Compose} &
\text{if} ~~x\not=x'
\end{array}
\right.
\end{equation}
Here, the \emph{null operator box},
\begin{equation}
\begin{Compose}{0}{0}\setsecondfont{\mathtt}
\crectangledouble{A}{0.6}{0.7}{0,0} \csymbolthird{0} \relpoint{A}{-2.5,0}{AL} \csymbolalt[-18,0]{x}
\relpoint{A}{2.5,0}{AR} \csymbolalt[15,0]{x}
\joinlrnoarrowthick{A}{0}{AL}{0}
\joinrlnoarrowthick{A}{0}{AR}{0}
\end{Compose}
\end{equation}
has the property that, if it is added to any operator circuit, that operator circuit evaluates to 0.

We can expand the operator readout box in terms of fiducial operators
\begin{equation}\label{xboxexpansionbwop}
\begin{Compose}{0}{-0.1}\setsecondfont{\mathtt}
\crectangledouble{A}{0.6}{0.7}{0,0} \csymbolthird{x} \relpoint{A}{-2.5,0}{AL} \csymbolalt[-18,0]{x}
\relpoint{A}{2.5,0}{AR} \csymbolalt[15,0]{x}
\joinlrnoarrowthick{A}{0}{AL}{0}
\joinrlnoarrowthick{A}{0}{AR}{0}
\end{Compose}
~~=~~
\begin{Compose}{0}{0}\setsecondfont{\mathtt}
\crectangle{AL}{0.7}{0.8}{-6,0} \csymbolthird{x}
\fidopleftwire{AL}{0}{x}
\fidoprightwirebw{AL}{0}{x}
\end{Compose}
\end{equation}
Following the same reasoning as in Sec.\ \ref{sec:duotensorsforreadoutboxes}, we can show that
\begin{equation}\label{xboxbwop}
\begin{Compose}{0}{0}\setsecondfont{\mathtt}
\crectangle{AL}{0.7}{0.8}{-6,0} \csymbolthird{x}
\blacksquarerightwire{AL}{0}{x}
\whitesquareleftwire{AL}{0}{x}
\end{Compose}
~=~
\left(
\begin{array}{ccccc}
0 &   & & &  \\
  & 0 & & &  \\
  &   & \ddots &   &  \\
  &   &        & 1 &  \\
  &   &        &   & \ddots
\end{array}
\right)
\end{equation}
where the $1$ is in position $x$ along the diagonal.
We can now verify the condition on the right of \eqref{flatrequirementoperators} by explicit calculation.

\subsection{Maximal operators}\label{sec:maximaloperators}

By analogy with maximal operations, we define maximal operators as operators
\begin{equation}\label{maximaloperators}
\begin{Compose}{0}{0}\setdefaultfont{\hat}\setthirdfont{\mathsfb}\setsecondfont{\mathtt}
\Crectangledouble{X}{1}{1}{0,0}
\thispoint{p}{0,3} \csymbolthird[0,20]{x} \jointbnoarrow[left]{X}{0}{p}{0}
\thispoint{pp}{-3,0} \csymbolalt[-20,0]{x} \joinlrnoarrowthick{X}{0}{pp}{0}
\end{Compose}
~~~~~~~~~~~~~~~~~~~~~~~~
\begin{Compose}{0}{0}\setdefaultfont{\hat}\setsecondfont{\mathtt}\setthirdfont{\mathsfb}
\Crectangledouble{X}{1}{1}{0,3}
\thispoint{p}{0,0} \csymbolthird[0,-20]{x} \jointbnoarrow[left]{p}{0}{X}{0}
\thispoint{pp}{3,3} \csymbolalt[20,0]{x} \joinlrnoarrowthick[left]{pp}{0}{X}{0}
\end{Compose}
\end{equation}
for which
\begin{equation}\label{doublemaximalityoperators}
\begin{Compose}{0}{0}\setdefaultfont{\hat}\setsecondfont{\mathtt} \setthirdfont{\mathsfb}
\Crectangledouble{X}{1}{1}{0,-2} \relpoint{X}{-4,0}{XL} \csymbolalt[-20,0]{x} \joinrlnoarrowthick{XL}{0}{X}{0}
\crectangledouble{Y}{1}{1}{0,2}\csymbol{X} \relpoint{Y}{4,0}{YR} \csymbolalt[20,0]{x}\joinlrnoarrowthick{YR}{0}{Y}{0}
\jointbnoarrow[left]{X}{0}{Y}{0}\csymbolthird[0,5]{x}
\end{Compose}
~~~~ = ~~~~~~~
\begin{Compose}{0}{0}\setsecondfont{\mathtt}
\thispoint{X}{-4,0}
\thispoint{Y}{4,0}
\joinrlnoarrowthick[above]{X}{0}{Y}{0}\csymbolalt[0,5]{x}
\end{Compose}
\end{equation}
where, also, $N_\mathtt{x}=N_\mathsfb{x}$.   As we will see, maximal operators are associated with projectors on to a basis set of the corresponding Hilbert space.  However, there are some subtleties that arise in the time symmetric situation that are not evident in the standard time asymmetric setting.  We can add fiducial elements to the ends of \eqref{doublemaximalityoperators} to obtain
\begin{equation}\label{doublemaximalityduotensors}
\begin{Compose}{0}{0}\setdefaultfont{\hat} \setsecondfont{\mathsfb}
\Crectangledouble{X}{1}{1}{0,-2}
\crectangledouble{Y}{1}{1}{0,2}\csymbol{X}
\whitesquareleftwire{X}{0}{x}
\blacksquarerightwire{Y}{0}{x}
\jointbnoarrow[left]{X}{0}{Y}{0}\csymbolalt{x}
\end{Compose}
= ~~
\begin{Compose}{0}{0}\setsecondfont{\mathtt}
\whitedotsq{X}{-1.75,0}
\blackdotsq{Y}{1.75,0}
\joinrlnoarrow[above]{X}{0}{Y}{0}\csymbolthird[0,5]{x}
\end{Compose}
~~~~~~~~~~~~~~~
\begin{Compose}{0}{0}\setdefaultfont{\hat} \setsecondfont{\mathsfb}
\Crectangledouble{X}{1}{1}{0,-2}
\crectangledouble{Y}{1}{1}{0,2}\csymbol{X}
\blacksquareleftwire{X}{0}{x}
\whitesquarerightwire{Y}{0}{x}
\jointbnoarrow[left]{X}{0}{Y}{0}\csymbolalt{x}
\end{Compose}
= ~~
\begin{Compose}{0}{0}\setsecondfont{\mathtt}
\blackdotsq{X}{-1.75,0}
\whitedotsq{Y}{1.75,0}
\joinrlnoarrow[above]{X}{0}{Y}{0}\csymbolthird[0,5]{x}
\end{Compose}
\end{equation}
(this is really the same equation twice with the dot colours changed).
Now consider the following four objects.
\begin{equation}
\begin{Compose}{0}{0}\setdefaultfont{\hat}\setfourthfont{\mathsfb}\setsecondfont{\mathtt}
\Crectangledouble{X}{1}{1}{0,0}
\thispoint{p}{0,3} \csymbolfourth[0,20]{x} \jointbnoarrow[left]{X}{0}{p}{0}
\blacksquareleftwire{X}{0}{x}
\end{Compose}
~~~~~~~~
\begin{Compose}{0}{0}\setdefaultfont{\hat}\setfourthfont{\mathsfb}\setsecondfont{\mathtt}
\Crectangledouble{X}{1}{1}{0,0}
\thispoint{p}{0,3} \csymbolfourth[0,20]{x} \jointbnoarrow[left]{X}{0}{p}{0}
\whitesquareleftwire{X}{0}{x}
\end{Compose}
~~~~~~~~~~~~
\begin{Compose}{0}{0}\setdefaultfont{\hat}\setsecondfont{\mathtt}\setfourthfont{\mathsfb}
\Crectangledouble{X}{1}{1}{0,3}
\thispoint{p}{0,0} \csymbolfourth[0,-20]{x} \jointbnoarrow[left]{p}{0}{X}{0}
\blacksquarerightwire{X}{0}{x}
\end{Compose}
~~~~~~~~
\begin{Compose}{0}{0}\setdefaultfont{\hat}\setsecondfont{\mathtt}\setfourthfont{\mathsfb}
\Crectangledouble{X}{1}{1}{0,3}
\thispoint{p}{0,0} \csymbolfourth[0,-20]{x} \jointbnoarrow[left]{p}{0}{X}{0}
\whitesquarerightwire{X}{0}{x}
\end{Compose}
\end{equation}
We want each of these objects to correspond to a set of projection operators onto a basis of $\mathcal{H}^{\mathsfb{x}_1}$ (for the ones with an output) and $\mathcal{H}_{\mathsfb{x}_1}$ (for the ones with an input).   We start by putting
\begin{equation}\label{forwardchoiceforwXx}
\begin{Compose}{0}{0}\setdefaultfont{\hat}\setfourthfont{\mathsfb}\setsecondfont{\mathtt}
\Crectangledouble{X}{1}{1}{0,0}
\thispoint{p}{0,3} \csymbolfourth[0,20]{x} \jointbnoarrow[left]{X}{0}{p}{0}
\whitesquareleftwire{X}{0}{x}
\end{Compose}
~=~
\alpha_\mathsfb{x}\sqrt{N_\mathsfb{x}} |x\rangle^{\mathsfb{x}_1}\negs\negs\langle x|
\end{equation}
where $\alpha_\mathsfb{x}$ is a real constant whose significance we will discuss below and $|x\rangle^{\mathsfb{x}_1}\negs\negs\langle x|$ is a set of projection operators projecting on to a orthonormal basis set, $|x\rangle^{\mathsfb{x}_1}$, for the Hilbert space, $\mathcal{H}^{\mathsfb{x}_1}$.
By application of the hopping matrix, $\bbdotsnoarrowsq$, we obtain
\begin{equation}
\begin{Compose}{0}{0}\setdefaultfont{\hat}\setfourthfont{\mathsfb}\setsecondfont{\mathtt}
\Crectangledouble{X}{1}{1}{0,0}
\thispoint{p}{0,3} \csymbolfourth[0,20]{x} \jointbnoarrow[left]{X}{0}{p}{0}
\blacksquareleftwire{X}{0}{x}
\end{Compose}
=
\frac{\alpha_\mathsfb{a}}{\sqrt{N_\mathsfb{a}}}
|x\rangle^{\mathsfb{x}_1}\negs\negs\langle x|
\end{equation}
In order to satisfy the equation on the left of \eqref{doublemaximalityduotensors} we require
\begin{equation}\label{xhatXb}
\begin{Compose}{0}{-0.6}\setdefaultfont{\hat}\setsecondfont{\mathtt}\setfourthfont{\mathsfb}
\Crectangledouble{X}{1}{1}{0,3}
\thispoint{p}{0,0} \csymbolfourth[0,-20]{x} \jointbnoarrow[left]{p}{0}{X}{0}
\blacksquarerightwire{X}{0}{x}
\end{Compose}
= \frac{1}{\alpha_\mathsfb{a}\sqrt{N_\mathsfb{a}}}|x\rangle_{\mathsfb{x}_1} \negs\negs\negs\langle x|
\end{equation}
$|x\rangle_{\mathsfb{x}_1}\negs\negs\negs\langle x|$ is a set of projection operators projecting on to a orthonormal basis set, $|x\rangle_{\mathsfb{x}_1}$, for the Hilbert space, $\mathcal{H}_{\mathsfb{x}_1}$.
In order to satisfy the equation on the right of \eqref{doublemaximalityduotensors} we require
\begin{equation}\label{xhatXw}
\begin{Compose}{0}{-0.6}\setdefaultfont{\hat}\setsecondfont{\mathtt}\setfourthfont{\mathsfb}
\Crectangledouble{X}{1}{1}{0,3}
\thispoint{p}{0,0} \csymbolfourth[0,-20]{x} \jointbnoarrow[left]{p}{0}{X}{0}
\whitesquarerightwire{X}{0}{x}
\end{Compose}
= \frac{\sqrt{N_\mathtt{x}}}{\alpha_\mathsfb{a}}|x\rangle_{\mathsfb{x}_1} \negs\negs\negs\langle x|
\end{equation}
As a consistency check, we can see that applying $\bbdotsnoarrowsq$ to \eqref{xhatXw} gives \eqref{xhatXb}.

The constant, $\alpha_\mathsfb{a}$ drops out of any real calculations so it is a gauge parameter.  However, it is interesting to see how it relates to the usual way of doing Quantum Theory.  We will discuss this in Sec.\ \ref{sec:chasingdowntheoneoverN}.

\subsection{Ignore operators}

We define ignore operators by analogy with ignore operations.  One way of writing an ignore operation is to write them as the sum over a maximal set
\begin{equation}
\begin{Compose}{0}{0}\setsecondfont{\mathtt}\setthirdfont{\mathsfb}
\Crectangle{I}{1}{1}{0,0}
\thispoint{p}{0,3.5}  \csymbolthird[0,20]{x} \jointbnoarrow{I}{0}{p}{0}
\end{Compose}
~\equiv~
\begin{Compose}{0}{0}\setsecondfont{\mathtt}\setthirdfont{\mathsfb}
\Crectangle{X}{1}{1}{0,0}
\thispoint{p}{0,3.5}  \csymbolthird[0,20]{x} \jointbnoarrow{X}{0}{p}{0}
\Rincome{X}{0}{x}
\end{Compose}
~~~~~~~~~~~~
\begin{Compose}{0}{0}\setsecondfont{\mathtt}\setthirdfont{\mathsfb}
\Crectangle{I}{1}{1}{0,0}
\thispoint{p}{0,-3.5}  \csymbolthird[0,-20]{x} \joinbtnoarrow{I}{0}{p}{0}
\end{Compose}
~\equiv~
\begin{Compose}{0}{0}\setsecondfont{\mathtt}\setthirdfont{\mathsfb}
\Crectangle{X}{1}{1}{0,0}
\thispoint{p}{0,-3.5}  \csymbolthird[0,-20]{x} \joinbtnoarrow{X}{0}{p}{0}
\Routcome{X}{0}{x}
\end{Compose}
\end{equation}
Hence, by analogy, we will write
\begin{equation}
\begin{Compose}{0}{0}\setdefaultfont{\hat}\setsecondfont{\mathtt}\setthirdfont{\mathsfb}
\Crectangledouble{I}{1}{1}{0,0}
\thispoint{p}{0,3.5}  \csymbolthird[0,20]{x} \jointbnoarrow{I}{0}{p}{0}
\end{Compose}
~\equiv~
\begin{Compose}{0}{0} \setdefaultfont{}\setsecondfont{\mathtt}\setthirdfont{\mathsfb}
\crectangledouble{X}{1}{1}{0,0}\csymbol{\hat{X}}
\thispoint{p}{0,3.5}  \csymbolthird[0,20]{x} \jointbnoarrow{X}{0}{p}{0}
\Rincomedouble{X}{0}{x}
\end{Compose}
~~~~~~~~~~~~
\begin{Compose}{0}{0}\setdefaultfont{\hat}\setsecondfont{\mathtt}\setthirdfont{\mathsfb}
\Crectangledouble{I}{1}{1}{0,0}
\thispoint{p}{0,-3.5}  \csymbolthird[0,-20]{x} \joinbtnoarrow{I}{0}{p}{0}
\end{Compose}
~=~
\begin{Compose}{0}{0}\setdefaultfont{}\setsecondfont{\mathtt}\setthirdfont{\mathsfb}
\crectangledouble{X}{1}{1}{0,0} \csymbol{\hat{X}}
\thispoint{p}{0,-3.5}  \csymbolthird[0,-20]{x} \joinbtnoarrow{X}{0}{p}{0}
\Routcomedouble{X}{0}{x}
\end{Compose}
\end{equation}
Using the wire expansion rule \eqref{wirepointertomlocoperator} we obtain
\begin{equation}
\begin{Compose}{0}{0}\setdefaultfont{\hat}\setsecondfont{\mathtt}\setthirdfont{\mathsfb}
\Crectangledouble{I}{1}{1}{0,0}
\thispoint{p}{0,3.5}  \csymbolthird[0,20]{x} \jointbnoarrow{I}{0}{p}{0}
\end{Compose}
~\equiv~
\begin{Compose}{0}{0} \setdefaultfont{}\setsecondfont{\mathtt}\setthirdfont{\mathsfb}
\crectangledouble{X}{1}{1}{0,0}\csymbol{\hat{X}}
\thispoint{p}{0,3.5}  \csymbolthird[0,20]{x} \jointbnoarrow{X}{0}{p}{0}
\Crectangle{R}{0.7}{0.9}{-4,0} \joinrlnoarrowbw[above]{R}{0}{X}{0} \csymbol[0,5]{x}
\end{Compose}
~~~~~~~~~~~~
\begin{Compose}{0}{0}\setdefaultfont{\hat}\setsecondfont{\mathtt}\setthirdfont{\mathsfb}
\Crectangledouble{I}{1}{1}{0,0}
\thispoint{p}{0,-3.5}  \csymbolthird[0,-20]{x} \joinbtnoarrow{I}{0}{p}{0}
\end{Compose}
~=~
\begin{Compose}{0}{0}\setdefaultfont{}\setsecondfont{\mathtt}\setthirdfont{\mathsfb}
\crectangledouble{X}{1}{1}{0,0} \csymbol{\hat{X}}
\thispoint{p}{0,-3.5}  \csymbolthird[0,-20]{x} \joinbtnoarrow{X}{0}{p}{0}
\Crectangle{R}{0.7}{0.9}{4,0} \joinlrnoarrowbw[above]{R}{0}{X}{0} \csymbol[0,5]{x}
\end{Compose}
\end{equation}
Using the expressions from Sec.\ \ref{sec:maximaloperators} and Sec.\ \ref{sec:flatdistributionoperators} we obtain
\begin{equation}
\begin{Compose}{0}{0}\setdefaultfont{\hat}\setsecondfont{\mathtt}\setthirdfont{\mathsfb}
\Crectangledouble{I}{1}{1}{0,0}
\thispoint{p}{0,3.5}  \csymbolthird[0,20]{x} \jointbnoarrow{I}{0}{p}{0}
\end{Compose}
~=~ \frac{\alpha_\mathsfb{x}}{\sqrt{N_\mathsfb{x}}} \hat{\mathds{1}}^{\mathsfb{x}_1}
~~~~~~~~~~~~
\begin{Compose}{0}{0}\setdefaultfont{\hat}\setsecondfont{\mathtt}\setthirdfont{\mathsfb}
\Crectangledouble{I}{1}{1}{0,0}
\thispoint{p}{0,-3.5}  \csymbolthird[0,-20]{x} \joinbtnoarrow{I}{0}{p}{0}
\end{Compose}
~=~ \frac{1}{\alpha_\mathsfb{x}\sqrt{N_\mathsfb{x}}} \hat{\mathds{1}}_{\mathsfb{x}_1}
\end{equation}
where
\begin{equation}
\hat{\mathds{1}}^{\mathsfb{x}_1} = \sum_x |x\rangle^{\mathsfb{x}_1}\negs\negs\langle x| ~~~~~~~~~~~~~ \hat{\mathds{1}}_{\mathsfb{x}_1} = \sum_x |x\rangle_{\mathsfb{x}_1} \negs\negs\negs\langle x|
\end{equation}
is the identity operator (as originally defined in Sec.\ \ref{sec:identityoperators}.  We will discuss in Sec.\ \ref{sec:chasingdowntheoneoverN} different choices of $\alpha_\mathsfb{a}$.

Given these results we clearly have the factorization results (analogous to \eqref{IequalsII})
\begin{equation}\label{IequalsIIop}
\begin{Compose}{0}{0}\setdefaultfont{\hat}\setsecondfont{\mathtt}\setthirdfont{\mathsfb}
\crectangledouble{X}{0.9}{0.9}{0,2} \csymbol{I}
\thispoint{Y}{0,-1.5}\csymbolthird[0,-20]{ab} \joinbtnoarrow[right]{X}{0}{Y}{0}
\end{Compose}
~~\equiv~~
\begin{Compose}{0}{0}\setdefaultfont{\hat}\setsecondfont{\mathtt}\setthirdfont{\mathsfb}
\crectangledouble{X}{0.9}{0.9}{0,2} \csymbol{I}
\thispoint{Y}{0,-1.5}\csymbolthird[0,-20]{a} \joinbtnoarrow[right]{X}{0}{Y}{0}
\end{Compose}
~~\begin{Compose}{0}{0}\setdefaultfont{\hat}\setsecondfont{\mathtt}\setthirdfont{\mathsfb}
\crectangledouble{X}{0.9}{0.9}{0,2} \csymbol{I}
\thispoint{Y}{0,-1.5}\csymbolthird[0,-20]{b} \joinbtnoarrow[right]{X}{0}{Y}{0}
\end{Compose}
~~~~~~~~~~~~~~~~~~~~
\begin{Compose}{0}{0}\setdefaultfont{\hat}\setsecondfont{\mathtt}\setthirdfont{\mathsfb}
\crectangledouble{X}{0.9}{0.9}{0,-2} \csymbol{I}
\thispoint{Y}{0,1.5}\csymbolthird[0,20]{ab} \jointbnoarrow[right]{X}{0}{Y}{0}
\end{Compose}
~~\equiv~~
\begin{Compose}{0}{0}\setdefaultfont{\hat}\setsecondfont{\mathtt}\setthirdfont{\mathsfb}
\crectangledouble{X}{0.9}{0.9}{0,-2} \csymbol{I}
\thispoint{Y}{0,1.5}\csymbolthird[0,20]{a} \jointbnoarrow[right]{X}{0}{Y}{0}
\end{Compose}
~~\begin{Compose}{0}{0}\setdefaultfont{\hat}\setsecondfont{\mathtt}\setthirdfont{\mathsfb}
\crectangledouble{X}{0.9}{0.9}{0,-2} \csymbol{I}
\thispoint{Y}{0,1.5}\csymbolthird[0,20]{b} \jointbnoarrow[right]{X}{0}{Y}{0}
\end{Compose}
\end{equation}

\subsection{Chasing down the $\frac{1}{N_\mathsfb{x}}$ factor} \label{sec:chasingdowntheoneoverN}

In standard Quantum Theory the maximally mixed normalised state is written as $\mathds{1}/N$ and the deterministic effect is written as $\mathds{1}$ (this latter object is related to the imposition of normalisation on a density matrix, $\rho$, by imposing $\text{tr}(\mathds{1}\rho)=\text{tr}(\rho)=1$.  The deterministic state is not unique in the standard formulation. However, even if it were unique and equal to $\mathds{1}/N$, there appears to be an additional time asymmetric residue in that we have $\mathds{I}/N$ and $\mathds{I}$ for objects that ought to be the time reverses of one another.   In the time symmetric operational formalism presented here this symmetry is restored because we work with operator tensors which are abstract objects. However, if we want to actually represent these objects then we need to make a choice for the constant $\alpha_\mathsfb{x}$.   We will review four possible choices for this constant that relate to different points of view. Interestingly, we can make choices for these constants such that we have explicit time symmetry in the concrete representation.  Whatever choice we make, the constants drop out of the calculations so the formalism remains time symmetric so far as actual predictions are concerned.  Before discussing choices for these constants we collect some previous results. We have
\begin{equation}
\begin{Compose}{0}{-0.07}\setdefaultfont{\hat}\setfourthfont{\mathsfb}\setsecondfont{\mathtt}
\Crectangledouble{X}{1}{1}{0,0}
\thispoint{p}{0,3} \csymbolfourth[0,20]{x} \jointbnoarrow[left]{X}{0}{p}{0}
\whitesquareleftwire{X}{0}{x}
\end{Compose}
= \alpha_\mathsfb{x}\sqrt{N_\mathsfb{x}}\mathbb{P}^{\mathsfb{x}_1}
~~~~~
\begin{Compose}{0}{-0.5}\setdefaultfont{\hat}\setsecondfont{\mathtt}\setfourthfont{\mathsfb}
\Crectangledouble{X}{1}{1}{0,3}
\thispoint{p}{0,0} \csymbolfourth[0,-20]{x} \jointbnoarrow[left]{p}{0}{X}{0}
\blacksquarerightwire{X}{0}{x}
\end{Compose}
\hspace{-2mm}
= \frac{1}{\alpha_\mathsfb{x}\sqrt{N_\mathsfb{x}}}\mathbb{P}_{\mathsfb{x}_1}
\end{equation}
(where $\mathbb{P}=|x\rangle\langle x|$).
This pair of operator tensors above correspond to a forward point of view (white dot going in means we condition on this income and black dot going out means we look at the probability for this outcome). We have
\begin{equation}
\begin{Compose}{0}{-0.5}\setdefaultfont{\hat}\setsecondfont{\mathtt}\setfourthfont{\mathsfb}
\Crectangledouble{X}{1}{1}{0,3}
\thispoint{p}{0,0} \csymbolfourth[0,-20]{x} \jointbnoarrow[left]{p}{0}{X}{0}
\whitesquarerightwire{X}{0}{x}
\end{Compose}
\hspace{-2mm}
=\frac{\sqrt{N_\mathsfb{x}}}{\alpha_\mathsfb{x}}\mathbb{P}^{\mathsfb{x}_1}
~~~~
\begin{Compose}{0}{-0.07}\setdefaultfont{\hat}\setfourthfont{\mathsfb}\setsecondfont{\mathtt}
\Crectangledouble{X}{1}{1}{0,0}
\thispoint{p}{0,3} \csymbolfourth[0,20]{x} \jointbnoarrow[left]{X}{0}{p}{0}
\blacksquareleftwire{X}{0}{x}
\end{Compose}
= \frac{\alpha_\mathsfb{x}}{\sqrt{N_\mathsfb{x}}} \mathbb{P}^{\mathsfb{x}_1}
\end{equation}
This pair of operator tensor correspond to a backwards point of view.  We also have ignore operators
\begin{equation}\label{identityalphaNstuff}
\begin{Compose}{0}{-0.07}\setdefaultfont{\hat}\setsecondfont{\mathtt}\setthirdfont{\mathsfb}
\Crectangledouble{I}{1}{1}{0,0}
\thispoint{p}{0,3.5}  \csymbolthird[0,20]{x} \jointbnoarrow{I}{0}{p}{0}
\end{Compose}
~=~ \frac{\alpha_\mathsfb{x}}{\sqrt{N_\mathsfb{x}}} \hat{\mathds{1}}^{\mathsfb{x}_1}
~~~~~~~~~~~~
\begin{Compose}{0}{-0.07}\setdefaultfont{\hat}\setsecondfont{\mathtt}\setthirdfont{\mathsfb}
\Crectangledouble{I}{1}{1}{0,0}
\thispoint{p}{0,-3.5}  \csymbolthird[0,-20]{x} \joinbtnoarrow{I}{0}{p}{0}
\end{Compose}
~=~ \frac{1}{\alpha_\mathsfb{x}\sqrt{N_\mathsfb{x}}} \hat{\mathds{1}}_{\mathsfb{x}_1}
\end{equation}
Now we will look at four particular choices for the gauge.
\begin{description}
\item[Forward gauge.]  For this we set $\alpha_\mathsfb{a}=\frac{1}{\sqrt{N_\mathsfb{a}}}$.  Then we obtain
\begin{equation*}
\begin{Compose}{0}{-0.07}\setdefaultfont{\hat}\setfourthfont{\mathsfb}\setsecondfont{\mathtt}
\Crectangledouble{X}{1}{1}{0,0}
\thispoint{p}{0,3} \csymbolfourth[0,20]{x} \jointbnoarrow[left]{X}{0}{p}{0}
\whitesquareleftwire{X}{0}{x}
\end{Compose}
= \mathbb{P}^{\mathsfb{x}_1}
~~~~~
\begin{Compose}{0}{-0.5}\setdefaultfont{\hat}\setsecondfont{\mathtt}\setfourthfont{\mathsfb}
\Crectangledouble{X}{1}{1}{0,3}
\thispoint{p}{0,0} \csymbolfourth[0,-20]{x} \jointbnoarrow[left]{p}{0}{X}{0}
\blacksquarerightwire{X}{0}{x}
\end{Compose}
\hspace{-2mm}
= \mathbb{P}_{\mathsfb{x}_1}
~~~~~~~~~~~~
\begin{Compose}{0}{-0.5}\setdefaultfont{\hat}\setsecondfont{\mathtt}\setfourthfont{\mathsfb}
\Crectangledouble{X}{1}{1}{0,3}
\thispoint{p}{0,0} \csymbolfourth[0,-20]{x} \jointbnoarrow[left]{p}{0}{X}{0}
\whitesquarerightwire{X}{0}{x}
\end{Compose}
\hspace{-2mm}
= N_\mathsfb{x}\mathbb{P}^{\mathsfb{x}_1}
~~~~
\begin{Compose}{0}{-0.07}\setdefaultfont{\hat}\setfourthfont{\mathsfb}\setsecondfont{\mathtt}
\Crectangledouble{X}{1}{1}{0,0}
\thispoint{p}{0,3} \csymbolfourth[0,20]{x} \jointbnoarrow[left]{X}{0}{p}{0}
\blacksquareleftwire{X}{0}{x}
\end{Compose}
= \frac{1}{N_\mathsfb{x}} \mathbb{P}^{\mathsfb{x}_1}
\end{equation*}
We see that the forward point of view operators are associated with a cooefficient 1.  This is the usual way of thinking about maximal preparations and maximal measurements. We also have
\begin{equation}
\begin{Compose}{0}{-0.07}\setdefaultfont{\hat}\setsecondfont{\mathtt}\setthirdfont{\mathsfb}
\Crectangledouble{I}{1}{1}{0,0}
\thispoint{p}{0,3.5}  \csymbolthird[0,20]{x} \jointbnoarrow{I}{0}{p}{0}
\end{Compose}
~=~ \frac{1}{N_\mathsfb{x}} \hat{\mathds{1}}^{\mathsfb{x}_1}
~~~~~~~~~~~~
\begin{Compose}{0}{-0.07}\setdefaultfont{\hat}\setsecondfont{\mathtt}\setthirdfont{\mathsfb}
\Crectangledouble{I}{1}{1}{0,0}
\thispoint{p}{0,-3.5}  \csymbolthird[0,-20]{x} \joinbtnoarrow{I}{0}{p}{0}
\end{Compose}
~=~ \hat{\mathds{1}}_{\mathsfb{x}_1}
\end{equation}
These are the same as the standard representations for the maximally mixed state and the deterministic effect in time asymmetric Quantum Theory as discussed at the beginning of this subsection.
\item[Backward gauge.] For this we set $\alpha_\mathsfb{a}=\sqrt{N_\mathsfb{a}}$.  Then we obtain
\begin{equation*}
\begin{Compose}{0}{-0.07}\setdefaultfont{\hat}\setfourthfont{\mathsfb}\setsecondfont{\mathtt}
\Crectangledouble{X}{1}{1}{0,0}
\thispoint{p}{0,3} \csymbolfourth[0,20]{x} \jointbnoarrow[left]{X}{0}{p}{0}
\whitesquareleftwire{X}{0}{x}
\end{Compose}
= N_\mathsfb{x}\mathbb{P}^{\mathsfb{x}_1}
~~~~~
\begin{Compose}{0}{-0.5}\setdefaultfont{\hat}\setsecondfont{\mathtt}\setfourthfont{\mathsfb}
\Crectangledouble{X}{1}{1}{0,3}
\thispoint{p}{0,0} \csymbolfourth[0,-20]{x} \jointbnoarrow[left]{p}{0}{X}{0}
\blacksquarerightwire{X}{0}{x}
\end{Compose}
\hspace{-2mm}
= \frac{1}{N_\mathsfb{x}}\mathbb{P}_{\mathsfb{x}_1}
~~~~~~~~~~~~
\begin{Compose}{0}{-0.5}\setdefaultfont{\hat}\setsecondfont{\mathtt}\setfourthfont{\mathsfb}
\Crectangledouble{X}{1}{1}{0,3}
\thispoint{p}{0,0} \csymbolfourth[0,-20]{x} \jointbnoarrow[left]{p}{0}{X}{0}
\whitesquarerightwire{X}{0}{x}
\end{Compose}
\hspace{-2mm}
=\mathbb{P}^{\mathsfb{x}_1}
~~~~
\begin{Compose}{0}{-0.07}\setdefaultfont{\hat}\setfourthfont{\mathsfb}\setsecondfont{\mathtt}
\Crectangledouble{X}{1}{1}{0,0}
\thispoint{p}{0,3} \csymbolfourth[0,20]{x} \jointbnoarrow[left]{X}{0}{p}{0}
\blacksquareleftwire{X}{0}{x}
\end{Compose}
= \mathbb{P}^{\mathsfb{x}_1}
\end{equation*}
We see that the backwards point of view operators are associated with a cooefficient 1.  This is the time reverse of the usual forward way of thinking about maximal preparations and maximal measurements. We also have
\begin{equation}
\begin{Compose}{0}{-0.07}\setdefaultfont{\hat}\setsecondfont{\mathtt}\setthirdfont{\mathsfb}
\Crectangledouble{I}{1}{1}{0,0}
\thispoint{p}{0,3.5}  \csymbolthird[0,20]{x} \jointbnoarrow{I}{0}{p}{0}
\end{Compose}
~=~ \hat{\mathds{1}}^{\mathsfb{x}_1}
~~~~~~~~~~~~
\begin{Compose}{0}{-0.07}\setdefaultfont{\hat}\setsecondfont{\mathtt}\setthirdfont{\mathsfb}
\Crectangledouble{I}{1}{1}{0,0}
\thispoint{p}{0,-3.5}  \csymbolthird[0,-20]{x} \joinbtnoarrow{I}{0}{p}{0}
\end{Compose}
~=~ \frac{1}{N_\mathsfb{x}}\hat{\mathds{1}}_{\mathsfb{x}_1}
\end{equation}
These are the time reverse of the standard representations for the maximally mixed state and the deterministic effect in time asymmetric Quantum Theory.  Now the $1/N_\mathsfb{x}$ cooefficient is on the ignore result.
\item[Symmetric gauge.] For this we set $\alpha_\mathsfb{a}=1$.  Then we obtain
\begin{equation*}
\begin{Compose}{0}{-0.07}\setdefaultfont{\hat}\setfourthfont{\mathsfb}\setsecondfont{\mathtt}
\Crectangledouble{X}{1}{1}{0,0}
\thispoint{p}{0,3} \csymbolfourth[0,20]{x} \jointbnoarrow[left]{X}{0}{p}{0}
\whitesquareleftwire{X}{0}{x}
\end{Compose}
= \sqrt{N_\mathsfb{x}}\mathbb{P}^{\mathsfb{x}_1}
~~~~~
\begin{Compose}{0}{-0.5}\setdefaultfont{\hat}\setsecondfont{\mathtt}\setfourthfont{\mathsfb}
\Crectangledouble{X}{1}{1}{0,3}
\thispoint{p}{0,0} \csymbolfourth[0,-20]{x} \jointbnoarrow[left]{p}{0}{X}{0}
\blacksquarerightwire{X}{0}{x}
\end{Compose}
\hspace{-2mm}
= \frac{\mathbb{P}_{\mathsfb{x}_1}}{\sqrt{N_\mathsfb{x}}}
~~~~~~~
\begin{Compose}{0}{-0.5}\setdefaultfont{\hat}\setsecondfont{\mathtt}\setfourthfont{\mathsfb}
\Crectangledouble{X}{1}{1}{0,3}
\thispoint{p}{0,0} \csymbolfourth[0,-20]{x} \jointbnoarrow[left]{p}{0}{X}{0}
\whitesquarerightwire{X}{0}{x}
\end{Compose}
\hspace{-2mm}
= \sqrt{N_\mathsfb{x}}\mathbb{P}^{\mathsfb{x}_1}
~~~~
\begin{Compose}{0}{-0.07}\setdefaultfont{\hat}\setfourthfont{\mathsfb}\setsecondfont{\mathtt}
\Crectangledouble{X}{1}{1}{0,0}
\thispoint{p}{0,3} \csymbolfourth[0,20]{x} \jointbnoarrow[left]{X}{0}{p}{0}
\blacksquareleftwire{X}{0}{x}
\end{Compose}
=  \frac{\mathbb{P}^{\mathsfb{x}_1}}{\sqrt{N_\mathsfb{x}}}
\end{equation*}
Now the white dot maximal operators have coefficient $\sqrt{N_\mathsfb{x}}$ and the black dot maximal operators have coefficient $\frac{1}{\sqrt{N_\mathsfb{x}}}$.  This is explicitly time symmetric.  We also have
\begin{equation}
\begin{Compose}{0}{-0.07}\setdefaultfont{\hat}\setsecondfont{\mathtt}\setthirdfont{\mathsfb}
\Crectangledouble{I}{1}{1}{0,0}
\thispoint{p}{0,3.5}  \csymbolthird[0,20]{x} \jointbnoarrow{I}{0}{p}{0}
\end{Compose}
~=~ \frac{1}{\sqrt{N_\mathsfb{x}}} \hat{\mathds{1}}^{\mathsfb{x}_1}
~~~~~~~~~~~~
\begin{Compose}{0}{-0.07}\setdefaultfont{\hat}\setsecondfont{\mathtt}\setthirdfont{\mathsfb}
\Crectangledouble{I}{1}{1}{0,0}
\thispoint{p}{0,-3.5}  \csymbolthird[0,-20]{x} \joinbtnoarrow{I}{0}{p}{0}
\end{Compose}
~=~ \frac{1}{\sqrt{N_\mathsfb{x}}} \hat{\mathds{1}}_{\mathsfb{x}_1}
\end{equation}
This is also explicitly time symmetric.
\end{description}
Probably the symmetric gauge is the most natural.  However, there is much historical baggage associated with Quantum Theory and so the forward gauge may be the most familiar of these three choices for $\alpha_\mathsfb{x}$.

When calculations are conducted at the abstract level of operator tensors, we do not need to make an explicit choice for $\alpha_\mathsfb{x}\sqrt{N_\mathsfb{x}}$.  The fact that we do not see these gauge parameters explicitly in the operator tensor formalism is evidence that this approach is more suited to a proper formulation of Quantum Theory than the objects used in the standard formalism.  However, when we want to actually do a calculation, we need to represent the operator tensors and then we need to be explicit about $\alpha_\mathsfb{x}$.

\subsection{Chasing down the $\frac{1}{N_\mathtt{x}}$ factor}\label{sec:chasingdowntheoneoverNfactorforpointers}

There is an exactly analogous story to be told to the previous section, but concerning pointer types instead of system types.  In Sec.\ \ref{sec:vectorspaceforincomesandoutcomes} we associated a fiducial vectors $\presub{x_2}{\vec{P}}^{\mathtt{x}_1}$ and ${\vec{P}}_{\mathtt{x}_1}^{x_2}$
with incomes and outcomes, these being represented diagrammatically by
\begin{equation}
\begin{Compose}{0}{0} \setsecondfont{\mathtt}
\blackdotsq{b}{-2.5,0}\csymbolfourth[-20,0]{x} \sleftvecfid{P}{0,0} \thispoint{p}{2.5,0} \csymbolalt[20,0]{x}
\joinrlnoarrow{b}{0}{P}{0} \joinrlnoarrowthick{P}{0}{p}{0}
\end{Compose}
~~~~~~~\text{and}~~~~~~~
\begin{Compose}{0}{0} \setsecondfont{\mathtt}
\blackdotsq{b}{2.5,0}\csymbolfourth[20,0]{x} \srightvecfid{P}{0,0} \thispoint{p}{-2.5,0} \csymbolalt[-20,0]{x}
\joinlrnoarrow{b}{0}{P}{0} \joinlrnoarrowthick{P}{0}{p}{0}
\end{Compose}
\end{equation}
and demanded that the scalar product of these is $\frac{1}{N_\mathtt{x}}$ times  a delta function (see \eqref{vechoppingsymbolic}).  Diagrammatically, this is
\begin{equation}
\begin{Compose}{0}{0} \setsecondfont{\mathtt}
\blackdotsq{FL}{-2.5,0} \csymbolthird[-24,0]{x}
\sleftvecfid{F}{0,0}
\srightvecfid{rF}{3,0}
\blackdotsq{rFR}{5.5,0} \csymbolthird[24,0]{x}
 \joinrlnoarrow[above]{FL}{0}{F}{0}
\joinrlnoarrowthick[above]{F}{0}{rF}{0} \csymbolalt{x}
\joinrlnoarrow[above]{rF}{0}{rFR}{0}
\end{Compose}
~~=~~ \frac{1}{N_\mathtt{x}}\left(\begin{array}{cccc}
                                    1 &   &        &  \\
                                      & 1 &        &  \\
                                      &   & \ddots &  \\
                                      &   &        & 1
                                  \end{array}\right)
\end{equation}
However, we haven't said how we actually represent these fiducial vectors.  We can put
\begin{equation}
\begin{Compose}{0}{0} \setsecondfont{\mathtt}
\blackdotsq{b}{-2.5,0}\csymbolfourth[-20,0]{x} \sleftvecfid{P}{0,0} \thispoint{p}{2.5,0} \csymbolalt[20,0]{x}
\joinrlnoarrow{b}{0}{P}{0} \joinrlnoarrowthick{P}{0}{p}{0}
\end{Compose}
~~=~~
\frac{\beta_\mathtt{x}}{\sqrt{N_\mathtt{x}}}
\left( \begin{array}{c}
         0 \\
         0 \\
         \vdots \\
         1 \\
         \vdots \\
         0
       \end{array} \right)
~~~~~~~\text{and}~~~~~~~
\begin{Compose}{0}{0} \setsecondfont{\mathtt}
\blackdotsq{b}{2.5,0}\csymbolfourth[20,0]{x} \srightvecfid{P}{0,0} \thispoint{p}{-2.5,0} \csymbolalt[-20,0]{x}
\joinlrnoarrow{b}{0}{P}{0} \joinlrnoarrowthick{P}{0}{p}{0}
\end{Compose}
~~=~~
\frac{1}{\beta_\mathtt{x}\sqrt{N_\mathtt{x}}}
\left( \begin{array}{c}
         0 \\
         0 \\
         \vdots \\
         1 \\
         \vdots \\
         0
       \end{array} \right)
\end{equation}
where the $1$ is in the $x$ position.  This gives the correct scalar product.  We have the gauge constant, $\beta_\mathtt{x}$.  We can consider three choices for this.
\begin{description}
\item[Forward choice.] In the forward time direction it is natural to regard the income preparation as a probability distribution so we set $\beta_\mathtt{x}=\sqrt{N_\mathtt{x}}$ so that
\begin{equation}
\begin{Compose}{0}{0} \setsecondfont{\mathtt}
\blackdotsq{b}{-2.5,0}\csymbolfourth[-20,0]{x} \sleftvecfid{P}{0,0} \thispoint{p}{2.5,0} \csymbolalt[20,0]{x}
\joinrlnoarrow{b}{0}{P}{0} \joinrlnoarrowthick{P}{0}{p}{0}
\end{Compose}
~~=~~
\frac{1}{{N_\mathtt{x}}}
\left( \begin{array}{c}
         0 \\
         0 \\
         \vdots \\
         1 \\
         \vdots \\
         0
       \end{array} \right)
~~~~~~~\text{and}~~~~~~~
\begin{Compose}{0}{0} \setsecondfont{\mathtt}
\blackdotsq{b}{2.5,0}\csymbolfourth[20,0]{x} \srightvecfid{P}{0,0} \thispoint{p}{-2.5,0} \csymbolalt[-20,0]{x}
\joinlrnoarrow{b}{0}{P}{0} \joinlrnoarrowthick{P}{0}{p}{0}
\end{Compose}
~~=~~
\left( \begin{array}{c}
         0 \\
         0 \\
         \vdots \\
         1 \\
         \vdots \\
         0
       \end{array} \right)
\end{equation}
\item[Backwards choice.] In the backwards time direction it is natural to regard the outcome as a probability distribution so we set       $\beta_\mathtt{x}=\frac{1}{\sqrt{N_\mathtt{x}}}$ so that
\begin{equation}
\begin{Compose}{0}{0} \setsecondfont{\mathtt}
\blackdotsq{b}{-2.5,0}\csymbolfourth[-20,0]{x} \sleftvecfid{P}{0,0} \thispoint{p}{2.5,0} \csymbolalt[20,0]{x}
\joinrlnoarrow{b}{0}{P}{0} \joinrlnoarrowthick{P}{0}{p}{0}
\end{Compose}
~~=~~
\left( \begin{array}{c}
         0 \\
         0 \\
         \vdots \\
         1 \\
         \vdots \\
         0
       \end{array} \right)
~~~~~~~\text{and}~~~~~~~
\begin{Compose}{0}{0} \setsecondfont{\mathtt}
\blackdotsq{b}{2.5,0}\csymbolfourth[20,0]{x} \srightvecfid{P}{0,0} \thispoint{p}{-2.5,0} \csymbolalt[-20,0]{x}
\joinlrnoarrow{b}{0}{P}{0} \joinlrnoarrowthick{P}{0}{p}{0}
\end{Compose}
~~=~~
\frac{1}{{N_\mathtt{x}}}
\left( \begin{array}{c}
         0 \\
         0 \\
         \vdots \\
         1 \\
         \vdots \\
         0
       \end{array} \right)
\end{equation}
\item[Symmetric choice.] The symmetric choice is where we put $\beta_\mathtt{x}=1$.  Then we have
\begin{equation}
\begin{Compose}{0}{0} \setsecondfont{\mathtt}
\blackdotsq{b}{-2.5,0}\csymbolfourth[-20,0]{x} \sleftvecfid{P}{0,0} \thispoint{p}{2.5,0} \csymbolalt[20,0]{x}
\joinrlnoarrow{b}{0}{P}{0} \joinrlnoarrowthick{P}{0}{p}{0}
\end{Compose}
~~=~~
\frac{1}{\sqrt{N_\mathtt{x}}}
\left( \begin{array}{c}
         0 \\
         0 \\
         \vdots \\
         1 \\
         \vdots \\
         0
       \end{array} \right)
~~~~~~~\text{and}~~~~~~~
\begin{Compose}{0}{0} \setsecondfont{\mathtt}
\blackdotsq{b}{2.5,0}\csymbolfourth[20,0]{x} \srightvecfid{P}{0,0} \thispoint{p}{-2.5,0} \csymbolalt[-20,0]{x}
\joinlrnoarrow{b}{0}{P}{0} \joinlrnoarrowthick{P}{0}{p}{0}
\end{Compose}
~~=~~
\frac{1}{\sqrt{N_\mathtt{x}}}
\left( \begin{array}{c}
         0 \\
         0 \\
         \vdots \\
         1 \\
         \vdots \\
         0
       \end{array} \right)
\end{equation}
\end{description}
We need to be explicit about $\beta_\mathtt{x}$ when we write down explicit expressions for operator tensors.

\part{Time Symmetric Operational Quantum Theory}\label{TSOQT}

We are now in a position to bring together all the strands developed in this paper to formulate Quantum Theory in a time symmetric operational manner.  First we set up a correspondence between operations and operators so that the operator circuit corresponding to any circuit (comprised of operations) gives the probability associated with that circuit.  This is achieved is by choosing fiducial elements for the operations and for the operators so we get the same hopping metrics for the operator theory as we do for the operation theory.  Then circuits comprised of operations will have the same probability as circuits comprised of operators if operations correspond to operators having the same duotensors.  This is explained in Sec.\ \ref{sec:correspondencebetweenoperationsandoperators}.  Next we need to establish what the conditions on allowed operators are. These conditions follow from the conditions on allowed operations - namely the physicality conditions. We will discuss this in Sec.\ \ref{sec:physicalityconditionsoperators}.  After developing these ideas we can state a simple mathematical axiom for quantum theory in the language of operations and operator tensors.  In Sec.\ \ref{sec:extensiontheorem} we will show that there is an kind of extension theorem like that of Stinespring \cite{stinespring1955positive} except adapted to this time symmetric.  We will also prove that Quantum Theory, formulated in this way, is explicitly time symmetric according to the definition in Sec.\ \ref{sec:definitionoftimesymmetricoperationaltheory}.

\section{Correspondence between operations and operators}\label{sec:correspondencebetweenoperationsandoperators}

So far we have set up the theory of operations showing how to calculate probabilities using duotensors.  We have also set up the theory of operators showing how to do calculations by means of duotensors.  We will now make operations and operators talk to each other by setting up an appropriate correspondence. In particular, we define such a correspondence in the following way.
\begin{quote}
\textbf{Operation-operator correspondence.}  We say that we have an operation-operator correspondence if there is an operator corresponding to every operation such that, if we start with any circuit comprised of operations then replace each operation by the corresponding operator, then the operator circuit so formed is equal to the probability associated with the circuit.
\end{quote}
Given the forgoing theory of operations and operators, it is in fact easy to set up such an operation-operator correspondence.   We state this in the form of the correspondence theorem given below.  The basic idea is that we choose fiducial sets from the operation approach and fiducial sets from the operator approach such that they give the same hopping matrices.   Then we can show that the probability for a circuit composed of operations is equal to the corresponding operator circuit whereby each operation is replaced by an operator \emph{having the same duotensor}.
\begin{quote}
\textbf{Correspondence Theorem}
If we have operation fiducials and operator fiducials such that
\begin{equation}\label{pointerhoppingcorrespondence}
\begin{Compose}{0}{0} \setsecondfont{\mathtt}
\blackdotsq{FL}{-2.5,0} \csymbolthird[-24,0]{x}
\sleftdatafid{F}{0,0}
\srightdatafid{rF}{3,0}
\blackdotsq{rFR}{5.5,0} \csymbolthird[24,0]{x}
 \joinrlnoarrow[above]{FL}{0}{F}{0}
\joinrlnoarrowthick[above]{F}{0}{rF}{0} \csymbolalt{x}
\joinrlnoarrow[above]{rF}{0}{rFR}{0}
\end{Compose}
~~\equiv~~
\begin{Compose}{0}{0} \setsecondfont{\mathtt}
\blackdotsq{FL}{-2.5,0} \csymbolthird[-24,0]{x}
\sleftvecfid{F}{0,0}
\srightvecfid{rF}{3,0}
\blackdotsq{rFR}{5.5,0} \csymbolthird[24,0]{x}
 \joinrlnoarrow[above]{FL}{0}{F}{0}
\joinrlnoarrowthick[above]{F}{0}{rF}{0} \csymbolalt{x}
\joinrlnoarrow[above]{rF}{0}{rFR}{0}
\end{Compose}
~:=~ \begin{Compose}{0}{0}
\bbmatrixnoarrowsq{h}{0,0} \csymbolthird{x}
\end{Compose}
\end{equation}
and
\begin{equation}\label{systemhoppingcorrespondence}
\begin{Compose}{0}{0} \setsecondfont{\mathsfb} \setfourthfont{\mathpzc}
\blackdot{u}{0,3.5} \csymbolfourth[0,20]{x}\sdownsysfid{up}{0,1.5} \supsysfid{do}{0,-1.5} \blackdot{d}{0,-3.5}\csymbolfourth[0,-20]{x}
\jointbnoarrow[left]{d}{0}{do}{0} \jointbnoarrow[left]{do}{0}{up}{0} \csymbolalt{x} \jointbnoarrow[left]{up}{0}{u}{0}
\end{Compose}
~~\equiv~~
\begin{Compose}{0}{0} \setsecondfont{\mathsfb} \setfourthfont{\mathpzc}
\blackdot{u}{0,3.5} \csymbolfourth[0,20]{x}\sdownopfid{up}{0,1.5} \supopfid{do}{0,-1.5} \blackdot{d}{0,-3.5}\csymbolfourth[0,-20]{x}
\jointbnoarrow[left]{d}{0}{do}{0} \jointbnoarrow[left]{do}{0}{up}{0} \csymbolalt{x} \jointbnoarrow[left]{up}{0}{u}{0}
\end{Compose}
~~:=~~
\begin{Compose}{0}{0} \setfourthfont{\mathpzc}
\blackdot{u}{0,1.5}\blackdot{d}{0,-1.5} \jointbnoarrow[left]{d}{0}{u}{0}\csymbolfourth{x}
\end{Compose}
\end{equation}
(so that the hopping matrices are the same for the operation theory and the operator theory)
then corresponding to an arbitrary the operation (expanded as a fiducial weighted sum over operation fiducials)
\begin{equation}\label{Boperationexpansion}
\begin{Compose}{0}{0}\setsecondfont{\mathsfb} \setthirdfont{\mathpzc}  
\Crectangle{B}{4}{4}{0,0}
\cupwire{B}{-3.5}{d} \cupwire{B}{-1}{e} \thispoint{down}{1.05,5.8}\csymbol{\dots} \cupwire{B}{3.5}{f}
\cdownwire{B}{-3.5}{a} \cdownwire{B}{-1}{b} \thispoint{down}{1.05,-5.8}\csymbol{\dots} \cdownwire{B}{3.5}{c}
\cincomewire{B}{-3.5}{x} \cincomewire{B}{-1}{y} \thispoint{down}{-5.8, 1.15}\csymbol{\vdots} \cincomewire{B}{3.5}{z}
\coutcomewire{B}{-3.5}{u} \coutcomewire{B}{-1}{v} \thispoint{down}{5.8, 1.15}\csymbol{\vdots} \coutcomewire{B}{3.5}{w}
\end{Compose}
~~~\equiv~~~
\begin{Compose}{0}{0} \setdefaultfont{}
\Crectangle{B}{4}{4}{0,0}
\setsecondfont{\mathtt} \setthirdfont{}
\cfiddyleftwire{B}{-3.5}{x}\cfiddyleftwire{B}{-1}{y} \thispoint{down}{-7, 1.25}\csymbol{\vdots} \cfiddyleftwire{B}{3.5}{z}
\cfiddyrightwire{B}{-3.5}{u} \cfiddyrightwire{B}{-1}{v} \thispoint{down}{7, 1.25}\csymbol{\vdots} \cfiddyrightwire{B}{3.5}{w}
\setsecondfont{\mathsfb} \setthirdfont{\mathpzc}
\cfiddyupwire{B}{-3.5}{d} \cfiddyupwire{B}{-1}{e} \thispoint{down}{1.25, 7}\csymbol{\dots} \cfiddyupwire{B}{3.5}{f}
\cfiddydownwire{B}{-3.5}{a} \cfiddydownwire{B}{-1}{b} \thispoint{down}{1.25,-7}\csymbol{\dots} \cfiddydownwire{B}{3.5}{c}
\end{Compose}
\end{equation}
is the operator
\begin{equation}\label{Boperatorexpansion}
\begin{Compose}{0}{0}\setdefaultfont{\hat}\setsecondfont{\mathsfb} \setthirdfont{\mathpzc}  
\Crectangledouble{B}{4}{4}{0,0}
\setdefaultfont{}
\cupwire{B}{-3.5}{d} \cupwire{B}{-1}{e} \thispoint{down}{1.05,5.8}\csymbol{\dots} \cupwire{B}{3.5}{f}
\cdownwire{B}{-3.5}{a} \cdownwire{B}{-1}{b} \thispoint{down}{1.05,-5.8}\csymbol{\dots} \cdownwire{B}{3.5}{c}
\cincomewire{B}{-3.5}{x} \cincomewire{B}{-1}{y} \thispoint{down}{-5.8, 1.15}\csymbol{\vdots} \cincomewire{B}{3.5}{z}
\coutcomewire{B}{-3.5}{u} \coutcomewire{B}{-1}{v} \thispoint{down}{5.8, 1.15}\csymbol{\vdots} \coutcomewire{B}{3.5}{w}
\end{Compose}
~~~=~~~
\begin{Compose}{0}{0} \setdefaultfont{}
\Crectangle{B}{4}{4}{0,0}
\setsecondfont{\mathtt} \setthirdfont{}
\cfidopleftwire{B}{-3.5}{x}\cfidopleftwire{B}{-1}{y} \thispoint{down}{-7, 1.25}\csymbol{\vdots} \cfidopleftwire{B}{3.5}{z}
\cfidoprightwire{B}{-3.5}{u} \cfidoprightwire{B}{-1}{v} \thispoint{down}{7, 1.25}\csymbol{\vdots} \cfidoprightwire{B}{3.5}{w}
\setsecondfont{\mathsfb} \setthirdfont{\mathpzc}
\cfidopupwire{B}{-3.5}{d} \cfidopupwire{B}{-1}{e} \thispoint{down}{1.25, 7}\csymbol{\dots} \cfidopupwire{B}{3.5}{f}
\cfidopdownwire{B}{-3.5}{a} \cfidopdownwire{B}{-1}{b} \thispoint{down}{1.25,-7}\csymbol{\dots} \cfidopdownwire{B}{3.5}{c}
\end{Compose}
\end{equation}
expanded as a duotensor weighted sum over operator fiducials.  Importantly, the duotensor in \eqref{Boperatorexpansion} is the same as the one in \eqref{Boperationexpansion}.
\end{quote}
It is easy to prove that this is an operation-operator correspondence - so that the probability associated with a circuit is given by the corresponding operator circuit under this correspondence.  Consider the circuit
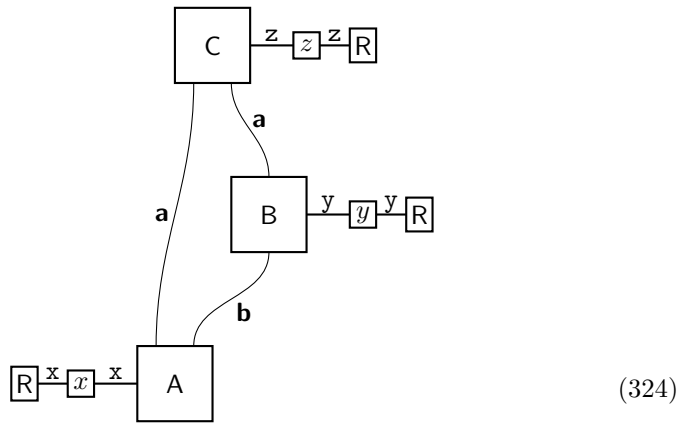
\begin{equation}\label{ABCcircuitagain}
\begin{Compose}{0}{0} \setsecondfont{\mathsfb}
\crectangle{A}{2}{2}{0,0} \csymbol{A}  \crectangle{B}{2}{2}{5,9} \csymbol{B} \crectangle{C}{2}{2}{2,18} \csymbol{C}
\jointbnoarrow[left]{A}{-1}{C}{-1} \csymbolalt[-5,0]{a}
\jointbnoarrow[below right]{A}{1}{B}{0}  \csymbolalt[5,-5]{b}
\jointbnoarrow[above right]{B}{0}{C}{1}  \csymbolalt{a}
\Rxboxincome{A}{0}{x}
\Rxboxoutcome{B}{0}{y}
\Rxboxoutcome{C}{0}{z}
\end{Compose}
\end{equation}
If, following the steps outlined in Sec.\ \ref{sec:calculatingprobabilitiesforcircuits}, we replace each operation by its expansion interms of duotensors (using  \eqref{Boperationexpansion}) and replace matched fiducial pairs with the corresponding hopping matrix, we obtain the equivalent diagram
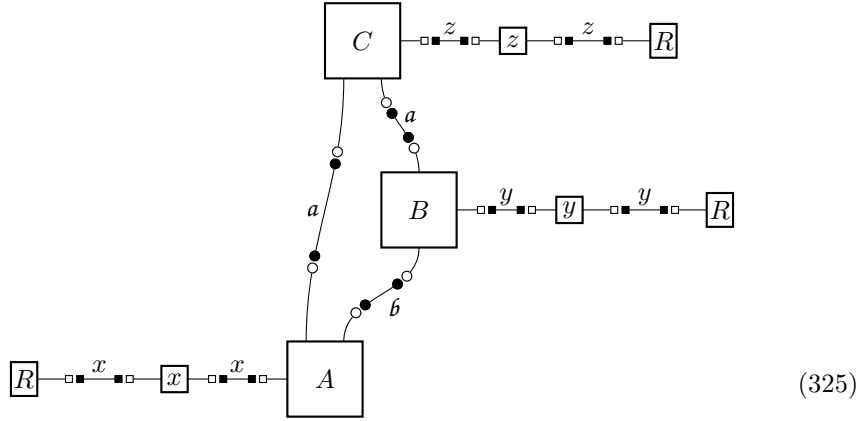
\begin{equation}\label{ABCduotensoragain}
\begin{Compose}{0}{0} \setsecondfont{\mathpzc}
\crectangle{A}{2}{2}{0,0} \csymbolfourth{A}  \crectangle{B}{2}{2}{5,9} \csymbolfourth{B} \crectangle{C}{2}{2}{2,18} \csymbolfourth{C}
\jointbnoarrowdotswbbw[left]{A}{-1}{C}{-1} \csymbolalt[-5,0]{a}
\jointbnoarrowdotswbbw[below right]{A}{1}{B}{0}  \csymbolalt[5,-5]{b}
\jointbnoarrowdotswbbw[above right]{B}{0}{C}{1}  \csymbolalt{a}
\Rxboxincomewbbw{A}{0}{x}
\Rxboxoutcomewbbw{B}{0}{y}
\Rxboxoutcomewbbw{C}{0}{z}
\end{Compose}
\end{equation}
Now, given that have equated hopping matrices between the operation theory and the operator theory as in \eqref{pointerhoppingcorrespondence} and \eqref{systemhoppingcorrespondence} the diagram in \eqref{ABCduotensoragain} is equal to
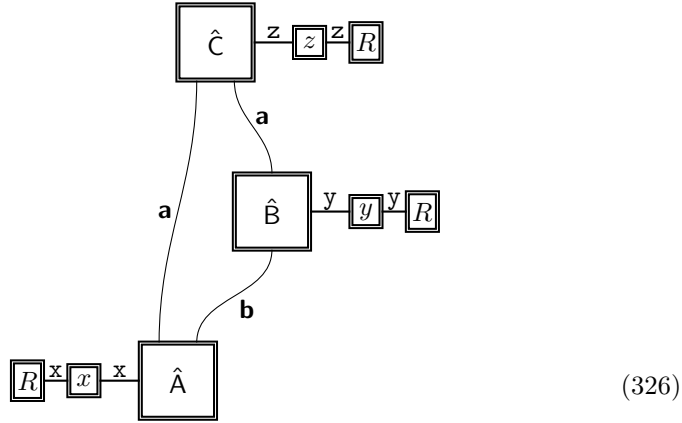
\begin{equation}
\begin{Compose}{0}{0}  \setsecondfont{\mathsfb}
\crectangledouble{A}{2}{2}{0,0} \csymbol{\hat{A}}  \crectangledouble{B}{2}{2}{5,9} \csymbol{\hat{B}} \crectangledouble{C}{2}{2}{2,18} \csymbol{\hat{C}}
\jointbnoarrow[left]{A}{-1}{C}{-1} \csymbolalt[-5,0]{a}
\jointbnoarrow[below right]{A}{1}{B}{0}  \csymbolalt[5,-5]{b}
\jointbnoarrow[above right]{B}{0}{C}{1}  \csymbolalt{a}
\Rxboxincomedouble{A}{0}{x}
\Rxboxoutcomedouble{B}{0}{y}
\Rxboxoutcomedouble{C}{0}{z}
\end{Compose}
\end{equation}
We can see this by following the steps in Sec.\ \ref{sec:operatorcircuits} backwards.   This proves that we have correspondence for this example.  It is clear that this will work for any circuit. Hence, we have proven the correspondence theorem.

To complete this subsection we remark that, if physics is described by Quantum Theory, we can choose fiducials so that the hopping matrices from the operation theory are equal to those from the operator theory.   We need $N_\mathsfb{x}^2$ linearly independent fiducials.  One choice is the following fiducial preparations
\begin{align}
&|x\rangle^{\mathsfb{x}_1}\negs\negs \langle x| ~~~~~~~~~~~\text{for} ~x=1~\text{to} ~N_\mathsfb{x}    \\
&|xx'\rangle^{\mathsfb{x}_1}\negs\negs \langle xx'|  ~~~~~~\text{for}~ x>x'\\
&|xx'i\rangle^{\mathsfb{x}_1}\negs\negs \langle xx'i|   ~~~~\text{for}~ x>x'
\end{align}
where
\begin{equation}
|xx'\rangle^{\mathsfb{x}_1} = \frac{1}{\sqrt{2}}
(|x\rangle^{\mathsfb{x}_1} + |x'\rangle^{\mathsfb{x}_1} )
\end{equation}
and
\begin{equation}
|xx'i\rangle^{\mathsfb{x}_1} = \frac{1}{\sqrt{2}}
(|x\rangle^{\mathsfb{x}_1} + i|x'\rangle^{\mathsfb{x}_1} )
\end{equation}
We know, from standard Quantum Theory, that these projectors correspond to preparations in the laboratory.  We choose similar projectors for the results (with subscripts rather than superscripts).  We have, here, used the forward gauge for $\alpha_\mathsfb{x}$.

\section{Physicality conditions}\label{sec:physicalityconditionsoperators}

In Sec.\ \ref{sec:physicalityconstraints} we saw how to impose physicality conditions. These ensured that (i) probabilities are non-negative and (ii) that the double bisumming property is satisfied.   Since we have a correspondence from operations to operators, it is a simple matter to transcribe these conditions from being conditions on operations to being conditions on operators.

\subsection{Complete positivity}\label{sec:completeoperatorpositivity}

Quantum theory satisfies double maximal purity.  Hence, every maximal preparation is pure parallel and every maximal result is pure parallel.  Hence, the tester positivity condition corresponding to \eqref{GeneralcircuitwithBwithpureEF} for operator tensors is
\begin{equation}\label{GeneralcircuitwithBwithpureEFoperators}
0~~\leq ~~
\begin{Compose}{0}{0}\setdefaultfont{}  \setsecondfont{\mathsfb}
\crectangledouble{A}{6}{2}{0,-6}  \csymbol{\hat{D}}
\crectangledouble{C}{6}{2}{0,6}  \csymbol{\hat{E}}
\crectangledouble{B}{2}{2}{5,0} \csymbol{\hat{B}}
\jointbnoarrow[left]{A}{-5}{C}{-5} \csymbolalt{h}
\jointbnoarrow[left]{A}{5}{B}{0}  \csymbolalt{a}
\jointbnoarrow[left]{B}{0}{C}{5}  \csymbolalt{b}
\RxBoxincomedouble{B}{2}{0}{x} \RxBoxoutcomedouble{B}{2}{0}{y}
\RxBoxincomedouble{A}{6}{0}{u} 
\RxBoxoutcomedouble{C}{6}{0}{v}
\end{Compose}
\end{equation}
for all maximal $\hat{D}$ and $\hat{E}$.  We will write this condition as
\begin{equation}\label{operatorpositivity}
0~\underset{T}{\leq}
\begin{Compose}{0}{0}\setdefaultfont{}\setsecondfont{\mathsfb}\setthirdfont{\mathtt}
\crectangledouble{B}{1.5}{1.5}{0,0} \csymbol{\hat{B}}
\thispoint{d}{0,-3} \csymbolalt[0,-20]{a}
\thispoint{u}{0,3}  \csymbolalt[0,20]{b}
\thispoint{l}{-3,0} \csymbolthird[-20,0]{x}
\thispoint{r}{3,0}  \csymbolthird[20,0]{y}
\jointbnoarrow{d}{0}{B}{0}
\jointbnoarrow{B}{0}{u}{0}
\joinrlnoarrowthick{l}{0}{B}{0}
\joinrlnoarrowthick{B}{0}{r}{0}
\end{Compose}
\end{equation}
where this means that $\hat{B}$ is positive with respect to application of any operator tester of the form
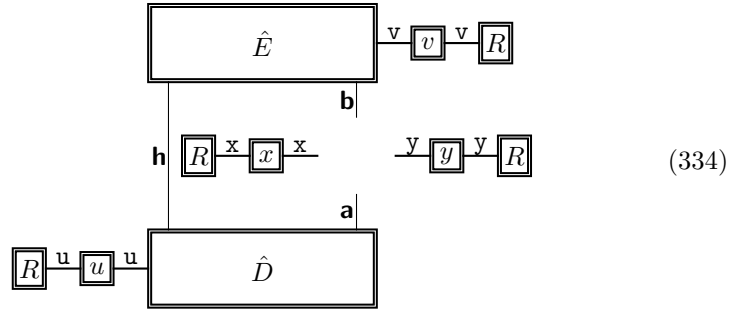
\begin{equation}\label{operatortester}
\begin{Compose}{0}{0}\setdefaultfont{} \setsecondfont{\mathsfb}
\crectangledouble{A}{6}{2}{0,-6}  \csymbol{\hat{D}}
\crectangledouble{C}{6}{2}{0,6}  \csymbol{\hat{E}}
\Crectangledouble[white]{B}{2}{2}{5,0}
\jointbnoarrow[left]{A}{-5}{C}{-5} \csymbolalt{h}
\jointbnoarrow[left]{A}{5}{B}{0}  \csymbolalt{a}
\jointbnoarrow[left]{B}{0}{C}{5}  \csymbolalt{b}
\RxBoxincomedouble{B}{2}{0}{x} \RxBoxoutcomedouble{B}{2}{0}{y}
\RxBoxincomedouble{A}{6}{0}{u} 
\RxBoxoutcomedouble{C}{6}{0}{v}
\end{Compose}
\end{equation}
for all maximal $\hat{D}$ and $\hat{E}$.  This is a different notion of positivity than the standard notion for operators.  We call this condition $T$-positivity (This is equivalent to the notion of $T$-positivity already defined in Sec.\ \ref{sec:Tpositiveoperatortensors} as will be explained below in Sec.\ \ref{sec:Tpositivityequivalenttoinputtransposepositivity}).

\subsection{$T$-positivity equivalent to input transpose positivity}\label{sec:Tpositivityequivalenttoinputtransposepositivity}

[This subsection can be skipped on a first reading.]~
We will see that $T$-positivity is equivalent to the condition that $\hat B$ is positive after taking the transpose over the input space (or equivalently, over the output space).  To explore the meaning of this, we will first absorb the $R$ and readout boxes into the main operators writing \eqref{GeneralcircuitwithBwithpureEFoperators} as
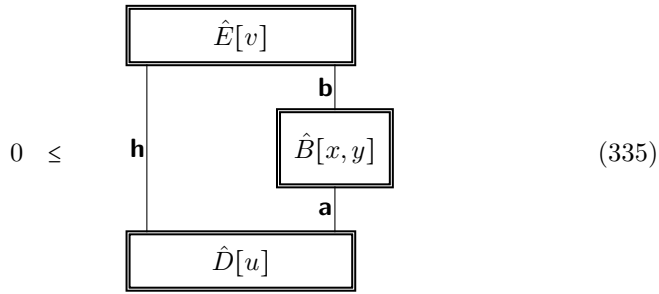
\begin{equation}\label{GeneralcircuitwithEFoperatorssimplified}
0~~\leq ~~~~~~
\begin{Compose}{0}{0}\setdefaultfont{}  \setsecondfont{\mathsfb}
\crectangledouble{A}{6}{1.5}{0,-6}  \csymbol{\hat{D}[u]}
\crectangledouble{C}{6}{1.5}{0,6}  \csymbol{\hat{E}[v]}
\crectangledouble{B}{3}{2}{5,0} \csymbol{\hat{B}[x,y]}
\jointbnoarrow[left]{A}{-5}{C}{-5} \csymbolalt{h}
\jointbnoarrow[left]{A}{5}{B}{0}  \csymbolalt{a}
\jointbnoarrow[left]{B}{0}{C}{5}  \csymbolalt{b}
\end{Compose}
\end{equation}
Now we can double this up using the notation of Sec.\ \ref{sec:hilbertspacesoperatorsanddiagrammatics}.
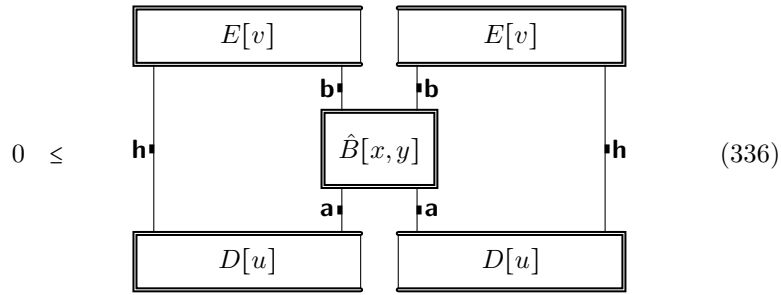
\begin{equation}
0~~\leq ~~~~~~
\begin{Compose}{0}{0}\setdefaultfont{}  \setsecondfont{\mathsfb}
\crectangledouble{B}{3}{2}{0,0} \csymbol{\hat{B}[x,y]}
\crectangledleft{A}{6}{1.5}{-7,-6}  \csymbol{{D}[u]}
\crectangledleft{C}{6}{1.5}{-7,6}  \csymbol{{E}[v]}
\jointbleft[left]{A}{-5}{C}{-5} \csymbolalt{h}
\jointbleft[left]{A}{5}{B}{-2}  \csymbolalt{a}
\jointbleft[left]{B}{-2}{C}{5}  \csymbolalt{b}
\crectangledright{A}{6}{1.5}{7,-6}  \csymbol{{D}[u]}
\crectangledright{C}{6}{1.5}{7,6}  \csymbol{{E}[v]}
\jointbright[right]{A}{5}{C}{5} \csymbolalt{h}
\jointbright[right]{A}{-5}{B}{2}  \csymbolalt{a}
\jointbright[right]{B}{2}{C}{-5}  \csymbolalt{b}
\end{Compose}
\end{equation}
We can expand the $D$ and $E$ boxes with respect to a basis for $\mathsfb h$ to give
\begin{equation}
0~~\leq ~~~~~~
\begin{Compose}{0}{0}\setdefaultfont{}  \setsecondfont{\mathsfb}
\crectangledouble{B}{3}{2}{0,0} \csymbol{\hat{B}[x,y]}
\crectangledleft{A}{2}{1.5}{-3.3,-6}  \csymbol{A[u]}
\crectangledleft{C}{2}{1.5}{-3.3,6}  \csymbol{C[v]}
\crectangledleft{nd}{0.8}{0.8}{-8,-9.5}  \joinrrnoarrow[right]{A}{0}{A}{-3.5} \thispoint{Ap}{-1.3,-9.5} \joinlrnoarrow{Ap}{0}{nd}{0}  \csymbol{n}
\crectangledleft{nu}{0.8}{0.8}{-8,9.5}   \joinrrnoarrow[right]{C}{0}{C}{3.5}  \thispoint{Cp}{-1.3,9.5}  \joinlrnoarrow{Cp}{0}{nu}{0}  \csymbol{n}
\jointbleft[left]{A}{0}{B}{-2}  \csymbolalt{a}
\jointbleft[left]{B}{-2}{C}{0}  \csymbolalt{b}
\jointbleft[left]{nd}{0}{nu}{0} \csymbolalt{h}
\crectangledright{A}{2}{1.5}{3.3,-6}  \csymbol{A[u]}
\crectangledright{C}{2}{1.5}{3.4,6}  \csymbol{C[v]}
\crectangledright{nd}{0.8}{0.8}{8,-9.5}  \joinllnoarrow[right]{A}{0}{A}{-3.5} \thispoint{Ap}{1.3,-9.5} \joinrlnoarrow{Ap}{0}{nd}{0}  \csymbol{n}
\crectangledright{nu}{0.8}{0.8}{8,9.5}   \joinllnoarrow[right]{C}{0}{C}{3.5}  \thispoint{Cp}{1.3,9.5}  \joinrlnoarrow{Cp}{0}{nu}{0}  \csymbol{n}
\jointbright[right]{A}{0}{B}{2}  \csymbolalt{a}
\jointbright[right]{B}{2}{C}{0}  \csymbolalt[5,0]{b}
\jointbright[right]{nd}{0}{nu}{0} \csymbolalt{h}
\end{Compose}
\end{equation}
using the orthogonality relations in \eqref{diagrammaticorthog}, we get
\begin{equation} \label{diagramBxywithtester}
0~~\leq ~~~~~~
\begin{Compose}{0}{0}\setdefaultfont{}  \setsecondfont{\mathsfb}
\crectangledouble{B}{3}{2}{0,0} \csymbol{\hat{B}[x,y]}
\crectangledleft{A}{2}{1.5}{-3.3,-6}  \csymbol{A[u]}
\crectangledleft{C}{2}{1.5}{-3.3,6}  \csymbol{C[v]}
\thispoint{nd}{-6,-9.5} \thispoint{ndd}{-8,-7.5} \joinrrnoarrow[right]{A}{0}{A}{-3.5} \thispoint{Ap}{-1.3,-9.5} \joinlrnoarrow{Ap}{0}{nd}{0}
\thispoint{nu}{-6,9.5} \thispoint{nuu}{-8, 7.5}  \joinrrnoarrow[right]{C}{0}{C}{3.5}  \thispoint{Cp}{-1.3,9.5}  \joinlrnoarrow{Cp}{0}{nu}{0}
\jointbleft[left]{A}{0}{B}{-2}  \csymbolalt{a}
\jointbleft[left]{B}{-2}{C}{0}  \csymbolalt{b}
\joinlbnoarrow{nd}{0}{ndd}{0} \joinltnoarrow{nu}{0}{nuu}{0}
\jointbnoarrow[left]{ndd}{0}{nuu}{0} \csymbol{n}
\crectangledright{A}{2}{1.5}{3.3,-6}  \csymbol{A[u]}
\crectangledright{C}{2}{1.5}{3.4,6}  \csymbol{C[v]}
\thispoint{nd}{6,-9.5} \thispoint{ndd}{8,-7.5}   \joinllnoarrow[right]{A}{0}{A}{-3.5} \thispoint{Ap}{1.3,-9.5} \joinrlnoarrow{Ap}{0}{nd}{0}
\thispoint{nu}{6,9.5} \thispoint{nuu}{8, 7.5}  \joinllnoarrow[right]{C}{0}{C}{3.5}  \thispoint{Cp}{1.3,9.5}  \joinrlnoarrow{Cp}{0}{nu}{0}
\jointbright[right]{A}{0}{B}{2}  \csymbolalt{a}
\jointbright[right]{B}{2}{C}{0}  \csymbolalt[5,0]{b}
\joinrbnoarrow{nd}{0}{ndd}{0} \joinrtnoarrow{nu}{0}{nuu}{0}
\jointbnoarrow[right]{ndd}{0}{nuu}{0} \csymbol{n}
\end{Compose}
\end{equation}
Thus, we are testing $\hat{B}$ with the tester
\begin{equation}\label{diagrammaticdoubletester}
\begin{Compose}{0}{0}\setdefaultfont{}  \setsecondfont{\mathsfb}
\crectangledouble[white]{B}{3}{2}{0,0}
\crectangledleft{A}{2}{1.5}{-3.3,-6}  \csymbol{A[u]}
\crectangledleft{C}{2}{1.5}{-3.3,6}  \csymbol{C[v]}
\thispoint{nd}{-6,-9.5} \thispoint{ndd}{-8,-7.5} \joinrrnoarrow[right]{A}{0}{A}{-3.5} \thispoint{Ap}{-1.3,-9.5} \joinlrnoarrow{Ap}{0}{nd}{0}
\thispoint{nu}{-6,9.5} \thispoint{nuu}{-8, 7.5}  \joinrrnoarrow[right]{C}{0}{C}{3.5}  \thispoint{Cp}{-1.3,9.5}  \joinlrnoarrow{Cp}{0}{nu}{0}
\jointbleft[left]{A}{0}{B}{-3.3}  \csymbolalt{a}
\jointbleft[left]{B}{-3.3}{C}{0}  \csymbolalt{b}
\joinlbnoarrow{nd}{0}{ndd}{0} \joinltnoarrow{nu}{0}{nuu}{0}
\jointbnoarrow[left]{ndd}{0}{nuu}{0} \csymbol{n}
\crectangledright{A}{2}{1.5}{3.3,-6}  \csymbol{A[u]}
\crectangledright{C}{2}{1.5}{3.3,6}  \csymbol{C[v]}
\thispoint{nd}{6,-9.5} \thispoint{ndd}{8,-7.5}   \joinllnoarrow[right]{A}{0}{A}{-3.5} \thispoint{Ap}{1.3,-9.5} \joinrlnoarrow{Ap}{0}{nd}{0}
\thispoint{nu}{6,9.5} \thispoint{nuu}{8, 7.5}  \joinllnoarrow[right]{C}{0}{C}{3.5}  \thispoint{Cp}{1.3,9.5}  \joinrlnoarrow{Cp}{0}{nu}{0}
\jointbright[right]{A}{0}{B}{3.3}  \csymbolalt{a}
\jointbright[right]{B}{3.3}{C}{0}  \csymbolalt[5,0]{b}
\joinrbnoarrow{nd}{0}{ndd}{0} \joinrtnoarrow{nu}{0}{nuu}{0}
\jointbnoarrow[right]{ndd}{0}{nuu}{0} \csymbol{n}
\end{Compose}
\end{equation}
The object on the left in \eqref{diagrammaticdoubletester} can represent any element of $\mathcal{H}^{\mathsfb{a}_1}\mathcal{H}_{\mathsfb{b}_2}$ (up to an over all factor) and, similarly, the object on the right can represent any element of $\bar{\mathcal{H}}^{\mathsfb{a}_1}\bar{\mathcal{H}}_{\mathsfb{b}_2}$ (up to an over all factor).  This explains equivalence to $T$-positivity as defined in Sec.\ \ref{sec:Tpositiveoperatortensors} since we can expand $\hat{B}[x,y]$ in terms of eigenvectors and then pick out each eigenvector with a corresponding tester to demonstrate that the corresponding eigenvalue must be positive.

It is useful to look at these same steps in symbolic (rather than diagrammatic) notation.  We  put
\begin{equation}
\hat{D}^\mathsfb{a_1h_3}[u] = |D[u]\rangle^\mathsfb{a_1h_3}\otimes \presup{\mathsfb{a_1h_3}\negs} \langle D[u]|
~~~~~~~~~~~~~
\hat{E}_\mathsfb{b_2h_3}[v] = |E[v]\rangle_\mathsfb{b_eh_3}\otimes \presub{\mathsfb{b_eh_3} \negs} \langle E[v]|
\end{equation}
These are proportional to projection operators. We have inserted the tensor product symbol, $\otimes$, since this will aid the exposition below.   We can write $|D[u]\rangle^\mathsfb{a_1h_3}$ as
\begin{equation}
|D[u]\rangle^\mathsfb{a_1h_3} = \sum_n  |A_n[u]\rangle^\mathsfb{a_1} \otimes |n\rangle^\mathsfb{h_3}
\end{equation}
where $|n\rangle^\mathsfb{h_3}$ are a basis.  Also, we can write $\presup{\mathsfb{b_2h_3}}\langle E|$ with respect to the basis $\presub{\mathsfb{h_3}}\langle n|$ as follows
\begin{equation}
\presub{\mathsfb{b_2h_3}}\langle E[v]| = \sum_n \presub{\mathsfb{b_2}}\langle C_n[v]|\otimes \presub{\mathsfb{h_3}}\langle n|
\end{equation}
If $n$ does not run over $N_\mathsfb{h}$ values, we can complete the basis.  Note that the relationship between $|n\rangle^\mathsfb{h_3}$ and $\presub{\mathsfb{h_3}}\langle n|$ is elucidated in Sec.\ \ref{sec:inputandoutputhilbertspaces}.
We obtain
\begin{align}
\hat{D}^\mathsfb{a_1h_3}[u] \hat{E}_\mathsfb{b_2h_3}[v] &= \presub{\mathsfb{b_2h_3}\negs}\langle E[v]|D[u]\rangle^\mathsfb{a_1h_3} ~ \presup{\mathsfb{a_1h_3}}\langle D[u]|E[v]\rangle_\mathsfb{b_2h_3}  \\
&=
\left( \sum_n   |A_n[u]\rangle^\mathsfb{a_1} \otimes \presub{\mathsfb{b_2}\negs}\langle C_n[v]| \right)
\otimes
\left( \sum_n  \presup{\mathsfb{a_1}}\langle A_n[u]| \otimes |C_n[v]\rangle_\mathsfb{b_1} \right)
\label{DEexpansion}
\end{align}
This is the symbolic version of the diagram shown in \eqref{diagrammaticdoubletester}.   It is clear that this can be equal to any rank one (Hermitian) operator in $\mathcal{V}^\mathsfb{a}\mathcal{V}_\mathsfb{b}$ up to an overall normalisation constant.   Hence, the system part of the operator tester is checking for positivity of $\hat B$ with respect to all rank one operators that can act on it.

The bra's and ket's in \eqref{DEexpansion} are mixed up from what we would usually expect.  In the usual language of operators, an operator, $\hat Q$, is positive if $\langle \psi|\hat{Q}|\psi\rangle\geq 0$.  To convert to this kind of statement, we can take the partial transpose over the input part of the Hilbert space as we will now explain.  To take the partial transpose with respect to the $\mathsfb{a}_2$ space we need to choose a basis.  We choose the $|a\rangle^\mathsfb{a_1}$ basis and write
\begin{equation}
|A_n[u]\rangle^\mathsfb{a_1} = \sum_a c_a[u] |a\rangle^\mathsfb{a_1}
\end{equation}
Substituting this into \eqref{DEexpansion} gives
\begin{equation}
\hat{D}^\mathsfb{a_1h_3}[u] \hat{E}_\mathsfb{b_2h_3}[v]
=
\left( \sum_{n'a'} c_{a'}[u]    |a'\rangle^\mathsfb{a_1} \otimes \presub{\mathsfb{b_2}}\langle C_{n'}[v]| \right)
\otimes
\left( \sum_{na} c_a^*[u]  \presup{\mathsfb{a_1}}\langle a| \otimes |C_n[v]\rangle_\mathsfb{b_1} \right)
\end{equation}
We have put primes on the dummy indices in the second parenthesis to aid discussion of taking the partial transpose.  We can now take the partial transpose in the $|a\rangle^\mathsfb{a_2}$ basis.  This means swapping elements on the diagonal in this basis, or replacing $|a\rangle_\mathsfb{a_1}\negs\langle a'|$ with $|a'\rangle_\mathsfb{a_1}\negs\langle a|$ which gives
\begin{equation}\label{DEexpansiontranspose}
\hat{D}^{\mathsfb{a_1^\mathsf{T}}\mathsfb{h_3}}[u] \hat{E}_\mathsfb{b_2h_3}[v]
=
\left( \sum_{n'a'} c_{a'}    \presub{\mathsfb{a_1}}\langle a'| \otimes \presub{\mathsfb{b_2}}\langle C_{n'}[v]| \right)
\otimes
\left( \sum_{na} c_a^*        |a\rangle_\mathsfb{a_1} \otimes |C_[v]\rangle_\mathsfb{b_1} \right)
\end{equation}
We have denoted the partial transpose in the expression on the left by a $\mathsf T$ superscript on the $\mathsfb{a}_1$.  By appropriate choice of $c_n$ and $|C_[v]\rangle_\mathsfb{b_1}$, the expression on the right can be made into any rank one projector onto $\mathcal{H}_\mathsfb{a_1}\otimes\mathcal{H}_\mathsfb{b_2}$ (up to an over all constant).

It is easy to show that if we take the partial transpose on both sides of a closed wire an expression is unchanged.  This is because the closed wire corresponds to taking the partial trace and
\begin{equation}
\text{trace}(|a\rangle_\mathsfb{a_1}\negs\langle a'| |a''\rangle^\mathsfb{a_1}\negs\langle a'''|) =
\text{trace}(|a'\rangle_\mathsfb{a_1}\negs\langle a| |a'''\rangle^\mathsfb{a_1}\negs\langle a''|)
\end{equation}
Hence,
\begin{equation}
\hat{D}^{\mathsfb{a_1}\mathsfb{h_3}}[u] \hat{E}_\mathsfb{b_2h_3}[v] \hat{B}_\mathsfb{a_1}^\mathsfb{b_2}[x,y]
=
\hat{D}^{\mathsfb{a_1^\mathsf{T}}\mathsfb{h_3}}[u] \hat{E}_\mathsfb{b_2h_3}[v] \hat{B}_\mathsfb{a_1^\mathsf{T}}^\mathsfb{b_2}[x,y]
\end{equation}
The left hand side can be read as being equal to the expression on the right of the tester positivity condition in \eqref{GeneralcircuitwithBwithpureEFoperators}.
Given that the expression on the right of equation \eqref{DEexpansiontranspose} can equal any rank one projector, we must have that
\begin{equation}
0\leq
\hat{B}_\mathsfb{a_1^\mathsf{T}}^\mathsfb{b_2}[x,y]
\end{equation}
This condition is time symmetric since we can equivalently write it as positivity under output transpose.  Also, note that, although we need to choose a particular basis to take the input (or output) transpose, the condition is independent of this choice (a different choice will yield the same result for the condition).   This follows from the fact that the condition is equivalent to that shown in \eqref{GeneralcircuitwithBwithpureEFoperators} where there is no need to choose a basis.

These manipulations elucidate the mathematical meaning of \eqref{operatorpositivity} - namely that for each given $x$ and $y$ on the income and outcome respectively, the operator must be positive under input transpose. However, there is a certain unnaturalness to taking input transposes.  All they do really is put matrix entries in a different place so we can apply the usual notion of positivity.  It is arguably more appropriate to take $T$-positivity as a more natural notion (without invoking any transposes). In support of this claim, it is worth noting that $T$-positivity is essentially the same as the notion positivity used due to Selinger \cite{selinger2007dagger} (called $\otimes$-positivity in the book of Coecke and Kissinger \cite{coecke2017picturing}).  This notion arises naturally in that work.

We can illustrate taking the input transpose diagrammatically.  The transpose operation is given by
\begin{equation}\label{transposediagrams}
\begin{Compose}{0}{0}\setdefaultfont{}  \setsecondfont{\mathsfb}
\ctransposeboxS{TZ}{0.8}{1.2}{0,0}
\thispoint{d}{0,-4}\thispoint{u}{0,4}
\jointbright[right]{d}{0}{TZ}{0}\csymbolalt{a}
\jointbleft[left]{TZ}{0}{u}{0}\csymbolalt{a}
\end{Compose}
~~:=~~
\begin{Compose}{0}{0}\setdefaultfont{} \setsecondfont{\mathsfb}
\crectangledleft{B}{0.8}{0.8}{0,1.8}
\crectangledright{A}{0.8}{0.8}{0,-1.8}
\thispoint{ml}{-0.8,0} \thispoint{mr}{0.8,0} \joinrlnoarrow[above]{ml}{0}{mr}{0}
\joinllnoarrow[right]{A}{0}{ml}{0} \joinrrnoarrow[right]{B}{0}{mr}{0}\csymbol{a}
\thispoint{d}{0,-4}\thispoint{u}{0,4}
\jointbright[left]{d}{0}{A}{0}\csymbolalt{a}
\jointbleft[right]{B}{0}{u}{0}\csymbolalt{a}
\end{Compose}
~~~~~~~~~~~~~~~
\begin{Compose}{0}{0}\setdefaultfont{}  \setsecondfont{\mathsfb}
\ctransposeboxZ{TZ}{0.8}{1.2}{0,0}
\jointbleft[left]{d}{0}{TZ}{0}\csymbolalt{a}
\jointbright[right]{TZ}{0}{u}{0}\csymbolalt{a}
\end{Compose}
~~:=~~
\begin{Compose}{0}{0}\setdefaultfont{} \setsecondfont{\mathsfb}
\crectangledleft{B}{0.8}{0.8}{0,-1.8}
\crectangledright{A}{0.8}{0.8}{0,1.8}
\thispoint{ml}{-0.8,0} \thispoint{mr}{0.8,0} \joinrlnoarrow[above]{ml}{0}{mr}{0}
\joinllnoarrow[right]{A}{0}{ml}{0} \joinrrnoarrow[right]{B}{0}{mr}{0}\csymbol{a}
\thispoint{d}{0,-4}\thispoint{u}{0,4}
\jointbleft[left]{d}{0}{B}{0}\csymbolalt{a}
\jointbright[right]{A}{0}{u}{0}\csymbolalt{a}
\end{Compose}
\end{equation}
[As an aside, note that if we rotate the lower box on the right hand side of either of these equations so it is pointing up then we obtain a maximally entangled state (though not normalised).  This shows how taking the partial transpose is related to the Choi-Jamio\l kowski isomorphism.]  Clearly, if we apply this transpose operation twice we implement the identity.  We take the partial transpose twice on the $\mathsf{a}$ wires so that \eqref{diagramBxywithtester} becomes
\begin{equation} \label{diagramBxywithtesterwithtransposes}
0~~\leq ~~~~~~
\begin{Compose}{0}{0}\setdefaultfont{}  \setsecondfont{\mathsfb}
\crectangledouble{B}{3}{2}{0,0} \csymbol{\hat{B}[x,y]}
\crectangledleft{A}{2}{1.5}{-3.3,-14}  \csymbol{A[u]}
\crectangledleft{C}{2}{1.5}{-3.3,6}  \csymbol{C[v]}
\thispoint{nd}{-6,-17.5} \thispoint{ndd}{-8,-15.5} \joinrrnoarrow[right]{A}{0}{A}{-3.5} \thispoint{Ap}{-1.3,-17.5} \joinlrnoarrow{Ap}{0}{nd}{0}
\thispoint{nu}{-6,9.5} \thispoint{nuu}{-8, 7.5}  \joinrrnoarrow[right]{C}{0}{C}{3.5}  \thispoint{Cp}{-1.3,9.5}  \joinlrnoarrow{Cp}{0}{nu}{0}
\jointbleft[left]{B}{-2}{C}{0}  \csymbolalt{b}
\joinlbnoarrow{nd}{0}{ndd}{0} \joinltnoarrow{nu}{0}{nuu}{0}
\jointbnoarrow[left]{ndd}{0}{nuu}{0} \csymbol{n}
\ctransposeboxS{Tu}{0.8}{1.2}{-2.7,-5} \ctransposeboxZ{Td}{0.8}{1.2}{-2.7,-9}
\jointbleft[left]{Tu}{0}{B}{-2}  \csymbolalt{a} \jointbright[right]{Td}{0}{Tu}{0}  \csymbolalt{a} \jointbleft[left]{A}{0}{Td}{0}  \csymbolalt{a}
\crectangledright{A}{2}{1.5}{3.3,-14}  \csymbol{A[u]}
\crectangledright{C}{2}{1.5}{3.4,6}  \csymbol{C[v]}
\thispoint{nd}{6,-17.5} \thispoint{ndd}{8,-15.5}   \joinllnoarrow[right]{A}{0}{A}{-3.5} \thispoint{Ap}{1.3,-17.5} \joinrlnoarrow{Ap}{0}{nd}{0}
\thispoint{nu}{6,9.5} \thispoint{nuu}{8, 7.5}  \joinllnoarrow[right]{C}{0}{C}{3.5}  \thispoint{Cp}{1.3,9.5}  \joinrlnoarrow{Cp}{0}{nu}{0}
\jointbright[right]{B}{2}{C}{0}  \csymbolalt[5,0]{b}
\joinrbnoarrow{nd}{0}{ndd}{0} \joinrtnoarrow{nu}{0}{nuu}{0}
\jointbnoarrow[right]{ndd}{0}{nuu}{0} \csymbol{n}
\ctransposeboxZ{Tu}{0.8}{1.2}{2.7,-5} \ctransposeboxS{Td}{0.8}{1.2}{2.7,-9}
\jointbright[right]{Tu}{0}{B}{2}  \csymbolalt{a} \jointbleft[left]{Td}{0}{Tu}{0}  \csymbolalt{a} \jointbright[right]{A}{0}{Td}{0}  \csymbolalt{a}
\end{Compose}
\end{equation}
We can read this diagram as stating that the input transpose of $\hat{B}_{\mathsfb{a}_1}^{\mathsfb{b}_2}[x,y]$ is positive in the usual sense of positivity for operators. This is because the transpose boxes attached to the $A$'s flips ket to bra and vice versa so the tester with these transpose boxes included has only kets on the left and bras on the right.

Any operator circuit in which all operators satisfy tester positivity will be equal to a non-negative number (essential if it is to be a probability).  One way to see this is to use the correspondence theorem and convert operator circuits into operation circuits and use the complete positivity theorem in Sec.\ \ref{sec:completepositivity}.  Alternatively, we can see directly that it is true by showing that, if we combine operators that satisfy the condition then we get a new operator that satisfies the condition.   The proof of this mirrors the proof in Sec.\ \ref{sec:completepositivity} so we will not repeat it here.

\subsection{Double bisummation from double causality}

In Sec.\ \ref{sec:doublebisummationfromdoublecausality} we proved the double causality theorem - that the given double causality conditions imply double bisummation.  It follow from that theorem and the correspondence theorem that, if operators
\begin{equation}
\begin{Compose}{0}{0}\setsecondfont{\mathtt} \setthirdfont{\mathsfb}
\crectangledouble{U}{2}{2}{0,0} \csymbol{B} \thispoint{AL}{-4,0} \csymbolalt[-20,0]{x} \thispoint{AR}{4,0} \csymbolalt[20,0]{y}
\joinlrnoarrowthick{U}{0}{AL}{0} \joinrlnoarrowthick{U}{0}{AR}{0}
\thispoint{a}{0,-4} \csymbolthird[0,-20]{a} \jointbnoarrow{a}{0}{U}{0}
\thispoint{d}{0,4} \csymbolthird[0,20]{b} \joinbtnoarrow{d}{0}{U}{0}
\end{Compose}
\end{equation}
satisfy the following two conditions
\begin{equation}\label{forwardunitaloperator}
\begin{Compose}{0}{0}\setdefaultfont{\hat}\setsecondfont{\mathtt}\setthirdfont{\mathsfb}
\crectangledouble{C}{2}{2}{0,0} \csymbol{B}
\crectangledouble{CL}{0.9}{0.9}{-4.5,0}\csymbolfourth{R} \joinlrnoarrowthick{C}{0}{CL}{0} \csymbolalt{x}
\thispoint{CR}{4,0} \csymbolalt[20,0]{y} \joinrlnoarrowthick{C}{0}{CR}{0}
\crectangledouble{A}{0.9}{0.9}{0,-5} \csymbol{I} \jointbnoarrow[left]{A}{0}{C}{0} \csymbolthird{a} 
\thispoint{B}{0,4}\csymbolthird[0,20]{b} \jointbnoarrow[left]{C}{0}{B}{0}
\end{Compose}
~\equiv ~
\begin{Compose}{0}{0}\setdefaultfont{\hat}\setsecondfont{\mathtt}\setthirdfont{\mathsfb}
\crectangledouble{B}{0.9}{0.9}{-0.3,0} \csymbol{I} \thispoint{BR}{-0.3,4} \csymbolthird[20,0]{b} \jointbnoarrow{B}{0}{BR}{0}
\crectangledouble{BD}{0.9}{0.9}{2,0} \csymbolfourth{R} \thispoint{BRD}{4,0} \csymbolalt[20,0]{y} \joinrlnoarrowthick{BD}{0}{BRD}{0}
\end{Compose}
\end{equation}
and
\begin{equation}\label{backwardunitaloperator}
\begin{Compose}{0}{0}\setdefaultfont{\hat}\setsecondfont{\mathtt}\setthirdfont{\mathsfb}
\crectangledouble{C}{2}{2}{0,0} \csymbol{B}
\thispoint{CL}{-4,0} \csymbolalt[-20,0]{x} \joinlrnoarrowthick{C}{0}{CL}{0}
\crectangledouble{CR}{0.9}{0.9}{4.5,0} \csymbolfourth{R} \joinrlnoarrowthick{C}{0}{CR}{0} \csymbolalt[0,5]{y}
\thispoint{A}{0,-4} \csymbolthird[0,-20]{a} \jointbnoarrow{A}{0}{C}{0}
\crectangledouble{B}{0.9}{0.9}{0,5} \csymbol{I} \jointbnoarrow[left]{C}{0}{B}{0}\csymbolthird{b}
\end{Compose}
~\equiv ~
\begin{Compose}{0}{0}\setdefaultfont{\hat}\setsecondfont{\mathtt}\setthirdfont{\mathsfb}
\crectangledouble{B}{0.9}{0.9}{0.3,0} \csymbol{I} \thispoint{BR}{0.3,-4} \csymbolthird[0,-20]{a} \joinbtnoarrow{B}{0}{BR}{0}
\crectangledouble{BD}{0.9}{0.9}{-2,0} \csymbolfourth{R} \thispoint{BRD}{-4,0} \csymbolalt[-20,0]{x} \joinlrnoarrowthick{BD}{0}{BRD}{0}
\end{Compose}
\end{equation}
then the double summation conditions hold for operator circuits.   Rather than relying on correspondence, we can prove this directly by a proof with the same steps as the proof in Sec.\ \ref{sec:doublebisummationfromdoublecausality}.   Many versions of the double causality conditions for operators can be obtained by treating some of the types as null (i.e.\ let the associated Hilbert space have dimension 1 for systems and let the associated vector space have dimension 1 for pointer types).  Some of these versions are shown in Table \ref{table:causalityoperators} (this is the operator version of Table \ref{table:causality}).
\begin{table}
\begin{center}
{\tabulinesep=1mm
\begin{tabu}{|c|c|c|}
  \hline
  1 & $\Funitaldoublexyab[0.2]{1}{1}{1}{1}$ &  $\Bunitaldoubleyxba[-0.2]{1}{1}{1}{1}$   \\ \hline
  2 & $\Funitaldoublexyab[-0.4]{1}{1}{0}{1}$ &  $\Bunitaldoubleyxba[0.4]{1}{1}{0}{1}$   \\ \hline
  3 & $\Funitaldoublexyab[0.4]{1}{1}{1}{0}$ &  $\Bunitaldoubleyxba[-0.4]{1}{1}{1}{0}$   \\ \hline
  4 & $\Funitaldoublexyab{0}{1}{1}{1}$ &  $\Bunitaldoubleyxba{0}{1}{1}{1}$   \\ \hline
  5 & $\Funitaldoublexyab[0.2]{1}{0}{1}{1}$ &  $\Bunitaldoubleyxba[-0.2]{1}{0}{1}{1}$   \\ \hline
  6 & $\Funitaldoublexyab[0.2]{0}{0}{1}{1}$ &  $\Bunitaldoubleyxba[-0.2]{0}{0}{1}{1}$   \\ \hline
  7 & $\Funitaldoublexyab{1}{1}{0}{0}$ &  $\Bunitaldoubleyxba{1}{1}{0}{0}$   \\ \hline
  8 & $\Funitaldoublexyab[-0.4]{0}{0}{0}{1}$ &  $\Bunitaldoubleyxba[0.4]{0}{0}{0}{1}$   \\ \hline
  9 & $\Funitaldoublexyab{0}{1}{0}{0}$ &  $\Bunitaldoubleyxba{0}{1}{0}{0}$   \\ \hline
\end{tabu} }
\end{center}
\caption{This tabulates many of the different versions of forward causality (left column) and backward causality (right column) for operators. }
\label{table:causalityoperators}
\end{table}
Two more applications of the double causality properties (not explicitly shown in the table) are
\begin{equation}\label{operatornormalisation}
\Funitaldoublexyab[0.4]{0}{0}{1}{0} 1  ~~~~~~~~~~~~~~~~  \Bunitaldoubleyxba[-0.6]{0}{0}{1}{0}  1
\end{equation}
The 1 on the right hand side of these equations follows since a box with no wires in or out is equal to 1.  The equation on the right is the normalisation condition for states.   In standard Quantum Theory the normalisation condition is $\text{trace}(\rho)=\text{trace}(\hat{\mathds{1}}\rho)=1$.  However, this form for the normalisation condition assumes we have adopted the forward gauge, wherein $\alpha_\mathsfb{a}\sqrt{N_\mathsfb{a}}=1$ for all system types (see Sec.\ \ref{sec:chasingdowntheoneoverN}).  In the forward gauge we represent the ignore result, $\hat{I}_{\mathsfb{b}_2}$ by $\hat{\mathds{1}}_\mathsfb{b_2}$ and the ignore preparation, $\hat{I}^{\mathsfb{a}_1}$ by $\hat{\mathds{1}}^{\mathsfb{a}_1}/N_{\mathsfb{a}_1}$.  However, if we use a different gauge then the normalisation will be different.  For example, in the time symmetric gauge, we represent $\hat{I}_{\mathsfb{b}_2}$ by $\hat{\mathds{1}}_{\mathsfb{b}_2}/\sqrt{N_\mathsfb{b}}$.  In this gauge, the equation on the right of \eqref{operatornormalisation} becomes $\hat{\mathbb{B}}^{\mathsfb{b}_2} \hat{\mathds{1}}_{\mathsfb{b}_2}/\sqrt{N_\mathsfb{b}} = 1$.  These considerations mean that the sum of the entries along the diagonal of $\hat{\mathbb{B}}^{\mathsfb{b}_2}$ must equal $\sqrt{N_\mathsfb{b}}$.  In fact, from row 8 of Table \ref{table:causalityoperators} we know that $\hat{\mathsfb{B}}^{\mathsfb{a}_1}=\hat{\mathsf{I}}^{\mathsfb{a}_1}$.  In the time symmetric gauge this means that $\hat{\mathbb{B}}^{\mathsfb{a}_1}=\hat{\mathds{1}}^{\mathsfb{a}_1}/\sqrt{N_\mathsfb{b}}$.  This does, indeed, have the property that the diagonal entries sum to $\sqrt{N_\mathsfb{b}}$.    These considerations reveal that there is an extra layer of representation when we go from abstract operator tensors to explicit representation of these operator tensors wherein we must specify $\alpha_\mathsfb{a}$.

\section{Mathematical axiom for Quantum Theory}\label{sec:mathematicalaxiomforquantumtheory}

We say that an operator is physical if (i) it is $T$-positive and (ii) it satisfies double causality.  We can state the following simple mathematical axiom for Time Symmetric Operational Quantum Theory.
\begin{quote}
\textbf{Quantum Axiom.}
Every physical operator has an operation corresponding to it.
\end{quote}
That's it.  To support the reasonableness of this axiom we will provide an extension theorem (basically this the analogue of Stinespring's theorem adjusted to apply to this time symmetric situation) and also show this is equivalent to standard (time asymmetric) operational quantum theory.

Note that this is not a reconstruction in terms of reasonable axioms in the spirit of \cite{hardy2001quantum, clifton2003characterizing, d2008probabilistic, wilce2009four, rau2009quantum, rau2010measurement, goyal2008information, dakic2009quantum, masanes2010derivation, goyal2010origin, helland2009steps, fuchs2010quantum, fivel2010derivation, chiribella2010informational, selby2018reconstructing}.  This axiom needs, up front, Hilbert spaces and the usual mathematical machinery of Quantum Theory.

\section{Extension theorem}\label{sec:extensiontheorem}

We will prove the following
\begin{quote}
\textbf{Extension theorem.}  All physical operators can be written in the (extended) form
\begin{equation}\label{extensiontheoremequation}
\begin{Compose}{0}{0}\setdefaultfont{} \setsecondfont{\mathtt} \setthirdfont{\mathsfb}
\crectangledouble{U}{2}{2}{0,0} \csymbol{B} \thispoint{AL}{-4,0} \csymbolalt[-20,0]{x} \thispoint{AR}{4,0} \csymbolalt[20,0]{y}
\joinlrnoarrowthick{U}{0}{AL}{0} \joinrlnoarrowthick{U}{0}{AR}{0}
\thispoint{a}{0,-4} \csymbolthird[0,-20]{a} \jointbnoarrow{a}{0}{U}{0}
\thispoint{d}{0,4} \csymbolthird[0,20]{b} \joinbtnoarrow{d}{0}{U}{0}
\end{Compose}
~~=~~
\begin{Compose}{0}{0} \setdefaultfont{}\setsecondfont{\mathsfb}\setthirdfont{\mathtt}
\crectangledouble{U}{3}{2}{0,0} \csymbol{\hat{U}_\mathsf{B}}
\relpoint{U}{-0.3,4.5}{Iupos}  \crectangledouble{Iu}{0.9}{0.9}{Iupos} \csymbol{\hat{I}}  \jointbnoarrow[left]{U}{-0.3}{Iu}{0}  \csymbolalt{z}
\relpoint{U}{0.3,-4.5}{Idpos}  \crectangledouble{Id}{0.9}{0.9}{Idpos} \csymbol{\hat{I}} \jointbnoarrow[left]{Id}{0}{U}{0.3} \csymbolalt[-3,0]{w}
\relpoint{U}{2.3,4.5}{Ypos}  \crectangledouble{Y}{0.9}{0.9}{Ypos} \csymbol{\hat{Y}}  \jointbnoarrow[left]{U}{2.3}{Y}{0}  \csymbolalt{y}
\relpoint{Y}{4,0}{r} \joinlrnoarrowthick[above]{r}{0}{Y}{0} \csymbolthird[0,3]{y}
\relpoint{U}{-2.3,-4.5}{Xpos}  \crectangledouble{X}{0.9}{0.9}{Xpos} \csymbol{\hat{X}}  \jointbnoarrow[left]{X}{0}{U}{-2.3} \csymbolalt{x} \relpoint{X}{-4,0}{l} \joinrlnoarrowthick[above]{l}{0}{X}{0} \csymbolthird{x}
\thispoint{down}{0,-7} \thispoint{up}{0,7}
\jointbnoarrow[right]{down}{2.3}{U}{2.3} \csymbolalt{a}
\jointbnoarrow[left]{U}{-2.3}{up}{-2.3} \csymbolalt{b}
\end{Compose}
\end{equation}
where $\hat{U}$ is unitary, and $\hat{X}$ and $\hat{Y}$ are maximal.
\end{quote}
The proof of this theorem is as follows.  First, we note that we can write
\begin{equation}
\begin{Compose}{0}{0}\setdefaultfont{\hat}\setsecondfont{\mathtt} \setthirdfont{\mathsfb}
\crectangledouble{U}{2}{2}{0,0} \csymbol{B} \thispoint{AL}{-4,0} \csymbolalt[-20,0]{x} \thispoint{AR}{4,0} \csymbolalt[20,0]{y}
\joinlrnoarrowthick{U}{0}{AL}{0} \joinrlnoarrowthick{U}{0}{AR}{0}
\thispoint{a}{0,-4} \csymbolthird[0,-20]{a} \jointbnoarrow{a}{0}{U}{0}
\thispoint{d}{0,4} \csymbolthird[0,20]{b} \joinbtnoarrow{d}{0}{U}{0}
\end{Compose}
~~=~~
\begin{Compose}{0}{0} \setdefaultfont{\hat} \setsecondfont{\mathtt} \setthirdfont{\mathsfb}
\crectangledouble{B}{2}{2}{0,0} \csymbol{B}
\crectangledouble{X}{0.9}{0.9}{-6,0} \csymbol{X} \crectangledouble{Y}{0.9}{0.9}{6,0} \csymbol{Y}
\crectangledouble{XD}{0.9}{0.9}{-6,-6} \csymbol{X} \crectangledouble{YU}{0.9}{0.9}{6,6} \csymbol{Y}
\relpoint{XD}{-3,0}{l} \relpoint{YU}{3,0}{r}
\joinrlnoarrowthick[above]{X}{0}{B}{0} \csymbolalt{x}  \joinrlnoarrowthick[above]{B}{0}{Y}{0} \csymbolalt{y}
\jointbnoarrow[left]{XD}{0}{X}{0} \csymbolthird{x}  \jointbnoarrow[left]{Y}{0}{YU}{0} \csymbolthird{y}
\joinrlnoarrowthick[above]{l}{0}{XD}{0} \csymbolalt{x}  \joinrlnoarrowthick[above]{YU}{0}{r}{0} \csymbolalt{y}
\thispoint{Bd}{0,-7} \thispoint{Bu}{0,7}
\jointbnoarrow[left]{Bd}{0}{B}{0} \csymbolthird{a}  \jointbnoarrow[left]{B}{0}{Bu}{0} \csymbolthird{b}
\end{Compose}
~~=~~
\begin{Compose}{0}{0} \setdefaultfont{\hat} \setsecondfont{\mathtt} \setthirdfont{\mathsfb}
\crectangledmark[-1]{B}{3}{2}{0,0} \csymbol{V}
\crectangledouble{XD}{0.9}{0.9}{-2,-6} \csymbol{X} \crectangledouble{YU}{0.9}{0.9}{2,6} \csymbol{Y}
\relpoint{XD}{-3,0}{l} \relpoint{YU}{3,0}{r}
\jointbnoarrow[left]{XD}{0}{B}{-2} \csymbolthird{x}  \jointbnoarrow[right]{B}{2}{YU}{0} \csymbolthird{y}
\joinrlnoarrowthick[above]{l}{0}{XD}{0} \csymbolalt{x}  \joinrlnoarrowthick[above]{YU}{0}{r}{0} \csymbolalt{y}
\thispoint{Bd}{0,-7} \thispoint{Bu}{0,7}
\jointbnoarrow[right]{Bd}{0}{B}{0} \csymbolthird{a}  \jointbnoarrow[left]{B}{0}{Bu}{0} \csymbolthird{b}
\end{Compose}
\end{equation}
using double maximality.  Now look at $\hat{V}$. Physicality requires $T$-positivity. Hence, following the steps in Sec.\ \ref{sec:Tpositiveoperatortensors},  we can write
\begin{equation}
\begin{Compose}{0}{0} \setdefaultfont{\hat} \setsecondfont{\mathtt} \setthirdfont{\mathsfb}
\crectangledmark[-1]{B}{3}{2.5}{0,0} \csymbol{V}
\thispoint{Bd}{0,-7} \thispoint{Bu}{0,7}
\jointbnoarrow[right]{Bd}{0}{B}{0} \csymbolthird{a}  \jointbnoarrow[left]{B}{0}{Bu}{0} \csymbolthird{b}
\jointbnoarrow[right]{Bd}{-2}{B}{-2} \csymbolthird{x}  \jointbnoarrow[left]{B}{2}{Bu}{2} \csymbolthird{y}
\end{Compose}
~~=~~
\begin{Compose}{0}{0}\setdefaultfont{} \setsecondfont{\mathsfb}
\crectangledleft[-1]{A}{1.5}{2.5}{-4,0} \csymbol{V}
\relpoint{A}{0,-7}{Ad} \jointbleft[left]{Ad}{0}{A}{0} \csymbolalt[-8,0]{xa}
\relpoint{A}{0,7}{Cd} \jointbleft[left]{A}{0}{Cd}{0} \csymbolalt[-8,0]{by}
\crectangledright[-1]{AA}{1.5}{2.5}{4,0} \csymbol{V}
\relpoint{AA}{0,-7}{Ad} \jointbright[right]{Ad}{0}{AA}{0} \csymbolalt[8,0]{xa}
\relpoint{AA}{0,7}{Cd} \jointbright[right]{AA}{0}{Cd}{0} \csymbolalt[8,0]{y}
%
\joinrlnoarrow[above]{A}{1.7}{AA}{1.7}\csymbol{xa}
\joinrlnoarrow[above]{A}{-1.7}{AA}{-1.7}\csymbol[0,3]{by}
\end{Compose}
~~=~~
\begin{Compose}{0}{0}\setdefaultfont{} \setsecondfont{\mathsfb}
\crectangledouble{Iup}{1.5}{1.5}{0,8}\csymbolthird{\hat{\mathds{1}}} \crectangledouble{Idown}{1.5}{1.5}{0,-8} \csymbolthird{\hat{\mathds{1}}}
\crectangledleft[-1]{A}{1.5}{2.5}{-6,0} \csymbol{V}
\relpoint{A}{0,-7}{Ad} \jointbleft[left]{Ad}{0}{A}{0} \csymbolalt[-8,0]{xa}
\relpoint{A}{0,7}{Cd} \jointbleft[left]{A}{0}{Cd}{0} \csymbolalt[-8,0]{by}
%
\crectangledleft{XA}{0.8}{0.8}{-2,3} \crectangledleft{BY}{0.8}{0.8}{-2,-3}
\relpoint{XA}{0.8,-1.8}{XAA} \joinrrnoarrow{XA}{0}{XAA}{0}
\joinrlnoarrow[below]{A}{1.2}{XAA}{0} \csymbol{xa}
\relpoint{BY}{0.8,1.8}{BYY} \joinrrnoarrow{BY}{0}{BYY}{0}
\joinrlnoarrow[above]{A}{-1.2}{BYY}{0} \csymbol{by}
\jointbleft[left]{XA}{0}{Iup}{-1} \csymbolalt[-8,6]{xa} \jointbleft[left]{Idown}{-1}{BY}{0} \csymbolalt[-8,-6]{by}
\crectangledright[-1]{AA}{1.5}{2.5}{6,0} \csymbol{V}
\relpoint{AA}{0,-7}{Ad} \jointbright[right]{Ad}{0}{AA}{0} \csymbolalt[8,0]{xa}
\relpoint{AA}{0,7}{Cd} \jointbright[right]{AA}{0}{Cd}{0} \csymbolalt[8,0]{by}
%
\crectangledright{XA}{0.8}{0.8}{2,3} \crectangledleft{BY}{0.8}{0.8}{2,-3}
\relpoint{XA}{-0.8,-1.8}{XAA} \joinllnoarrow{XA}{0}{XAA}{0}
\joinlrnoarrow[below]{AA}{1.2}{XAA}{0} \csymbol{xa}
\relpoint{BY}{-0.8,1.8}{BYY} \joinllnoarrow{BY}{0}{BYY}{0}
\joinlrnoarrow[above]{AA}{-1.2}{BYY}{0} \csymbol{by}
\jointbright[right]{XA}{0}{Iup}{1} \csymbolalt[8,6]{xa} \jointbright[right]{Idown}{1}{BY}{0} \csymbolalt[8,-6]{by}
\end{Compose}
\end{equation}
We have grouped $x$ and $a$ together into a single wire so the diagram does not become too messy (and similarly with $b$ and $y$).  We have included the small black squares so we can track taking the adjoint in the following.   Using the relationships between $\hat{I}$ and $\mathds{1}$ in \eqref{identityalphaNstuff}, we obtain
\begin{equation}
\begin{Compose}{0}{0} \setdefaultfont{\hat} \setsecondfont{\mathtt} \setthirdfont{\mathsfb}
\crectangledmark[-1]{B}{3}{2.5}{0,0} \csymbol{V}
\thispoint{Bd}{0,-7} \thispoint{Bu}{0,7}
\jointbnoarrow[right]{Bd}{0}{B}{0} \csymbolthird{a}  \jointbnoarrow[left]{B}{0}{Bu}{0} \csymbolthird{b}
\jointbnoarrow[right]{Bd}{-2}{B}{-2} \csymbolthird{x}  \jointbnoarrow[left]{B}{2}{Bu}{2} \csymbolthird{y}
\end{Compose}
~~=~~
\begin{Compose}{0}{0}\setdefaultfont{} \setsecondfont{\mathsfb}
\crectangle[thin]{coef}{3.2}{2}{-12,0} \csymbol{\frac{\alpha_\mathsfb{xa}\sqrt{N_\mathsfb{xa}N_\mathsfb{by}}}{\alpha_\mathsfb{by}} }
\crectangledouble{Iup}{1.5}{1.5}{0,8}\csymbolthird{\hat{I}} \crectangledouble{Idown}{1.5}{1.5}{0,-8} \csymbolthird{\hat{I}}
\crectangledleft[-1]{A}{1.5}{2.5}{-6,0} \csymbol{V}
\relpoint{A}{0,-7}{Ad} \jointbleft[left]{Ad}{0}{A}{0} \csymbolalt[-8,0]{xa}
\relpoint{A}{0,7}{Cd} \jointbleft[left]{A}{0}{Cd}{0} \csymbolalt[-8,0]{by}
%
\crectangledleft{XA}{0.8}{0.8}{-2,3} \crectangledleft{BY}{0.8}{0.8}{-2,-3}
\relpoint{XA}{0.8,-1.8}{XAA} \joinrrnoarrow{XA}{0}{XAA}{0}
\joinrlnoarrow[below]{A}{1.2}{XAA}{0} \csymbol{xa}
\relpoint{BY}{0.8,1.8}{BYY} \joinrrnoarrow{BY}{0}{BYY}{0}
\joinrlnoarrow[above]{A}{-1.2}{BYY}{0} \csymbol{by}
\jointbleft[left]{XA}{0}{Iup}{-1} \csymbolalt[-8,6]{xa} \jointbleft[left]{Idown}{-1}{BY}{0} \csymbolalt[-8,-6]{by}
\crectangledright[-1]{AA}{1.5}{2.5}{6,0} \csymbol{V}
\relpoint{AA}{0,-7}{Ad} \jointbright[right]{Ad}{0}{AA}{0} \csymbolalt[8,0]{xa}
\relpoint{AA}{0,7}{Cd} \jointbright[right]{AA}{0}{Cd}{0} \csymbolalt[8,0]{by}
%
\crectangledright{XA}{0.8}{0.8}{2,3} \crectangledleft{BY}{0.8}{0.8}{2,-3}
\relpoint{XA}{-0.8,-1.8}{XAA} \joinllnoarrow{XA}{0}{XAA}{0}
\joinlrnoarrow[below]{AA}{1.2}{XAA}{0} \csymbol{xa}
\relpoint{BY}{-0.8,1.8}{BYY} \joinllnoarrow{BY}{0}{BYY}{0}
\joinlrnoarrow[above]{AA}{-1.2}{BYY}{0} \csymbol{by}
\jointbright[right]{XA}{0}{Iup}{1} \csymbolalt[8,6]{xa} \jointbright[right]{Idown}{1}{BY}{0} \csymbolalt[8,-6]{by}
\end{Compose}
\end{equation}
where the expression in the box is a multiplicative factor.   Hence, if we can show that
\begin{equation}\label{candidateunitary}
\begin{Compose}{0}{0}\setdefaultfont{} \setsecondfont{\mathsfb}
\crectangle[thin]{coef}{3.2}{2}{-12,0} \csymbol{\frac{\alpha_\mathsfb{xa}\sqrt{N_\mathsfb{xa}N_\mathsfb{by}}}{\alpha_\mathsfb{by}} }\thispoint{Iup}{0,7} \thispoint{Idown}{0,-7}
\crectangledleft[-1]{A}{1.5}{2.5}{-6,0} \csymbol{V}
\relpoint{A}{0,-7}{Ad} \jointbleft[left]{Ad}{0}{A}{0} \csymbolalt[-8,0]{xa}
\relpoint{A}{0,7}{Cd} \jointbleft[left]{A}{0}{Cd}{0} \csymbolalt[-8,0]{by}
%
\crectangledleft{XA}{0.8}{0.8}{-2,3} \crectangledleft{BY}{0.8}{0.8}{-2,-3}
\relpoint{XA}{0.8,-1.8}{XAA} \joinrrnoarrow{XA}{0}{XAA}{0}
\joinrlnoarrow[below]{A}{1.2}{XAA}{0} \csymbol{xa}
\relpoint{BY}{0.8,1.8}{BYY} \joinrrnoarrow{BY}{0}{BYY}{0}
\joinrlnoarrow[above]{A}{-1.2}{BYY}{0} \csymbol{by}
\jointbleft[left]{XA}{0}{Iup}{-2} \csymbolalt[-8,6]{xa} \jointbleft[left]{Idown}{-2}{BY}{0} \csymbolalt[-8,-6]{by}
\crectangledright[-1]{AA}{1.5}{2.5}{6,0} \csymbol{V}
\relpoint{AA}{0,-7}{Ad} \jointbright[right]{Ad}{0}{AA}{0} \csymbolalt[8,0]{xa}
\relpoint{AA}{0,7}{Cd} \jointbright[right]{AA}{0}{Cd}{0} \csymbolalt[8,0]{by}
%
\crectangledright{XA}{0.8}{0.8}{2,3} \crectangledleft{BY}{0.8}{0.8}{2,-3}
\relpoint{XA}{-0.8,-1.8}{XAA} \joinllnoarrow{XA}{0}{XAA}{0}
\joinlrnoarrow[below]{AA}{1.2}{XAA}{0} \csymbol{xa}
\relpoint{BY}{-0.8,1.8}{BYY} \joinllnoarrow{BY}{0}{BYY}{0}
\joinlrnoarrow[above]{AA}{-1.2}{BYY}{0} \csymbol{by}
\jointbright[right]{XA}{0}{Iup}{2} \csymbolalt[8,6]{xa} \jointbright[right]{Idown}{2}{BY}{0} \csymbolalt[8,-6]{by}
\end{Compose}
\end{equation}
is a unitary operator tensor then we have shown we can write $\hat{B}$ in the form in \eqref{extensiontheoremequation}. To prove this unitary property we use double causality (as required by physicality).   Applying backward causality to \eqref{candidateunitary} gives
\begin{equation}
\begin{Compose}{0}{0}\setdefaultfont{} \setsecondfont{\mathsfb}
\crectangle[thin]{coef}{3.2}{2}{-12,0} \csymbol{\frac{\alpha_\mathsfb{xa}\sqrt{N_\mathsfb{xa}N_\mathsfb{by}}}{\alpha_\mathsfb{by}} } \crectangledouble{Iup}{1.5}{1.5}{0,8}\csymbolthird{\hat{I}}\crectangledouble{Iupper}{1.5}{1.5}{0,14}\csymbolthird{\hat{I}}  \thispoint{Idown}{0,-7}
\crectangledleft[-1]{A}{1.5}{2.5}{-6,0} \csymbol{V}
\relpoint{A}{0,-7}{Ad} \jointbleft[left]{Ad}{0}{A}{0} \csymbolalt[-8,0]{xa}
\relpoint{A}{0,7}{Cd} \jointbleft[left]{A}{0}{Iupper}{-1} \csymbolalt[-8,0]{by}
%
\crectangledleft{XA}{0.8}{0.8}{-2,3} \crectangledleft{BY}{0.8}{0.8}{-2,-3}
\relpoint{XA}{0.8,-1.8}{XAA} \joinrrnoarrow{XA}{0}{XAA}{0}
\joinrlnoarrow[below]{A}{1.2}{XAA}{0} \csymbol{xa}
\relpoint{BY}{0.8,1.8}{BYY} \joinrrnoarrow{BY}{0}{BYY}{0}
\joinrlnoarrow[above]{A}{-1.2}{BYY}{0} \csymbol{by}
\jointbleft[left]{XA}{0}{Iup}{-1} \csymbolalt[-8,6]{xa} \jointbleft[left]{Idown}{-2}{BY}{0} \csymbolalt[-8,-6]{by}
\crectangledright[-1]{AA}{1.5}{2.5}{6,0} \csymbol{V}
\relpoint{AA}{0,-7}{Ad} \jointbright[right]{Ad}{0}{AA}{0} \csymbolalt[8,0]{xa}
\relpoint{AA}{0,7}{Cd} \jointbright[right]{AA}{0}{Iupper}{1} \csymbolalt[8,0]{by}
%
\crectangledright{XA}{0.8}{0.8}{2,3} \crectangledleft{BY}{0.8}{0.8}{2,-3}
\relpoint{XA}{-0.8,-1.8}{XAA} \joinllnoarrow{XA}{0}{XAA}{0}
\joinlrnoarrow[below]{AA}{1.2}{XAA}{0} \csymbol{xa}
\relpoint{BY}{-0.8,1.8}{BYY} \joinllnoarrow{BY}{0}{BYY}{0}
\joinlrnoarrow[above]{AA}{-1.2}{BYY}{0} \csymbol{by}
\jointbright[right]{XA}{0}{Iup}{1} \csymbolalt[8,6]{xa} \jointbright[right]{Idown}{2}{BY}{0} \csymbolalt[8,-6]{by}
\end{Compose}
~~=~~
\begin{Compose}{0}{0} \setsecondfont{\mathsfb}
\crectangledouble{Iup}{1.5}{1.5}{0,0}\csymbolthird{\hat{I}}
\relpoint{Iup}{0,-7}{Ad}
\jointbleft[left]{Ad}{-1}{Iup}{-1} \csymbolalt[-8,0]{xa}
\jointbright[right]{Ad}{1}{Iup}{1} \csymbolalt[8,0]{xa}
\crectangledouble{Iup}{1.5}{1.5}{7,0}\csymbolthird{\hat{I}}
\relpoint{Iup}{0,-7}{Ad}
\jointbleft[left]{Ad}{-1}{Iup}{-1} \csymbolalt[-8,0]{by}
\jointbright[right]{Ad}{1}{Iup}{1} \csymbolalt[8,0]{by}
\end{Compose}
\end{equation}
If we replace the $\hat{I}$'s by $\hat{\mathds{1}}$'s using \eqref{identityalphaNstuff}, we obtain
\begin{equation}
\begin{Compose}{0}{0}\setdefaultfont{} \setsecondfont{\mathsfb}
\crectangle[thin]{coef}{3.2}{2}{-12,0} \csymbol{\frac{\alpha_\mathsfb{xa}\sqrt{N_\mathsfb{xa}N_\mathsfb{by}}}{\alpha_\mathsfb{by}} }
\thispoint{Iup}{0,7} \thispoint{Idown}{0,-7}
\crectangledleft[-1]{A}{1.5}{2.5}{-6,0} \csymbol{V}
\relpoint{A}{0,-7}{Ad} \jointbleft[left]{Ad}{0}{A}{0} \csymbolalt[-8,0]{xa}
\relpoint{A}{0,7}{Cd} \jointbleft[left]{A}{0}{Cd}{0} \csymbolalt[-8,0]{by}
%
\crectangledleft{XA}{0.8}{0.8}{-2,3} \crectangledleft{BY}{0.8}{0.8}{-2,-3}
\relpoint{XA}{0.8,-1.8}{XAA} \joinrrnoarrow{XA}{0}{XAA}{0}
\joinrlnoarrow[below]{A}{1.2}{XAA}{0} \csymbol{xa}
\relpoint{BY}{0.8,1.8}{BYY} \joinrrnoarrow{BY}{0}{BYY}{0}
\joinrlnoarrow[above]{A}{-1.2}{BYY}{0} \csymbol{by}
\jointbleft[left]{XA}{0}{Iup}{-2} \csymbolalt[-8,6]{xa} \jointbleft[left]{Idown}{-2}{BY}{0} \csymbolalt[-8,-6]{by}
\crectangledright[-1]{AA}{1.5}{2.5}{6,0} \csymbol{V}
\relpoint{AA}{0,-7}{Ad} \jointbright[right]{Ad}{0}{AA}{0} \csymbolalt[8,0]{xa}
\relpoint{AA}{0,7}{Cd} \jointbright[right]{AA}{0}{Cd}{0} \csymbolalt[8,0]{by}
%
\crectangledright{XA}{0.8}{0.8}{2,3} \crectangledleft{BY}{0.8}{0.8}{2,-3}
\relpoint{XA}{-0.8,-1.8}{XAA} \joinllnoarrow{XA}{0}{XAA}{0}
\joinlrnoarrow[below]{AA}{1.2}{XAA}{0} \csymbol{xa}
\relpoint{BY}{-0.8,1.8}{BYY} \joinllnoarrow{BY}{0}{BYY}{0}
\joinlrnoarrow[above]{AA}{-1.2}{BYY}{0} \csymbol{by}
\jointbright[right]{XA}{0}{Iup}{2} \csymbolalt[8,6]{xa} \jointbright[right]{Idown}{2}{BY}{0} \csymbolalt[8,-6]{by}
\jointtnoarrow{Iup}{-2}{Iup}{2} \jointtnoarrow{Iup}{-6}{Iup}{6}
\end{Compose}
~~=~~
\begin{Compose}{0}{0} \setdefaultfont{} \setsecondfont{\mathsfb}
\thispoint{b}{0,-7}\thispoint{t}{0,0}
\jointbleft[left]{b}{-6}{t}{-6} \csymbolalt[-8,0]{xa} \jointbleft[left]{b}{-2}{t}{-2} \csymbolalt[-8,0]{by}
\jointbright[right]{b}{6}{t}{6} \csymbolalt[8,0]{xa} \jointbright[right]{b}{2}{t}{2} \csymbolalt[8,0]{by}
\jointtnoarrow{t}{-2}{t}{2} \jointtnoarrow{t}{-6}{t}{6}
\end{Compose}
\end{equation}
Note, we have cancelled out the common factor arising from converting from $\hat{I}$'s to $\hat{\mathds{1}}$ (this is clearly the same factor on both sides of the equation).  Also, note we now use the caps notation for the $\mathds{1}$'s from Sec.\ \ref{sec:identityoperators}.  We see that this is exactly one of the two conditions for unitarity given in \eqref{Udefinitioncupcapversion} for the operator tensor given in \eqref{candidateunitary}.  We obtain the other condition by applying backwards causality and following similar steps (flipped in the vertical direction).  This proves the extension theorem.

\section{Proof that formulation is time symmetric}\label{sec:proofthatformulationistimesymmetric}

The definition of a time symmetric operational theory was provided in Sec.\ \ref{sec:definitionoftimesymmetricoperationaltheory}.  Given the correspondence from operations to operators, this means (i) we need a map from any operator to the time reverse of that operator such that (ii) the time reversed operator circuit equates to the same value (the probability).  The obvious map between an operator and its time reverse is the adjoint map discussed in Sec.\ \ref{sec:adjoint} (see equation \eqref{optensoradjoint}).  However, there is a subtlety here because of the gauge associated with the choice of the gauge parameters $\alpha_\mathsfb{a}$ (introduced in Sec.\ \ref{sec:chasingdowntheoneoverN}) and $\beta_\mathsfb{a}$ (introduced in Sec.\ \ref{sec:chasingdowntheoneoverNfactorforpointers}).   If we choose the symmetric gauges (whereby $\alpha_\mathsfb{a}=\beta_\mathtt{a}=1$ for all systems) then the adjoint does the job.  However, it is more interesting to allow freedom in the choice of gauge (especially since the standard presentation of quantum theory is in the forward gauge).  We define the \emph{time reverse cups and caps} which act as normal cups and caps except that they also introduce an overall factor.  Using these instead of the standard cups and caps we can define the \emph{time reverse operator} which is just the adjoint with but with these extra factors.

Let us begin by defining time reverse cups and caps.  For systems these are
\begin{equation}
\begin{Compose}{0}{0} \setsecondfont{\mathsfb}
\thispoint{a}{0,-1}
\ccapTR{t}{2.25}{1}{0,1}
\jointbnoarrow[left]{a}{-2.25}{t}{-2.25}  \csymbolalt{a}
\jointbnoarrow[right]{a}{2.25}{t}{2.25}  \csymbolalt{a}
\end{Compose}
~~=~~
\begin{Compose}{0}{0} \setsecondfont{\mathsfb}
\crectangle[thin]{A}{1.2}{1.5}{-5,0}\csymbolthird{\frac{1}{\alpha_\mathsfb{a}^2}}
\thispoint{a}{0,-1}
\ccap{t}{2.25}{1}{0,1}
\jointbnoarrow[left]{a}{-2.25}{t}{-2.25}  \csymbolalt{a}
\jointbnoarrow[right]{a}{2.25}{t}{2.25}  \csymbolalt{a}
\end{Compose}
~~~~~~~~~~~~~~~
\begin{Compose}{0}{0} \setsecondfont{\mathsfb}
\thispoint{a}{0,1}
\ccupTR{t}{2.25}{1}{0,-1}
\jointbnoarrow[left]{t}{-2.25}{a}{-2.25}  \csymbolalt{a}
\jointbnoarrow[right]{t}{2.25}{a}{2.25}  \csymbolalt{a}
\end{Compose}
~~=~~
\begin{Compose}{0}{0} \setsecondfont{\mathsfb}
\crectangle[thin]{A}{1.2}{1.1}{-5,0}\csymbolthird{\alpha_\mathsfb{a}^2}
\thispoint{a}{0,1}
\ccup{t}{2.25}{1}{0,-1}
\jointbnoarrow[left]{t}{-2.25}{a}{-2.25}  \csymbolalt{a}
\jointbnoarrow[right]{t}{2.25}{a}{2.25}  \csymbolalt{a}
\end{Compose}
\end{equation}
These defined in terms of the caps and cups introduced in Sec.\ \ref{sec:identityoperators} except that they introduce a factor (given inside the box).  It can be verified from Sec.\ \ref{sec:chasingdowntheoneoverN} that this is the correct factor to convert outputs to inputs and vice versa.  For pointers we have
\begin{equation}
\begin{Compose}{0}{0} \setsecondfont{\mathtt}
\thispoint{l}{-1,0} \thispoint{c}{0,0}
\chcapTR{t}{2.25}{0}{0,0}
\joinrlnoarrowthick[below]{l}{-2.25}{c}{-2.25}  \csymbolalt{a}
\joinrlnoarrowthick[above]{l}{2.25}{c}{2.25}  \csymbolalt{a}
\end{Compose}
~~=~~
\begin{Compose}{0}{0} \setsecondfont{\mathtt}
\crectangle[thin]{A}{1.2}{1.5}{-5,0}\csymbolthird{\frac{1}{\beta_\mathtt{a}^2}}
\thispoint{l}{-1,0} \thispoint{c}{0,0}
\chcap{t}{2.25}{0}{0,0}
\joinrlnoarrowthick[below]{l}{-2.25}{c}{-2.25}  \csymbolalt{a}
\joinrlnoarrowthick[above]{l}{2.25}{c}{2.25}  \csymbolalt{a}
\end{Compose}
~~~~~~~~~~~~~~~
\begin{Compose}{0}{0} \setsecondfont{\mathtt}
\thispoint{l}{1,0} \thispoint{c}{0,0}
\chcupTR{t}{2.25}{0}{0,0}
\joinlrnoarrowthick[below]{l}{-2.25}{c}{-2.25}  \csymbolalt{a}
\joinlrnoarrowthick[above]{l}{2.25}{c}{2.25}  \csymbolalt{a}
\end{Compose}
~~=~~
\begin{Compose}{0}{0} \setsecondfont{\mathtt}
\crectangle[thin]{A}{1.2}{1.1}{-5,0}\csymbolthird{\beta_\mathtt{a}^2}
\thispoint{l}{1,0} \thispoint{c}{0,0}
\chcup{t}{2.25}{0}{0,0}
\joinlrnoarrowthick[below]{l}{-2.25}{c}{-2.25}  \csymbolalt{a}
\joinlrnoarrowthick[above]{l}{2.25}{c}{2.25}  \csymbolalt{a}
\end{Compose}
\end{equation}
That these are correct can be seen by applying them to the fiducial elements in Sec.\ \ref{sec:chasingdowntheoneoverNfactorforpointers}.

Given these time reverse cups and caps, we can define the time reverse operator as follows
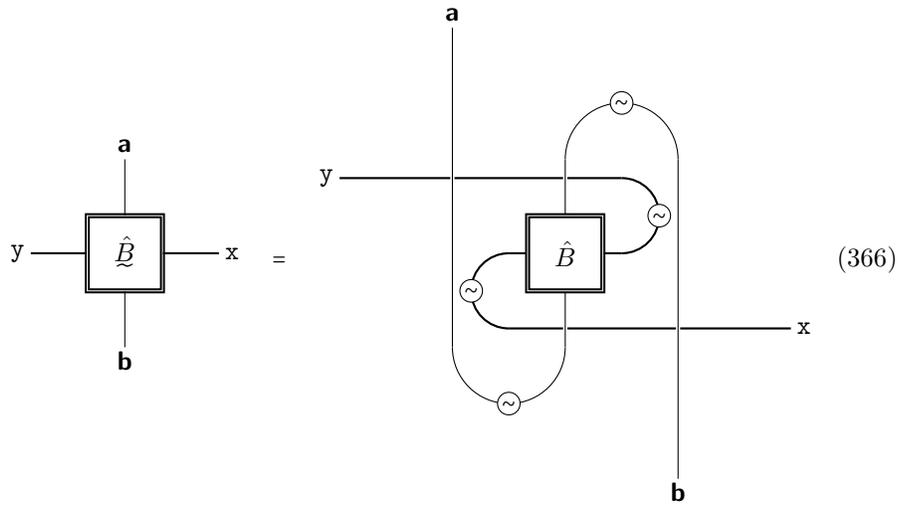
\begin{equation}\label{timereverseofB}
\begin{Compose}{0}{0} \setdefaultfont{} \setsecondfont{\mathsfb} \setthirdfont{\mathtt}
\crectangledouble{B}{2}{2}{0,0} \csymbol{\utilde{\hat{B}}}
\thispoint{l}{-5,0} \csymbolthird[-20,0]{y}  \thispoint{r}{5,0} \csymbolthird[20,0]{x}
\thispoint{d}{0,-5} \csymbolalt[0,-20]{b}  \thispoint{u}{0,5} \csymbolalt[0,20]{a}
\joinrlnoarrowthick{l}{0}{B}{0} \joinlrnoarrowthick{r}{0}{B}{0}
\jointbnoarrow{d}{0}{B}{0} \joinbtnoarrow{u}{0}{B}{0}
\end{Compose}
~~=~~
\begin{Compose}{0}{0} \setdefaultfont{} \setsecondfont{\mathsfb} \setthirdfont{\mathtt}
\crectangledouble[white]{B}{2}{2}{0,0}
\thispoint{l}{-3,0} \thispoint{ll}{-12,0} \thispoint{lll}{-12,4}\csymbolthird[-20,0]{y}
\thispoint{r}{3,0} \thispoint{rr}{12,0} \thispoint{rrr}{12,-4} \csymbolthird[20,0]{x}
\chcupTR{a}{2}{0}{-3,-2} \joinrlnoarrowthick{l}{0}{B}{0} \joinrlnoarrowthick{l}{-4}{rr}{-4}
\chcapTR{a}{2}{0}{3,2} \joinlrnoarrowthick{r}{0}{B}{0} \joinlrnoarrowthick{r}{4}{ll}{4}
\thispoint{d}{0,-5} \thispoint{dd}{0,-12} \thispoint{ddd}{6,-12} \csymbolalt[0,-20]{b}
\thispoint{u}{0,5} \thispoint{uu}{0,12} \thispoint{uuu}{-6,12} \csymbolalt[0,20]{a}
\ccupTR{a}{3}{0}{-3,-5}
\jointbnoarrowthickwhite{d}{0}{B}{0} \jointbnoarrowthickwhite{d}{-6}{uu}{-6}
\jointbnoarrow{d}{0}{B}{0} \jointbnoarrow{d}{-6}{uu}{-6}
\ccapTR{a}{3}{0}{3,5}
\joinbtnoarrowthickwhite{u}{0}{B}{0} \joinbtnoarrowthickwhite{u}{6}{dd}{6}
\joinbtnoarrow{u}{0}{B}{0} \joinbtnoarrow{u}{6}{dd}{6}
\crectangledouble{B}{2}{2}{0,0} \csymbol{\hat{B}}
\end{Compose}
\end{equation}
If we take the time reverse of the time reverse we get back to the original operation because the over all factors associated with the time reverse caps and cups are reciprocals of one another.   The adjoint is defined similarly - as
\begin{equation}
\begin{Compose}{0}{0} \setdefaultfont{} \setsecondfont{\mathsfb} \setthirdfont{\mathtt}
\crectangledouble{B}{2}{2}{0,0} \csymbol{\hat{B}^\dagger}
\thispoint{l}{-5,0} \csymbolthird[-20,0]{y}  \thispoint{r}{5,0} \csymbolthird[20,0]{x}
\thispoint{d}{0,-5} \csymbolalt[0,-20]{b}  \thispoint{u}{0,5} \csymbolalt[0,20]{a}
\joinrlnoarrowthick{l}{0}{B}{0} \joinlrnoarrowthick{r}{0}{B}{0}
\jointbnoarrow{d}{0}{B}{0} \joinbtnoarrow{u}{0}{B}{0}
\end{Compose}
~~=~~
\begin{Compose}{0}{0} \setdefaultfont{} \setsecondfont{\mathsfb} \setthirdfont{\mathtt}
\crectangledouble[white]{B}{2}{2}{0,0}
\thispoint{l}{-3,0} \thispoint{ll}{-12,0} \thispoint{lll}{-12,4}\csymbolthird[-20,0]{y}
\thispoint{r}{3,0} \thispoint{rr}{12,0} \thispoint{rrr}{12,-4} \csymbolthird[20,0]{x}
\chcup{a}{2}{0}{-3,-2} \joinrlnoarrowthick{l}{0}{B}{0} \joinrlnoarrowthick{l}{-4}{rr}{-4}
\chcap{a}{2}{0}{3,2} \joinlrnoarrowthick{r}{0}{B}{0} \joinlrnoarrowthick{r}{4}{ll}{4}
\thispoint{d}{0,-5} \thispoint{dd}{0,-12} \thispoint{ddd}{6,-12} \csymbolalt[0,-20]{b}
\thispoint{u}{0,5} \thispoint{uu}{0,12} \thispoint{uuu}{-6,12} \csymbolalt[0,20]{a}
\ccup{a}{3}{0}{-3,-5}
\jointbnoarrowthickwhite{d}{0}{B}{0} \jointbnoarrowthickwhite{d}{-6}{uu}{-6}
\jointbnoarrow{d}{0}{B}{0} \jointbnoarrow{d}{-6}{uu}{-6}
\ccap{a}{3}{0}{3,5}
\joinbtnoarrowthickwhite{u}{0}{B}{0} \joinbtnoarrowthickwhite{u}{6}{dd}{6}
\joinbtnoarrow{u}{0}{B}{0} \joinbtnoarrow{u}{6}{dd}{6}
\crectangledouble{B}{2}{2}{0,0} \csymbol{\hat{B}}
\end{Compose}
\end{equation}
Consequently, we have
\begin{equation}\label{TRtoAdjointwithconversionfactor}
\begin{Compose}{0}{0} \setdefaultfont{} \setsecondfont{\mathsfb} \setthirdfont{\mathtt}
\crectangledouble{B}{2}{2}{0,0} \csymbol{\utilde{\hat{B}}}
\thispoint{l}{-5,0} \csymbolthird[-20,0]{y}  \thispoint{r}{5,0} \csymbolthird[20,0]{x}
\thispoint{d}{0,-5} \csymbolalt[0,-20]{b}  \thispoint{u}{0,5} \csymbolalt[0,20]{a}
\joinrlnoarrowthick{l}{0}{B}{0} \joinlrnoarrowthick{r}{0}{B}{0}
\jointbnoarrow{d}{0}{B}{0} \joinbtnoarrow{u}{0}{B}{0}
\end{Compose}
~~=~~
\begin{Compose}{0}{0} \setdefaultfont{} \setsecondfont{\mathsfb} \setthirdfont{\mathtt}
\crectangle[thin]{A}{2}{1.5}{-3,4}\csymbolthird{\frac{\alpha_\mathsfb{a}^2\beta_\mathtt{x}^2}{\alpha_\mathsfb{b}^2\beta_\mathtt{y}^2}}
\crectangledouble{B}{2}{2}{0,0} \csymbol{\hat{B}^\dagger}
\thispoint{l}{-5,0} \csymbolthird[-20,0]{y}  \thispoint{r}{5,0} \csymbolthird[20,0]{x}
\thispoint{d}{0,-5} \csymbolalt[0,-20]{b}  \thispoint{u}{0,5} \csymbolalt[0,20]{a}
\joinrlnoarrowthick{l}{0}{B}{0} \joinlrnoarrowthick{r}{0}{B}{0}
\jointbnoarrow{d}{0}{B}{0} \joinbtnoarrow{u}{0}{B}{0}
\end{Compose}
\end{equation}
where the conversion factor is inside the rectangle.  Note that if we are in the time symmetric gauge (where $\alpha_\mathsfb{a}=\alpha_\mathsfb{b}=\beta_\mathtt{x}=\beta_\mathtt{y}=1$ then the conversion factor is equal to 1 so the time reverse is just equal to the adjoint.  Also, in the case where the input and output are the same type and also the income and outcome are the same type then the conversion factor is equal to 1.  Note that if an operator is physical then its time reverse is also physical.  This is clear since (a) $T$-positivity is a time symmetric notion that does not depend on any overall factors and (b) double causality is also a time symmetric notion but where we need to look after the overall factors as is done by the time reverse caps and cups.

If we have an operator circuit we can replace each operator by the expression involving its time reverse.  Consider the following example of this procedure:
\begin{equation}\label{ABCreversea}
\begin{Compose}{0}{0} \setdefaultfont{} \setsecondfont{\mathsfb} \setthirdfont{\mathtt}
\crectangledouble{A}{1.5}{1.5}{0,-6}\csymbol{\hat{A}} \crectangledouble{B}{1.5}{1.5}{0,0}\csymbol{\hat{B}} \crectangledouble{C}{1.5}{1.5}{0,6}\csymbol{\hat{C}}
\jointbnoarrow[left]{A}{0}{B}{0} \csymbolalt{a} \jointbnoarrow[left]{B}{0}{C}{0} \csymbolalt{b}
\end{Compose}
~~=~~
\begin{Compose}{0}{0} \setdefaultfont{} \setsecondfont{\mathsfb} \setthirdfont{\mathtt}
\crectangledouble{A}{1.5}{1.5}{0,-6}\csymbol{\hat{A}} \crectangledouble{B}{1.5}{1.5}{0,0}\csymbol{\hat{B}} \crectangledouble{C}{1.5}{1.5}{0,6}\csymbol{\hat{C}}
\jointbnoarrow[left]{A}{0}{B}{0} \csymbolalt{a} \jointbnoarrow[left]{B}{0}{C}{0} \csymbolalt{b}
\crectangledouble{Id}{0.9}{0.9}{0,-11}\csymbol{\hat{I}}  \crectangledouble{Iu}{0.9}{0.9}{0,11}\csymbol{\hat{I}}
\jointbnoarrow[left]{Id}{0}{A}{0} \csymbolalt{0} \jointbnoarrow[left]{C}{0}{Iu}{0} \csymbolalt{0}
\end{Compose}
~~=~~
\begin{Compose}{0}{0} \setdefaultfont{} \setsecondfont{\mathsfb} \setthirdfont{\mathtt}
\crectangledouble{A}{1.5}{1.5}{5,-6}\csymbol{\utilde{\hat{A}}} \crectangledouble{B}{1.5}{1.5}{10,0}\csymbol{\utilde{\hat{B}}} \crectangledouble{C}{1.5}{1.5}{15,6}\csymbol{\utilde{\hat{C}}}
\crectangledouble{Id}{0.9}{0.9}{2.5,-11}\csymbol{\hat{I}}  \crectangledouble{Iu}{0.9}{0.9}{17.5,11}\csymbol{\hat{I}}
\relpoint{A}{-1.25,1.5}{Aul} \ccapTR{IA}{1.25}{0}{Aul} \jointbnoarrow[left]{Id}{0}{Aul}{-1.25} \csymbolalt{0}
\relpoint{A}{1.25,-1.5}{Adr}  \relpoint{B}{-1.25,1.5}{Bul} \ccupTR{AB}{1.25}{0}{Adr} \jointbnoarrow[left]{Adr}{1.25}{Bul}{-1.25} \csymbolalt{a} \ccapTR{BC}{1.25}{0}{Bul}
\relpoint{B}{1.25,-1.5}{Bdr} \relpoint{C}{-1.25,1.5}{Cul} \ccupTR{BC}{1.25}{0}{Bdr} \jointbnoarrow[left]{Bdr}{1.25}{Cul}{-1.25} \csymbolalt{b} \ccapTR{BC}{1.25}{0}{Cul}
\relpoint{C}{1.25,-1.5}{Cdr}  \ccupTR{AB}{1.25}{0}{Cdr} \jointbnoarrow[right]{Cdr}{1.25}{Iu}{0} \csymbolalt{0}
\end{Compose}
\end{equation}
Note that we introduce the ignore operators with the null system (this is associated with a Hilbert space of dimension 1 so this changes nothing).  Also note that each time reverse cup is matched by a time reverse cap. Hence, the overall factors cancel (since they are reciprocals).  Hence, \eqref{ABCreversea} continues as
\begin{equation}
=~~
\begin{Compose}{0}{0} \setdefaultfont{} \setsecondfont{\mathsfb} \setthirdfont{\mathtt}
\crectangledouble{A}{1.5}{1.5}{5,-6}\csymbol{\utilde{\hat{A}}} \crectangledouble{B}{1.5}{1.5}{10,0}\csymbol{\utilde{\hat{B}}} \crectangledouble{C}{1.5}{1.5}{15,6}\csymbol{\utilde{\hat{C}}}
\crectangledouble{Id}{0.9}{0.9}{2.5,-11}\csymbol{\hat{I}}  \crectangledouble{Iu}{0.9}{0.9}{17.5,11}\csymbol{\hat{I}}
\relpoint{A}{-1.25,1.5}{Aul} \ccap{IA}{1.25}{0}{Aul} \jointbnoarrow[left]{Id}{0}{Aul}{-1.25} \csymbolalt{0}
\relpoint{A}{1.25,-1.5}{Adr}  \relpoint{B}{-1.25,1.5}{Bul} \ccup{AB}{1.25}{0}{Adr} \jointbnoarrow[left]{Adr}{1.25}{Bul}{-1.25} \csymbolalt{a} \ccap{BC}{1.25}{0}{Bul}
\relpoint{B}{1.25,-1.5}{Bdr} \relpoint{C}{-1.25,1.5}{Cul} \ccup{BC}{1.25}{0}{Bdr} \jointbnoarrow[left]{Bdr}{1.25}{Cul}{-1.25} \csymbolalt{b} \ccap{BC}{1.25}{0}{Cul}
\relpoint{C}{1.25,-1.5}{Cdr}  \ccup{AB}{1.25}{0}{Cdr} \jointbnoarrow[right]{Cdr}{1.25}{Iu}{0} \csymbolalt{0}
\end{Compose}
~~=~~
\begin{Compose}{0}{0} \setdefaultfont{} \setsecondfont{\mathsfb} \setthirdfont{\mathtt}
\crectangledouble{C}{1.5}{1.5}{0,-6}\csymbol{\utilde{\hat{C}}} \crectangledouble{B}{1.5}{1.5}{0,0}\csymbol{\utilde{\hat{B}}} \crectangledouble{A}{1.5}{1.5}{0,6}\csymbol{\utilde{\hat{A}}}
\jointbnoarrow[left]{C}{0}{B}{0} \csymbolalt{a} \jointbnoarrow[left]{B}{0}{A}{0} \csymbolalt{b}
\crectangledouble{Id}{0.9}{0.9}{-3.5,3}\csymbol{\hat{I}}  \crectangledouble{Iu}{0.9}{0.9}{3.5,-3}\csymbol{\hat{I}}
\relpoint{A}{-1.75,1.5}{Aul} \ccap{IA}{1.75}{0}{Aul} \jointbnoarrow[left]{Id}{0}{Aul}{-1.75} \csymbolalt{0}
\relpoint{C}{1.75,-1.5}{Cdr}  \ccup{AB}{1.75}{0}{Cdr} \jointbnoarrow[right]{Cdr}{1.75}{Iu}{0} \csymbolalt{0}
\end{Compose}
~~=~~
\begin{Compose}{0}{0} \setdefaultfont{} \setsecondfont{\mathsfb} \setthirdfont{\mathtt}
\crectangledouble{C}{1.5}{1.5}{0,-6}\csymbol{\utilde{\hat{C}}} \crectangledouble{B}{1.5}{1.5}{0,0}\csymbol{\utilde{\hat{B}}} \crectangledouble{A}{1.5}{1.5}{0,6}\csymbol{\utilde{\hat{A}}}
\jointbnoarrow[left]{C}{0}{B}{0} \csymbolalt{a} \jointbnoarrow[left]{B}{0}{A}{0} \csymbolalt{b}
\end{Compose}
\end{equation}
Thus, in this example, the operator circuit formed by replacing elements by their time reverse and inverting the circuit has the same probability.  This works for any circuit - even if it has some boxes in parallel.  One way to see this is to imagine foliating the circuit and then sliding one layer at a time over a cap.  This deals with the case where we have system (not pointer) types.  It clearly works when there are pointer types as well.  One way to see this is to bend the pointer types so they point up and down and apply the same reasoning.  However, for the simple case where we have $R$ and $x$ boxes attached to each operator, we  can see we have the required behaviour by considering the following.
\begin{equation}
\begin{Compose}{0}{0} \setdefaultfont{} \setsecondfont{\mathsfb} \setthirdfont{\mathtt}
\crectangledouble{B}{2}{2}{0,0} \csymbol{\hat{B}}
\thispoint{l}{-5,0}   \thispoint{r}{5,0}
\thispoint{d}{0,-5} \csymbolalt[0,-20]{a}  \thispoint{u}{0,5} \csymbolalt[0,20]{b}
\RxBoxincomedouble{B}{1.5}{0}{x} \RxBoxoutcomedouble{B}{1.5}{0}{y}
\jointbnoarrow{d}{0}{B}{0} \joinbtnoarrow{u}{0}{B}{0}
\end{Compose}
~~=~~
\begin{Compose}{0}{0} \setdefaultfont{} \setsecondfont{\mathsfb} \setthirdfont{\mathtt}
\crectangledouble[white]{B}{2}{2}{0,0}
\thispoint{l}{-3,0} \thispoint{ll}{-5,0} \thispoint{lll}{-12,4}
\thispoint{r}{3,0} \thispoint{rr}{5,0} \thispoint{rrr}{12,-4}
\chcupTR{a}{2}{0}{-3,-2} \joinrlnoarrowthick{l}{0}{B}{0} \joinrlnoarrowthick[below]{l}{-4}{rr}{-4} \csymbolthird[60,-4]{y}
\GenxdifBoxoutcomedouble{rr}{0}{-4}{}{y}{R} 
\chcapTR{a}{2}{0}{3,2} \joinlrnoarrowthick{r}{0}{B}{0}\joinlrnoarrowthick[above]{r}{4}{ll}{4} \csymbolthird[-60,0]{x}
\GenxdifBoxincomedouble{ll}{0}{4}{}{x}{R} 
\thispoint{d}{0,-5} \thispoint{dd}{0,-12} \thispoint{ddd}{6,-12} \csymbolalt[0,-20]{a}
\thispoint{u}{0,5} \thispoint{uu}{0,12} \thispoint{uuu}{-6,12} \csymbolalt[0,20]{b}
\ccupTR{a}{3}{0}{-3,-5}
\jointbnoarrowthickwhite{d}{0}{B}{0} \jointbnoarrowthickwhite{d}{-6}{uu}{-6}
\jointbnoarrow{d}{0}{B}{0} \jointbnoarrow{d}{-6}{uu}{-6}
\ccapTR{a}{3}{0}{3,5}
\joinbtnoarrowthickwhite{u}{0}{B}{0} \joinbtnoarrowthickwhite{u}{6}{dd}{6}
\joinbtnoarrow{u}{0}{B}{0} \joinbtnoarrow{u}{6}{dd}{6}
\crectangledouble{B}{2}{2}{0,0} \csymbol{\utilde{\hat{B}}}
\end{Compose}
\end{equation}
If we now go through the rest of the steps inserting cups and caps in the horizontal direction then we obtain the following:
\begin{equation}
\begin{Compose}{0}{0} \setdefaultfont{} \setsecondfont{\mathsfb} \setthirdfont{\mathtt}
\crectangledouble{B}{2}{2}{0,0} \csymbol{\hat{B}}
\thispoint{l}{-5,0}   \thispoint{r}{5,0}
\thispoint{d}{0,-5} \csymbolalt[0,-20]{a}  \thispoint{u}{0,5} \csymbolalt[0,20]{b}
\RxBoxincomedouble{B}{1.5}{0}{x} \RxBoxoutcomedouble{B}{1.5}{0}{y}
\jointbnoarrow{d}{0}{B}{0} \joinbtnoarrow{u}{0}{B}{0}
\end{Compose}
~~=~~
\begin{Compose}{0}{0} \setdefaultfont{} \setsecondfont{\mathsfb} \setthirdfont{\mathtt}
\crectangledouble[white]{B}{2}{2}{0,0}
\thispoint{l}{-5,0}   \thispoint{r}{5,0}
\thispoint{d}{0,-5} 
\thispoint{u}{0,5}  
\RxBoxincomedouble{B}{1.5}{0}{y} \RxBoxoutcomedouble{B}{1.5}{0}{x}
%
\thispoint{d}{0,-5} \thispoint{dd}{0,-12} \thispoint{ddd}{6,-12}
\thispoint{u}{0,5} \thispoint{uu}{0,12} \thispoint{uuu}{-6,12} 
\ccupTR{a}{5}{0}{-5,-5}
\jointbnoarrowthickwhite{d}{0}{B}{0} 
\jointbnoarrow{d}{0}{B}{0} \jointbnoarrow[right]{d}{-10}{uu}{-10} \csymbolalt{a}
\ccapTR{a}{5}{0}{5,5}
\joinbtnoarrowthickwhite{u}{0}{B}{0} 
\joinbtnoarrow{u}{0}{B}{0} \joinbtnoarrow[left]{u}{10}{dd}{10} \csymbolalt{b}
\crectangledouble{B}{2}{2}{0,0} \csymbol{\utilde{\hat{B}}}
\end{Compose}
\end{equation}
It is clear that, in general, we can invert circuits as required. This completes the proof that the we have a time symmetric formulation of Quantum Theory as defined in Sec.\ \ref{sec:proofthatformulationistimesymmetric}.

When we take the time reverse of an operator we pick up a gauge dependent factor in the representation of this operator (as seen in \eqref{TRtoAdjointwithconversionfactor}).  There is a natural time symmetric choice for $\alpha_\mathsfb{a}$ and $\beta_\mathtt{x}$ so this factor is equal to 1. However, the proof that the formulation is time symmetric presented in this section goes through regardless of what values we choose for these gauge parameters.

The extension theorem provides a nice way to take the time reverse of an operation.  By the above techniques, it is easy to prove that the time reverse of
\begin{equation}
\begin{Compose}{0}{0}\setsecondfont{\mathtt} \setthirdfont{\mathsfb}
\crectangledouble{U}{2}{2}{0,0} \csymbol{B} \thispoint{AL}{-4,0} \csymbolalt[-20,0]{x} \thispoint{AR}{4,0} \csymbolalt[20,0]{y}
\joinlrnoarrowthick{U}{0}{AL}{0} \joinrlnoarrowthick{U}{0}{AR}{0}
\thispoint{a}{0,-4} \csymbolthird[0,-20]{a} \jointbnoarrow{a}{0}{U}{0}
\thispoint{d}{0,4} \csymbolthird[0,20]{b} \joinbtnoarrow{d}{0}{U}{0}
\end{Compose}
~~=~~
\begin{Compose}{0}{0} \setdefaultfont{}\setsecondfont{\mathsfb}\setthirdfont{\mathtt}
\crectangledmark[-1]{U}{3}{2}{0,0} \csymbol{\hat{U}_\mathsf{B}}
\relpoint{U}{-0.3,4.5}{Iupos}  \crectangledouble{Iu}{0.9}{0.9}{Iupos} \csymbol{\hat{I}}  \jointbnoarrow[left]{U}{-0.3}{Iu}{0}  \csymbolalt{z}
\relpoint{U}{0.3,-4.5}{Idpos}  \crectangledouble{Id}{0.9}{0.9}{Idpos} \csymbol{\hat{I}} \jointbnoarrow[left]{Id}{0}{U}{0.3} \csymbolalt[-3,0]{w}
\relpoint{U}{2.3,4.5}{Ypos}  \crectangledouble{Y}{0.9}{0.9}{Ypos} \csymbol{\hat{Y}}  \jointbnoarrow[left]{U}{2.3}{Y}{0}  \csymbolalt{y}
\relpoint{Y}{4,0}{r} \joinlrnoarrowthick[above]{r}{0}{Y}{0} \csymbolthird[0,3]{y}
\relpoint{U}{-2.3,-4.5}{Xpos}  \crectangledouble{X}{0.9}{0.9}{Xpos} \csymbol{\hat{X}}  \jointbnoarrow[left]{X}{0}{U}{-2.3} \csymbolalt{x} \relpoint{X}{-4,0}{l} \joinrlnoarrowthick[above]{l}{0}{X}{0} \csymbolthird{x}
\thispoint{down}{0,-7} \thispoint{up}{0,7}
\jointbnoarrow[right]{down}{2.3}{U}{2.3} \csymbolalt{a}
\jointbnoarrow[left]{U}{-2.3}{up}{-2.3} \csymbolalt{b}
\end{Compose}
\end{equation}
is
\begin{equation}
\begin{Compose}{0}{0}\setdefaultfont{}\setsecondfont{\mathtt} \setthirdfont{\mathsfb}
\crectangledouble{U}{2}{2}{0,0} \csymbol{\utilde{B}} \thispoint{AL}{-4,0} \csymbolalt[-20,0]{y} \thispoint{AR}{4,0} \csymbolalt[20,0]{x}
\joinlrnoarrowthick{U}{0}{AL}{0} \joinrlnoarrowthick{U}{0}{AR}{0}
\thispoint{a}{0,-4} \csymbolthird[0,-20]{b} \jointbnoarrow{a}{0}{U}{0}
\thispoint{d}{0,4} \csymbolthird[0,20]{a} \joinbtnoarrow{d}{0}{U}{0}
\end{Compose}
~~=~~
\begin{Compose}{0}{0} \setdefaultfont{}\setsecondfont{\mathsfb}\setthirdfont{\mathtt}
\crectangle[thin]{coeff}{3}{2}{-7.5,0}
\csymbol{
\frac{\alpha_\mathsfb{a}^2\beta_\mathtt{x}^2}{\alpha_\mathsfb{b}^2\beta_\mathtt{y}^2}
}
\crectangledmark[1]{U}{3}{2}{0,0} \csymbol{\hat{U}_\mathsf{B}}
\relpoint{U}{-0.3,4.5}{Iupos}  \crectangledouble{Iu}{0.9}{0.9}{Iupos} \csymbol{\hat{I}}  \jointbnoarrow[left]{U}{-0.3}{Iu}{0}  \csymbolalt[-3,0]{w}
\relpoint{U}{0.3,-4.5}{Idpos}  \crectangledouble{Id}{0.9}{0.9}{Idpos} \csymbol{\hat{I}} \jointbnoarrow[left]{Id}{0}{U}{0.3} \csymbolalt{z}
\relpoint{U}{2.3,4.5}{Ypos}  \crectangledouble{Y}{0.9}{0.9}{Ypos} \csymbol{\hat{X}}  \jointbnoarrow[left]{U}{2.3}{Y}{0}  \csymbolalt{x}
\relpoint{Y}{4,0}{r} \joinlrnoarrowthick[above]{r}{0}{Y}{0} \csymbolthird{x}
\relpoint{U}{-2.3,-4.5}{Xpos}  \crectangledouble{X}{0.9}{0.9}{Xpos} \csymbol{\hat{Y}}  \jointbnoarrow[left]{X}{0}{U}{-2.3} \csymbolalt{y} \relpoint{X}{-4,0}{l} \joinrlnoarrowthick[above]{l}{0}{X}{0} \csymbolthird[0,3]{y}
\thispoint{down}{0,-7} \thispoint{up}{0,7}
\jointbnoarrow[right]{down}{2.3}{U}{2.3} \csymbolalt{b}
\jointbnoarrow[left]{U}{-2.3}{up}{-2.3} \csymbolalt{a}
\end{Compose}
\end{equation}
The black dot position indicates the adjoint.

\section{Equivalence to standard operational formulation}\label{sec:equivalencetostandardoperationalformulation}

We will now show that the time symmetric operational formulation of Quantum Theory presented here is equivalent to the standard (time asymmetric) operational formulation of Quantum Theory which is construed in a forward time direction.  First we note that, by the extension theorem presented in Sec.\ \ref{sec:extensiontheorem}, we know we can model every time symmetric operation by a corresponding operator that can be expressed in the language of standard quantum theory.  This proves that the standard operational formulation can model the time-symmetric formulation.   To prove equivalence we also need to prove the converse - that the time symmetric formulation can model the standard (time asymmetric) operational formulation of Quantum Theory.  In fact, we prove this at the more general level of operational theories since the pertinent properties ($T$-positivity and double causality) can be discussed at that level.

In the standard (time forward) operational formulation all incomes are regarded as settings - we condition on particular values of them.  Consider the circuit
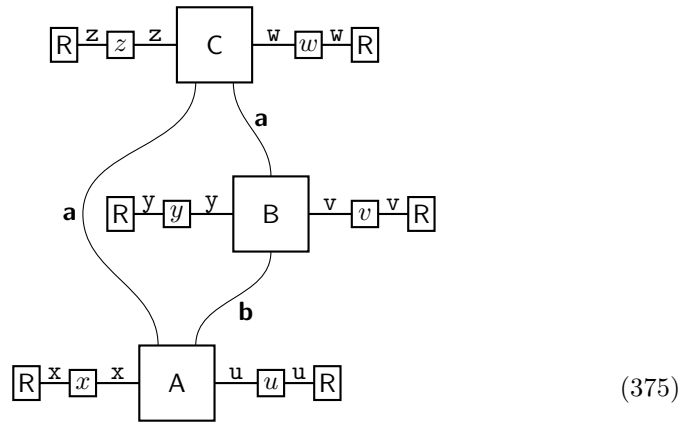
\begin{equation}\label{ABCxyxuvwcircuit}
\begin{Compose}{0}{0} \setsecondfont{\mathsfb}
\crectangle{A}{2}{2}{0,0} \csymbol{A}  \crectangle{B}{2}{2}{5,9} \csymbol{B} \crectangle{C}{2}{2}{2,18} \csymbol{C}
\thispoint{mid}{-5,9}\csymbolalt[-20,0]{a} \jointbnoarrow[left]{A}{-1}{mid}{0}  \jointbnoarrow[left]{mid}{0}{C}{-1}
\jointbnoarrow[below right]{A}{1}{B}{0}  \csymbolalt[5,-5]{b}
\jointbnoarrow[above right]{B}{0}{C}{1}  \csymbolalt{a}
\Rxboxincome{A}{0}{x} \Rxboxoutcome{A}{0}{u}
\Rxboxincome{B}{0}{y} \Rxboxoutcome{B}{0}{v}
\Rxboxincome{C}{0}{z} \Rxboxoutcome{C}{0}{w}
\end{Compose}
\end{equation}
We are interested in the probability
\begin{equation}
\text{prob}(uvw|xyz)~~ = ~~
\frac{ \text{prob}\left(
\begin{Compose}{0}{-2.3} \setsecondfont{\mathsfb}
\crectangle{A}{2}{2}{0,0} \csymbol{A}  \crectangle{B}{2}{2}{5,9} \csymbol{B} \crectangle{C}{2}{2}{2,18} \csymbol{C}
\thispoint{mid}{-5,9}\csymbolalt[-20,0]{a} \jointbnoarrow[left]{A}{-1}{mid}{0}  \jointbnoarrow[left]{mid}{0}{C}{-1}
\jointbnoarrow[below right]{A}{1}{B}{0}  \csymbolalt[5,-5]{b}
\jointbnoarrow[above right]{B}{0}{C}{1}  \csymbolalt{a}
\Rxboxincome{A}{0}{x} \Rxboxoutcome{A}{0}{u}
\Rxboxincome{B}{0}{y} \Rxboxoutcome{B}{0}{v}
\Rxboxincome{C}{0}{z} \Rxboxoutcome{C}{0}{w}
\end{Compose}\right)
}{ \text{prob} \left(~~
\begin{Compose}{0}{-2.3} \setsecondfont{\mathsfb}
\crectangle{A}{2}{2}{0,0} \csymbol{A}  \crectangle{B}{2}{2}{5,9} \csymbol{B} \crectangle{C}{2}{2}{2,18} \csymbol{C}
\thispoint{mid}{-5,9}\csymbolalt[-20,0]{a} \jointbnoarrow[left]{A}{-1}{mid}{0}  \jointbnoarrow[left]{mid}{0}{C}{-1}
\jointbnoarrow[below right]{A}{1}{B}{0}  \csymbolalt[5,-5]{b}
\jointbnoarrow[above right]{B}{0}{C}{1}  \csymbolalt{a}
\Rxboxincome{A}{0}{x} \Routcome{A}{0}{u}
\Rxboxincome{B}{0}{y} \Routcome{B}{0}{v}
\Rxboxincome{C}{0}{z} \Routcome{C}{0}{w}
\end{Compose} ~~ \right)
}
\end{equation}
We know from backwards summation (or backwards flatness) that the denominator is equal to $\frac{1}{N_\mathtt{x} N_\mathtt{y} N_\mathtt{z}}$.  This means we can write
\begin{equation}
\text{prob}(uvw|xyz)~~ = ~~
\begin{Compose}{0}{0} \setsecondfont{\mathsfb}
\crectangle{A}{2}{2}{0,-9} \csymbol{A}  \crectangle{B}{2}{2}{5,0} \csymbol{B} \crectangle{C}{2}{2}{2,9} \csymbol{C}
\thispoint{mid}{-5,0}\csymbolalt[-20,0]{a} \jointbnoarrow[left]{A}{-1}{mid}{0}  \jointbnoarrow[left]{mid}{0}{C}{-1}
\jointbnoarrow[below right]{A}{1}{B}{0}  \csymbolalt[5,-5]{b}
\jointbnoarrow[above right]{B}{0}{C}{1}  \csymbolalt{a}
\cxconditionincome[1.3]{A}{0}{x} \cxconditionoutcome[1.3]{A}{0}{u}
\cxconditionincome[1.3]{B}{0}{y} \cxconditionoutcome[1.3]{B}{0}{v}
\cxconditionincome[1.3]{C}{0}{z} \cxconditionoutcome[1.3]{C}{0}{w}
\end{Compose}
\end{equation}
where
\begin{equation}
\begin{Compose}{0}{0}
\thispoint{p}{0,0} \cxconditionincome{p}{0}{x}
\end{Compose}
~~:=~~
\begin{Compose}{0}{0}
\crectangle[thin]{N}{1}{1}{-8,0} \csymbolfourth{N_\mathtt{x}}
\thispoint{p}{0,0} \RxBoxincome[0.8]{p}{0}{0}{x}
\end{Compose}
~~~~~~~~~~
\begin{Compose}{0}{0}
\thispoint{p}{0,0} \cxconditionoutcome{p}{0}{x}
\end{Compose}
~~:=~~
\begin{Compose}{0}{0}
\thispoint{p}{0,0} \RxBoxoutcome[0.8]{p}{0}{0}{x}
\end{Compose}
\end{equation}
The $N_\mathtt{x}$ factor in the box is a multiplicative factor.  The interpretation of this is that we prepare $x$.  The $N_\mathtt{x}$ can be thought of as canceling the $\frac{1}{N_\mathtt{x}}$ from the $\mathsf R$ box.

What happens to the axiom of Quantum Theory when we go into this income conditioned picture?  In particular, how is the notion of physicality affected?  Physicality has two parts (i) $T$-positivity and (ii) double causality.  It is clear that $T$-positivity goes through unaltered (since multiplication by $N_\mathtt{x}$ factors does not change this property).  However, double causality needs to be revisited.  Double causality consists of the following two properties.
\begin{equation}\label{causalityconditionsagain}
\Funitalxyab{1}{1}{1}{1}    ~~~~~~~~~~~~~~~  \Bunitalyxba{1}{1}{1}{1}
\end{equation}
The forward causality property on the left becomes unapplicable since we are always conditioning on an income - except in the special case where there is no income when we get
\begin{equation}
\Funitalxyab{0}{1}{1}{1}
\end{equation}
This corresponds preparing the maximally mixed state. In this case, the outcome gives no information.  The backwards causality property (on the right of \eqref{causalityconditionsagain}) is applicable.  We can rewrite it assuming that we have conditioned on a particular value of $x$ in which case it becomes
\begin{equation}
\begin{Compose}{0}{0} \setsecondfont{\mathsfb}
\Crectangle{B}{1.3}{1.3}{0,0}
\Crectangle{I}{0.8}{0.8}{0,4.4} \jointbnoarrow[left]{B}{0}{I}{0} \csymbolalt{b}
\Routcome{B}{0}{y}
\cxconditionincome{B}{0}{x}
\thispoint{b}{0,-4}\jointbnoarrow[left]{b}{0}{B}{0} \csymbolalt{a}
\end{Compose}
~~=~~
\begin{Compose}{0}{0} \setsecondfont{\mathsfb}
\Crectangle{I}{0.8}{0.8}{0,0.3}
\thispoint{b}{0,-4}\jointbnoarrow[left]{b}{0}{I}{0} \csymbolalt{a}
\end{Compose}
\end{equation}
This corresponds to the causality condition from standard Quantum Theory \cite{chiribella2010informational}(see diagrammatic notation in \cite{hardy2011reformulating}) whereby the deterministic effect is unique.

In the context of Quantum Theory, these considerations mean that we have, in this time asymmetric picture, the usual physicality conditions (as discussed in \cite{hardy2011reformulating} for example).  These are
\begin{equation}
0\underset{T}{\leq} ~~~
\begin{Compose}{0}{-0.07} \setdefaultfont{\hat} \setsecondfont{\mathsfb} \setthirdfont{\mathtt}
\Crectangledouble{B}{1.3}{1.3}{0,0}
\thispoint{b}{0,-4}\thispoint{t}{0,4} \thispoint{l}{-4,0} \thispoint{r}{4,0}
\jointbnoarrow[left]{b}{0}{B}{0} \csymbolalt{a}
\jointbnoarrow[left]{B}{0}{t}{0} \csymbolalt{b}
\joinrlnoarrowthick[above]{l}{0}{B}{0} \csymbolthird{x}
\joinrlnoarrowthick[above]{B}{0}{r}{0} \csymbolthird{y}
\end{Compose}
~~~~~~~~~~~~~~~~~~~
\begin{Compose}{0}{0} \setdefaultfont{} \setsecondfont{\mathsfb}
\crectangledouble{B}{1.3}{1.3}{0,0} \csymbol{\hat{B}}
\crectangledouble{I}{0.9}{0.9}{0,4.4} \csymbol{\hat{I}} \jointbnoarrow[left]{B}{0}{I}{0} \csymbolalt{b}
\Routcomedouble{B}{0}{y}
\cxconditionincomedouble{B}{0}{x}
\thispoint{b}{0,-4}\jointbnoarrow[left]{b}{0}{B}{0} \csymbolalt{a}
\end{Compose}
~~=~~
\begin{Compose}{0}{0} \setdefaultfont{\hat} \setsecondfont{\mathsfb}
\Crectangledouble{I}{0.9}{0.9}{0,0.3}
\thispoint{b}{0,-4}\jointbnoarrow[left]{b}{0}{I}{0} \csymbolalt{a}
\end{Compose}
\end{equation}
In the usual parlance, the condition on the left is called complete positivity and the condition on the right tells us that, the trace is preserved if we sum over the outcomes.

We can also prove equivalence to a backwards formulation wherein we calculate probabilities for incomes conditioned on outcomes.  To do this, we would set
\begin{equation}
\begin{Compose}{0}{0}
\thispoint{p}{0,0} \cxconditionincome{p}{0}{x}
\end{Compose}
~~:=~~
\begin{Compose}{0}{0}
\thispoint{p}{0,0} \RxBoxincome[0.8]{p}{0}{0}{x}
\end{Compose}
~~~~~~~~~~
\begin{Compose}{0}{0}
\thispoint{p}{0,0} \cxconditionoutcome{p}{0}{x}
\end{Compose}
~~:=~~
\begin{Compose}{0}{0}
\crectangle[thin]{N}{1}{1}{8,0} \csymbolfourth{N_\mathtt{x}}
\thispoint{p}{0,0} \RxBoxoutcome[0.8]{p}{0}{0}{x}
\end{Compose}
\end{equation}
This is discussed further in Sec.\ \ref{sec:generalisedtheoryofconditionalframes}.

\section{Proof that Quantum Theory is doubly summing}\label{sec:proofthatquantumtheoryisdoublysumming}

Consider a circuit like that in \eqref{ABCxyxuvwcircuit}.  The corresponding operator circuit is
\begin{equation}\label{ABCxyxuvwcircuitoperator}
\begin{Compose}{0}{0} \setdefaultfont{} \setsecondfont{\mathsfb}
\crectangledouble{A}{2}{2}{0,0} \csymbol{\hat{A}}  \crectangledouble{B}{2}{2}{5,9} \csymbol{\hat{B}} \crectangledouble{C}{2}{2}{2,18} \csymbol{\hat{C}}
\thispoint{mid}{-5,9}\csymbolalt[-20,0]{a} \jointbnoarrow[left]{A}{-1}{mid}{0}  \jointbnoarrow[left]{mid}{0}{C}{-1}
\jointbnoarrow[below right]{A}{1}{B}{0}  \csymbolalt[5,-5]{b}
\jointbnoarrow[above right]{B}{0}{C}{1}  \csymbolalt{a}
\Rxboxincomedouble{A}{0}{x} \Rxboxoutcomedouble{A}{0}{u}
\Rxboxincomedouble{B}{0}{y} \Rxboxoutcomedouble{B}{0}{v}
\Rxboxincomedouble{C}{0}{z} \Rxboxoutcomedouble{C}{0}{w}
\end{Compose}
\end{equation}
If we substitute each operation for its extended version (as in \eqref{extensiontheoremequation}) we obtain
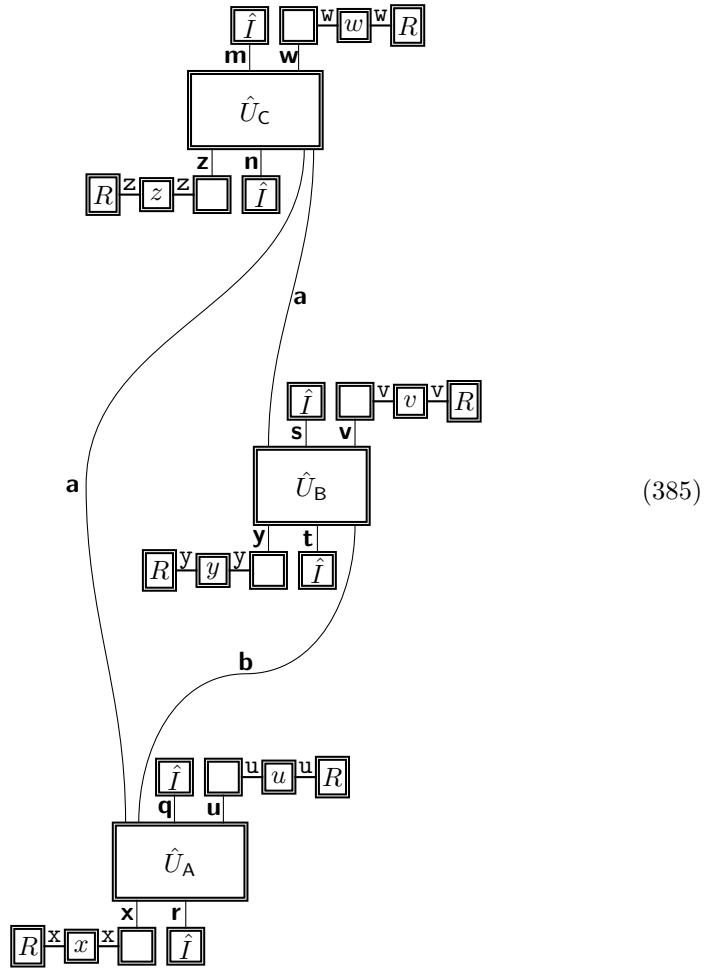
\begin{equation}
\begin{Compose}{0}{0} \setdefaultfont{} \setsecondfont{\mathsfb}
\crectangledouble{U}{3.5}{2}{0,-20} \csymbol{\hat{U}_\mathsf{A}}
\relpoint{U}{-0.3,4.5}{Iupos}  \crectangledouble{Iu}{0.9}{0.9}{Iupos} \csymbol{\hat{I}}  \jointbnoarrow[left]{U}{-0.3}{Iu}{0}  \csymbolalt{q}
\relpoint{U}{0.3,-4.5}{Idpos}  \crectangledouble{Id}{0.9}{0.9}{Idpos} \csymbol{\hat{I}} \jointbnoarrow[left]{Id}{0}{U}{0.3} \csymbolalt{r}
\relpoint{U}{2.3,4.5}{Ypos}  \crectangledouble{Y}{0.9}{0.9}{Ypos}   \jointbnoarrow[left]{U}{2.3}{Y}{0}  \csymbolalt{u} \RxBoxoutcomedouble[0.5]{Y}{0.9}{0}{u}
\relpoint{U}{-2.3,-4.5}{Xpos}  \crectangledouble{X}{0.9}{0.9}{Xpos}  \jointbnoarrow[left]{X}{0}{U}{-2.3} \csymbolalt{x} \RxBoxincomedouble[0.5]{X}{0.9}{0}{x}
\relpoint{U}{-2.3,2.1}{outputA}
\relpoint{U}{2.3,-2.1}{inputA}
\crectangledouble{U}{3}{2}{7,0} \csymbol{\hat{U}_\mathsf{B}}
\relpoint{U}{-0.3,4.5}{Iupos}  \crectangledouble{Iu}{0.9}{0.9}{Iupos} \csymbol{\hat{I}}  \jointbnoarrow[left]{U}{-0.3}{Iu}{0}  \csymbolalt{s}
\relpoint{U}{0.3,-4.5}{Idpos}  \crectangledouble{Id}{0.9}{0.9}{Idpos} \csymbol{\hat{I}} \jointbnoarrow[left]{Id}{0}{U}{0.3} \csymbolalt{t}
\relpoint{U}{2.3,4.5}{Ypos}  \crectangledouble{Y}{0.9}{0.9}{Ypos}   \jointbnoarrow[left]{U}{2.3}{Y}{0}  \csymbolalt{v} \RxBoxoutcomedouble[0.5]{Y}{0.9}{0}{v}
\relpoint{U}{-2.3,-4.5}{Xpos}  \crectangledouble{X}{0.9}{0.9}{Xpos}  \jointbnoarrow[left]{X}{0}{U}{-2.3} \csymbolalt{y} \RxBoxincomedouble[0.5]{X}{0.9}{0}{y}
\relpoint{U}{-2.3,2.1}{outputB}
\relpoint{U}{2.3,-2.1}{inputB}
\crectangledouble{U}{3.5}{2}{4,20} \csymbol{\hat{U}_\mathsf{C}}
\relpoint{U}{-0.3,4.5}{Iupos}  \crectangledouble{Iu}{0.9}{0.9}{Iupos} \csymbol{\hat{I}}  \jointbnoarrow[left]{U}{-0.3}{Iu}{0}  \csymbolalt[-7,0]{m}
\relpoint{U}{0.3,-4.5}{Idpos}  \crectangledouble{Id}{0.9}{0.9}{Idpos} \csymbol{\hat{I}} \jointbnoarrow[left]{Id}{0}{U}{0.3} \csymbolalt{n}
\relpoint{U}{2.3,4.5}{Ypos}  \crectangledouble{Y}{0.9}{0.9}{Ypos}   \jointbnoarrow[left]{U}{2.3}{Y}{0}  \csymbolalt{w} \RxBoxoutcomedouble[0.5]{Y}{0.9}{0}{w}
\relpoint{U}{-2.3,-4.5}{Xpos}  \crectangledouble{X}{0.9}{0.9}{Xpos}  \jointbnoarrow[left]{X}{0}{U}{-2.3} \csymbolalt{z} \RxBoxincomedouble[0.5]{X}{0.9}{0}{z}
\relpoint{U}{-2.3,2.1}{outputC}
\relpoint{U}{2.3,-2.1}{inputC}
\thispoint{AB}{$0.5*(outputA)+0.5*(inputB)$} \csymbolalt[0,20]{b}
\jointlnoarrow{outputA}{0.1}{AB}{0} \joinrbnoarrow{AB}{0}{inputB}{0}
\jointbnoarrow[right]{outputB}{0}{inputC}{0.8}  \csymbolalt{a}
\thispoint{AC}{-5,0}\csymbolalt[-20,0]{a} \jointbnoarrow{outputA}{-0.6}{AC}{0} \jointbnoarrow{AC}{0}{inputC}{0.3}
\end{Compose}
\end{equation}
The blank boxes are maximal elements.  We can stretch wires and move boxes so this is the same circuit as
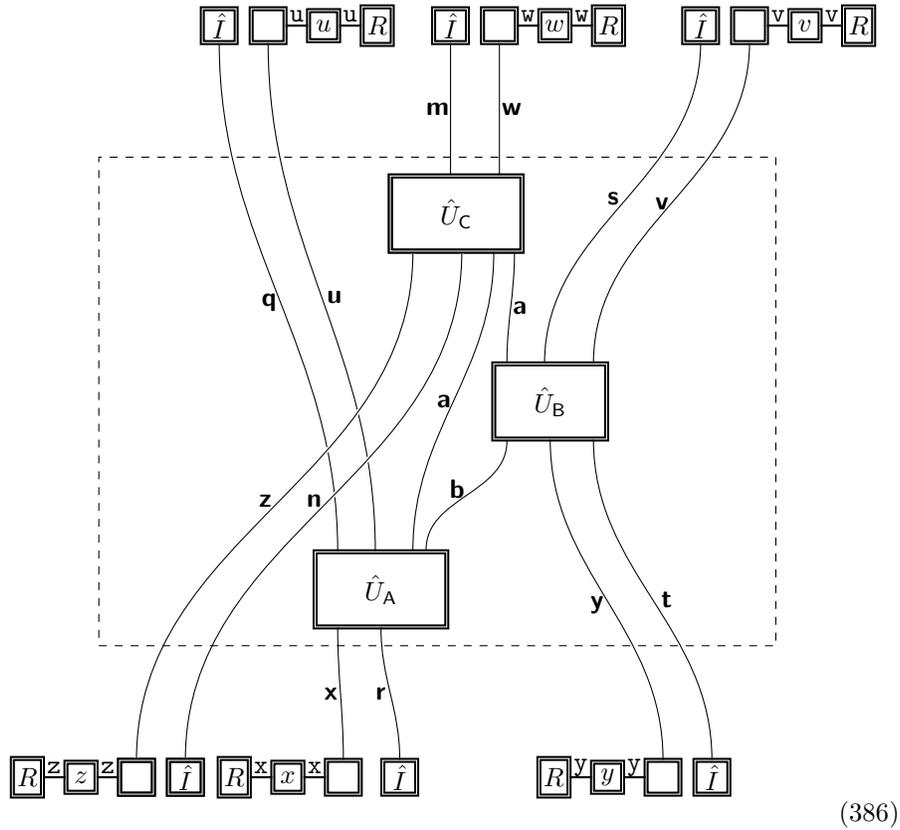
\begin{equation}
\begin{Compose}{0}{0} \setdefaultfont{} \setsecondfont{\mathsfb}
\crectangledouble{U}{3.5}{2}{0,-10} \csymbol{\hat{U}_\mathsf{A}}
\thispoint{Iupos}{-8.6,20}  \crectangledouble{Iu}{0.9}{0.9}{Iupos} \csymbol{\hat{I}}  \jointbnoarrow[left]{U}{-2.3}{Iu}{0}  \csymbolalt[0,-8]{q}
\thispoint{Idpos}{1,-20}  \crectangledouble{Id}{0.9}{0.9}{Idpos} \csymbol{\hat{I}} \jointbnoarrow[left]{Id}{0}{U}{0} \csymbolalt{r}
\thispoint{Ypos}{-6,20}  \crectangledouble{Y}{0.9}{0.9}{Ypos}   \jointbnoarrow[right]{U}{-0.3}{Y}{0}  \csymbolalt[5,0]{u} \RxBoxoutcomedouble[0.5]{Y}{0.9}{0}{u}
\thispoint{Xpos}{-2,-20}  \crectangledouble{X}{0.9}{0.9}{Xpos}  \jointbnoarrow[left]{X}{0}{U}{-2.3} \csymbolalt{x} \RxBoxincomedouble[0.5]{X}{0.9}{0}{x}
\relpoint{U}{2.3,2.1}{outputA}
\relpoint{U}{2.3,-2.1}{inputA}
\crectangledouble{U}{3}{2}{9,0} \csymbol{\hat{U}_\mathsf{B}}
\relpoint{U}{8,20}{Iupos}  \crectangledouble{Iu}{0.9}{0.9}{Iupos} \csymbol{\hat{I}}  \jointbnoarrow[left]{U}{-0.3}{Iu}{0}  \csymbolalt[0,7]{s}
\relpoint{U}{8.6,-20}{Idpos}  \crectangledouble{Id}{0.9}{0.9}{Idpos} \csymbol{\hat{I}} \jointbnoarrow[right]{Id}{0}{U}{2.3} \csymbolalt[7,0]{t}
\relpoint{U}{10.6,20}{Ypos}  \crectangledouble{Y}{0.9}{0.9}{Ypos}   \jointbnoarrow[left]{U}{2.3}{Y}{0}  \csymbolalt{v} \RxBoxoutcomedouble[0.5]{Y}{0.9}{0}{v}
\relpoint{U}{6,-20}{Xpos}  \crectangledouble{X}{0.9}{0.9}{Xpos}  \jointbnoarrow[left]{X}{0}{U}{0} \csymbolalt[0,-7]{y} \RxBoxincomedouble[0.5]{X}{0.9}{0}{y}
\relpoint{U}{-2.3,2.1}{outputB}
\relpoint{U}{-2.3,-2.1}{inputB}
\crectangledouble{U}{3.5}{2}{4,10} \csymbol{\hat{U}_\mathsf{C}}
\relpoint{U}{-0.3,10}{Iupos}  \crectangledouble{Iu}{0.9}{0.9}{Iupos} \csymbol{\hat{I}}  \jointbnoarrow[left]{U}{-0.3}{Iu}{0}  \csymbolalt[-5,0]{m}
\relpoint{U}{-14.4,-30}{Idpos}  \crectangledouble{Id}{0.9}{0.9}{Idpos} \csymbol{\hat{I}}
\jointbnoarrowthickwhite[left]{Id}{0}{U}{0.3}\jointbnoarrow[left]{Id}{0}{U}{0.3} \csymbolalt[0,7]{n}
\relpoint{U}{2.3,10}{Ypos}  \crectangledouble{Y}{0.9}{0.9}{Ypos}   \jointbnoarrow[right]{U}{2.3}{Y}{0}  \csymbolalt[5,0]{w} \RxBoxoutcomedouble[0.5]{Y}{0.9}{0}{w}
\relpoint{U}{-17,-30}{Xpos}  \crectangledouble{X}{0.9}{0.9}{Xpos}
\jointbnoarrowthickwhite[left]{X}{0}{U}{-2.3} \jointbnoarrow[left]{X}{0}{U}{-2.3} \csymbolalt[0,8]{z} \RxBoxincomedouble[0.5]{X}{0.9}{0}{z}
\relpoint{U}{-2.3,2.1}{outputC}
\relpoint{U}{2.3,-2.1}{inputC}
%
%
\jointbnoarrow[left]{outputA}{0.1}{inputB}{0}\csymbolalt[0,10]{b}
\jointbnoarrow[right]{outputB}{0}{inputC}{0.8}  \csymbolalt{a}
\jointbnoarrow[left]{outputA}{-0.6}{inputC}{-0.3} \csymbolalt{a}
\crectangledouble{U}{3.5}{2}{4,10} \crectangledouble{Id}{0.9}{0.9}{Idpos} \crectangledouble{X}{0.9}{0.9}{Xpos} 
\crectangle[thin, dashed]{dots}{18}{13}{3,0}
\end{Compose}
\end{equation}
(We have taken some liberties with the positions at which the wires join the boxes to minimise wires crossing).
Now we see that we have a big unitary transformation happening in the middle (a dashed box is drawn round this).   Thus, this is equivalent to
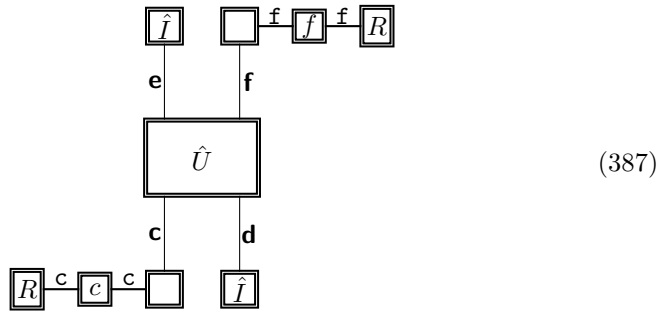
\begin{equation}
\begin{Compose}{0}{0} \setdefaultfont{} \setsecondfont{\mathsfb}
\crectangledouble{U}{3}{2}{0,0} \csymbol{\hat{U}}
\crectangledouble{Iu}{0.9}{0.9}{-2,7} \csymbol{\hat{I}} \crectangledouble{Yu}{0.9}{0.9}{2,7}
\crectangledouble{Xd}{0.9}{0.9}{-2,-7} \crectangledouble{Id}{0.9}{0.9}{2,-7} \csymbol{\hat{I}}
\jointbnoarrow[left]{U}{-2}{Iu}{0} \csymbolalt{e} \jointbnoarrow[right]{U}{2}{Yu}{0} \csymbolalt{f}
\jointbnoarrow[left]{Xd}{0}{U}{-2} \csymbolalt{c} \jointbnoarrow[right]{Id}{0}{U}{2} \csymbolalt{d}
\RxBoxincomedouble{Xd}{0.9}{0}{c} \RxBoxoutcomedouble{Yu}{0.9}{0}{f}
\end{Compose}
\end{equation}
where we have formed composite systems to combine the ignore elements and the maximal elements with $R$'s and readout boxes attached.  It is now clear that we have the double summation property.  If we sum over $c$ then we have two ignore preparations going into the unitary $\hat{U}$ and so we get the ignore preparation out.  Then we have a tensor product of an ignore result and a maximal measurement.  Thus, the circuit equates to $\frac{1}{N_\mathsfb{f}}$. This proves forwards summation. Backwards summation follows by similar reasoning.

\part{End remarks}

\section{Conclusions}

In this paper we have presented a time symmetric approach to operational theories in general and applied this to the particular case of Quantum Theory.  The starting point was a time symmetric notion of the basic idea of an operation. In addition to having inputs and outputs (for physical systems like electrons for example) this also has incomes as well as the usual outcomes.  We have presented many double properties (all of which hold for Quantum Theory).  Most importantly, we obtained a double causality principle that goes beyond the causality principle of Chiribella, D'Ariano, and Perinotti.

We have shown how to formulate Quantum Theory in a time symmetric operational way.  To do this we set up the theory of operator tensors with additional structure corresponding to the incomes and outcomes.  We have provided an extension theorem showing how an arbitrary operator (to which an operation corresponds) can be modeled as a unitary behaviour on a larger system (similar to the Stinespring extension theorem).

There remains much work to be done on this project.  I will discuss some future directions in the next section.

\section{Future directions}

\subsection{Interpretation}\label{sec:interpretation}

I have not discussed the interpretation of this formalism.  In part this is because the mathematical results stand for themselves and interpretation is better seen as additional discussion.  Nevertheless, there are some important questions.

We know, as a contingent fact, that the world is not time symmetric.  It is possible, as a matter of principle, for a theory to have some symmetry while the stuff that makes up the universe does not. For example, having a fundamental theory that is Lorentz invariant does not preclude the existence of some given oak table even though the oak table itself picks out a preferred frame of reference.  However, in the present case, it is not so simple.  In Sec.\ \ref{sec:doublepointercausalityanddoubleflatness} we argued that distributions over incomes (regarded forward in time and outcomes (regarded backwards in time) must be flat.  However, this does not appear to be true of the actual distributions we see.  What should we make of this discrepancy?  One attitude is the following. The formalism presented here provides a way of calculating any probabilities (including conditional probabilities) in terms of probabilities for circuits.  In the case of conditional probabilities we divide one circuit probability by another.  Furthermore, when we use the mathematical machinery in this paper to calculate the probability for these circuits, we get the same answer whether we do a calculation forwards or backwards in time.  This is a formal time symmetry at the level of calculations.

How should we interpret double causality?  If we look at the right column of row 8 of Table \ref{table:causality} we see that results with input only (for a given system type) are all equivalent.  This means that different choices of result do not change the probability of a circuit.  Consequently we cannot signal backward in time (this is the Pavia causality condition \cite{chiribella2010informational}).  Similarly, the left column of row 8 in the same table tells us that there is a unique preparation having an output only.  This tells us that we cannot signal forward in time.  Of course, we are quite used to communicating forward in time so this is rather odd.  In fact, we can steer information forwards (or backwards) in time in the sense that an earlier income can be correlated to a later outcome just as measurement outcomes on an entangled state can be correlated.   This still does not explain why, typically, we appear to be able to signal to the future but not the past.  A discussion of exactly this situation is provided by Di Biagio, Don\`a, and Rovelli \cite{di2020quantum} which appeals to the second law.  For the current project, a fully satisfactory resolution would involve showing how this apparent time asymmetry can appear within the framework described in this paper.  Some steps in this direction are discussed below in Sec.\ \ref{sec:portinginformationarround}.   Such a project may also require modeling the agent themselves.  The discussion of time asymmetry from the perspective of a causal agent by Evans, Milburn, and Shrapnel in \cite{evans2021causal} is a good starting point for this project.

\subsection{Generalised theory of conditional frames}\label{sec:generalisedtheoryofconditionalframes}

In the time symmetric formalism we calculate joint probabilities
\begin{equation}
p(\text{incomes},\text{outcomes})
\end{equation}
In Sec.\ \ref{sec:proofthatformulationistimesymmetric} we defined the following object
\begin{equation}\label{transtoforwardpicture}
\begin{Compose}{0}{0}
\thispoint{p}{0,0} \cxconditionincome{p}{0}{x}
\end{Compose}
~~:=~~
\begin{Compose}{0}{0}
\crectangle[thin]{N}{1}{1}{-8,0} \csymbolfourth{N_\mathtt{x}}
\thispoint{p}{0,0} \RxBoxincome[0.8]{p}{0}{0}{x}
\end{Compose}
~~~~~~~~~~
\begin{Compose}{0}{0}
\thispoint{p}{0,0} \cxconditionoutcome{p}{0}{x}
\end{Compose}
~~:=~~
\begin{Compose}{0}{0}
\thispoint{p}{0,0} \RxBoxoutcome[0.8]{p}{0}{0}{x}
\end{Compose}
\end{equation}
We could regard this as a transformation between different \emph{conditional frames of reference}.  In the original frame we have a time symmetric formulation.  In the new frame we have a time forward formulation. In this new conditional frame of reference we calculate probabilities of the form
\begin{equation}
p(\text{outcomes}|\text{incomes})
\end{equation}
as illustrated by the example in Sec.\ \ref{sec:proofthatformulationistimesymmetric}.
We could also define a time backwards frame of reference as follows
\begin{equation}
\begin{Compose}{0}{0}
\thispoint{p}{0,0} \cxconditionincome{p}{0}{x}
\end{Compose}
~~:=~~
\begin{Compose}{0}{0}
\thispoint{p}{0,0} \RxBoxincome[0.8]{p}{0}{0}{x}
\end{Compose}
~~~~~~~~~~
\begin{Compose}{0}{0}
\thispoint{p}{0,0} \cxconditionoutcome{p}{0}{x}
\end{Compose}
~~:=~~
\begin{Compose}{0}{0}
\crectangle[thin]{N}{1}{1}{8,0} \csymbolfourth{N_\mathtt{x}}
\thispoint{p}{0,0} \RxBoxoutcome[0.8]{p}{0}{0}{x}
\end{Compose}
\end{equation}
In this frame of reference we would calculate probabilities of the form
\begin{equation}
p(\text{incomes}|\text{outcomes})
\end{equation}
All three conditional frames of reference (the time symmetric one, the forward one, and the backwards one) are useful and afford equivalent formulations of operational Quantum Theory (by virtue of the arguments in Sec.\ \ref{sec:proofthatformulationistimesymmetric}).

An interesting question is whether we can usefully define general conditional frames of reference that go beyond the three examples just discussed.  One thing that makes these three cases special is that the normalisation factor $N_\mathtt{x}$ does not depend on the circuit it pertains to.  In the forward frame, this is because, by double flatness, the probability $p(\text{incomes},-)$ (where we sum over outcomes) is equal to $\frac{1}{N_\text{incomes}}$ (where $N_\text{incomes}$ is the product of the $N$'s for the income types.  And consequently, we can attribute the appropriate $N$ factor to each of the incomes as in \eqref{transtoforwardpicture}. Similar remarks apply to the backwards frame.  There may be other examples that are more contingent on the form of the circuit in which we can do this (for example, see the example in \eqref{goodcircuitexampleZflatness} discussed in Sec.\ \ref{sec:doubleZflatness} on double $Z$-flatness).  However, in general, we the normalisation factor will depend on the circuit.  There is no reason we cannot deal consistently with such general conditional frames.

This project could be driven by analogy with the role of general coordinate systems in General Relativity. In that case the special case is when we have an inertial frame of reference and there are linear transformations between different inertial frames of reference. More generally, the transformations are nonlinear.

This project would inform the interpretational questions discussed in Sec.\ \ref{sec:interpretation}.  Each observer would carry with them beliefs/knowledge which they condition on. They would want to choose a conditional frame of reference consistent with these beliefs/knowledge.

\subsection{Porting income and outcome information around}\label{sec:portinginformationarround}

We introduces incomes as providing information that (within the confines of the operation) is available before the operation takes place but not afterwards.  At first sight this appears problematic because we need to have income and outcome information in one place if we are to analyse it (and deduce joint probabilities for example).  However, away from the confines of the operation we can allow the income information and outcome information to be processed.  In fact, we could imagine doing this by using classical operations.   Let us consider how we might go about this.   First consider the example discussed in Sec.\ \ref{sec:sectionasimpleclassicalsituation} wherein operations are wired together by their systems and then we simplify this to the situation
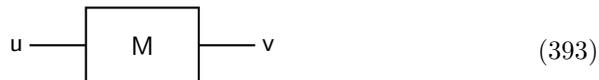
\begin{equation}
\begin{Compose}{0}{0}
\crectangle{M}{3}{2}{0,0} \csymbol{M}
\thispoint{u}{-6,0} \csymbol[-20,0]{u}
\thispoint{v}{6,0} \csymbol[20,0]{v}
\joinrlnoarrowthick[above]{u}{0}{M}{0}
\joinrlnoarrowthick[above]{M}{0}{v}{0}
\end{Compose}
\end{equation}
Now imagine we want to port the income readouts and outcome readouts to a single location so we can deduce joint probabilities.  We could attempt to do this in the following way
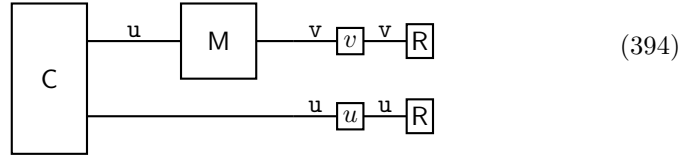
\begin{equation}
\begin{Compose}{0}{0} \setsecondfont{\mathtt}
\Crectangle{C}{2}{4}{-9,-2}
\crectangle{M}{2}{2}{0,0} \csymbol{M}
\thispoint{u}{4,-4} 
\thispoint{v}{4,0} 
\joinrlnoarrowthick[above]{C}{2}{M}{0} \csymbolalt{u}
\joinrlnoarrowthick[above]{C}{-2}{u}{0}
\joinrlnoarrowthick[above]{M}{0}{v}{0}
\RxBoxoutcome{u}{0}{0}{u}
\RxBoxoutcome{v}{0}{0}{v}
\end{Compose}
\end{equation}
The idea is that the $\mathsf C$ operation correlates its two outcomes so that they have the same $u$ value and then we can read off both $u$ and $v$ after $\mathsf M$.  However, forward flatness tells us that
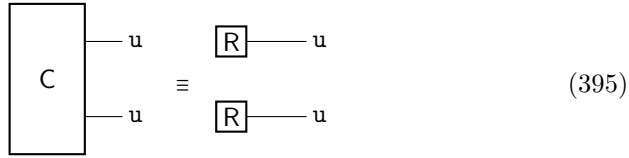
\begin{equation}
\begin{Compose}{0}{0} \setsecondfont{\mathtt}
\Crectangle{C}{2}{4}{0,0}
\thispoint{u}{4,-2}  \csymbolalt[20,0]{u}
\thispoint{v}{4,2}  \csymbolalt[20,0]{u}
\joinrlnoarrow{C}{-2}{u}{0}
\joinrlnoarrow{C}{2}{v}{0}
\end{Compose}
~~\equiv~~
\begin{Compose}{0}{0} \setsecondfont{\mathtt}
\Crectangle{R}{0.8}{0.8}{0,-2} \crectangle{RR}{0.8}{0.8}{0,2} \csymbol{R}
\thispoint{u}{4,-2}  \csymbolalt[20,0]{u}
\thispoint{v}{4,2}  \csymbolalt[20,0]{u}
\joinrlnoarrow{R}{0}{u}{0}
\joinrlnoarrow{RR}{0}{v}{0}
\end{Compose}
\end{equation}
and consequently it is impossible to arrange that $\mathsf C$ correlates its outcomes.  This strategy fails.  It seems that the only way we can bring information forward is if we have some information about the past. Thus, if we substitute $\mathsf C$ in the above for $\mathsf D$ which has an income
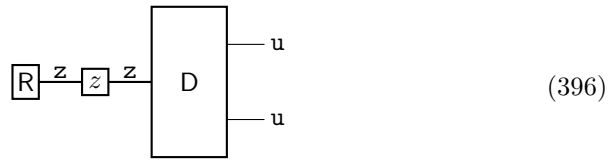
\begin{equation}
\begin{Compose}{0}{0} \setsecondfont{\mathtt}
\crectangle{C}{2}{4}{0,0} \csymbol{D}
\thispoint{u}{4,-2}  \csymbolalt[20,0]{u}
\thispoint{v}{4,2}  \csymbolalt[20,0]{u}
\joinrlnoarrow{C}{-2}{u}{0}
\joinrlnoarrow{C}{2}{v}{0}
\RxBoxincome{C}{2}{0}{z}
\end{Compose}
\end{equation}
then forward flatness does apply as it did with $\mathsf C$.  However, this still leaves the problem of how to get the information $z$ to the future.  We could attempt to copy it but that would give rise to the same problem we had with copying $u$.  Formally, we could choose to go into a forward conditional frame of reference just with respect to this instance of $\mathtt z$ such that
\begin{equation}
\begin{Compose}{0}{0}
\thispoint{p}{0,0} \cxconditionincome{p}{0}{z}
\end{Compose}
~~:=~~
\begin{Compose}{0}{0}
\crectangle[thin]{N}{1}{1}{-8,0} \csymbolfourth{N_\mathtt{x}}
\thispoint{p}{0,0} \RxBoxincome[0.8]{p}{0}{0}{z}
\end{Compose}
\end{equation}
In this case, we would be calculating the probability $p(u,v|z)$.  This enables us to set up the mathematics so we can port information around.  However, it does not resolve the interpretational issues of how we do it.

\subsection{Implications of gauge parameters $\alpha_\mathsfb{x}$ and $\beta_\mathtt{x}$}

In Sec.\ \ref{sec:chasingdowntheoneoverN} and Sec.\ \ref{sec:chasingdowntheoneoverNfactorforpointers} we discussed the gauge parameters $\alpha_\mathsfb{x}$ and $\beta_\mathtt{x}$. Different choices of these gauge parameters go naturally with different points of view (forward, backward, and time symmetric).  We saw, further, in Sec.\ \ref{sec:proofthatformulationistimesymmetric} that these parameters enter when we transform an operation to its time reverse.  What are the implications of having such gauge parameters in the theory?  Do they represent a fundamental symmetry?  In the context of the Lagrangian framework, gauge symmetries are associated with conserved quantities according to Noether's theorem.  We are not in the Lagrangian framework here but it is interesting to ask whether there are conserved quantities associated with these parameters.

Operator tensors are tensor-like objects. One might ask how they transform.  In \cite{hardy2012operator} I provided a brief discussion of this topic. General transformations of operator tensors may provide the appropriate framework in which to study $\alpha_\mathsfb{x}$ and $\beta_\mathtt{x}$.

\subsection{Reasonable axioms for Quantum Theory}

It was twenty years ago that I wrote the paper \emph{Can we obtain Quantum Theory from reasonable axioms?}  Since then there have been many papers on this subject (see Sec.\ \ref{sec:previouswork}).  All this work has been in the forward time picture (a notable exception being the reconstruction due to Ding Jia \cite{jia2018quantum}).  However, the time symmetric picture is arguably much more natural.  Consequently it is a better setting for such reconstruction efforts.  We might hold out the hope that, in this time symmetric setting, a really compelling set of axioms is possible (as has been advocated for by Fuchs \cite{fuchs2011some}).

To do this we need suitable axioms and a reconstruction.  In this paper we have presented many double properties and some subset of these might provide a suitable set of axioms.  The mathematical axiom provided in Sec.\ \ref{sec:mathematicalaxiomforquantumtheory} is not suitable as it stands since it refers to operators. It assumes the operator structure we would like to reconstruct.  However, the idea of correspondence from operations to some sort of mathematical object and physicality (so we have positive probabilities and causality) are reasonable.  Thus, one approach might be to start with the mathematical axiom and back fill so we do not assume the operator structure up front.  The approach in this paper is diagrammatic.  Other diagrammatic reconstructions may provide clues.  The reconstruction of Chiribella, D'Ariano, and Perinotti \cite{chiribella2010informational} invokes diagrams as does my reconstruction \cite{hardy2011reformulating}.   In the reconstruction of Selby, Scandolo, and Coecke \cite{selby2018reconstructing} the diagrammatics are more central to the proof.  Interestingly, notions of time reversal play an important role.  Further, Postulate 5 (\emph{symmetric purification}) in that work clearly has a connection with the extension theorem presented in Sec.\ \ref{sec:extensiontheorem}.

Di Biagio, Dona, and Rovelli also have some speculations on directions to take for a reconstruction program \cite{di2020quantum} (see also \cite{dibiagio2020can}).

\subsection{Conditional probabilities}

To formulate these principles we have introduced diagrammatic notation for the operational theory, for duotensors, and for operator tensors (as well as a version of the diagrammatic notation of Coecke and Kissinger for Hilbert space objects).  In so doing, we have added classical structure corresponding to the incomes and outcomes.  Interestingly, the duotensor notation can be thought of as implementing the chain rule for conditional probabilities in the classical (square dot) case.  There is a similar looking structure for the round dot duotensors. However, the inverse of the hopping matrix can be negative so there is not a simple interpretation of this structure as implementing the chain rule for conditional probabilities.  However, this may relate to attempts to provide objects like a conditional density matrix \cite{leifer2006quantum2, leifer2013towards}.

\subsection{Combs and other generalisations}

The quantum combs approach considers more general objects than operations called quantum combs.  These have inputs and outputs at multiple times (such that we could imagine, for example, feeding an output into an input on the same comb).  Associated with quantum combs is a generalization of the causality condition.  This generalisation is iterative.  However, it is in the time forward picture.  It would be useful to find the generalization of these conditions to the time symmetric picture presented in this work.

The approach to duotensors and operator tensors presented here goes beyond my previous work on these subjects.  This is because we add classical structure associated with incomes and outcomes.  In the duotensor formalism, for example, these are associated with square dots.  The work of Schmid, Selby, and Spekkens \cite{schmid2020unscrambling} does something similar adding inferential structure to a process framework.  In fact, this work is more general since it considers objects that might be regarded as inferential combs. In the time symmetric setting there are constraints on what can usefully be implemented (see the discussion of the attempt to construct a classical operation, $\mathsf C$, that provides two copies of an income in Sec.\ \ref{sec:portinginformationarround} for an example).  Thus there is an interesting project (suggested, in fact, by Di Biagio in \cite{dibiagio2020can}) in developing the interplay between the inferential structure of Schmid et al. and formulating operational  Quantum Theory in a time symmetric fashion.

\subsection{Time reverse experiments}

There are many experiments discussed in the foundations of Quantum Theory and Quantum Information Theory having conceptual import.  We can take any of these and consider the time reverse experiment.  We must get the same predictions for this time reverse experiment.  Consequently, any conceptual implications that apply in one time direction will apply in the reverse direction as well.  Obvious examples to look at are Bell's theorem \cite{bell1964einstein} and its variants (such as \cite{greenberger1989going, hardy1993nonlocality}), quantum teleportation \cite{bennett1993teleporting}, and the Pusey, Barrett, and Rudolph result \cite{pusey2012reality}.  Wood and Spekkens's investigation of Bell's theorem \cite{wood2015lesson} using causal discovery algorithms included retrocausal explanations on the same footing as forward in time causal explanations so this is of the right flavour.

To get a taste of how this might work. Here is the standard quantum teleportation experiment in the diagrammatic notation of this paper.
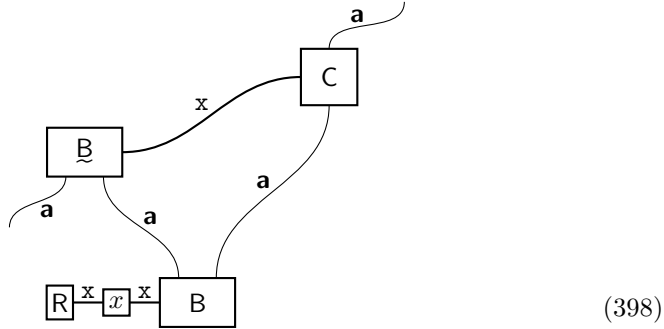
\begin{equation}
\begin{Compose}{0}{0} \setsecondfont{\mathsfb} \setthirdfont{\mathtt}
\crectangle{B}{2}{1.3}{0,0} \csymbol{B}
\crectangle{BB}{2}{1.3}{-6,8} \csymbol{\utilde{\mathsf{B}}} \Crectangle{C}{1.5}{1.5}{7,12}
\jointbnoarrow[above right]{B}{-1}{BB}{1} \csymbolalt{a} \jointbnoarrow[above left]{B}{1}{C}{0} \csymbolalt{a}
\relpoint{BB}{-4,-4}{pB} \jointbnoarrow[below right]{pB}{0}{BB}{-1} \csymbolalt{a}
\joinrlnoarrowthick[above left]{BB}{0}{C}{0} \csymbolthird{x}
\relpoint{C}{4,4}{pC} \jointbnoarrow[above left]{C}{0}{pC}{0} \csymbolalt{a}
\RxBoxincome[0.7]{B}{2}{0}{x}
\end{Compose}
\end{equation}
Note that we would normally condition on a given value of $x$. Then this implements the identity channel on $\mathsfb a$.  We could write $\mathtt{x}=\mathtt{aa}$.  The time reverse is the following
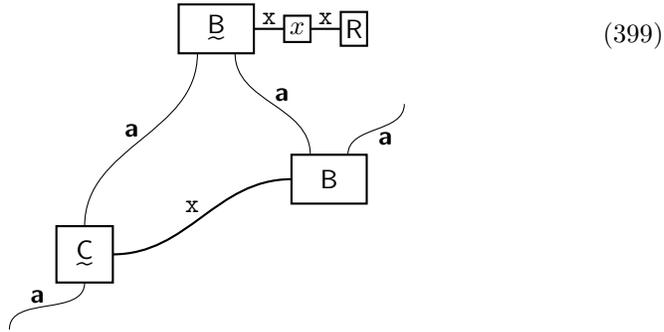
\begin{equation}
\begin{Compose}{0}{0} \setsecondfont{\mathsfb} \setthirdfont{\mathtt}
\crectangle{B}{2}{1.3}{0,0} \csymbol{\utilde{\mathsf{B}}}
\crectangle{BB}{2}{1.3}{6,-8} \csymbol{B} \crectangle{C}{1.5}{1.5}{-7,-12} \csymbol{\utilde{\mathsf{C}}}
\joinbtnoarrow[above right]{B}{1}{BB}{-1} \csymbolalt{a} \joinbtnoarrow[above left]{B}{-1}{C}{0} \csymbolalt{a}
\relpoint{BB}{4,4}{pB} \joinbtnoarrow[below right]{pB}{0}{BB}{1} \csymbolalt{a}
\joinlrnoarrowthick[above left]{BB}{0}{C}{0} \csymbolthird{x}
\relpoint{C}{-4,-4}{pC} \joinbtnoarrow[above left]{C}{0}{pC}{0} \csymbolalt{a}
\RxBoxoutcome[0.7]{B}{2}{0}{x}
\end{Compose}
\end{equation}
In the case where we post-select on the same given value of $x$, this will implement the identity channel on $\mathsfb a$.
We can model $\mathsf C$ in extended form as a controlled unitary implementing the standard operations in the quantum teleportation protocol.
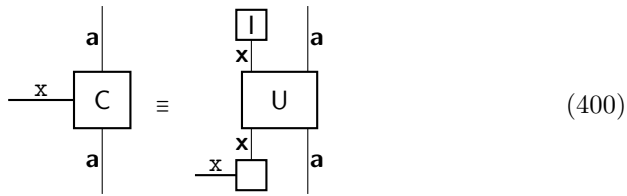
\begin{equation}
\begin{Compose}{0}{0} \setsecondfont{\mathsfb} \setthirdfont{\mathtt}
\crectangle{C}{1.5}{1.5}{0,0} \csymbol{C}
\thispoint{d}{0,-5} \thispoint{u}{0,5} \thispoint{h}{-5,0}
\jointbnoarrow[left]{d}{0}{C}{0} \csymbolalt{a}
\jointbnoarrow[left]{C}{0}{u}{0} \csymbolalt{a}
\joinrlnoarrowthick[above]{h}{0}{C}{0} \csymbolthird{x}
\end{Compose}
~~\equiv~~
\begin{Compose}{0}{0} \setsecondfont{\mathsfb} \setthirdfont{\mathtt}
\crectangle{U}{2}{1.5}{0,0} \csymbol{U}
\crectangle{X}{0.8}{0.8}{-1.5,-4}  \crectangle{I}{0.8}{0.8}{-1.5,4} \csymbol{I}
\thispoint{d}{1.5,-5} \thispoint{u}{1.5,5}
\jointbnoarrow[left]{X}{0}{U}{-1.5} \csymbolalt{x}
\jointbnoarrow[left]{U}{-1.5}{I}{0} \csymbolalt{x}
\jointbnoarrow[right]{d}{0}{U}{1.5} \csymbolalt{a}
\jointbnoarrow[right]{U}{1.5}{u}{0} \csymbolalt{a}
\relpoint{X}{-3,0}{Xp} \joinrlnoarrowthick[above]{Xp}{0}{X}{0} \csymbolthird{x}
\end{Compose}
\end{equation}
where the blank box is a maximal preparation.  The time reverse of this is
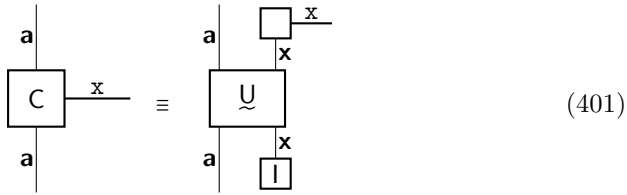
\begin{equation}
\begin{Compose}{0}{0} \setsecondfont{\mathsfb} \setthirdfont{\mathtt}
\crectangle{C}{1.5}{1.5}{0,0} \csymbol{C}
\thispoint{d}{0,5} \thispoint{u}{0,-5} \thispoint{h}{5,0}
\joinbtnoarrow[left]{d}{0}{C}{0} \csymbolalt{a}
\joinbtnoarrow[left]{C}{0}{u}{0} \csymbolalt{a}
\joinlrnoarrowthick[above]{h}{0}{C}{0} \csymbolthird{x}
\end{Compose}
~~\equiv~~
\begin{Compose}{0}{0} \setsecondfont{\mathsfb} \setthirdfont{\mathtt}
\crectangle{U}{2}{1.5}{0,0} \csymbol{\utilde{\mathsf{U}}}
\crectangle{X}{0.8}{0.8}{1.5,4}  \crectangle{I}{0.8}{0.8}{1.5,-4} \csymbol{I}
\thispoint{d}{-1.5,5} \thispoint{u}{-1.5,-5}
\joinbtnoarrow[right]{X}{0}{U}{1.5} \csymbolalt{x}
\joinbtnoarrow[right]{U}{1.5}{I}{0} \csymbolalt{x}
\joinbtnoarrow[left]{d}{0}{U}{-1.5} \csymbolalt{a}
\joinbtnoarrow[left]{U}{-1.5}{u}{0} \csymbolalt{a}
\relpoint{X}{3,0}{Xp} \joinlrnoarrowthick[above]{Xp}{0}{X}{0} \csymbolthird{x}
\end{Compose}
\end{equation}
This is still a controlled unitary but where noise is sent in, then an outcome is read out at the end. The time reverse teleportation protocol works as follows.  The state, on system $\mathsfb a$, to be teleported is sent into $\mathsf C$.  A classical outcome is extracted and system $\mathsfb a$ continues.  The outcome from $\mathsf C$ is used to select a Bell state.  Meanwhile, a Bell measurement is made on the system and one of the two outputs from the Bell state.  If the outcome of this Bell measurement is the right one then the state of the system is transferred onto the outgoing $\mathsfb a$ system.

\subsection{Connection to Operational Quantum Gravity program}

While this may not be apparent, the real motivation for the present work is the need to solve the problem of Quantum Gravity (that is to find a theory that limits to Quantum Theory and General Relativity under appropriate circumstances).   I have pursued an operational approach to this problem \cite{hardy2005probability, hardy2016operational, hardy2018construction} wherein the indefinite causal structure is regarded as representing the most basic challenge to making progress.  We expect to have indefinite causal structure in a theory of Quantum Gravity because causal structure (via the metric) is a dynamical variable in General Relativity while, in Quantum Theory, dynamical variables are typically subject to indefiniteness.  Indefinite causal structure is a problem because it makes it difficult to use standard concepts from physics.  In particular, how do we define causality if causal structure is indefinite.  It seems that indefinite causal structure needs to be \lq\lq tamed".  The best idea I have for taming it is the Quantum Equivalence Principle - this is the idea that it should be possible to find a quantum coordinate system such that, in the vicinity of any given point, we have definite causal structure \cite{hardy2020implementation} (see also \cite{giacomini2020einstein, marletto2020testability}).  Once we are in a quantum coordinate system where we have definite causal structure then we can attempt to impose causality.  However, the previous Pavia causality condition is time asymmetric which is at odds with the underlying symmetry of General Relativity.  This motivates the search for a time-symmetric approach which accounts for the present paper.

\subsection{LaTeX/Tikz package for diagrams}

In this paper I have drawn numerous diagrams. These were composed using the \verb+.sty+ packages that come with this arXiv submission. These packages define a number of commands built in terms of basic Tikz commands \cite{tantautikz}.  For example, the diagram
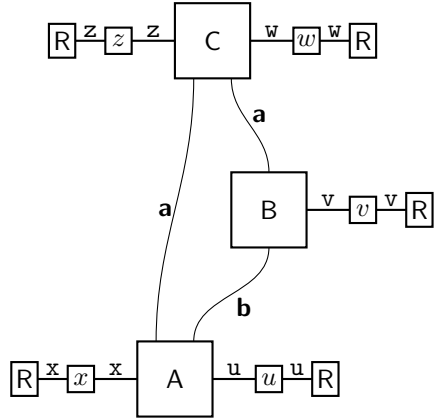
\[
\begin{Compose}{0}{0} \setsecondfont{\mathsfb}
\crectangle{A}{2}{2}{0,0} \csymbol{A}
\crectangle{B}{2}{2}{5,9} \csymbol{B}
\crectangle{C}{2}{2}{2,18} \csymbol{C}
\jointbnoarrow[left]{A}{-1}{C}{-1} \csymbolalt{a}
\jointbnoarrow[below right]{A}{1}{B}{0}  \csymbolalt[5,-5]{b}
\jointbnoarrow[above right]{B}{0}{C}{1}  \csymbolalt{a}
\Rxboxincome{A}{0}{x} \Rxboxoutcome{A}{0}{u}
\Rxboxoutcome{B}{0}{v}
\Rxboxincome{C}{0}{z} \Rxboxoutcome{C}{0}{w}
\end{Compose}
\]
is given by the code
\begin{verbatim}
\[
\begin{Compose}{0}{0} \setsecondfont{\mathsfb}
\crectangle{A}{2}{2}{0,0} \csymbol{A}
\crectangle{B}{2}{2}{5,9} \csymbol{B}
\crectangle{C}{2}{2}{2,18} \csymbol{C}
%
\jointbnoarrow[left]{A}{-1}{C}{-1} \csymbolalt{a}
\jointbnoarrow[below right]{A}{1}{B}{0}  \csymbolalt[5,-5]{b}
\jointbnoarrow[above right]{B}{0}{C}{1}  \csymbolalt{a}
%
\Rxboxincome{A}{0}{x} \Rxboxoutcome{A}{0}{u}
\Rxboxoutcome{B}{0}{v}
\Rxboxincome{C}{0}{z} \Rxboxoutcome{C}{0}{w}
\end{Compose}
\]
\end{verbatim}
However, the \verb+.sty+ packages, as they stand, have a lot of legacy structure in them having been used for multiple projects.  The ability to draw diagrams is basic to projects of this nature and some work on providing more user friendly packages is needed for the current project.

\section*{Acknowledgements}

This project began in ernest following Zoom discussions that formed part of the QISS project (see below).   I am grateful to Flaminia Giacomini, Laurent Freidel, Doreen Fraser, Francesca Vidotto, Marios Christodoulou and especially Carlo Rovelli.  Further, I am grateful to Andrea Di Biagio who gave a presentation to this small QISS discussion group on an early version of his work (as outlined in Sec.\ \ref{sec:previouswork}).

I am further grateful to Bob Coecke, Ognyan Oreshkov, Mauro D'Ariano, Paulo Perinotti, Giulio Chribella, Robert Oeckl, David Schmid, Robert Spekkens, Ding Jia, Nitica Sakharwade and many others on topics related to this work.

I am grateful to my family (co-workers) during this stay-at-home covid era, especially to Zivy for frequently reminding me, with simple examples, that the world is not time symmetric.

Research at Perimeter Institute is supported in part by the Government
of Canada through the Department of Innovation, Science and Industry Canada and by the Province of Ontario through the Ministry of Colleges and Universities

As emphasised above and in Sec.\ \ref{sec:previouswork}, this work was made possible through the support of a grant from the John TempletonFoundation (ID\# 61466) funding “The Quantum Information Structure of Spacetime (QISS)” project (qiss.fr).

\bibliography{QGbibMarch2021}{}
\bibliographystyle{plain}

\end{document}